\documentclass[a4paper,11pt]{article}
\usepackage{silence}
\usepackage{jheppub}
\usepackage{amsmath}
\usepackage{amsfonts}
\usepackage{amssymb}
\usepackage{xcolor}
\usepackage{comment}
\usepackage{booktabs}
\usepackage{makecell}
\usepackage{upgreek}
\usepackage{cancel}
\usepackage{multirow}
\usepackage{wasysym}
\usepackage[normalem]{ulem}
\usepackage{lscape}
\usepackage{rotating}
\usepackage{adjustbox}
\usepackage{caption}
\usepackage{subcaption}
\usepackage{hyperref}

\usepackage{graphicx}   
\graphicspath{{fig/}}

\allowdisplaybreaks

\newcommand{\vp}{\varphi}
\newcommand{\vpj }{\mbox{${\Phi^\dag\, \raisebox{1.5mm}{${}^\leftrightarrow$}\hspace{-4mm} D_\mu\,\Phi}$}}
\newcommand{\vpjt}{\mbox{${\Phi^\dag \tau^I\,\raisebox{1.5mm}{${}^\leftrightarrow$}\hspace{-4mm} D_\mu\,\Phi}$}}
\newcommand{\sw}{\sin\theta_w}
\newcommand{\cw}{\cos\theta_w}

\begin{document}

%%%%%%%%%%%%%%%%%%%%%%%%%%%

% \author{Xiaocong Ai (艾小聪)$^{1}$,} % zzu
\author{Xiaocong Ai$^{1}$,} % zzu
\affiliation{${}^{1}$School of Physics, Zhengzhou University, Zhengzhou 450001, China}
%\emailAdd{xiaocongai@zzu.edu.cn}

\author{Wolfgang Altmannshofer$^{2}$,} % ucsc
\affiliation{${}^{2}$Department of Physics and Santa Cruz Institute for Particle Physics, University of California, Santa Cruz, 95064, USA}
%\emailAdd{waltmann@ucsc.edu}

\author{Peter Athron$^{3}$,} % njnu
\affiliation{${}^{3}$Department of Physics and Institute of Theoretical Physics, Nanjing Normal University, Nanjing 210023, China}
%\emailAdd{peter.athron@njnu.edu.cn}

% \author{Xiaozhi Bai (白晓智)$^{4}$,} % ustc
\author{Xiaozhi Bai$^{4}$,} % ustc
\affiliation{${}^{4}$University of Science and Technology of China, Hefei 230026, China}
%\emailAdd{baixiaozhi@ustc.edu.cn}

\author{Lorenzo Calibbi$^{5*}$,} % nankai
\affiliation{${}^{5}$School of Physics, Nankai University, Tianjin 300071, China}
\emailAdd{calibbi@nankai.edu.cn}

% \author{Lu Cao (曹璐)${^{6,7}}$,} % fudan_0, desy
\author{Lu Cao${^{6,7}}$,} % fudan_0, desy
\affiliation{${}^{6}$Fudan University, Shanghai 200438, China}
\affiliation{${}^{7}$Deutsches Elektronen-Synchrotron (DESY), Hamburg 22607, Germany}
%\emailAdd{lu_cao@fudan.edu.cn}

% \author{Yuzhi Che (车逾之)$^{8,9}$,} % IHEP, UCAS
\author{Yuzhi Che$^{8,9}$,} % IHEP, UCAS
\affiliation{${}^{8}$Institute of High Energy Physics, Chinese Academy of Sciences, Beijing 100049, China}
\affiliation{${}^{9}$School of Physical Sciences, University of Chinese Academy of Sciences (UCAS), Beijing 100049, China}
%\emailAdd{cheyuzhi@ihep.ac.cn}

% \author{Chunhui Chen (陈春晖)$^{10}$,} % Iowa
\author{Chunhui Chen$^{10}$,} % Iowa
\affiliation{${}^{10}$Department of Physics and Astronomy, Iowa State University, Ames, IA 50011-1026, USA}
%\emailAdd{cchen23@iastate.edu}

% \author{Ji-Yuan Chen (陈纪元)$^{31}$,} % sjtu
\author{Ji-Yuan Chen$^{31}$,} % sjtu
%\emailAdd{jy_chen@sjtu.edu.cn}

% \author{Long Chen (陈龙)$^{11}$,} % sdu
\author{Long Chen$^{11}$,} % sdu
\affiliation{${}^{11}$School of Physics, Shandong University, Jinan, Shandong 250100, China}
%\emailAdd{longchen@sdu.edu.cn}

% \author{Mingshui Chen (陈明水)$^{8,9,77}$,} % IHEP, UCAS, henan_hep
\author{Mingshui Chen$^{8,9,77}$,} % IHEP, UCAS, henan_hep
%\emailAdd{chenms@ihep.ac.cn}

% \author{Shanzhen Chen (陈缮真)$^{8,9,77\dagger}$,} % IHEP, UCAS, henan_hep
\author{Shanzhen Chen$^{8,9,77\dagger}$,} % IHEP, UCAS, henan_hep
%\emailAdd{szchen@ihep.ac.cn}

% \author{Xuan Chen (陈暄)$^{11}$,} % sdu
\author{Xuan Chen$^{11}$,} % sdu
%\emailAdd{xuan.chen@sdu.edu.cn}

% \author{Shan Cheng (程山)$^{12}$,} % hnu
\author{Shan Cheng$^{12}$,} % hnu
\affiliation{${}^{12}$School of Physics and Electronics, Hunan University, Changsha 410082, China}
%\emailAdd{scheng@hnu.edu.cn}

% \author{Cheng-Wei Chiang (蔣正偉)$^{13}$,} % ntu.tw
\author{Cheng-Wei Chiang$^{13}$,} % ntu.tw
\affiliation{${}^{13}$Department of Physics and Center for Theoretical Physics, National Taiwan University, 10617, Taipei}
%\emailAdd{chengwei@phys.ntu.edu.tw}

\author{Andreas Crivellin$^{14,15}$,} % PSI, Physik-Institut
\affiliation{${}^{14}$Paul Scherrer Institute (PSI), Villigen, CH-5232, Switzerland}
\affiliation{${}^{15}$Physik-Institut, Universit\"at Z\"urich, Z\"urich, CH-8057, Switzerland}
%\emailAdd{andreas.crivellin@psi.ch}

% \author{Hanhua Cui (崔瀚化)$^{8,9}$,} % IHEP, UCAS
\author{Hanhua Cui$^{8,9}$,} % IHEP, UCAS
%\emailAdd{cuihanhua@ihep.ac.cn}

\author{Olivier Deschamps$^{16}$,} % clermont
\affiliation{${}^{16}$Universit\'e Clermont Auvergne, CNRS/IN2P3, LPC, Clermont-Ferrand, 63178, France}
%\emailAdd{olivier.deschamps@clermont.in2p3.fr}

\author{Sébastien Descotes-Genon$^{17}$,} % IJCLab
\affiliation{${}^{17}$Universit\'e Paris-Saclay, CNRS/IN2P3, IJCLab, Orsay, 91405, France}
%\emailAdd{sebastien.descotes-genon@ijclab.in2p3.fr}

% \author{Xiaokang Du (都小康)$^{18}$,} % hnas
\author{Xiaokang Du$^{18}$,} % hnas
\affiliation{${}^{18}$Institute of Physics, Henan Academy of Sciences, Zhengzhou 450046, China}
%\emailAdd{xkdu@hnas.ac.cn}

% \author{Shuangshi Fang (房双世)$^{8,9}$,} % IHEP, UCAS
\author{Shuangshi Fang$^{8,9}$,} % IHEP, UCAS
%\emailAdd{fangss@ihep.ac.cn}

% \author{Yu Gao (高宇)$^{8,9}$,} % IHEP, UCAS
\author{Yu Gao$^{8,9}$,} % IHEP, UCAS
%\emailAdd{gaoyu@ihep.ac.cn}

% \author{Yuanning Gao (高原宁)$^{46}$,} % pku
\author{Yuanning Gao$^{46}$,} % pku
%\emailAdd{yuanning.gao@pku.edu.cn}

% \author{Li-Sheng Geng (耿立升)$^{19}$,} % buaa
\author{Li-Sheng Geng$^{19}$,} % buaa
\affiliation{${}^{19}$School of Physics, Beihang University, Beijing 100191, China}
%\emailAdd{lisheng.geng@buaa.edu.cn}

\author{Pablo Goldenzweig$^{20}$,} % KIT
\affiliation{${}^{20}$Karlsruher Institut f\"ur Technologie (KIT), Karlsruhe, 76131, Germany}
%\emailAdd{pablo.goldenzweig@kit.edu}

% \author{Jiayin Gu (顾嘉荫)$^{21,22,23}$,} % fudan_phy, fudan_Theory, fudan_MOEKeyLab
\author{Jiayin Gu$^{21,22,23}$,} % fudan_phy, fudan_Theory, fudan_MOEKeyLab
\affiliation{${}^{21}$Department of Physics, Fudan University, Shanghai 200438, China}
\affiliation{${}^{22}$Center for Field Theory and Particle Physics, Fudan University, Shanghai 200438, China}
\affiliation{${}^{23}$Key Laboratory of Nuclear Physics and Ion-beam Application (MOE), Fudan University, Shanghai 200438, China}
%\emailAdd{jiayin\_gu@fudan.edu.cn}

% \author{Feng-Kun Guo (郭奉坤)$^{24,9,25\dagger}$,} % itp, UCAS, Peng Huanwu Beihang
\author{Feng-Kun Guo$^{24,9,25\dagger}$,} % itp, UCAS, Peng Huanwu Beihang
\affiliation{${}^{24}$CAS Key Laboratory of Theoretical Physics, Institute of Theoretical Physics, Chinese Academy of Sciences, Beijing 100190, China}
\affiliation{${}^{25}$Peng Huanwu Collaborative Center for Research and Education, Beihang University, Beijing 100191, China}
%\emailAdd{fkguo@itp.ac.cn}

% \author{Yuchen Guo (郭禹辰)$^{26,27}$,} % lnnu
\author{Yuchen Guo$^{26,27}$,} % lnnu
\affiliation{${}^{26}$Department of Physics, Liaoning Normal University, Dalian 116029, China}
\affiliation{${}^{27}$Center for Theoretical and Experimental High Energy Physics, Liaoning Normal University, Dalian 116029, China}
%\emailAdd{ycguo@lnnu.edu.cn}

% \author{Zhi-Hui Guo (郭志辉)$^{28\dagger}$,} % Hebei Normal
\author{Zhi-Hui Guo$^{28\dagger}$,} % Hebei Normal
\affiliation{${}^{28}$Department of Physics and Hebei Key Laboratory of Photophysics Research and Application, Hebei Normal University, Shijiazhuang 050024, China}
%\emailAdd{zhguo@hebtu.edu.cn}

% \author{Tao Han (韩涛)$^{29}$,} % University of Pittsburgh
\author{Tao Han$^{29}$,} % University of Pittsburgh
\affiliation{${}^{29}$Pittsburgh Particle physics, Astrophysics, and Cosmology Center, Department of Physics \& Astronomy, University of Pittsburgh, 3941 O'Hara St., Pittsburgh, PA 15260, USA}
%\emailAdd{than@pitt.edu}

% \author{Hong-Jian He (何红建)$^{30,31}$,} % TDL, sjtu
\author{Hong-Jian He$^{30,31}$,} % TDL, sjtu
\affiliation{${}^{30}$Tsung-Dao Lee Institute, Shanghai Jiao Tong University, Shanghai 200240, China}
\affiliation{${}^{31}$School of Physics and Astronomy, Shanghai Jiao Tong University, Shanghai 200240, China}
%\emailAdd{hjhe@sjtu.edu.cn}

% \author{Jibo He (何吉波)$^{9}$,} % UCAS
\author{Jibo He$^{9}$,} % UCAS
%\emailAdd{jibo.he@ucas.ac.cn}

% \author{Miao He (何苗)$^{8,9}$,} % IHEP, UCAS
\author{Miao He$^{8,9}$,} % IHEP, UCAS
%\emailAdd{hem@ihep.ac.cn}

% \author{Xiaogang He (何小刚)$^{30,31,65}$} % TDL, sjtu, sjtu_MOEKeyLab
\author{Xiaogang He$^{30,31,65}$} % TDL, sjtu, sjtu_MOEKeyLab
% \emailAdd{xghe@sjtu.edu.cn}

% \author{Yanping Huang (黄燕萍)$^{8,9}$,} % IHEP, UCAS
\author{Yanping Huang$^{8,9}$,} % IHEP, UCAS
%\emailAdd{huangyp@ihep.ac.cn}

\author{Gino Isidori$^{15}$,} % Physik-Institut
%\emailAdd{gino.isidori@uzh.ch}

% \author{Quan Ji (纪全)$^{8,9}$,} % IHEP, UCAS
\author{Quan Ji$^{8,9}$,} % IHEP, UCAS
%\emailAdd{jiq@ihep.ac.cn}

% \author{Jianfeng Jiang (江建锋)$^{8,9}$,} % IHEP, UCAS
\author{Jianfeng Jiang$^{8,9}$,} % IHEP, UCAS
%\emailAdd{jianfeng.jiang@ihep.ac.cn}

% \author{Xu-Hui Jiang (蒋旭辉)$^{8,32,33}$,} % IHEP, CCAST, HKUST
\author{Xu-Hui Jiang$^{8,32,33}$,} % IHEP, CCAST, HKUST
\affiliation{${}^{32}$China Center of Advanced Science and Technology, Beijing 100190, China}
\affiliation{${}^{33}$Department of Physics and Jockey Club Institute for Advanced Study, The Hong Kong University of Science and Technology, Hong Kong S.A.R., China}
%\emailAdd{jiangxh@ihep.ac.cn}

\author{Jernej F. Kamenik$^{34,35}$,}
\affiliation{${}^{34}$Jožef Stefan Institute, Jamova 39, Ljubljana, 1000, Slovenia}
\affiliation{${}^{35}$Faculty of Mathematics and Physics, University of Ljubljana, Jadranska 19, 1000 Ljubljana, Slovenia}
%\emailAdd{jernej.kamenik@cern.ch}

% \author{Tsz Hong Kwok (郭子康)$^{33\dagger}$,} % HKUST
\author{Tsz Hong Kwok$^{33\dagger}$,} % HKUST
%\emailAdd{thkwokae@connect.ust.hk}

% \author{Gang Li (李刚)$^{8,9}$,} % IHEP, UCAS
\author{Gang Li$^{8,9}$,} % IHEP, UCAS
%\emailAdd{ligang@ihep.ac.cn}

% \author{Geng Li (李更)$^{36}$,} % UCAS_Hangzhou
\author{Geng Li$^{36}$,} % UCAS_Hangzhou
\affiliation{${}^{36}$Hangzhou Institute for Advanced Study, University of Chinese Academy of Sciences, Hangzhou 310024, China}
%\emailAdd{ligeng@ucas.ac.cn}

% \author{Haibo Li (李海波)$^{8,9}$,} % IHEP, UCAS
\author{Haibo Li$^{8,9}$,} % IHEP, UCAS
%\emailAdd{haibo.li@ihep.ac.cn}

% \author{Haitao Li (李海涛)$^{11}$,} % sdu
\author{Haitao Li$^{11}$,} % sdu
%\emailAdd{haitao.li@sdu.edu.cn}

% \author{Hengne Li (李衡讷)$^{37}$,} % scnu
\author{Hengne Li$^{37}$,} % scnu
\affiliation{${}^{37}$Institute of Quantum Matter, South China Normal University, Guangzhou, Guangdong 510631, China}
%\emailAdd{hengne.li@m.scnu.edu.cn}

% \author{Honglei Li (李洪蕾)$^{38}$,} % ujn
\author{Honglei Li$^{38}$,} % ujn
\affiliation{${}^{38}$School of Physics and Technology, University of Jinan, Jinan 250022, China}
%\emailAdd{sps\_lihl@ujn.edu.cn}

% \author{Liang Li (李亮)$^{31,65,66}$,} % sjtu, sjtu_MOEKeyLab, sjtu_ShanghaiKeyLab
\author{Liang Li$^{31,65,66}$,} % sjtu, sjtu_MOEKeyLab, sjtu_ShanghaiKeyLab
%\emailAdd{liangliphy@sjtu.edu.cn}

% \author{Lingfeng Li (李凌风)$^{39,33*}$,} % Brown, HKUST
\author{Lingfeng Li$^{39,33*}$,} % Brown, HKUST
\affiliation{${}^{39}$Department of Physics, Brown University, Providence, 02912, USA}
\emailAdd{l.f.li165@gmail.com}

% \author{Qiang Li (李强)$^{40}$,} % nwpu
\author{Qiang Li$^{40}$,} % nwpu
\affiliation{${}^{40}$School of Physical Science and Technology, Northwestern Polytechnical University, Xi’an 710072, China}
%\emailAdd{liruo@nwpu.edu.cn}

% \author{Qiang Li (李强)$^{46}$,} % pku
\author{Qiang Li$^{46}$,} % pku
% \emailAdd{qliphy0@pku.edu.cn}

% \author{Shu Li (李数)$^{30,31}$,} % TDL, sjtu
\author{Shu Li$^{30,31}$,} % TDL, sjtu
%\emailAdd{shuli@sjtu.edu.cn}

% \author{Xiaomei Li (李笑梅)$^{41}$,} % ciae
\author{Xiaomei Li$^{41}$,} % ciae
\affiliation{${}^{41}$China Institute of Atomic Energy, Beijing 102413, China}
%\emailAdd{xiao_mei_li@foxmail.com}

% \author{Xin-Qiang Li (李新强)$^{42\dagger}$,} % ccnu
\author{Xin-Qiang Li$^{42\dagger}$,} % ccnu
\affiliation{${}^{42}$Institute of Particle Physics and Key Laboratory of Quark and Lepton Physics (MOE), Central China Normal University, Wuhan 430079, China}
%\emailAdd{xqli@mail.ccnu.edu.cn}

% \author{Yiming Li (李一鸣)$^{8,9}$,} % IHEP, UCAS
\author{Yiming Li$^{8,9}$,} % IHEP, UCAS
%\emailAdd{liyiming@ihep.ac.cn}

% \author{Yubo Li (李郁博)$^{43}$,} % xjtu
\author{Yubo Li$^{43}$,} % xjtu
\affiliation{${}^{43}$Xi'an Jiaotong University, Xi'an 710049, China}
%\emailAdd{Yubo.li@xjtu.edu.cn}

% \author{Yuji Li (李玉己)$^{6}$,} % fudan_0
\author{Yuji Li$^{6}$,} % fudan_0
%\emailAdd{yuji.li@cern.ch}

% \author{Zhao Li (李钊)$^{8,9}$,} % IHEP, UCAS
\author{Zhao Li$^{8,9}$,} % IHEP, UCAS
%\emailAdd{zhaoli@ihep.ac.cn}

% \author{Hao Liang (梁浩)$^{8,9}$,} % IHEP, UCAS
\author{Hao Liang$^{8,9}$,} % IHEP, UCAS
%\emailAdd{lihawl@hotmail.com}

% \author{Zhijun Liang (梁志均)$^{8,9}$,} % IHEP, UCAS
\author{Zhijun Liang$^{8,9}$,} % IHEP, UCAS
%\emailAdd{liangzj@ihep.ac.cn}

% \author{Libo Liao (廖立波)$^{44}$,} % sysu_shenzhen
\author{Libo Liao$^{44}$,} % sysu_shenzhen
\affiliation{${}^{44}$School of Science, Shenzhen Campus of Sun Yat-sen University, Shenzhen 518107, China}
%\emailAdd{liaolb3@mail2.sysu.edu.cn}

\author{Zoltan Ligeti$^{45}$,} % University of California, Berkeley
\affiliation{${}^{45}$Theory Group, Lawrence Berkeley National Laboratory and Berkeley Center for Theoretical Physics, University of California, Berkeley, 94720, USA}
%\emailAdd{zligeti@lbl.gov}

% \author{Jia Liu (刘佳)$^{46}$,} % pku
\author{Jia Liu$^{46}$,} % pku
\affiliation{${}^{46}$School of Physics, Peking University, Beijing 100871, China}
%\emailAdd{jialiu@pku.edu.cn}

% \author{Jianbei Liu (刘建北)$^{75,76}$,} % ustc_KeyLab, ustc_ModernPhy
\author{Jianbei Liu$^{75,76}$,} % ustc_KeyLab, ustc_ModernPhy
%\emailAdd{liujianb@ustc.edu.cn}

% \author{Tao Liu (刘滔)$^{33*}$,} % HKUST
\author{Tao Liu$^{33*}$,} % HKUST
\emailAdd{taoliu@ust.hk}

% \author{Yi Liu (刘义)$^{1}$,} % zzu
\author{Yi Liu$^{1}$,} % zzu
%\emailAdd{yiliu@zzu.edu.cn}

% \author{Yong Liu (刘勇)$^{8,9}$,} % IHEP, UCAS
\author{Yong Liu$^{8,9}$,} % IHEP, UCAS
%\emailAdd{liuyong@ihep.ac.cn}

% \author{Zhen Liu (刘真)$^{47}$,} % University of Minnesota
\author{Zhen Liu$^{47}$,} % University of Minnesota
\affiliation{${}^{47}$School of Physics and Astronomy, University of Minnesota, Minneapolis, MN 55455, USA}
%\emailAdd{zliuphys@umn.edu}

% \author{Xinchou Lou (娄辛丑)$^{8,77,78}$,} % IHEP, henan_hep, Texas
\author{Xinchou Lou$^{8,77,78}$,} % IHEP, henan_hep, Texas
%\emailAdd{xinchou@ihep.ac.cn}

% \author{Peng-Cheng Lu (路鹏程)$^{11}$,} % sdu
\author{Peng-Cheng Lu$^{11}$,} % sdu
%\emailAdd{pclu@sdu.edu.cn}

\author{Alberto Lusiani$^{48}$,}
\affiliation{${}^{48}$Scuola Normale Superiore and INFN sezione di Pisa, Pisa, 56126, Italy}
%\emailAdd{alberto.lusiani@pi.infn.it}

% \author{Hong-Hao Ma (马鸿浩)$^{49}$,} % gxnu
\author{Hong-Hao Ma$^{49}$,} % gxnu
\affiliation{${}^{49}$Department of Physics, Guangxi Normal University, Guilin 541004, China}
%\emailAdd{mahonghao@gxnu.edu.cn}

% \author{Kai Ma (马凯)$^{50}$,} % XianUAT
\author{Kai Ma$^{50}$,} % XianUAT
\affiliation{${}^{50}$Faculty of Science, Xi’an University of Architecture and Technology, Xi’an 710055, China}
%\emailAdd{makai@ucas.ac.cn}

\author{Farvah Mahmoudi$^{79,80,81}$,}
%\emailAdd{mahmoudi@in2p3.fr}

% \author{Yajun Mao (冒亚军)$^{46}$,} % pku
\author{Yajun Mao$^{46}$,} % pku
% \emailAdd{maoyj@pku.edu.cn}

% \author{Yaxian Mao (毛亚显)$^{42}$,} % ccnu
\author{Yaxian Mao$^{42}$,} % ccnu
%\emailAdd{yaxian.mao@mail.ccnu.edu.cn}

\author{David Marzocca$^{51}$,}
\affiliation{${}^{51}$INFN, Sezione di Trieste, SISSA, Via Bonomea 265, 34136, Trieste, Italy}
%\emailAdd{david.marzocca@ts.infn.it}

% \author{Juan-Juan Niu (牛娟娟)$^{49}$,} % gxnu
\author{Juan-Juan Niu$^{49}$,} % gxnu
%\emailAdd{niujj@gxnu.edu.cn}

\author{Soeren Prell$^{10}$,} % Iowa
%\emailAdd{prell@iastate.edu}

% \author{Huirong Qi (祁辉荣)$^{8,9}$,} % IHEP, UCAS
\author{Huirong Qi$^{8,9}$,} % IHEP, UCAS
%\emailAdd{qihr@ihep.ac.cn }

% \author{Sen Qian (钱森)$^{8,9}$,} % IHEP, UCAS
\author{Sen Qian$^{8,9}$,} % IHEP, UCAS
%\emailAdd{qians@ihep.ac.cn}

% \author{Zhuoni Qian (钱卓妮)$^{52}$,} % hznu
\author{Zhuoni Qian$^{52}$,} % hznu
\affiliation{${}^{52}$School of Physics, Hangzhou Normal University, Hangzhou 311121, China}
%\emailAdd{zhuoniqian@hznu.edu.cn}

% \author{Qin Qin (秦溱)$^{53\dagger}$,} % hust
\author{Qin Qin$^{53\dagger}$,} % hust
\affiliation{${}^{53}$School of Physics, Huazhong University of Science and Technology, Wuhan 430074, China}
%\emailAdd{qqin@hust.edu.cn}

\author{Ariel Rock$^{33}$,} % HKUST
%\emailAdd{iasarock@ust.hk}

\author{Jonathan L. Rosner$^{54,55}$,} % uchicago_Phy, uchicago_Fermi
\affiliation{${}^{54}$Department of Physics, University of Chicago, Chicago, IL 60637, USA}
\affiliation{${}^{55}$Enrico Fermi Institute, University of Chicago, Chicago, IL 60637, USA}
%\emailAdd{rosner@hep.uchicago.edu}

% \author{Manqi Ruan (阮曼奇)$^{8,9,77*}$,} % IHEP, UCAS, henan_hep
\author{Manqi Ruan$^{8,9,77*}$,} % IHEP, UCAS, henan_hep
\emailAdd{ruanmq@ihep.ac.cn}

% \author{Dingyu Shao (邵鼎煜)$^{6}$,} % fudan_0
\author{Dingyu Shao$^{6}$,} % fudan_0
%\emailAdd{dingyu.shao@cern.ch}

% \author{Chengping Shen (沈成平)$^{56,23}$,} % fudan_ModernPhy, fudan_MOEKeyLab
\author{Chengping Shen$^{56,23}$,} % fudan_ModernPhy, fudan_MOEKeyLab
\affiliation{${}^{56}$Institute of Modern Physics, Fudan University, Shanghai 200438, China}
%\emailAdd{shencp@fudan.edu.cn}

% \author{Xiaoyan Shen (沈肖雁)$^{8,9}$,} % IHEP, UCAS
\author{Xiaoyan Shen$^{8,9}$,} % IHEP, UCAS
%\emailAdd{shenxy@ihep.ac.cn}

% \author{Haoyu Shi (石澔玙)$^{8,9}$,} % IHEP, UCAS
\author{Haoyu Shi$^{8,9}$,} % IHEP, UCAS
%\emailAdd{shihy@ihep.ac.cn}

% \author{Liaoshan Shi (石辽珊)$^{57\dagger}$,} % UCL_London
\author{Liaoshan Shi$^{57\dagger}$,} % UCL_London
\affiliation{${}^{57}$University College London, London, WC1E 6BT, United Kingdom}
%\emailAdd{liaoshan.shi@ihep.ac.cn}

% \author{Zong-Guo Si (司宗国)$^{11}$,} % sdu
\author{Zong-Guo Si$^{11}$,} % sdu
%\emailAdd{zgsi@sdu.edu.cn}

\author{Cristian Sierra$^{3}$,} % njnu
%\emailAdd{cristian.sierra@njnu.edu.cn}

% \author{Huayang Song (宋华洋)$^{24}$,} % itp
\author{Huayang Song$^{24}$,} % itp
%\emailAdd{huayangs1990@ibs.re.kr}

% \author{Shufang Su (苏淑芳)$^{58}$,}
\author{Shufang Su$^{58}$,}
\affiliation{${}^{58}$Department of Physics, University of Arizona, 1118 E. 4th St., Tucson, AZ 85721, USA}
%\emailAdd{shufang@email.arizona.edu}

% \author{Wei Su (苏伟)$^{44}$,} % sysu_shenzhen
\author{Wei Su$^{44}$,} % sysu_shenzhen
%\emailAdd{suwei26@mail.sysu.edu.cn}

% \author{Zhijia Sun (孙志嘉)$^{8,9,62}$} % IHEP, UCAS, CSNS
\author{Zhijia Sun$^{8,9,62}$} % IHEP, UCAS, CSNS
% \emailAdd{sunzj@ihep.ac.cn}

\author{Michele Tammaro$^{59}$,}
\affiliation{${}^{59}$INFN Sezione di Firenze, Via G. Sansone 1, 1-50019 Sesto Fiorentino, Italy}
%\emailAdd{michele.tammaro@f.infn.it}

% \author{Dayong Wang (王大勇)$^{46}$,} % pku
\author{Dayong Wang$^{46}$,} % pku
% \emailAdd{dayong.wang@pku.edu.cn}

% \author{En Wang (王恩)$^{1}$,} % zzu
\author{En Wang$^{1}$,} % zzu
%\emailAdd{wangen@zzu.edu.cn}

% \author{Fei Wang (王飞)$^{1}$,} % zzu
\author{Fei Wang$^{1}$,} % zzu
%\emailAdd{feiwang@zzu.edu.cn}

% \author{Hengyu Wang (汪恒宇)$^{8,9}$,} % IHEP, UCAS
\author{Hengyu Wang$^{8,9}$,} % IHEP, UCAS
%\emailAdd{wanghengyu@ihep.ac.cn}

% \author{Jian Wang (王健)$^{11}$,} % sdu
\author{Jian Wang$^{11}$,} % sdu
%\emailAdd{j.wang@sdu.edu.cn}

% \author{Jianchun Wang (王建春)$^{8,9,77}$,} % IHEP, UCAS, henan_hep
\author{Jianchun Wang$^{8,9,77}$,} % IHEP, UCAS, henan_hep
%\emailAdd{jwang@ihep.ac.cn}

% \author{Kun Wang (王坤)$^{74}$,} % usst
\author{Kun Wang$^{74}$,} % usst
%\emailAdd{kwang@usst.edu.cn}

% \author{Lian-Tao Wang (王连涛)$^{54}$,} % uchicago_Phy
\author{Lian-Tao Wang$^{54}$,} % uchicago_Phy
%\emailAdd{liantaow@uchicago.edu}

% \author{Wei Wang (王伟)$^{31,60}$,} % sjtu, imp_Huizhou
\author{Wei Wang$^{31,60}$,} % sjtu, imp_Huizhou
\affiliation{${}^{60}$Southern Center for Nuclear-Science Theory, Institute of Modern Physics, Huizhou 516000, China}
%\emailAdd{wei.wang@sjtu.edu.cn}

% \author{Xiaolong Wang (王小龙)$^{56}$,} % fudan_ModernPhy
\author{Xiaolong Wang$^{56}$,} % fudan_ModernPhy
%\emailAdd{xiaolong@fudan.edu.cn}

% \author{Xiaoping Wang (王小平)$^{19}$,} % buaa
\author{Xiaoping Wang$^{19}$,} % buaa
%\emailAdd{hcwangxiaoping@buaa.edu.cn}

% \author{Yadi Wang (王雅迪)$^{61}$,} % ncepu
\author{Yadi Wang$^{61}$,} % ncepu
\affiliation{${}^{61}$North China Electric Power University, Beijing 102206, China}
%\emailAdd{wangyadi@ncepu.edu.cn}

% \author{Yifang Wang (王贻芳)$^{8,9,77}$,} % IHEP, UCAS, henan_hep
\author{Yifang Wang$^{8,9,77}$,} % IHEP, UCAS, henan_hep
%\emailAdd{yfwang@ihep.ac.cn}

% \author{Yuexin Wang (王悦心)$^{8,62\dagger}$,} % IHEP, Dongguan
\author{Yuexin Wang$^{8,62\dagger}$,} % IHEP, Dongguan
\affiliation{${}^{62}$China Spallation Neutron Source Science Center, Dongguan 523803, China}
%\emailAdd{wangyuexin@ihep.ac.cn}

% \author{Xing-Gang Wu (吴兴刚)$^{63}$,} % cqu
\author{Xing-Gang Wu$^{63}$,} % cqu
\affiliation{${}^{63}$Department of Physics, Chongqing Key Laboratory for Strongly Coupled Physics, Chongqing University, Chongqing 401331, China}
%\emailAdd{wuxg@cqu.edu.cn}

% \author{Yongcheng Wu (吴永成)$^{3}$,} % njnu
\author{Yongcheng Wu$^{3}$,} % njnu
%\emailAdd{ycwu@njnu.edu.cn}

% \author{Rui-Qing Xiao (肖瑞卿)$^{30,31,64}$,} % TDL, sjtu, King's College London
\author{Rui-Qing Xiao$^{30,31,64}$,} % TDL, sjtu, King's College London
\affiliation{${}^{64}$Department of Physics, King's College London, London, UK}
%\emailAdd{xiaoruiqing@sjtu.edu.cn}

% \author{Ke-Pan Xie (谢柯盼)$^{19}$,} % buaa
\author{Ke-Pan Xie$^{19}$,} % buaa
%\emailAdd{kpxie@buaa.edu.cn}

% \author{Yuehong Xie (谢跃红)$^{42}$,} % ccnu
\author{Yuehong Xie$^{42}$,} % ccnu
%\emailAdd{yuehong.xie@mail.ccnu.edu.cn}

% \author{Zijun Xu (徐子骏)$^{8,9}$,} % IHEP, UCAS
\author{Zijun Xu$^{8,9}$,} % IHEP, UCAS
%\emailAdd{xuzj@ihep.ac.cn}

% \author{Haijun Yang (杨海军)$^{30,31,65,66}$,} % TDL, sjtu, sjtu_MOEKeyLab, sjtu_ShanghaiKeyLab
\author{Haijun Yang$^{30,31,65,66}$,} % TDL, sjtu, sjtu_MOEKeyLab, sjtu_ShanghaiKeyLab
\affiliation{${}^{65}$Key Laboratory for Particle Astrophysics and Cosmology (Ministry of Education), Shanghai Jiao Tong University, Shanghai 200240, China}
\affiliation{${}^{66}$Shanghai Key Laboratory for Particle Physics and Cosmology, Shanghai Jiao Tong University, Shanghai 200240, China}
%\emailAdd{haijun.yang@sjtu.edu.cn}

% \author{Hongtao Yang (杨洪洮)$^{4}$,} % ustc
\author{Hongtao Yang$^{4}$,} % ustc
%\emailAdd{hongtaoyang@ustc.edu.cn}

% \author{Lin Yang (杨林)$^{30}$,} % TDL
\author{Lin Yang$^{30}$,} % TDL
%\emailAdd{young_forest@sjtu.edu.cn}

% \author{Shuo Yang (杨硕)$^{26,27}$,} % lnnu
\author{Shuo Yang$^{26,27}$,} % lnnu
%\emailAdd{shuoyanglnnu@163.com}

% \author{Zhongbao Yin (殷中宝)$^{42}$,} % ccnu
\author{Zhongbao Yin$^{42}$,} % ccnu
%\emailAdd{zbyin@ccnu.edu.cn}

% \author{Fusheng Yu (于福升)$^{67}$,} % lzu
\author{Fusheng Yu$^{67}$,} % lzu
\affiliation{${}^{67}$MOE Frontiers Science Center for Rare Isotopes, and School of Nuclear Science and Technology, Lanzhou University, Lanzhou 730000, China}
%\emailAdd{yufsh@lzu.edu.cn}

% \author{Changzheng Yuan (苑长征)$^{8,9}$,} % IHEP, UCAS
\author{Changzheng Yuan$^{8,9}$,} % IHEP, UCAS
%\emailAdd{yuancz@ihep.ac.cn}

% \author{Xing-Bo Yuan (袁兴博)$^{42}$,} % ccnu
\author{Xing-Bo Yuan$^{42}$,} % ccnu
%\emailAdd{y@ccnu.edu.cn}

% \author{Xuhao Yuan (袁煦昊)$^{8,9}$,} % IHEP, UCAS
\author{Xuhao Yuan$^{8,9}$,} % IHEP, UCAS
%\emailAdd{xuhao.yuan@ihep.ac.cn}

% \author{Chongxing Yue (岳崇兴)$^{26,27}$,} % lnnu
\author{Chongxing Yue$^{26,27}$,} % lnnu
%\emailAdd{cxyue@lnnu.edu.cn}

% \author{Xi-Jie Zhan (展希杰)$^{68}$,} % hbu
\author{Xi-Jie Zhan$^{68}$,} % hbu
\affiliation{${}^{68}$Department of Physics, Hebei University, Baoding 071002, China}
%\emailAdd{zhanxj@hbu.edu.cn}

% \author{Hong-Hao Zhang (张宏浩)$^{82}$} % sysu_guangzhou
\author{Hong-Hao Zhang$^{82}$} % sysu_guangzhou
% \emailAdd{zhh98@mail.sysu.edu.cn}

% \author{Kaili Zhang (张凯栗)$^{8,62}$,} % IHEP, Dongguan
\author{Kaili Zhang$^{8,62}$,} % IHEP, Dongguan
%\emailAdd{zhangkl@ihep.ac.cn}

% \author{Liming Zhang (张黎明)$^{69}$,} % tsinghua
\author{Liming Zhang$^{69}$,} % tsinghua
\affiliation{${}^{69}$Department of Engineering Physics, Tsinghua University, Beijing 100084, China}
%\emailAdd{liming\_zhang@tsinghua.edu.cn}

% \author{Xiaoming Zhang (张晓明)$^{42}$,} % ccnu
\author{Xiaoming Zhang$^{42}$,} % ccnu
%\emailAdd{xiaoming.zhang@mail.ccnu.edu.cn}

% \author{Yang Zhang (张阳)$^{1}$,} % zzu
\author{Yang Zhang$^{1}$,} % zzu
%\emailAdd{zhangyangphy@zzu.edu.cn}

% \author{Yanxi Zhang (张艳席)$^{46}$,} % pku
\author{Yanxi Zhang$^{46}$,} % pku
%\emailAdd{yanxi.zhang@pku.edu.cn}

% \author{Ying Zhang (张盈)$^{83}$} % xjtu_sci
\author{Ying Zhang$^{83}$} % xjtu_sci
% \emailAdd{hepzhy@mail.xjtu.edu.cn}

% \author{Yongchao Zhang (张永超)$^{70}$,} % seu
\author{Yongchao Zhang$^{70}$,} % seu
\affiliation{${}^{70}$School of Physics, Southeast University, Nanjing 211189, China}
%\emailAdd{zhangyongchao@seu.edu.cn}

% \author{Yu Zhang (张宇)$^{71}$,} % hfut
\author{Yu Zhang$^{71}$,} % hfut
\affiliation{${}^{71}$School of Physics, Hefei University of Technology, Hefei 230601, China}
%\emailAdd{dayu@hfut.edu.cn}

% \author{Zhen-Hua Zhang (张振华)$^{72}$,} % usc
\author{Zhen-Hua Zhang$^{72}$,} % usc
\affiliation{${}^{72}$University of South China, Hengyang 421001, China}
%\emailAdd{zhangzh@usc.edu.cn}

% \author{Zhong Zhang (张重)$^{57,70}$,} % UCL_London
\author{Zhong Zhang$^{57,70}$,} % UCL_London
%\emailAdd{zhong.zhang.19@ucl.ac.uk}

% \author{Mingrui Zhao (赵明锐)$^{41}$,} % ciae
\author{Mingrui Zhao$^{41}$,} % ciae
%\emailAdd{mingrui.zhao@cern.ch}

% \author{Qiang Zhao (赵强)$^{8,9}$,} % IHEP, UCAS
\author{Qiang Zhao$^{8,9}$,} % IHEP, UCAS
%\emailAdd{zhaoq@ihep.ac.cn}

% \author{Xu-Chang Zheng (郑绪昌)$^{63}$,} % cqu
\author{Xu-Chang Zheng$^{63}$,} % cqu
%\emailAdd{zhengxc@cqu.edu.cn}

% \author{Yangheng Zheng (郑阳恒)$^{9}$,} % UCAS
\author{Yangheng Zheng$^{9}$,} % UCAS
%\emailAdd{zhengyh@ucas.ac.cn}

% \author{Chen Zhou (周辰)$^{46}$,} % pku
\author{Chen Zhou$^{46}$,} % pku
%\emailAdd{czhouphy@pku.edu.cn}

% \author{Daicui Zhou (周代翠)$^{42}$,} % ccnu
\author{Daicui Zhou$^{42}$,} % ccnu
% \emailAdd{dczhou@ccnu.edu.cn}

% \author{Pengxuan Zhu (朱鹏轩)$^{24}$,} % itp
\author{Pengxuan Zhu$^{24}$,} % itp
%\emailAdd{zhupx99@icloud.com}

% \author{Yongfeng Zhu (朱永峰)$^{46}$,} % pku
\author{Yongfeng Zhu$^{46}$,} % pku
%\emailAdd{zhuyf@ihep.ac.cn}

% \author{Xuai Zhuang (庄胥爱)$^{8,9}$,} % IHEP, UCAS
\author{Xuai Zhuang$^{8,9}$,} % IHEP, UCAS
% \emailAdd{zhuangxa@ihep.ac.cn}

% \author{Xunwu Zuo (左训午)$^{20\dagger}$,} % KIT
\author{Xunwu Zuo$^{20\dagger}$,} % KIT
%\emailAdd{xunwu.zuo@cern.ch}

\author{Jure Zupan${^{73}}$} % University of Cincinnati
\affiliation{${}^{73}$Department of Physics, University of Cincinnati, Cincinnati, 45221, USA}
%\emailAdd{zupanje@ucmail.uc.edu}

%%%%%%%%%%%%%%%%%%
% Newly added author-> maillist
%%%%%%%%%%%%%%%%%%
% \author{author$^{0}$,}
% \affiliation{${}^{0}$affiliation}
% \emailAdd{}

%%%%%%%%%%%%%%%%%%
% Newly added aff.
%%%%%%%%%%%%%%%%%%
% usst
\affiliation{${}^{74}$College of Science, University of Shanghai for Science and Technology, Shanghai 200093, China}
% ustc_KeyLab
\affiliation{${}^{75}$State Key Laboratory of Particle Detection and Electronics, University of Science and Technology of China, Hefei 230026, China}
% ustc_ModernPhy
\affiliation{${}^{76}$Department of Modern Physics, University of Science and Technology of China, Hefei 230026, China}
% henan_hep
\affiliation{${}^{77}$Center for High Energy Physics, Henan Academy of Sciences, Zhengzhou 450046, China}
% Texas
\affiliation{${}^{78}$University of Texas at Dallas, Richardson, 75083, Texas, USA}

\affiliation{${}^{79}$Universit\'e Claude Bernard Lyon 1, CNRS/IN2P3, Institut de Physique des 2 Infinis de Lyon, UMR 5822, F-69622, Villeurbanne, France}
\affiliation{${}^{80}$Theoretical Physics Department, CERN, CH-1211 Geneva 23, Switzerland}
\affiliation{${}^{81}$Institut Universitaire de France (IUF), 75005 Paris, France}

% sysu_guangzhou
\affiliation{${}^{82}$School of Physics, Sun Yat-sen University, Guangzhou 510275, China}
% xjtu_sci
\affiliation{${}^{83}$School of Science, Xi'an Jiaotong University, Xi'an 710049, China}

%%%%%%%%%%%%%%%%%%
\note[*]{Corresponding author.}
\note[$\dagger$]{Primary contributor.}
%%%%%%%%%%%%%%%%%%

\title{Flavor Physics at the CEPC: a General Perspective}

\abstract{We discuss the landscape of flavor physics at the Circular Electron-Positron Collider (CEPC), based on the nominal luminosity outlined in its Technical Design Report. The CEPC is designed to operate in multiple modes to address a variety of tasks. At the $Z$ pole, the expected production of 4 Tera $Z$ bosons will provide unique and highly precise measurements of $Z$ boson couplings, while the substantial number of boosted heavy-flavored quarks and leptons produced in clean $Z$ decays will facilitate investigations into their flavor physics with unprecedented precision. We investigate the prospects of measuring various physics benchmarks and discuss their implications for particle theories and phenomenological models. Our studies indicate that, with its highlighted advantages and anticipated excellent detector performance, the CEPC can explore beauty and $\tau$ physics in ways that are superior to or complementary with the Belle~II and Large-Hadron-Collider-beauty experiments, potentially enabling the detection of new physics at energy scales of 10 TeV and above. This potential also extends to the observation of yet-to-be-discovered rare and exotic processes, as well as testing fundamental principles such as lepton flavor universality, lepton and baryon number conservation, etc., making the CEPC a vibrant platform for flavor physics research. The $WW$ threshold scan, Higgs-factory operation and top-pair productions of the CEPC further enhance its merits in this regard, especially for measuring the Cabibbo-Kobayashi-Maskawa matrix elements, and Flavor-Changing-Neutral-Current physics of Higgs boson and top quarks. We outline the requirements for detector performance and considerations for future development to achieve the anticipated scientific goals.  The role of machine learning for innovative detector design and advanced reconstruction algorithms is also stressed. The CEPC flavor physics program not only develops new capabilities for exploring flavor physics beyond existing projects but also enriches the physics opportunities of this machine. It should be remarked that, given the richness of the CEPC flavor physics, this manuscript is not meant to be a comprehensive survey, but rather an investigation of representative cases. Uncovering the full potential of flavor physics at the CEPC will require further dedicated explorations in the future.
}

\makeatletter
\gdef\@fpheader{}
\makeatother

\maketitle

%\newpage 

%\input{ReviewComments}

\section{Introduction}
\label{sec:intro}

\begin{figure}[b]
    \centering
    \includegraphics[width=0.5\textwidth]{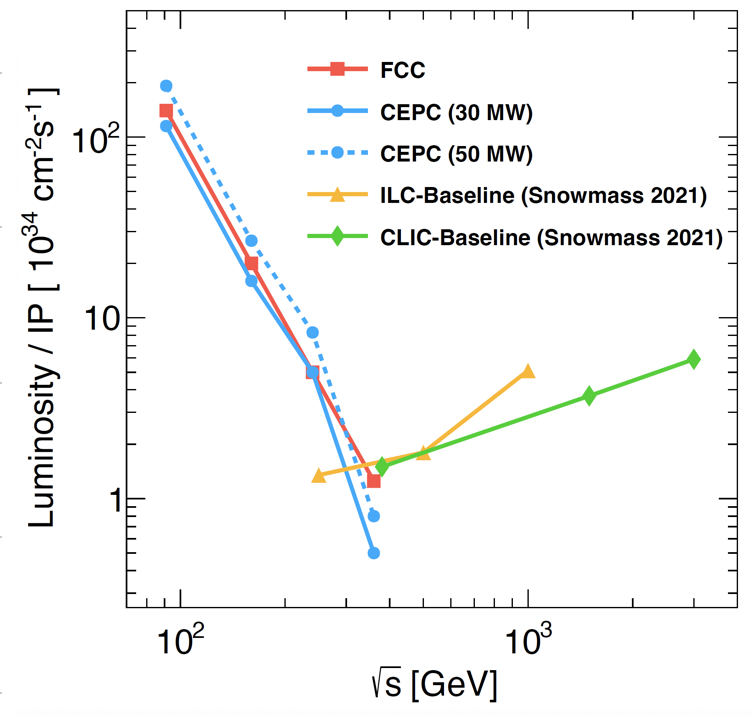}
    \caption{Designed luminosities of the CEPC at the $Z$ pole, Higgs, $WW$ and the $t\bar t$ thresholds operation modes with the baseline and upgrade shown in solid and dashed blue curves, respectively. Luminosities for several other proposals of $e^-e^+$ colliders are also shown for comparison. See Ref.~\cite{CEPCStudyGroup:2023quu} for details.}
    \label{fig:lumiplan}
\end{figure}

The Circular Electron-Positron Collider (CEPC)~\cite{CEPCStudyGroup:2023quu,CEPCStudyGroup:2018ghi} was proposed in 2012 by the Chinese high-energy physics community to function primarily as a Higgs factory  at a center-of-mass energy of 240 GeV. 
It is also set to operate as a $Z$ factory at the $Z$ pole, conduct precise $WW$ threshold scans, and potentially be upgraded to operate at a center-of-mass energy of 360 GeV, {\it i.e.}, above the $t \bar t $ threshold. In the proposed nominal operation scenario~\cite{CEPCStudyGroup:2023quu,Gao:2022lew}, the CEPC is anticipated to produce significant numbers of Higgs and $Z$ bosons, $W$ boson pairs and, potentially, top quarks. 
With respect to the accelerator design, the development of key technologies has led to a significant enhancement in the instantaneous luminosity per interaction point (IP) compared to those reported in the Conceptual Design Report (CDR), as shown in Figure~\ref{fig:lumiplan}.
Based on this progress, the CEPC study group proposes a new nominal operation scenario, shown in Table~\ref{tab:Yield_T1}, which would allow for precision measurements of Higgs boson couplings, electroweak (EW) observables, and QCD differential rates. It would also provide ample opportunities to search for rare decays and new physics (NP) signals. Moreover, the large quantities of bottom quarks, charm quarks, and tau leptons from the decays of $Z$ bosons create opportunities for numerous critical flavor physics measurements. 
It should be noted that the results presented here are based on the updated running scenario using a 50 MW synchrotron radiation (SR) power beam~\cite{CEPCStudyGroup:2023quu}.

Flavor physics, as a well-developed area within particle physics, has contributed substantially to the establishment of the Standard Model (SM) over recent decades. 
This was achieved through the examination of the properties of SM fermion flavors in a myriad of experiments, yielding significant findings and discoveries. 
The CEPC can serve as a flavor factory, and its flavor physics program enhances the CEPC's overarching physics objectives. 
The flavor sector provides substantial motivations for the CEPC operation, 
given the existing multitude of unknowns within the SM and beyond.

\begin{table}[t]
    \centering
    \resizebox{0.9\textwidth}{!}{
    \begin{tabular}[t]{ccccc}
        \toprule[1pt]
        Operation mode & $Z$ factory  & $WW$ threshold  & Higgs factory & $t\bar{t}$ \\
        \midrule
        $\sqrt{s}$ (GeV) & 91.2 & 160 & 240 & 360 \\
        Run time (year) & 2 & 1 & 10 & 5 \\
        \makecell{Instantaneous luminosity \\ ($10^{34} {\rm cm}^{-2}{\rm s}^{-1}$, per IP)} & 191.7 & 26.7 & 8.3 & 0.83 \\
        \makecell{Integrated luminosity \\ (${\rm ab}^{-1}$, 2~IPs)} & 100 & 6.9 & 21.6 & 1 \\
        
        Event yields & $4.1 \times 10^{12}$ & $2.1 \times 10^{8}$ & $4.3 \times 10^{6}$ & $0.6 \times 10^{6}$ \\
        
        % \midrule
        % \multicolumn{5}{c}{Particle yields} \\
        % \midrule
        % $Z$ & $4.3 \times 10^{12}$ & \YX{$2.8 \times 10^{7}$} & $1.5 \times 10^{8}$ &  \\
        % $W$ & - & $8.0 \times 10^{7}$ & $8.3 \times 10^{8}$ & \\
        % Higgs & - & - & $4.4 \times 10^{6}$ & \\
        % $t$ & - & - & - & $1.2 \times 10^{6}$ \\

        \bottomrule[1pt]
    \end{tabular}
    }
    \caption{Nominal CEPC operation scheme of four different modes. See \cite{CEPCStudyGroup:2023quu,Gao:2022lew} for details. }
    \label{tab:Yield_T1}
\end{table}

Understanding the flavor physics potential of the CEPC is not an isolated field of study, as it also influences other primary fields of explorations at the CEPC, including Higgs physics, EW precision observables (EWPOs), QCD, and Beyond the Standard Model (BSM) physics. 
For instance, within the SM the fermion mixing, specifically the Cabibbo-Kobayashi-Maskawa (CKM) matrix~\cite{Cabibbo:1963yz,Kobayashi:1973fv} and its hierarchical structure, originates from the Yukawa couplings of the Higgs field to the fermion gauge eigenstates.
While some of the diagonal Yukawa couplings will be pinned down by the direct Higgs measurements at CEPC~\cite{An:2018dwb}, studying the origin of the off-diagonal flavor mixing terms and their $CP$-violating phases remains mainly within the realm of flavor physics. 
Conversely, while most heavy-flavored particles decay via EW transitions at the tree level, many rare processes are only induced by EW one-loop effects, such as Flavor-Changing-Neutral-Current (FCNC) transitions. 
Their measurements may also serve as alternative tests of the EW sector at an energy scale lower than $Z$-pole measurements. 
Meanwhile, many EWPOs necessitate precise flavor tagging and high-precision reconstruction, {\it e.g.}, the forward-backward asymmetry of charm and bottom quarks. 
Furthermore, most flavor physics studies involve QCD since all quarks are colored and $\tau$ leptons can decay to hadronic final states. 
In fact, most flavor physics studies rely on the theory of QCD, both perturbatively and non-perturbatively, to provide insights into the corresponding production, spectroscopy, and decays of hadronic states. 
In turn, the plethora of flavor measurements could provide crucial inputs to, and calibration of, QCD theory in multiple ways. It is also noteworthy that flavor physics provides a set of probes sensitive to BSM physics. 
For instance, the decay of a heavy-flavored fermion is suppressed by EW scale, 
$G_F^2 m_f^4 \lesssim 10^{-7}$, and consequently $f$ becomes long-lived. 
Such a narrow width makes it possible to reveal even small BSM effects, which are not easily observable otherwise. 
Finally, the ambitious goals of flavor physics studies motivate developments on the instrumentation frontier, demanding enhanced detector performance in vertexing, tracking, particle identification (PID), and calorimetry.

\begin{table}[t]
    \centering
    \resizebox{1.\textwidth}{!}{\begin{tabular}[t]{cccccc}
    \toprule[1pt]
        Particle
        & BESIII & STCF (1 ab$^{-1}$) & Belle II (50 ab$^{-1}$ on $\Upsilon(4S)$)  & LHCb (300 fb$^{-1}$)  & CEPC (TDR) \\
    \midrule
        $B^{0}$, $\bar{B}^{0}$
        & - & - & $5.4 \times 10^{10}$   & $3 \times 10^{13}$    & $4.8 \times 10^{11}$ \\
        $B^{\pm}$
        & - & - & $5.7 \times 10^{10}$  & $3 \times 10^{13}$    & $4.8 \times 10^{11}$ \\
        $B^{0}_{s}$, $\bar{B}^{0}_{s}$
        & - & - & $6.0 \times 10^{8}$ (5 ab$^{-1}$ on $\Upsilon(5S)$)    & $1 \times 10^{13}$    & $1.2 \times 10^{11}$ \\
        $B^{\pm}_{c}$
        & - & - & -                                                      & $1 \times 10^{11}$    & $7.2 \times 10^{8}$ \\
        $\Lambda_{b}^{0}$, $\bar{\Lambda}_{b}^{0}$
        & - & - & -                                                      & $2 \times 10^{13}$    & $1 \times 10^{11}$ \\ 
    \midrule
        $D^0$, $\bar{D}^0$
        & $1.2 \times 10^{8}$ & $7.2 \times 10^{9}$ & $4.8 \times 10^{10}$ & $7 \times 10^{14}$ & $8.3 \times 10^{11}$ \\
        $D^{\pm}$
        & $1.2 \times 10^{8}$ & $5.6 \times 10^{9}$ & $4.8 \times 10^{10}$ & $3 \times 10^{14}$ & $4.9 \times 10^{11}$ \\
        $D^{\pm}_s$
        & $1 \times 10^{7}$ & $1.8 \times 10^{9}$ & $1.6 \times 10^{10}$ & $1 \times 10^{14}$ & $1.8 \times 10^{11}$ \\
        $\Lambda^{\pm}_c$
        & $0.3 \times 10^{7}$ & $1.1 \times 10^{9}$ & $1.6 \times 10^{10}$ & $1 \times 10^{14}$ & $6.2 \times 10^{10}$ \\

    \midrule
        $\tau^{+}\tau^{-}$
        & $3.6 \times 10^{8}$ & $3.6 \times 10^{9}$ & $4.5 \times 10^{10}$ &  & $1.2 \times 10^{11}$ \\
    \bottomrule[1pt]
    \end{tabular}}
    \caption{Expected yields of $b$-hadrons, $c$-hadrons, and $\tau$ leptons at BESIII, STCF, Belle II, LHCb Upgrade II, and CEPC (according to the TDR~\cite{CEPCStudyGroup:2023quu}, $4\times 10^{12}$ $Z$ bosons are expected). For $b$- and $c$-hadrons, their yields include both charge conjugates, while the yield of $\tau$ leptons refers to the $\tau^- \tau^+$ events, namely the number of $\tau$ pairs. 
    We take the cross sections for $b\bar b$ and $c\bar c$ productions at center-of-mass energies corresponding to $\Upsilon(4S)$ and $\Upsilon(5S)$ from Ref.~\cite{Belle-II:2018jsg}, and of the $b$ quark productions within LHCb detector acceptance from Ref.~\cite{LHCb_CX}. 
    To estimate the production fractions of $B^0$ and $B^{\pm}$ at LHCb, we utilize the $B^0_s$ and $\Lambda^0_b$ production fractions in Ref.~\cite{LHCb_bfraction} and assume $f_u + f_d + f_s + f_{\rm baryon} = 1$, with $f_u = f_d$, and $f_{\Lambda^0_b} = f_{\rm baryon}$.
    For $Z$ decays, the production fractions of $B^0$, $B^{\pm}$, $B^0_s$, and $\Lambda^0_b$ are presented in Ref.~\cite{HeavyFlavorAveragingGroup:2022wzx}. 
    The $B_c$ meson production fraction at LHCb is taken from Ref.~\cite{Aaij:2019ths}, while its production fraction at the $Z$ pole (including the contribution from $B_c^*$ decays) is taken from Ref.~\cite{Zpole_BcFraction}. 
    For inclusive charm meson productions at the $Z$ pole,  including the contribution from $b$-hadron decays, see Refs.~\cite{Gladilin:2014tba,DELPHI:1993gqe,ALEPH:1993sez,OPAL:1994cct,Boehrer:1996pr}.
    The yields of $\tau$ leptons at the CEPC are rescaled from Ref.~\cite{CEPCStudyGroup:2018ghi}.
    The particle yields at the STCF are taken from Ref.~\cite{Achasov:2023gey}.
    }
    \label{tab:BYield}
\end{table}
%~~~~~~~~~~~~~~~~~~~~~~~~~~~~~~~~~~~~~~~~~~~

The successful realization of the flavor physics program at the CEPC relies on a number of key factors:
\begin{itemize}
    \item An abundant luminosity of the data at the CEPC $Z$ pole, which yields substantial heavy flavor statistics. With a high integrated luminosity and the large cross section $\sigma (e^-e^+\to Z \to b\bar{b},c\bar{c},\tau^- \tau^+)$, the Tera-$Z$ will generate extensive statistics of heavy-flavored hadrons and $\tau$ leptons~\cite{CEPCStudyGroup:2018ghi}, rivaling other proposed flavor physics experiments. This is demonstrated by the expected yields of $b$-hadrons in Belle II, LHCb and a representative future $Z$ factory, as listed in Table~\ref{tab:BYield}. The Tera-$Z$ yields approximately $4.8\times 10^{11}$ $B^0/\bar B^0$ or $B^\pm$ mesons, which is one order of magnitude larger than that expected at Belle II~\cite{Belle-II:2018jsg}. Even though this yield is roughly two orders of magnitude lower compared to that of LHCb, studies at the Tera-$Z$ can benefit significantly from the clean experimental environment and the precisely known center-of-mass energy.

    \item The clean environment of $e^-e^+$ collisions constitutes another cornerstone, substantially diminishing the background level and systematic uncertainties associated with neutral particles. This environment is particularly beneficial to flavor physics studies involving heavy $b$-hadrons, especially given the significantly limited event reconstruction efficiency in the noisy data environment of the LHCb~\cite{Bediaga:2018lhg}.
    
    \item The scale separation $m_Z \gg m_{b,c,\tau} \gtrsim \Lambda_{\rm QCD}$ underpins the success of the project, as it facilitates the production of a wide array of particle species. 
    In addition, even decay products with low momentum in the center-of-mass frame of heavy-flavored particles are expected to be boosted to higher energies and larger displacements.
    % , augmenting measurement precision.
    The significantly higher boost at the $Z$ factory compared to the $B$ and $C$ factories offers substantial advantages for particle identification and measurement precision.
    
    \item  Lastly, state-of-the-art detector technologies and algorithms for data analysis under development today will be crucial when deployed in the CEPC era. These technologies will enhance the investigation of extremely rare decay modes that contain neutral or invisible particles, as the cleanliness of a lepton collider enables such studies. The evolving field of advanced algorithms, especially deep learning ones, could also benefit flavor physics at the CEPC in almost all aspects by fully utilizing the large amount of data recorded from the hardware.
\end{itemize}

While the flavor physics program at the CEPC benefits from the various advantages above, it confronts new challenges. 
The first of these challenges is related to the significant increase in event statistics at the CEPC, which is expected to be greater by a factor of $\gtrsim \mathcal{O}(10^5)$ than the LEP run at the $Z$ pole. 
Given the improved detector systems and electronics, the volume of data to be processed will increase substantially. 
Meanwhile, the precision goals of flavor physics, driven by theoretical interests, will also reach an elevated level in the CEPC era. 
Therefore, it becomes essential to improve the understanding of backgrounds and to control systematic effects to match statistical uncertainties,
thus to fully benefit from CEPC's luminosity.
% prevent the dominance by systematic uncertainties, which could potentially undermine the benefits of the CEPC’s high instantaneous luminosity.

A second challenge arises from the multitude of viable channels to be studied at the CEPC. 
Compared to the other proposed future flavor physics experiments (or the upgrades of the current ones), the improvement achievable at the CEPC varies significantly channel by channel. 
Initial studies indicate that while the CEPC could enhance the precision of measurements by orders of magnitude in many instances, the improvement could be marginal in others. 
Therefore, identifying the most valuable systems, or ``golden channels'' - those with the highest potential for significant progresses or even discovery potential - for investigation in the CEPC context could substantially reduce the allocation of future resources. 
As it stands, some of these golden modes at the CEPC may have been overlooked as they are not suited for the existing experiments.

Besides these aforementioned experimental challenges, control of theoretical uncertainties is critical for CEPC flavor physics measurements and their interpretation.
Theoretical inputs come in multiple forms, such as the non-perturbative theory of hadronization, perturbative QCD and EW corrections to fermion production, lattice extrapolations of heavy flavor form factors, the relation between the CKM matrix elements and the observed $CP$ asymmetries, as well as the proper modeling of the electron beam and detector system. To accurately scrutinize the SM and to search for NP, the precision of these theoretical tools must align with those of the experimental outputs.

The principal objective of this document is to present an general perspective on the discovery potential of flavor physics at the CEPC, through Monte Carlo (MC) simulations and relevant phenomenological analyses. During the compilation of this white paper, simultaneous efforts were dedicated to promoting flavor physics programs at other future lepton colliders, such as the Future Circular Collider (FCC-$ee$)~\cite{Abada:2019lih,Bernardi:2022hny} and the International Linear Collider (ILC)~\cite{Fujii:2019zll}, both of which also include a $Z$ factory phase and higher energy operations. In particular, the FCC-$ee$ $Z$ pole run has a similar integrated luminosity (180 ab$^{-1}$) to the current CEPC proposal, and the higher-energy runs are likewise comparable. Since both proposals share similar detector performances~\cite{Bacchetta:2019fmz,CEPCStudyGroup:2018ghi}, and both adopt a particle flow algorithm (PFA)-oriented detector design~\cite{CEPCStudyGroup:2018ghi} and IDEA (Innovative Detector for Electron-positron Accelerator) detector design~\cite{Tassielli:2021rjk}, some relevant FCC-$ee$ studies were also incorporated into the current summaries, with only minimal rescaling applied as necessary. For the same reason, many physical discussions and conclusions in this white paper could be also applied to the FCC-$ee$ project.

This document is structured as follows. 
In Section~\ref{sec:CEPC}, we provide an overview of the CEPC facility, delineating key features of the collider and the detector that are crucial for flavor physics. Additionally, the simulation methods utilized at the CEPC are explained. 
Section~\ref{sec:FCCC} delves into Flavor-Changing-Charged-Current (FCCC) semileptonic and leptonic $b$ decays, discussing their theoretical framework, recent progress and future research directions. 
Rare $b$ decays mediated by FCNC are explored in Section~\ref{sec:FCNC}, featuring a preliminary theoretical interpretation and discussion of di-leptonic, neutrino and radiative modes. Section~\ref{sec:CPV} is dedicated to the discussions on the measurements of $CP$ asymmetries. 
Sections~\ref{sec:charmstrange} and~\ref{sec:tau} focus on charm/strange and $\tau$ physics respectively. 
Flavor physics measurements via leptonic or hadronic $Z$ decays are discussed in Section~\ref{sec:Zdecay}.
Section~\ref{sec:beyondZ} extends the discussions to flavor physics at higher energies, including $|V_{cb}|$ measurements through on-shell $W$ boson decays, Higgs exotic and FCNC decays, as well as touching upon other possibilities. 
Prospects of hadron spectroscopy and exotic states are covered in Section~\ref{sec:exotic}.
The production of light BSM particles by heavy flavor interactions forms the central theme of Section~\ref{sec:BSM}. The detector performance requirements for a successful flavor physics program at the CEPC are discussed in Section~\ref{sec:detectorrequirement}.
Finally, we summarize the topics covered in this document and provide an outlook for future explorations in Section~\ref{sec:summary}.

\section{Description of CEPC Facility}  
\label{sec:CEPC}

\subsection{Key Collider Features for Flavor Physics}

As an $e^-e^+$ collider operating around the EW scale, flavor physics studies at the CEPC are affected by three major features. 
Firstly, as $\sqrt{s}\gg m_{b,c,\tau}$, the CEPC produces highly relativistic heavy-flavored quarks or leptons. 
Their boosted decay products allow for precise momentum and lifetime measurements. 
This is in contradistinction to the situations at low energy $e^-e^+$ colliders such as Belle II~\cite{Belle-II:2018jsg}, BaBar~\cite{Aubert:2001tu}, BESIII~\cite{Ablikim:2009aa}, and other future proposals, such as the Super Tau-Charm Factory (STCF)~\cite{Achasov:2023gey}. 
Secondly, as an $e^-e^+$ collider, the CEPC provides a clean environment for flavor physics studies with low QCD backgrounds, negligible pileup events, and an almost fixed $E_{cm}$. 
Compared to hadron collider experiments, such as the LHCb~\cite{Alves:2008zz}, the CEPC enables more effective identification and reconstruction of final states that include neutral or invisible particles.
The above arguments show the uniqueness of CEPC flavor physics studies. 
Thanks to advanced accelerator design, the large instantaneous luminosity will allow to collect $\mathcal{O}(10^5)$ times more statistics than the LEP $Z$ pole run~\cite{Alcaraz:2009jr}. 
As a consequence, the search and analysis strategies may differ significantly from those employed in the relevant studies at LEP. 
For instance, high signal statistics allows sharper cuts to reduce backgrounds. 
At the same time, one needs to carefully address other systematic uncertainty sources using the plethora of data. 
Hence, the large luminosity of the CEPC brings new challenges and existing projections based on LEP must be reconsidered. 
Such challenges are especially severe for precision measurements. According to the CEPC CDR~\cite{CEPCStudyGroup:2018ghi}, the beam energy spread could typically be controlled to the level of 0.1\%. This, together with a detector that can reconstruct precisely hadronic events -- allowing for precise determination of missing energy/momentum -- thus enables relevant physics measurements with high precision; for instance, tagging semileptonic heavy quark decay and searching for dark matter candidates in hadronic events, especially at the $Z$ factory mode.    

The CEPC uses a nano beam scenario and therefore the typical beam spot sizes are of order $\upmu$m in the $x$ direction, order nm in the $y$ direction, and correspondingly of order a few hundred $\upmu$m in the $z$ direction.
The beam sizes at different operation modes of the CEPC are summarized in Table~\ref{tab:BeamSize}.
The accelerator will provide a collision area with a typical size of order $\upmu$m in the transverse direction and of order $\sim\mathcal{O}$(100) $\upmu$m along the beam direction. 
The spatial uncertainty of the interaction point can therefore be limited, enabling high precision measurements with $\tau$ final states -- for example, in dark matter searches with $Z\to\tau^- \tau^+$ events at $Z$ factory.

\begin{table}[t]
    \centering
    \resizebox{1.\textwidth}{!}{
    \begin{tabular}[t]{ccccc}
    \toprule[1pt]
        Operation mode & $Z$ factory  & $WW$ threshold  & Higgs factory & $t\bar{t}$ \\
        \midrule
        $\sqrt{s}$ (GeV)
        & 91.2 & 160 & 240 & 360 \\
        Beam size $\sigma_x$ ($\upmu$m)
        & 6 & 13 & 14 & 39 \\
        Beam size $\sigma_y$ ($\upmu$m)
        & 0.035 & 0.042 & 0.036 & 0.113 \\
        Bunch length (total, mm)
        & 8.7 & 4.9 & 4.1 & 2.9 \\
        \midrule
        Crossing angle at IP (mrad)
        & \multicolumn{4}{c}{33} \\
    \bottomrule[1pt]
    \end{tabular}
    }
    \caption{Beam size, bunch length, and crossing angle at different operation modes of the CEPC~\cite{CEPCStudyGroup:2023quu,Gao:2022lew}.}
    \label{tab:BeamSize}
\end{table}

\subsection{Key Detector Features for Flavor Physics}
\label{sec:DetectorFeature}

Flavor physics program at Tera-$Z$ is enormously rich and extremely demanding on detector performance.
In general, a Tera-$Z$ detector would have a large acceptance with a solid angle coverage up to $|\cos\theta| < 0.99$.
This detector would also have low energy and momentum thresholds at the 100 MeV level to record and recognize low energy objects that characterize certain hadron decays, {\it e.g.},~soft photons and pions generated from excited heavy hadrons, as well as some low energy hadrons that are essential for understanding relevant QCD processes~\cite{Pythia_2015}.

\begin{figure}[t]
    \centering
    \includegraphics[width=0.45\textwidth]{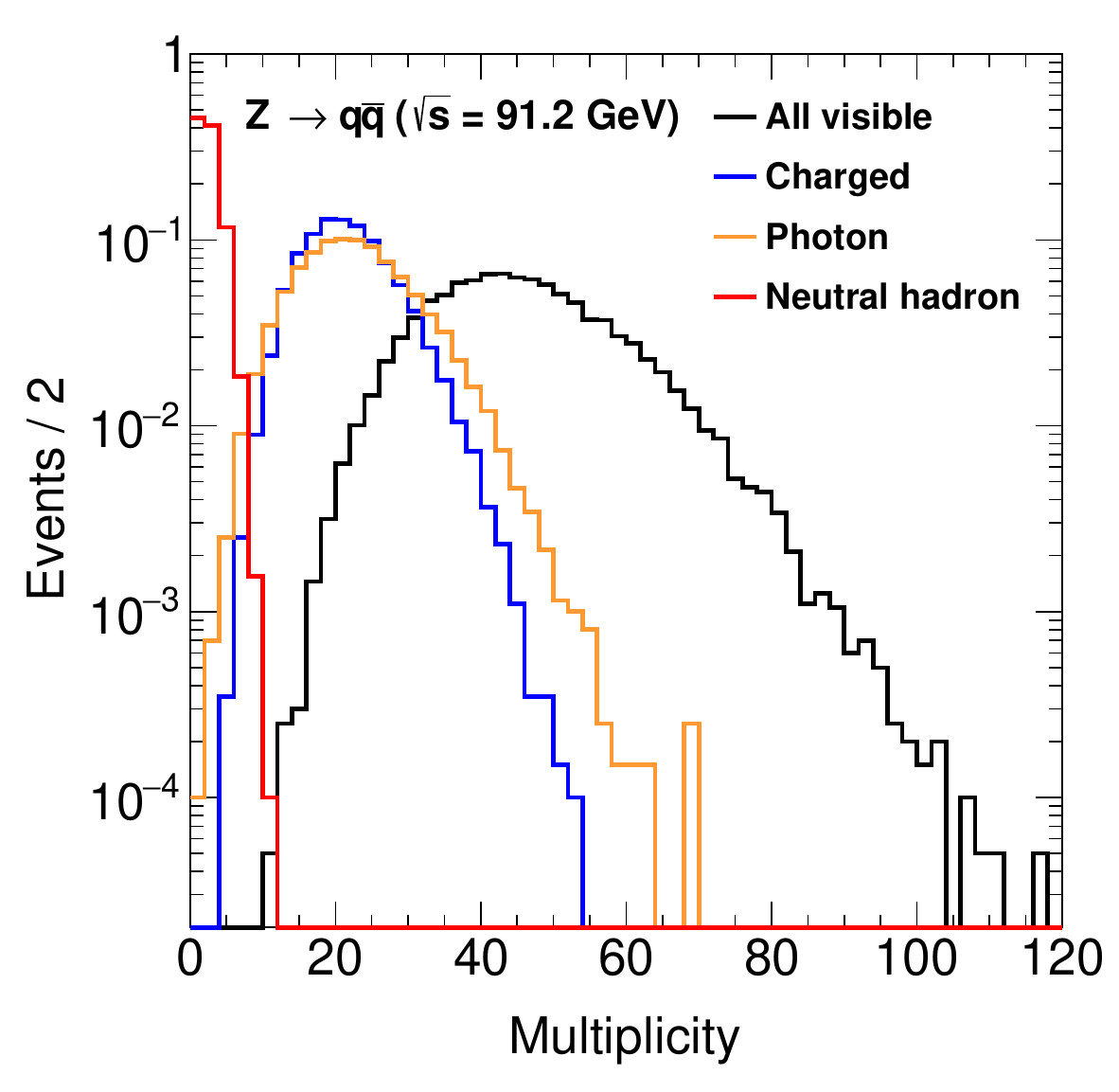}
    \includegraphics[width=0.45\textwidth]{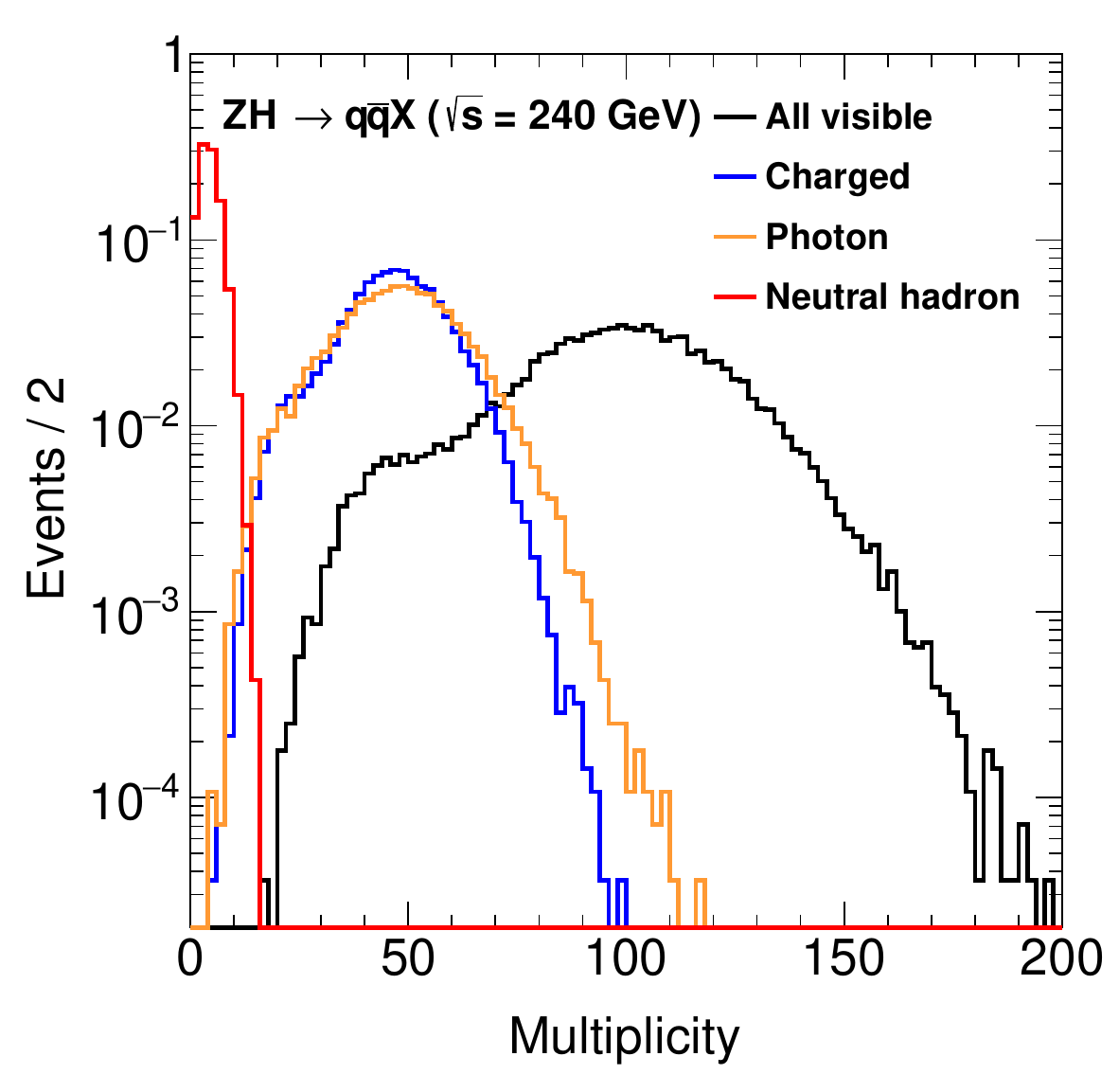}
    \caption{Multiplicities of different types of final state particles in $Z\to q\bar{q}$ (91.2 GeV) and $Z(\to q\bar{q})H(\to {\rm inclusive})$ (240 GeV) events.}
    \label{fig:Multiplicity}
\end{figure}

To efficiently separate signal events from background, it is essential to identify the relevant physics objects and to precisely reconstruct their properties --- especially their energies and momenta. For a Tera-$Z$ detector, a typical benchmark is to reconstruct the intermediate particles, such as $\pi^0\to\gamma\gamma$, $K^0_S\to\pi^+\pi^-$, $\phi\to K^+K^-$, $\Lambda\to p\pi^-$, etc., inside hadronic $Z$ events. 
A more challenging case would be to identify the decay products of a target heavy-flavored hadron which may decay into $\mathcal{O}(10)$ particles with a complicated and rich decay cascading order inside a jet. 
These decay products include not only charged final state particles (leptons and charged hadrons), but also photons, neutral hadrons, and the missing energy/momentum induced by neutrinos. A hadronic $Z$ event could have up to 100 final state particles, 
as shown in Figure~\ref{fig:Multiplicity}. 
To successfully separate and reconstruct the decay products of the target particle is a key challenge for measurements performed in hadronic $Z$ events, for which it is necessary to employ the particle flow method~\cite{Ruan:2013rkk,Marshall:2013bda}. Such a method emphasizes the separation of final state particles and has been proven capable of providing better reconstruction of both the hadronic system and the missing energy/momentum.

In addition, good intrinsic resolution of subdetectors, ({\it i.e.}, momentum reconstruction by the tracker and energy measurement by the calorimeter), is always critical for flavor physics measurements. It not only enables the precise reconstruction of physics properties, such as particle masses, but also significantly reduces the combinatorial background, which is especially present in physics measurements involving narrow resonances. In particular, achieving excellent electromagnetic (EM) energy resolution with a particle-flow-oriented, high-granularity calorimeter is both challenging and necessary for the flavor physics program, as photons and neutral pions are common decay products in many fundamental flavor physics measurements. A benchmark analysis~\cite{Wang:2022nrm} of the measurement of the standard CKM unitarity triangle angle $\alpha$ via $B^0\to\pi^0\pi^0$ suggests an EM resolution of approximately $\mathcal{O}(3\%/\sqrt{E\text{(GeV)}})$ 
to fulfill the requirement of 3\,$\sigma$ separation between $B^0$ and $B^0_s$ with a 30 MeV $B$-meson mass resolution.

Most flavor physics measurements at the CEPC will involve hadronic events, particularly di-jet events at the $Z$ pole. It is essential to identify the origin of a jet, {\it i.e.}, to determine whether it originates from a quark, an anti-quark, or even a gluon. The jet origin identification~\cite{Liang:2023wpt}, to a certain extent, shall be regarded as a natural extension of jet flavor tagging, quark-gluon jet separation, and jet charge measurements, which is indispensable in flavor physics measurements such as CKM and $CP$ violation measurements.

\begin{figure}[t]
    \centering
    \includegraphics[width=0.48\textwidth]{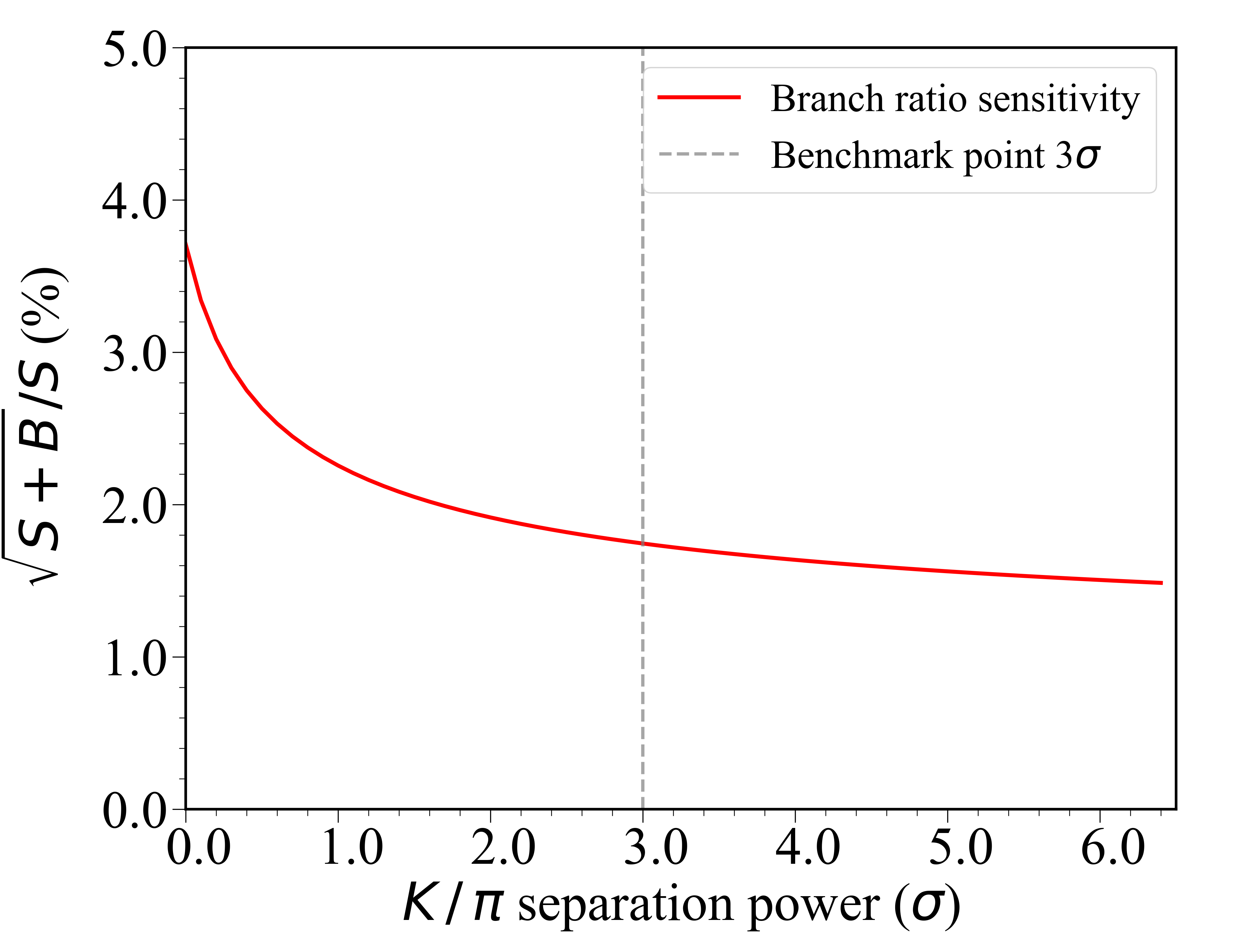}
    \includegraphics[width=0.48\textwidth]{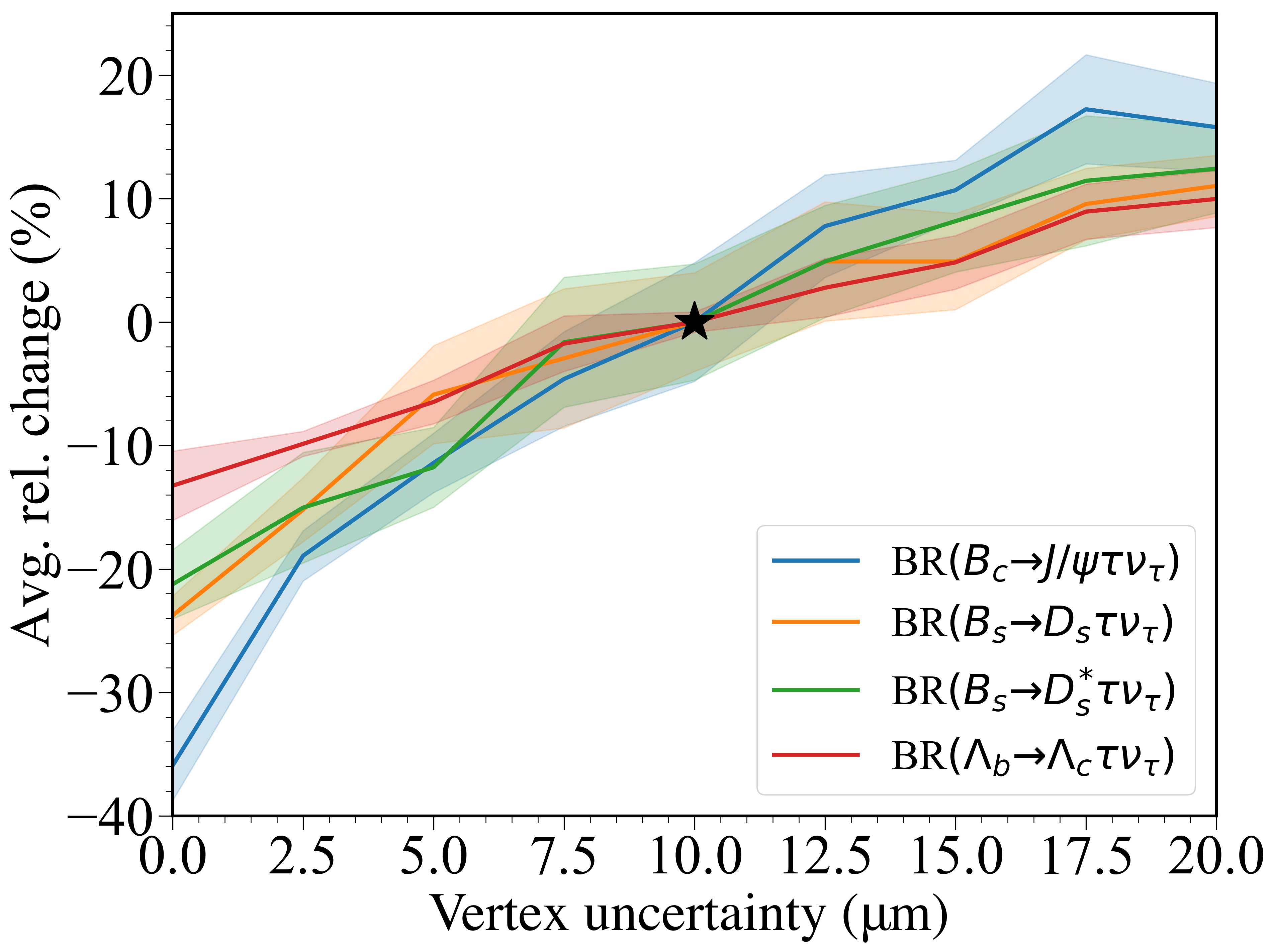}
    \caption{\textbf{LEFT:} Precision of measuring BR($B_s^0\to\phi\nu\bar{\nu}$) as a function of PID performance, parameterized by the $K/\pi$ separation power~\cite{Li:2022tov}. \textbf{RIGHT:} Precision variance of measuring $\text{BR}(H_b\to H_c \tau\nu_\tau )$ as a function of detector vertex uncertainties~\cite{Ho:2022ipo}, with starred reference point set by a vertex uncertainty of 10 $\mu$m.}
    \label{fig:Phivv_accuracy_pid}
\end{figure}

A successful flavor physics program also needs high efficiency/purity PID. An efficient PID not only suppresses the combinatorial background, induced by misidentified particles, but also separates decays with similar topologies in the final states, such as $B^0_{(s)}\to\pi^+\pi^-$, $B^0_{(s)}\to K^+K^-$, and $B^0_{(s)}\to K^{\pm}\pi^{\pm}$~\cite{LHCb:2012ihl}. A decent PID is also critical for the jet origin identification~\cite{Liang:2023wpt,Cui:2023kqb} and relevant physics measurements such as the Higgs rare/exotic decay measurement~\cite{Liang:2023wpt}.
The benchmark analysis of $B_s^0\to\phi\nu\bar{\nu}$~\cite{Li:2022tov} shows that a relative uncertainty of BR($B_s^0\to\phi\nu\bar{\nu}$) less than 2\% at a Tera-$Z$ collider requires a 3\,$\sigma$ $K^\pm/\pi^\pm$ separation for the identification of charged hadrons, see the left panel of Figure~\ref{fig:Phivv_accuracy_pid}.
This requirement can be addressed by multiple PID technologies. For instance, the CEPC CDR detector~\cite{CEPCStudyGroup:2018ghi} can separate different species of hadrons using $dE/dx$ information measured by the time projection chamber (TPC) and time-of-flight (TOF) information provided by either a dedicated TOF device, or by combining TOF and EM calorimeter (ECAL) together.
Detector optimization studies~\cite{Zhu:2022hyy} suggests that $dE/dx$ needs to reach 3\% in combination with a TOF resolution of 50 ps to statisfy this PID requirement.
In addition, the $dN/dx$ cluster-counting method proposed by the IDEA drift chamber~\cite{Cuna:2021sho} is promising to further improve the PID performance.

\begin{figure}[t]
    \centering
    \includegraphics[width=0.9\textwidth]{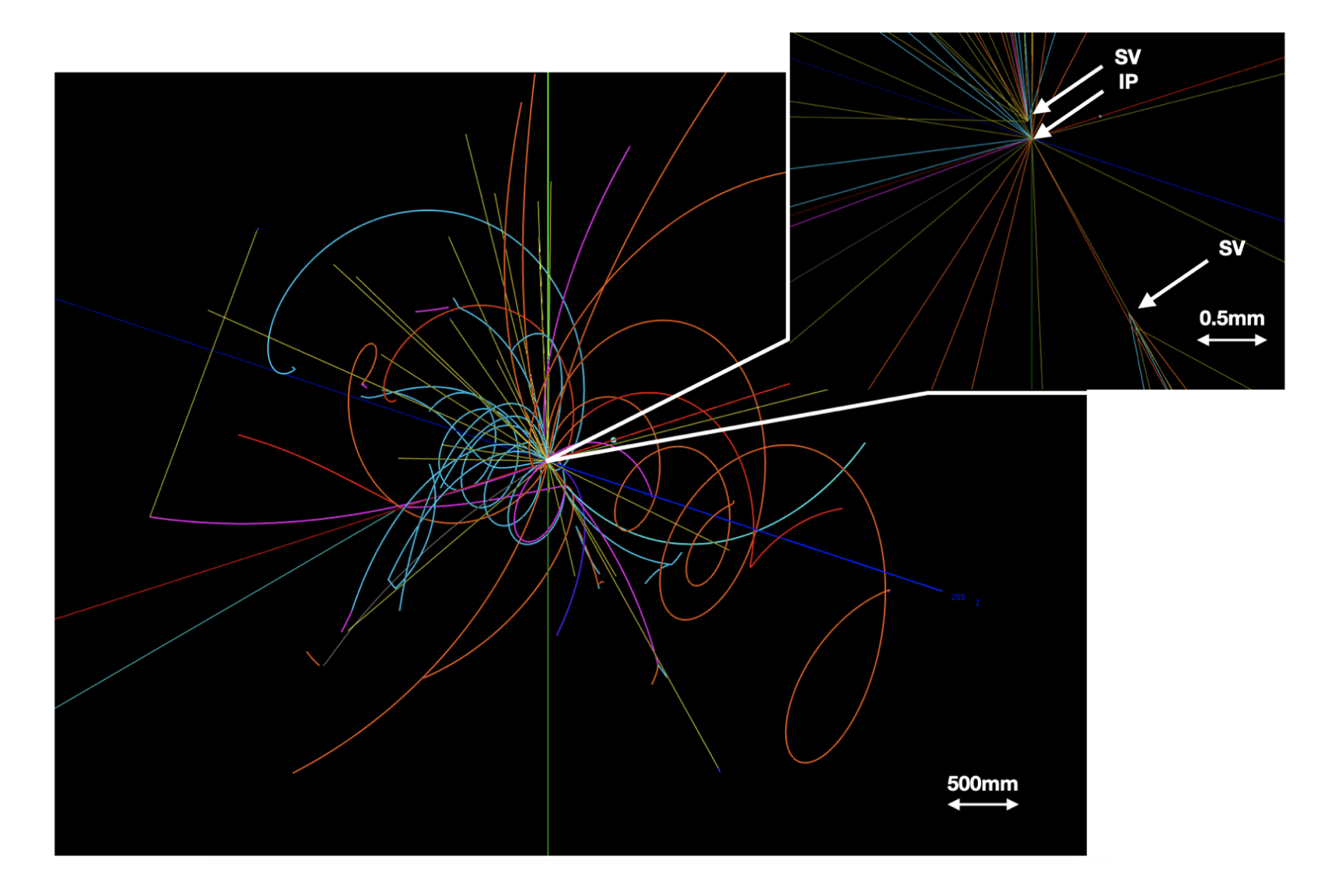}
    \caption{Display of a $Z\to b\bar{b}$ event with typical secondary vertices (SV).}
    \label{fig:Zbb_Vertex_Display}
\end{figure}

A high-precision and low-material vertex system is vital for the CEPC flavor physics program. Precise vertex measurements provide pivotal information to distinguish the species of the initial quark that fragments into a jet, namely the jet origin identification. Precise vertex information is also critical for determining the decay time or lifetime of heavy-flavored hadrons with high precision. To match the characteristic timescales such as those of $B_s^0-\bar{B}_s^0$ mixing ($\sim 56$~fs), of $D_s$ decay ($\sim 500$~fs), and of $\tau$ decay ($\sim 290$~fs), the decay time resolution is required to reach order $\mathcal{O}(10)$~fs. These accurate lifetime measurements will also benefit flavor tagging and time-dependent $CP$ violation measurements. In addition, a high-performance vertex system can provide a precise reconstruction of the secondary vertices that characterize some of the heavy-flavored hadron decays, such as the example shown in Figure~\ref{fig:Zbb_Vertex_Display}. Such a system can also help to suppress the backgrounds, especially from the IP. One concrete application can be the measurements of FCCC $\text{BR}(H_b\to H_c \tau\nu_\tau )$, where the reconstruction of the $b$ hadron $H_b$ can significantly rely on the determination of the decay vertex of the charmed hadron $H_c$ and on the measurement of the muon track originating from the $\tau$ decay~\cite{Ho:2022ipo}. As shown in the right panel of  
Figure~\ref{fig:Phivv_accuracy_pid}, the improved resolution  of vertex system can uniformly benefit these  measurements, yielding an improvement in precision of $\mathcal {O}(10\%)$ level.

The above-mentioned requirements are also highly beneficial for the physics programs at higher center-of-mass energies, {\it i.e.}, the 160 GeV $W^+W^-$ threshold scan, the 240 GeV Higgs run, and the 360 GeV top-pair operation.
On top of their core physics programs, such as $W$ mass and precise Higgs/top properties measurements, the data samples and key detector features also support an intensive flavor physics program, see Section~\ref{sec:beyondZ}.

To address these physics requirements, intensive efforts have been devoted to the detector conceptual design, physics performance studies, and key technology R\&D.
We refer to two benchmark detector concepts considered in the CDR study~\cite{CEPCStudyGroup:2018ghi}.
These concepts are used in the simulations presented in this manuscript, providing reference performance for relevant physics potential studies.

\begin{figure}[t]
    \centering
    \includegraphics[width=0.38\textwidth]{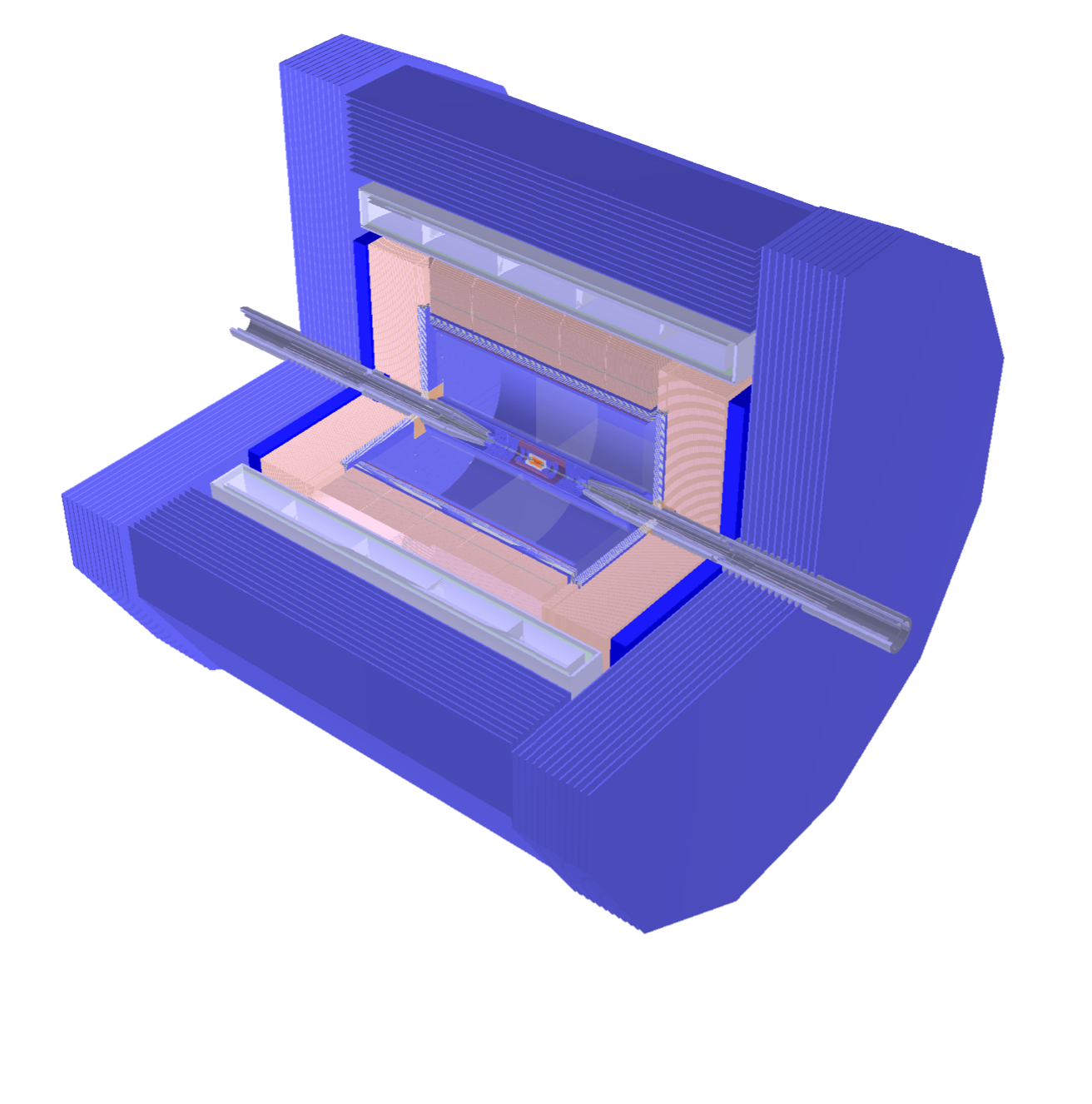}
    \includegraphics[width=0.42\textwidth]{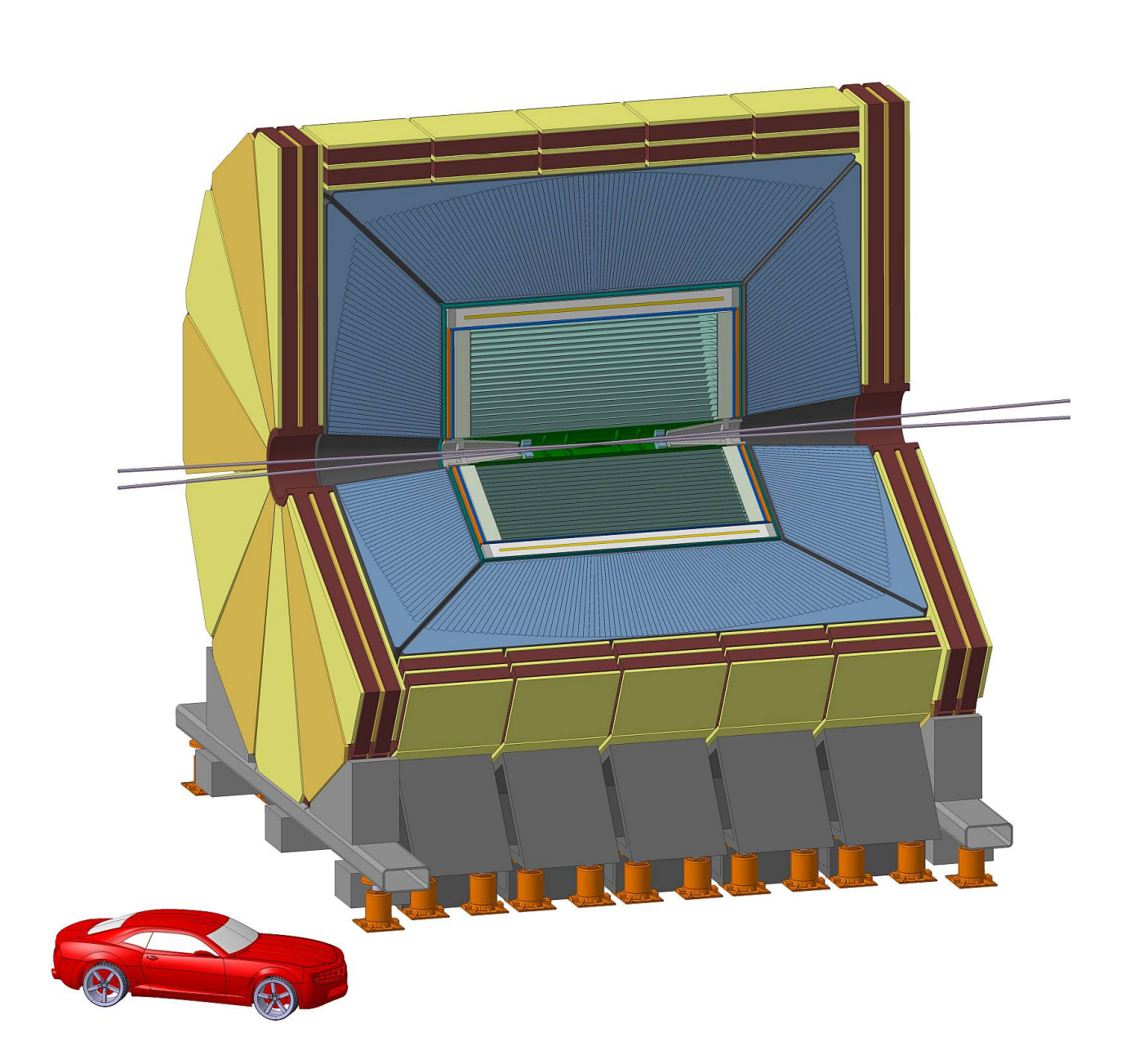}
    \caption{Schematic layouts of the CEPC CDR detector~\cite{CEPCStudyGroup:2018ghi} (\textbf{LEFT}) and the IDEA detector~\cite{IDEA_Snowmass_2021} (\textbf{RIGHT}).}
    \label{fig:CEPC_Geo_CDR}
\end{figure}

The starting point of our discussion is the particle-flow-oriented detector design in the CEPC CDR~\cite{CEPCStudyGroup:2018ghi}.
As the majority of the full simulation studies uses this detector design, we will refer to it as the CDR detector for simplicity.
Guided by the particle flow principle, the CDR detector features a high-precision tracking system, a high-granularity calorimeter system, and a high magnetic field.
As shown in detail in Figure~\ref{fig:CEPC_Geo_CDR}, from inside to outside, the CDR detector consists of a silicon pixel vertex detector, a silicon tracker, a TPC, a silicon-tungsten sampling EM calorimeter (Si-W ECAL), a steel-glass Resistive Plate Chambers (RPC) sampling hadronic calorimeter (SDHCAL), a superconducting solenoid magnet providing a magnetic field of 2--3 Tesla, and a flux return yoke embedded with a muon detector.
Additionally, the Si-W ECAL could also be instrumented with a few timing layers to enable TOF measurements with a precision of 50 ps or even better~\cite{CEPCStudyGroup:2018ghi,Che:2022dig}.

Alongside the CDR detector, an alternative detector concept known as IDEA~\cite{IDEA_Snowmass_2021} is also utilized in various studies covered in this white paper. In comparison to the CDR detector, the IDEA detector incorporates a dual readout calorimeter system to attain superior energy resolution for both EM and hadronic showers. Moreover, the IDEA detector operates with a reduced magnetic field of 2 Tesla while compensating for this reduction by offering a larger tracking volume. The overall structure of both the detectors can be seen in Figure~\ref{fig:CEPC_Geo_CDR}.

\begin{table}[t]
\begin{center}
\resizebox{1.\textwidth}{!}{\begin{tabular}{cccc}
    \toprule[1pt]
    Item & CDR~\cite{CEPCStudyGroup:2018ghi} & $4^{\rm th}$ concept~\cite{4th_Jianchun_talk} & Comments \\
    \midrule
    \multicolumn{4}{c}{Basic Performance}  \\
    \midrule
    Acceptance & $|\cos\theta|<0.99$~\cite{CEPCStudyGroup:2018ghi} & & \\
    Threshold & 200 MeV~\cite{Manqi_Talk_IAS_2019,Shen:2019yhf} & 100 MeV & For tracks \& photons \\
    
    Beam energy spread  & $\mathcal{O}(0.1\%)$~\cite{CEPCStudyGroup:2018ghi} &  & \\
    Tracker momentum resolution & $\mathcal{O}(0.1\%)$~\cite{CEPCStudyGroup:2018ghi} &  & \\
    ECAL energy resolution & $17\%/\sqrt{E\text{(GeV)}} \oplus 1\%$~\cite{CEPCStudyGroup:2018ghi} & $3\%/\sqrt{E\text{(GeV)}}$~\cite{Wang:2022nrm} & \\
    HCAL energy resolution & $60\%/\sqrt{E\text{(GeV)}} \oplus 1\%$~\cite{CEPCStudyGroup:2018ghi} & $30\%/\sqrt{E\text{(GeV)}}$~\cite{Hu:2023dbm} & \\
    Vertex resolution & 10--200 $\upmu$m~\cite{CEPCStudyGroup:2018ghi} & 5--100 $\upmu$m & \\
    Jet energy resolution & 3--5\%~\cite{CEPCStudyGroup:2018ghi,Lai:2021rko} &  &  For 20--100 GeV \\

    $\ell-\pi$ mis-ID & $<1\%$~\cite{Yu:2021pxc} &  & In jet, $|\vec{p}| > 2$~GeV \\
    $\pi-K$ separation & $> 2\sigma$~\cite{CEPCStudyGroup:2018ghi}  & $> 3\sigma$~\cite{Li:2022tov} & In jet, $|\vec{p}| > 1$~GeV, TOF+$dE/dx$ \\
    \midrule
    \multicolumn{4}{c}{Flavor Physics Benchmarks (Depending on the Above)}  \\
    \midrule
    $\sigma(m_{H,W,Z})$ & 3.7\%~\cite{CEPCStudyGroup:2018ghi} &  & Hadronic decays\\

    $b$-jet efficiency$\times$purity & $\sim 86\%$~\cite{Liang:2023wpt} & & In $Z$ hadronic decays \\
    $c$-jet efficiency$\times$purity & $\sim 64\%$~\cite{Liang:2023wpt} & & In $Z$ hadronic decays \\
    $b$-jet charge tagging $\epsilon_{\rm eff}=\epsilon(1-2\omega)^2$ & $\sim 37\%$~\cite{Liang:2023wpt} & &  \\
    $c$-jet charge tagging $\epsilon_{\rm eff}=\epsilon(1-2\omega)^2$ & $\sim 58\%$~\cite{Liang:2023wpt} & &  \\
    
    $\pi^0$ efficiency$\times$purity & $\gtrsim 70\%$~\cite{Shen:2019yhf} & $\gtrsim 80\%$~\cite{Wang:2022nrm} & In $Z$ hadronic decays, $|\vec{p}_{\pi^0}|>5$~GeV\\
    $K_S^0$, $\Lambda$ efficiency & 60\%-85\%~\cite{Zheng:2020qyh} &  & In $Z$ hadronic decays, all tracks\\
    $\tau$ efficiency$\times$purity & 70\%~\cite{Yu:2020bxh} &  & In $WW\to\tau\nu q\bar{q}^\prime$, inclusive\\
    $\tau$ mis-ID & $\mathcal{O}(1\%)$~\cite{Yu:2020bxh} &  & In $WW\to\tau\nu q\bar{q}^\prime$, inclusive\\
    \bottomrule[1pt]
\end{tabular}}
\end{center}
\caption{Performance of the CEPC CDR detector and some suggested objectives.}
\label{tab:performance}
\end{table}

\begin{figure}[t]
    \centering
    \includegraphics[width=.88\textwidth]{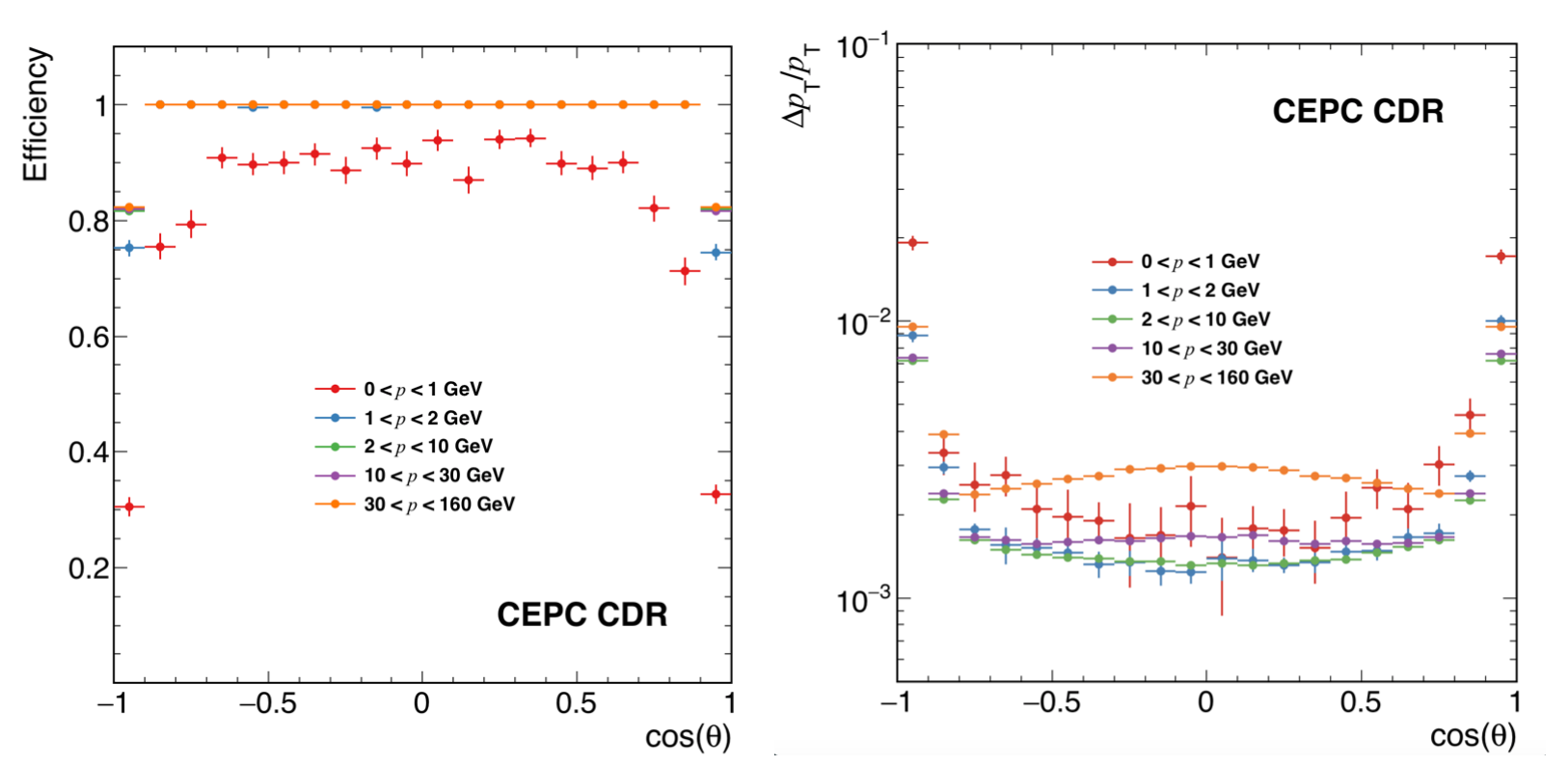}
    \caption{Single track reconstruction efficiency (\textbf{LEFT}) and momentum resolution (\textbf{RIGHT}) of the CEPC CDR detector~\cite{CEPCStudyGroup:2018ghi}.}
    \label{fig:CDR_Track_Performance}
\end{figure}

\begin{figure}[t]
    \centering
    \includegraphics[width=0.45\textwidth]{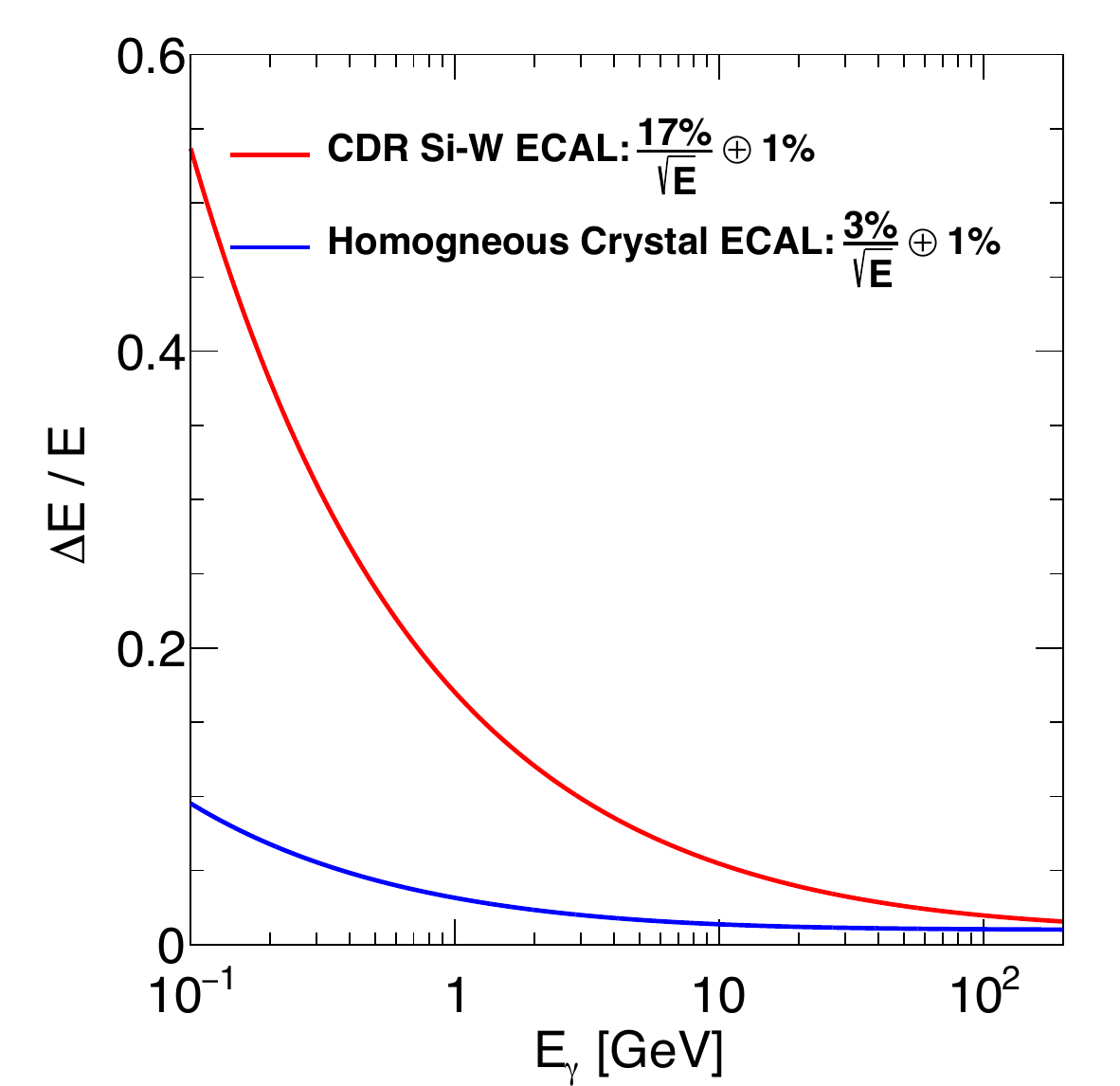}
    \includegraphics[width=0.45\textwidth]{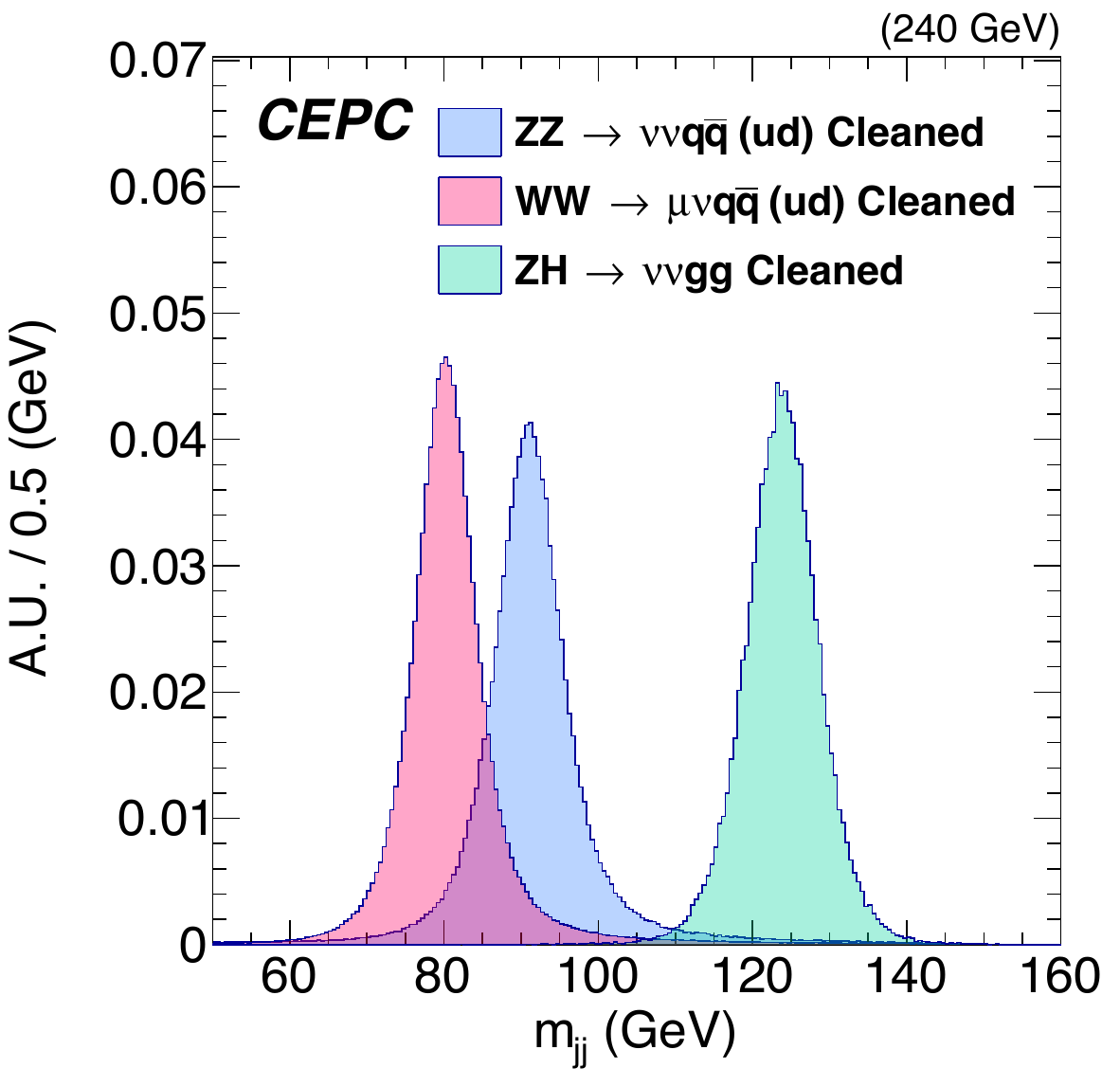}
    \caption{\textbf{LEFT:} Comparison of the CEPC CDR photon energy resolution achieved by the sampling Si-W ECAL~\cite{CEPCStudyGroup:2018ghi} and expected photon energy resolution of homogeneous crystal ECAL. \textbf{RIGHT:} Reconstructed boson masses of cleaned $\nu\bar{\nu}q\bar{q}$, $l\nu q\bar{q}$, and $\nu\bar{\nu}H,~H\to gg$ events~\cite{Lai:2021rko}.}
    \label{fig:EMReso_BMR}
\end{figure}

\begin{figure}[t]
    \centering
    \includegraphics[width=0.9\textwidth]{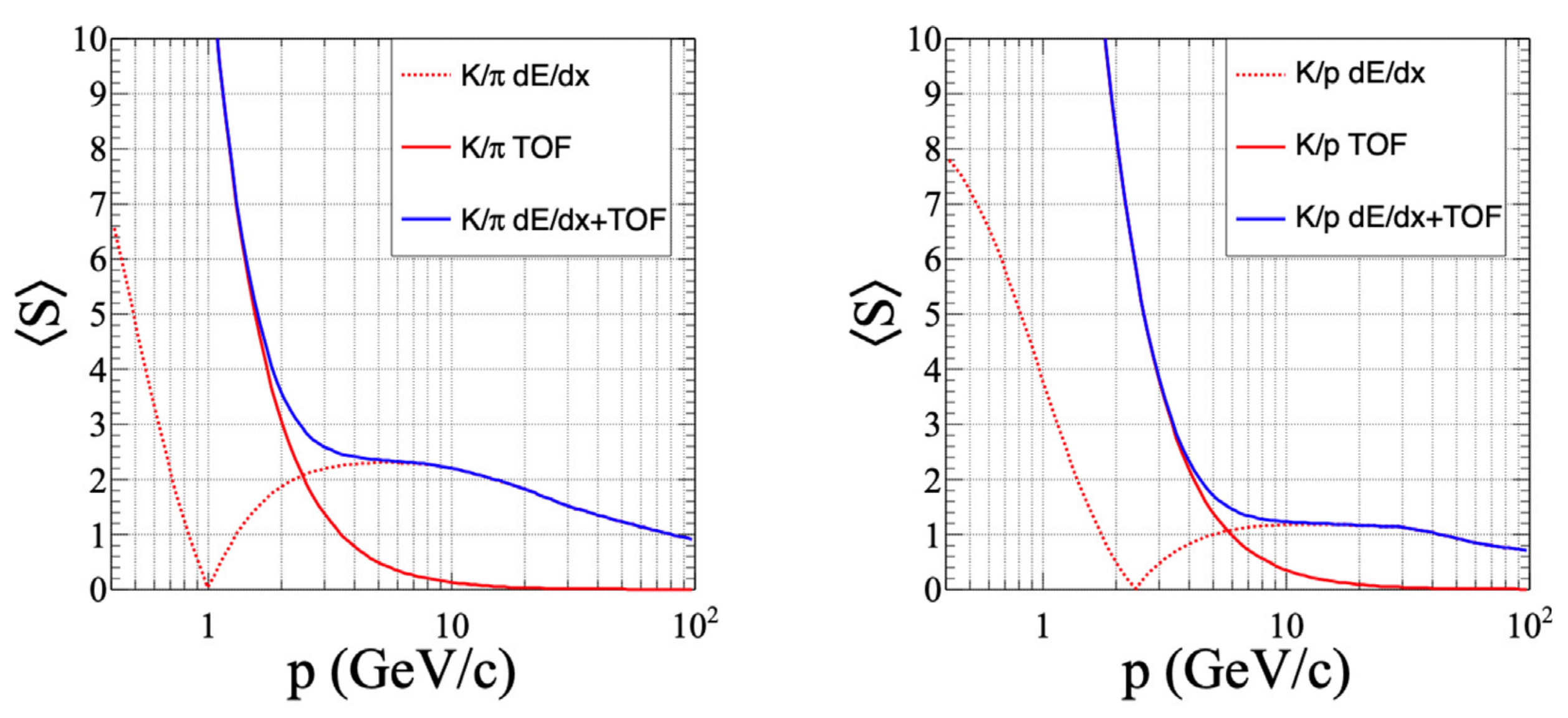}
    \caption{Separation power of $K/\pi$ (\textbf{LEFT}) and $K/p$ (\textbf{RIGHT}) using different techniques~\cite{Zhu:2022hyy}.}
    \label{fig:PID_Baseline}
\end{figure}

By virtue of the PFA-oriented design, the CEPC CDR detector performs well in efficient tracking, lepton identification, and precise reconstruction of hadronic systems. These excellent features of the CEPC CDR detector provide a solid basis for flavor physics studies. 
The expected performance of the CEPC CDR detector is summarized in Table~\ref{tab:performance}.
Notably, the CDR tracking system demonstrates an efficiency close to 90\% and a relative momentum resolution approaching $\mathcal{O}(10^{-3})$ for individual tracks with momenta exceeding 1 GeV within the barrel region, as illustrated in Figure~\ref{fig:CDR_Track_Performance}.
As depicted in left panel of Figure~\ref{fig:EMReso_BMR}, the CDR photon energy resolution is $17\%/\sqrt{E\text{(GeV)}} \oplus 1\%$, achieved by the sampling Si-W ECAL, which features the high granularity critical for particle flow reconstruction.
In terms of PID performance, the CEPC CDR design achieves a $K/\pi$ separation better than 2\,$\sigma$ in the momentum range up to 20 GeV by effectively combining TOF and $dE/dx$ information, as shown in Figure~\ref{fig:PID_Baseline}. The inclusive $Z\to q\bar{q}$ sample exhibits an overall $K^\pm$ identification efficiency and purity exceeding 95\%~\cite{Zhu:2022hyy}.
Regarding hadronic systems, the CEPC CDR detector attains a boson mass resolution (BMR) better than 4\% for hadronically decaying $W$, $Z$, and Higgs bosons, as illustrated in right panel of Figure~\ref{fig:EMReso_BMR}. This not only enables a separation exceeding 2\,$\sigma$ between $W$ and $Z$ bosons in their hadronic decays, but also enhances the precision of missing energy/momentum measurements, which are vital for flavor physics investigations.

After the release of the CEPC CDR, intensive detector R\&D efforts continue to address the CEPC physics requirements. 
These efforts have led to the development of the $4^{\rm th}$ detector concept~\cite{4th_Jianchun_talk}, which demonstrates significant improvements in EM energy resolution, intrinsic hadronic energy resolution, PID performance, and the vertexing. 
The $4^{\rm th}$ detector concept employs a PFA-compatible homogeneous crystal ECAL to enhance the EM resolution, achieving an energy resolution of $3\%/\sqrt{E\text{(GeV)}} \oplus 1\%$ (see the comparison in the left panel of Figure~\ref{fig:EMReso_BMR}). This resolution is crucial for the separation of $B^0$ and $B^0_s$ that decay into EM final states~\cite{Wang:2022nrm}.
It utilizes high-density glass-scintillator HCAL, which can improve the hadronic energy resolution by nearly a factor of two, consequently enhancing the BMR~\cite{Hu:2023dbm}. 
The $4^{\rm th}$ detector also features a pixelated TPC that provides precise $dE/dx$~\cite{Zhu:2022hyy,She:2023puo} or $dN/dx$~\cite{Cuna:2021sho} measurements, both of which are critical for PID. 
Furthermore, the $4^{\rm th}$ detector concept incorporates a vertex detector with stitching technology~\cite{4th_vtx_talk}, which has significantly lower material budget.

Another significant advancement is in the jet charge measurement.
The performance of jet charge measurement is typically characterized by the effective tagging efficiency (power) $\epsilon_{\rm eff} \equiv \epsilon_{\rm tag} (1-2\omega)^2$, where $\epsilon_{\rm tag}$ is the flavor tagging efficiency and $\omega$ is the wrong tag fraction.
The study~\cite{Cui:2023kqb} develops a Leading Particle Jet Charge method (LPJC) and combines it with a Weighted Jet Charge (WJC) method to form a Heavy Flavor Jet Charge method (HFJC).
This study evaluates the effective tagging power for $c$/$b$ jets at the CEPC $Z$ pole and finds it to be 39\%/20\%, respectively.
Additionally, by implementing benchmark impact parameter cuts of 0.02/0.04 mm to distinguish the origin of the leading charged particle (whether from the decay of the leading heavy hadron or QCD fragmentation), the effective tagging power for $c$/$b$ jets was found to be 39\%/27\%.
Furthermore, a dedicated $b$-jet charge tagging algorithm developed specifically for the study of $B^0_s \to J/\psi\phi$ at the CEPC~\cite{Li:2022tlo} achieved an effective tagging power of 20\%.

\begin{figure}[t]
    \centering
    \includegraphics[width=0.45\textwidth]{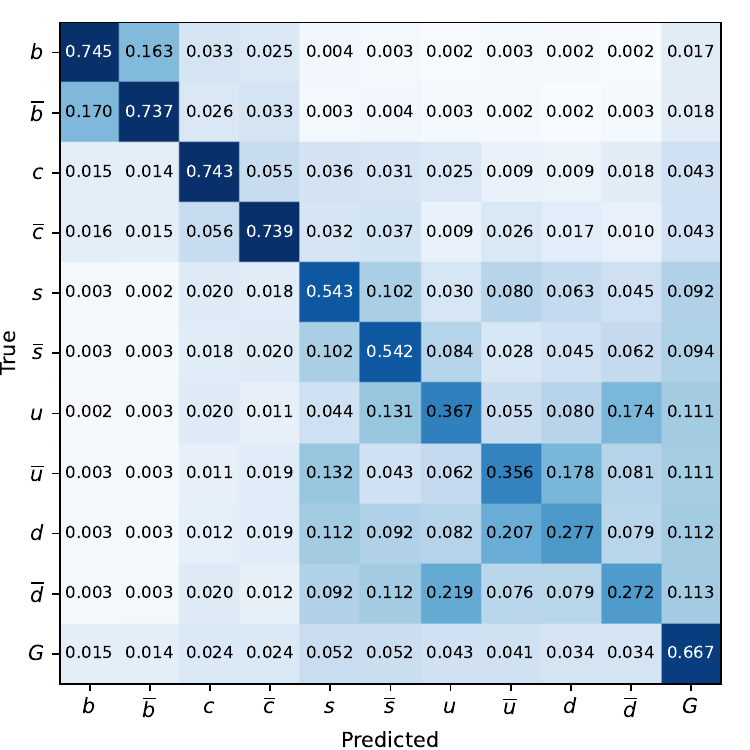}
    \includegraphics[width=0.45\textwidth]{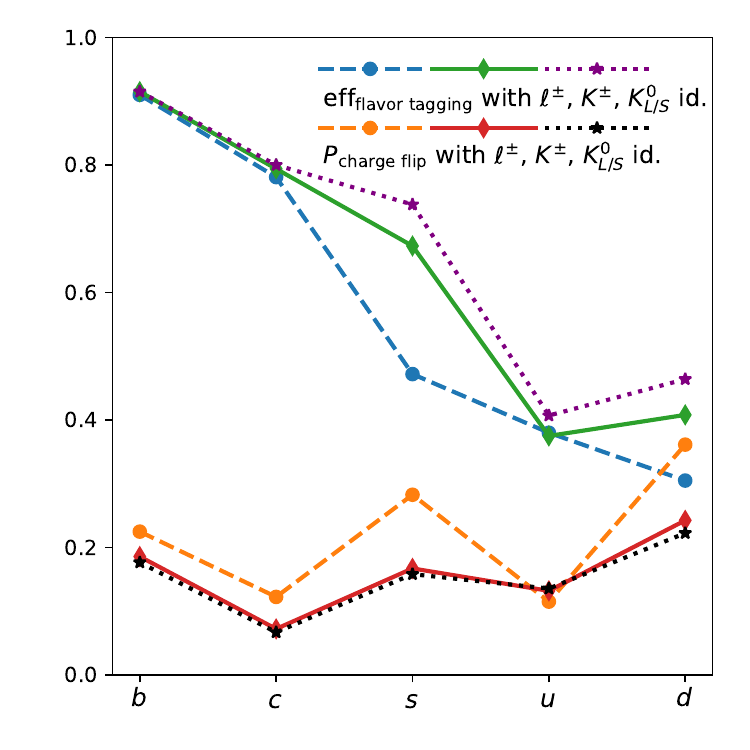}
    \caption{Jet origin identification performance~\cite{Liang:2023wpt} of full simulated Higgs/$Z$ to di-jet processes with CEPC conceptual detector. \textbf{LEFT:} The confusion matrix $M_{11}$ with perfect identification of leptons and charged hadrons. \textbf{RIGHT:} Jet flavor tagging efficiency and charge flip rates for quark jets with different scenarios of particle identification: with only lepton identification, plus identification of charged hadrons, plus identification of neutral kaons.}
    \label{fig:M11}
    
\end{figure}

Recently, the idea of jet origin identification has been proposed.
This idea aims at simultaneously identifying jets originating from eleven different colored particle species of the SM, namely five types of quarks ($u$, $d$, $s$, $c$, $b$), their corresponding anti-quarks, and gluons.
The jet origin identification combines the concepts of jet flavor tagging, jet charge measurement, strange jet and gluon jet identification together.
The idea of jet origin identification is then realized at the full simulated data of the CEPC CDR detector and using state-of-the-art reconstruction tools, including the Arbor particle flow reconstruction and 
the ParticleNet algorithm~\cite{Qu:2019gqs}, which simultaneously reaches jet flavor tagging efficiencies of 92\%, 79\%, 67\%, 37\%, and 41\%
and jet charge flip rates of 18\%, 7\%, 15\%, 15\%, and 19\% for $b$, $c$, $s$, $u$, and $d$ quarks, respectively, and meanwhile it could deliver a gluon jet identification efficiency of 66\%~\cite{Liang:2023wpt}, see Figure~\ref{fig:M11}.
These performances infer an effective tagging power of 37\%/58\% for $b$/$c$-jets, respectively, see Table~\ref{tab:performance}. 

The jet origin identification has significant impact on many physics measurements at the future electron-positron Higgs factories. 
For instance, the rare and exotic hadronic Higgs boson decays (see Section~\ref{sec:Higgsfcnc}), 
the determination of CKM matrix elements directly from $W$ boson decay (see Section~\ref{sec:Vcb_W}),
the time-dependent $CP$ measurements, the measurements of weak mixing angle, the differential measurements with multi-jet final states, etc.

\subsection{Simulation Method}

\begin{figure}[t]
    \centering
    \includegraphics[width=0.7\textwidth]{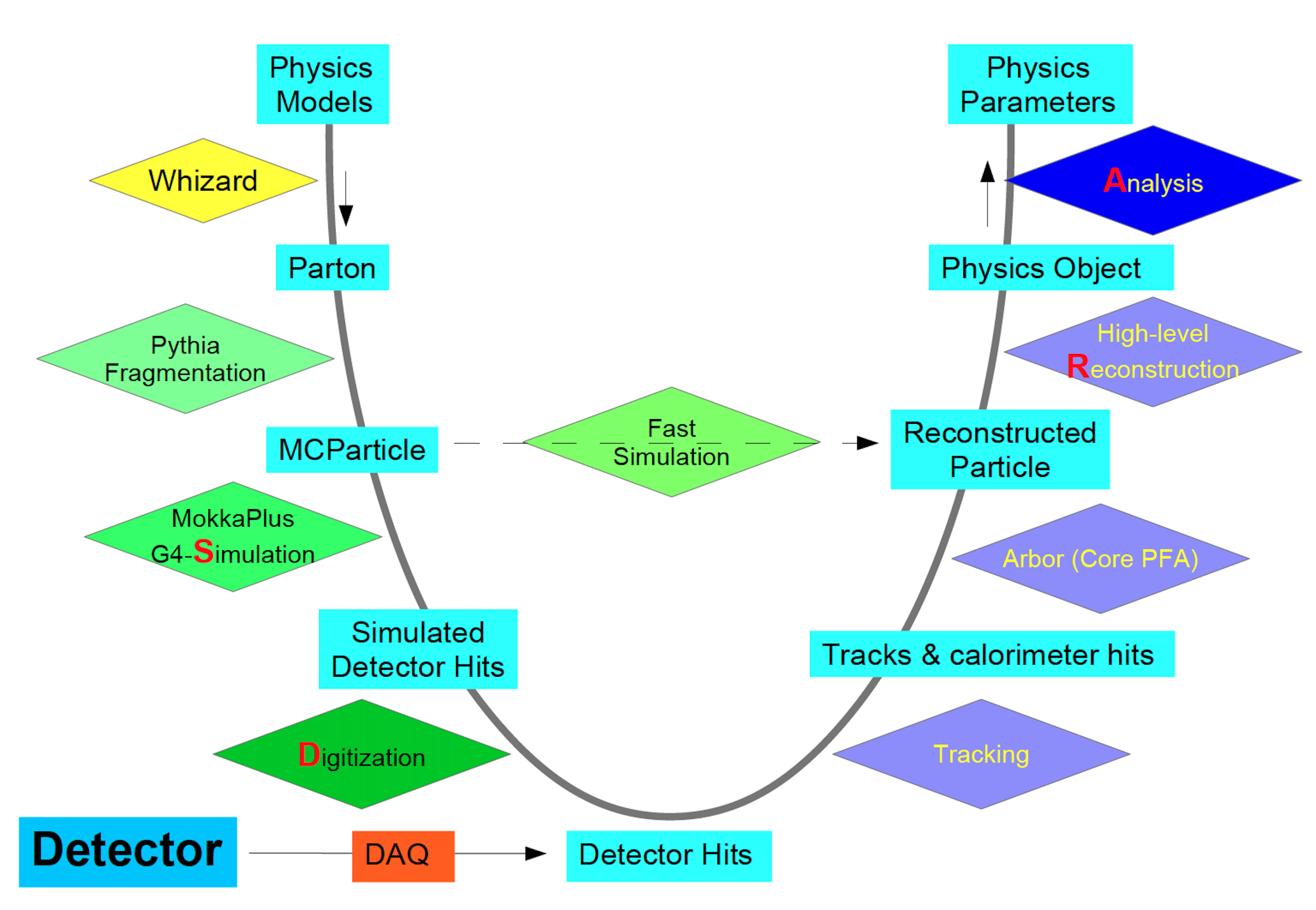}
    \caption{The CEPC official software chain and analysis flow~\cite{CEPCSoftWeb}. More detailed information can be found in the CEPC CDR~\cite{CEPCStudyGroup:2018ghi}.}
    \label{fig:CEPCSoft}
\end{figure}

To explore the flavor physics potential of the CEPC, various benchmark analyses that have been evaluated at the simulation level are covered in this manuscript.
Many of them are performed in the CEPC official software framework, illustrated in Figure~\ref{fig:CEPCSoft}, with full simulation and reconstruction of the CEPC CDR detector.
Limited by the available computing resource, a dataset of $\mathcal{O}(10^9)$ generator level inclusive $Z\to q\bar{q}$ events is generated for the physics potential studies at Tera-$Z$.
Since the full simulation of the whole dataset is computationally expensive and time-consuming, pre-selections are generally applied to refine the dataset into core subsets.
The analysis of $B_c \to \tau\nu_{\tau}$ in Section~\ref{sec:FCCC}, the study of $B_s^0 \to \phi\nu\bar{\nu}$ in Section~\ref{sec:FCNC}, and the $\phi_s$ measurement via $B_s^0 \to J/\psi\phi$ in Section~\ref{sec:CPV} are three typical examples.

For some studies, especially those that are oriented towards phenomenology and detector requirements, fast simulation is usually adopted.
Based on the understanding of detector responses and validated by the full simulation results, key detector performance is parameterized and modelled, and its effect on final physics observables is evaluated accordingly. This evaluation is used in studies such as the measurement of the $\alpha$ angle via $B^0_{(s)}\to\pi\pi$ channels discussed in Section~\ref{sec:CPV}. In this way, we can investigate the whole parameter space as much as possible with fast convergence.

To make the physics picture complete, we also list many benchmarks that have not been fully explored via simulation, but via first principle estimation, such as $\tau$ relevant studies in Section~\ref{sec:tau} and exclusive hadronic $Z$ decays in Section~\ref{sec:HadronicZ}.

%%%%%%%%%%%%%%%%%%%%%%%%%%%%%%%%%%%%%%%%%%%%%%%%%%%%%%%%%%%%%%%%%%%%%%%%%%%

\section{FCCC Semileptonic and Leptonic \texorpdfstring{$\boldsymbol{b}$}{b}-Hadron Decays}
\label{sec:FCCC}
Historically, $\beta$ decays, probably the best-known FCCC processes, have resulted in the discovery of weak interactions. While sensitivities to heavy-flavored leptonic and semileptonic FCCC decays in ongoing experiments are relatively limited, their explorations will continue to be significant for flavor physics in the CEPC era. 
 Firstly, measuring the signal rates of these channels can be used to determine the values of the CKM matrix elements such as $V_{cb}$ and $V_{ub}$~\cite{Bernlochner:2017jka}. Moreover, by performing these measurements, one can test lepton flavor universality (LFU), one of the most important predictions of the SM, see Refs.~\cite{Bryman:2021teu,Bifani:2018zmi,Bernlochner:2021vlv,Duan:2024ayo} for reviews. 
So the FCCC measurements can be an efficient way to probe NP that couples to leptons family-dependently.
For instance, given a relative deviation $\delta_{\rm SL}$ in the signal rate from the SM prediction, the energy scale probed can reach 
\begin{equation}
    \Lambda_{\rm NP}^{\rm SL} \sim (G_F|V_{cb}|\delta_{\rm SL})^{-\frac{1}{2}}  \sim (1.5~\text{TeV})\times \delta^{-\frac{1}{2}}_{\rm SL}~
\end{equation}
for $b\to c\ell\nu$ transitions and 
\begin{equation}
    \Lambda_{\rm NP}^{\rm SL} \sim(G_F|V_{ub}|\delta_{\rm SL})^{-\frac{1}{2}}  \sim (5~\text{TeV})\times \delta^{-\frac{1}{2}}_{\rm SL}~
\end{equation}
for $b\to u\ell\nu$ transitions. Notice that here the NP effective interactions have been assumed to be agnostic w.r.t. the SM flavor structure and have a strength of $\mathcal{O}(1)$.

The operation of the CEPC at the $Z$ pole enables the detector to access a full spectrum of $b$ hadrons with high statistics, including  multiple heavy-flavored mesons like $B_c$ and baryons like $\Lambda_b$, which are $b$-hadrons not accessible or  planned to produce at $B$-factories.  Measuring their (semi)leptonic decays would cross-validate our current understanding of FCCCs and further reveal hitherto unexplored physics. Particularly interesting among the list of expected measurements are the ones involving $\tau$ decays. These measurements are crucial for, {\it inter alia}, achieving a full test of LFU. However the multi-body decays of $\tau$ leptons complicate the event topology and kinematics. Even worse, the signature of neutrinos as missing momentum is hardly accessible at hadron colliders. The event reconstruction thus becomes a challenging task. In contrast, the reconstruction of these events including the $\tau$ leptons and other particles may greatly benefit from the excellent collider environment of the CEPC and the high-performance of its detector. These measurements thus define one of the ``golden'' channels for flavor physics at the CEPC.

\begin{figure}[t] 
\centering
       \includegraphics[width=\textwidth]{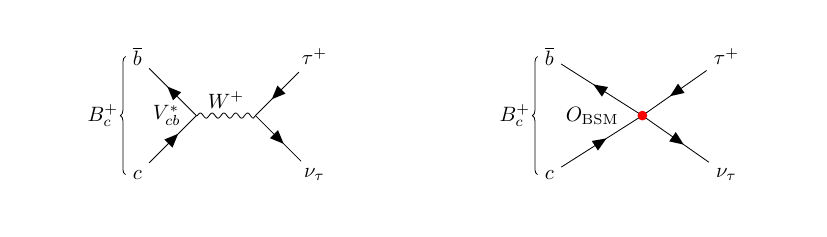}
    \caption{Illustrative Feynman diagrams for the decay $B_c^+\to \tau^+\nu_\tau$. \textbf{LEFT}: SM example. \textbf{RIGHT}: BSM example.}
    \label{fig:FCCC_Bc_Feynman}
\end{figure}

The above discussion can also be applied to the measurement of FCNC processes. Since such processes are forbidden at tree level and suppressed at loop level in the SM, these channels are capable of probing NP (see detailed discussions in Section~\ref{sec:FCNC}). The results obtained from both classes of measurements can be interpreted in various NP models. In a simplified NP model, these processes can arise from either colorless or colored mediators. The simplest colorless example might be a family non-universal $Z'$ boson with off-diagonal couplings to both quarks and leptons, thus yielding FCNC processes, see, {\it e.g.},~\cite{Langacker:2000ju,Barger:2009eq,Barger:2009qs,Langacker:2008yv,Altmannshofer:2014cfa}. This setup can be extended to a framework with an extra $SU(2)$ gauge triplet, where the additional $W'$ gauge bosons will contribute to the FCCC processes~\cite{Greljo:2015mma}. Another example is provided by leptoquarks,  namely scalar or vector bosons that couple to quarks and leptons simultaneously and therefore carry color. Leptoquarks are predicted by a wide range of ultraviolet (UV) theories such as grand unified theories, supersymmetry, composite Higgs models, etc. -- for a review see Ref.~\cite{Dorsner:2016wpm}. Such interpretations are model-dependent, and hence often limited in their applicability.

Alternatively, one can interpret the results in an Effective Field Theory (EFT) framework. The EFT is usually defined to parameterize the NP effects by integrating out the short distance physics. As a manifestation of physics at a low energy scale, the EFT is insensitive to the concrete format of UV physics. Here, let us consider the low-energy EFT (LEFT)~\cite{Jenkins:2017jig} with a natural cutoff at the EW-breaking scale. For $b\to c \ell \nu$ transitions, we have the dimension-6 LEFT Hamiltonian
\begin{equation}
    \mathcal{H}_{b\to c \ell \nu}^{\rm eff} = \frac{4G_F}{\sqrt{2}} V_{cb} \sum_i C_i O_i + \textrm{h.c.} \, , 
    \label{eq:FCCCLEFT}
\end{equation}
where $O_i$ denote the left(right)-handed scalar, vector, and tensor operators, namely 
\begin{eqnarray}
 O_{S_{L(R)}} &=& (\bar{c} P_{L(R)} b )(\bar{\ell} P_L \nu)\, ,  \nonumber  \\
    O_{V_{L(R)}} &=& (\bar{c} \gamma^\mu P_{L(R)} b )(\bar{\ell} \gamma_\mu P_L \nu)\, ,
   \\
    O_{T} &=& (\bar{c} \sigma^{\mu\nu} b )(\bar{\ell} \sigma_{\mu\nu} P_L \nu)\, ,    \nonumber 
\end{eqnarray}
and $C_i$ represent the corresponding Wilson coefficients. The SM can only contribute to $C_{V_{L}}$ via the exchange of a $W$ boson. Any deviation from this prediction will indicate the presence of NP, and the specific pattern of such deviation will carry crucial information on the nature of the underlying NP sector.

\begin{figure}[t]
    \centering
    \includegraphics[width=0.5\textwidth]{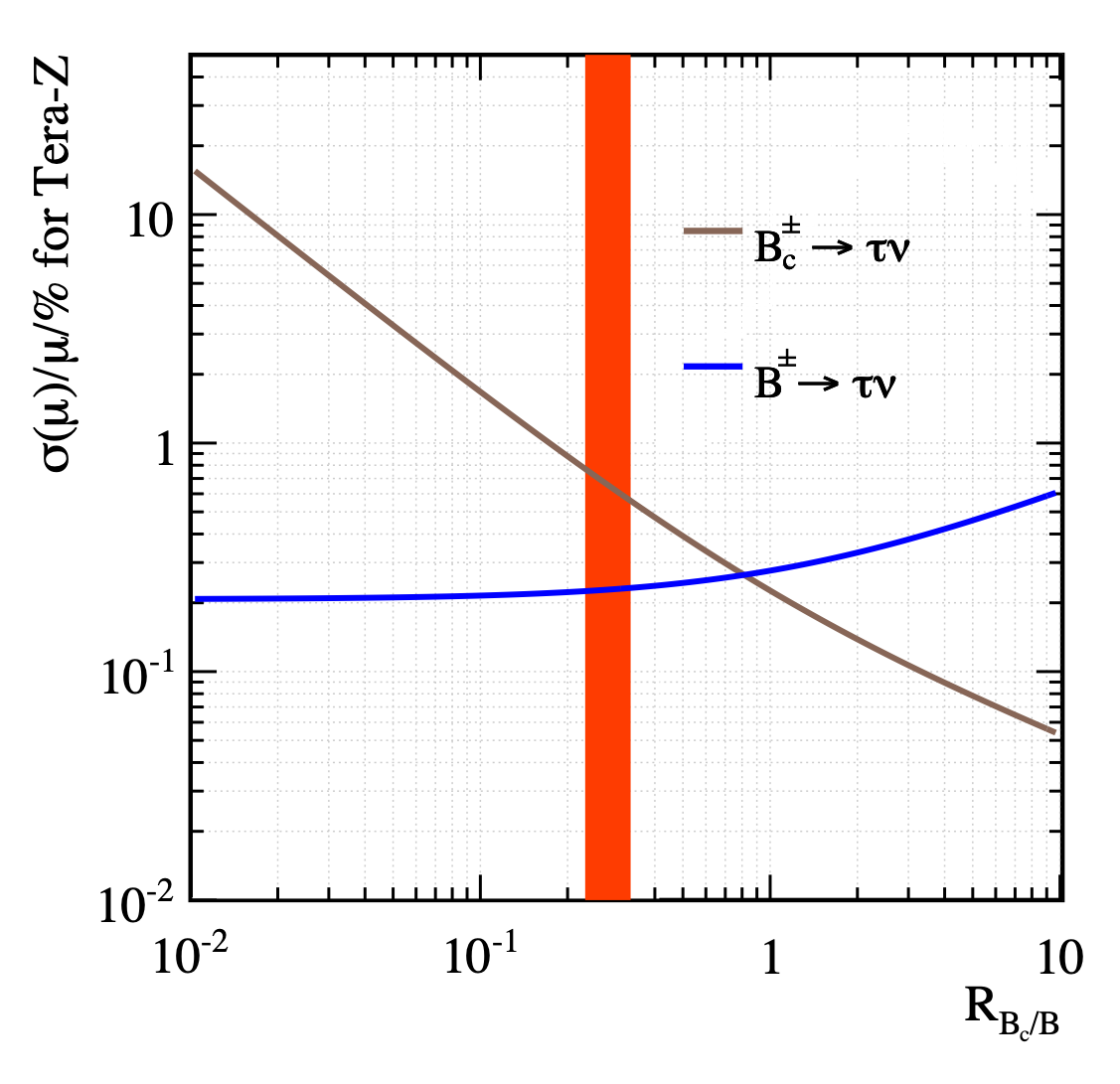}
    \caption{
    Relative precision of measuring the $B_{(c)}^\pm \to \tau\nu$ rate at the CEPC Tera-$Z$, as a function of $R_{B_{c}/B}\equiv N(B_c^\pm\to\tau\nu)/N(B^\pm\to\tau\nu)$~\cite{Zheng:2020ult}. Here the red band denotes the SM prediction for $R_{B_{c}/B}$. 
    }
    \label{fig:BcTauNu_Accuracy}
\end{figure}

\subsection{Leptonic Modes}

One important case regarding the $b\to c \ell \nu$ transitions is the leptonic decay of $B_q \to \tau\nu$ ($q=u,c$). As shown in Figure~\ref{fig:FCCC_Bc_Feynman}, this decay mode is sensitive to the axial vector $(C_{V_L} - C_{V_R} )$ and pseudoscalar $(C_{S_L} - C_{S_R})$ Wilson coefficients, with the branching ratio (BR) given by
\begin{align}
\label{eq:BRBqtaunu}
\mathrm{BR}(B_q^{+} \rightarrow \tau^+\nu_\tau)  =&~\tau_{B_q^+} \frac{G_F^2 |V_{q b}|^2 f_{B_q^+}^2 m_{B_q^+} m_\tau^2}{8 \pi}\left(1-\frac{m_\tau^2}{m_{B_q^+}^2}\right)^2 \nonumber \\
& \times \left|1+(C_{V_L}-C_{V_R})-\frac{m_{B_q^+}^2}{m_\tau\left(m_b+m_q\right)} (C_{S_L}-C_{S_R})\right|^2\,,
\end{align}
where $G_F$ is the Fermi constant, $m_\tau$ is the mass of the $\tau$ lepton, and $m_{B_q^+}$, $\tau_{B_q^+}$ and $f_{B_q^+}$ denote the $B_q^+$ mass, lifetime and decay constant, respectively. The SM prediction for the BR of the decay $B_c \to \tau \nu$ is rather large, $\sim 2.3\times 10^{-2}$~\cite{Fedele:2023gyi}, but the current constraint is relatively weak, BR$(B_c\to \tau\nu)\lesssim 30\%$. Detailed studies indicate that a Tera-$Z$ factory can measure this BR with a precision of $\mathcal{O}(10^{-4})$~\cite{Zheng:2020ult, Amhis:2021cfy,Fedele:2023gyi}. 
Specifically, the CEPC study in Ref.~\cite{Zheng:2020ult} employs a full simulation and incorporates leptonic $\tau$ decays $\tau^\pm\to\ell^\pm\nu\bar{\nu}$. The major features that differentiate $B_u^+$ from $B_c^+$ stem from their differing lifetimes and hadrons associated with their hadronization. As illustrated by Figure~\ref{fig:BcTauNu_Accuracy}, a measurement of the rate of $B_c \to \tau\nu$ with a relative precision $\sim \mathcal{O}(1\%)$ can be achieved at the Tera-$Z$ run of the CEPC. The study in Ref.~\cite{Fedele:2023gyi} instead focuses on the 3-prong $\tau$ decay, namely $\tau^\pm \to \pi^\pm \pi^\pm \pi^\mp \nu$. 
Within the considered analysis scenarios, the expected precision of the measurements of the rates ranges from 1.6\% to 2.3\% for $B_c^+ \to \tau^+\nu_\tau$ and from 1.8\% to 3.6\% for $B^+ \to \tau^+\nu_\tau$.

Within the SM, Eq.~(\ref{eq:BRBqtaunu}) can be further used to extract the $|V_{qb}|$ value by measuring the $B_q \to \tau \nu$ decay rates~\cite{Fedele:2023gyi}. 
Such a determination depends on precise inputs on the decay constants of the $B_q^+$ mesons $f_{B_q^+}$ as well as their production fractions.
Currently, the relative precision is $\sim 0.7\%$ for $f_{B_u^+}$~\cite{FlavourLatticeAveragingGroupFLAG:2024oxs} and $\sim 4.6\%$ for $f_{B_c^+}$~\cite{Narison:2019tym}, which could be improved in the coming decade. The $B^+$ production fraction is known with a precision $\sim 2\%$~\cite{HeavyFlavorAveragingGroup:2022wzx} and could be significantly improved in the CEPC era due to the abundant $Z\to b\bar{b}$ data. As for the $B_c^+$ production fraction, however, no information is available from any existing measurements or future projections.

With the high precision measurement of BR($B^+ \to \tau^+ \nu_\tau$) expected at Tera-$Z$ factories~\cite{Zheng:2020ult,Fedele:2023gyi} and the theoretical uncertainties described above, we expect that the $|V_{ub}|$ value can be determined with a relative precision of 1\% or better. In comparison, the Belle~II experiment is expected to perform a similar determination with a relative precision of 2-3\% employing the full integrated luminosity~\cite{Belle-II:2018jsg}. Notably, these measurements may cast new insights on the long-standing discrepancy of more than $3\sigma$ between the inclusive and exclusive determinations of $|V_{cb}|$~\cite{HeavyFlavorAveragingGroupHFLAV:2024ctg,HeavyFlavorAveragingGroup:2022wzx,Gambino:2007rp,UTfit:2022hsi,CKMfitter}. \footnote{Constraints on $|V_{cb}|$ can also be obtained from $W$ hadronic decays, where the $W$ bosons are produced at the $WW$ threshold or Higgs factory runs. See Section~\ref{sec:Vcb_W} for details.}

\subsection{Semileptonic Modes}

\begin{figure}[t]
\centering
\includegraphics[width=\textwidth]{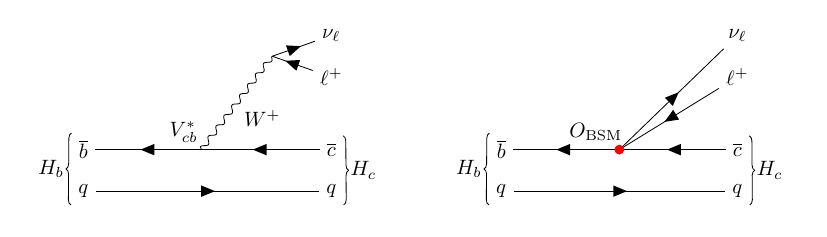}
    \caption{Illustrative Feynman diagrams for the transition $H_b\to H_c \ell^+\nu_\ell$. \textbf{LEFT}: SM example. \textbf{RIGHT}: BSM example.}
    \label{fig:FCCC_Hb_Feynman}
\end{figure}

The semileptonic decays induced by the $b \to c \ell\nu$ tansitions are often applied for the test of LFU. The LFU is predicted in the SM, because all three lepton families possess the same gauge charges. Consequently, any differences in decays involving different leptons can only arise from the Yukawa sector, in addition to any variations due to phase space. To highlight the special role of $\tau$ flavor, we introduce 
\begin{equation}
    R_{H_c} = \frac{\text{BR}(H_b\to H_c \tau\nu_\tau )}{\text{BR}(H_b\to H_c \ell'\nu_{\ell'})} 
    \label{eq:RHc}
\end{equation}
as an indicator for the LFU, where $H_{b(c)}$ represents a $b(c)$-hadron, and  $\ell'=e, \mu$ unless stated otherwise. Such an observable can be also defined for the decays of $B_c \to \tau \nu_\tau$ and $B_c \to \ell'\nu_{\ell'}$. For these observables, the systematics, such as the uncertainties from the CKM matrix elements and form factors, largely cancel. As an illustration, we show the Feynman diagrams for the SM and BSM contributions to the $H_b\to H_c \ell^+\nu_\ell$ transitions in Figure~\ref{fig:FCCC_Hb_Feynman}.

For the test of LFU at the $Z$ pole, a variety of $R_{H_c}$ observables ($R_{D_s}$, $R_{D_s^\ast}$, $R_{J/\psi}$, and $R_{\Lambda_c}$) have been recently investigated employing the fast simulation template of the CEPC~\cite{Ho:2022ipo}. The relative precisions that can be achieved, considering statistical errors only, are summarized in Table~\ref{tab:Rhc_sens}. Systematics in the $R_{H_c}$ measurements, as mentioned before, are expected to cancel largely since $R_{H_c}$ denotes a ratio of two aligned measurements.
This study indicates that at CEPC, a relative precision of $\lesssim 3\%$ for $R_{J/\psi}$, as well as $\lesssim 0.2\%$ and $\sim 0.05\%$ for $R_{D_s^{(\ast)}}$ and $R_{\Lambda_c}$, respectively, could be reached. 
Due to the complex topology and dynamics, these outcomes rely heavily on a vertex-based strategy for event reconstruction. 
They would benefit from a higher detector performance in general. Concretely, the $R_{J/\psi}$ measurement benefits the most from the improvement of tracker resolution, (see right panel of Figure~\ref{fig:Phivv_accuracy_pid} also), in reconstructing the $B_c^\pm$ vertex as well as in identifying the $J/\psi$ one, while the $R_{D_s^{(\ast)}}$ measurements gain more from the increase of soft photon identification efficiency in distinguishing the $D_s^{\ast}$ and $D_s $ modes via the decay $D_s^{\ast} \to D_s \gamma$.

% \begin{table}[t]
% \centering
% \fontsize{10pt}{12pt}\selectfont
% \begin{tabular}{ccccccc}
% \toprule[1pt]
% & $R_{H_c}$ & SM Value & Tera-$Z$ & 4 Tera-$Z$ & 10 Tera-$Z$   &\\
% \hline
% & $R_{J/\psi}$    & 0.289 & $4.3\times 10^{-2}$ & $2.1\times 10^{-2}$ & $1.4\times 10^{-2}$&\\
% \hline
% & $R_{D_s}$       & 0.393 & $4.1\times 10^{-3}$ & $2.1\times 10^{-3}$ & $1.3\times 10^{-3}$ &\\
% \hline
% & $R_{D_s^*}$     & 0.303 & $3.3\times 10^{-3}$ & $1.6\times 10^{-3}$ & $1.0\times 10^{-3}$ &\\
% \hline
% & $R_{\Lambda_c}$ & 0.334 & $9.8\times 10^{-4}$ & $4.9\times 10^{-4}$ & $3.1\times 10^{-4}$ &\\
% \bottomrule[1pt]
% \end{tabular}
% \caption{SM predictions for the $R_{H_c}$ observables and relative precision for their measurements at Tera-$Z$, $4\times$Tera-$Z$, and $10\times$Tera-$Z$, considering statistical uncertainties only~\cite{Ho:2022ipo}. }
% \label{tab:Rhc_sens}
% \end{table}

\begin{table}[t]
\centering
\fontsize{10pt}{12pt}\selectfont
\begin{tabular}{cccc}
\toprule[1pt]
& $R_{H_c}$ & SM Value & 4 Tera-$Z$ \\
\hline
& $R_{J/\psi}$    & 0.289 & $2.1\times 10^{-2}$ \\
\hline
& $R_{D_s}$       & 0.393 & $2.1\times 10^{-3}$ \\
\hline
& $R_{D_s^*}$     & 0.303 & $1.6\times 10^{-3}$ \\
\hline
& $R_{\Lambda_c}$ & 0.334 & $4.9\times 10^{-4}$ \\
\bottomrule[1pt]
\end{tabular}
\caption{SM predictions for the $R_{H_c}$ observables and relative precision for their measurements at 4 Tera-$Z$, considering statistical uncertainties only~\cite{Ho:2022ipo}. }
\label{tab:Rhc_sens}
\end{table}

\begin{figure}[t]
    \centering
    \includegraphics[width=0.8\textwidth]{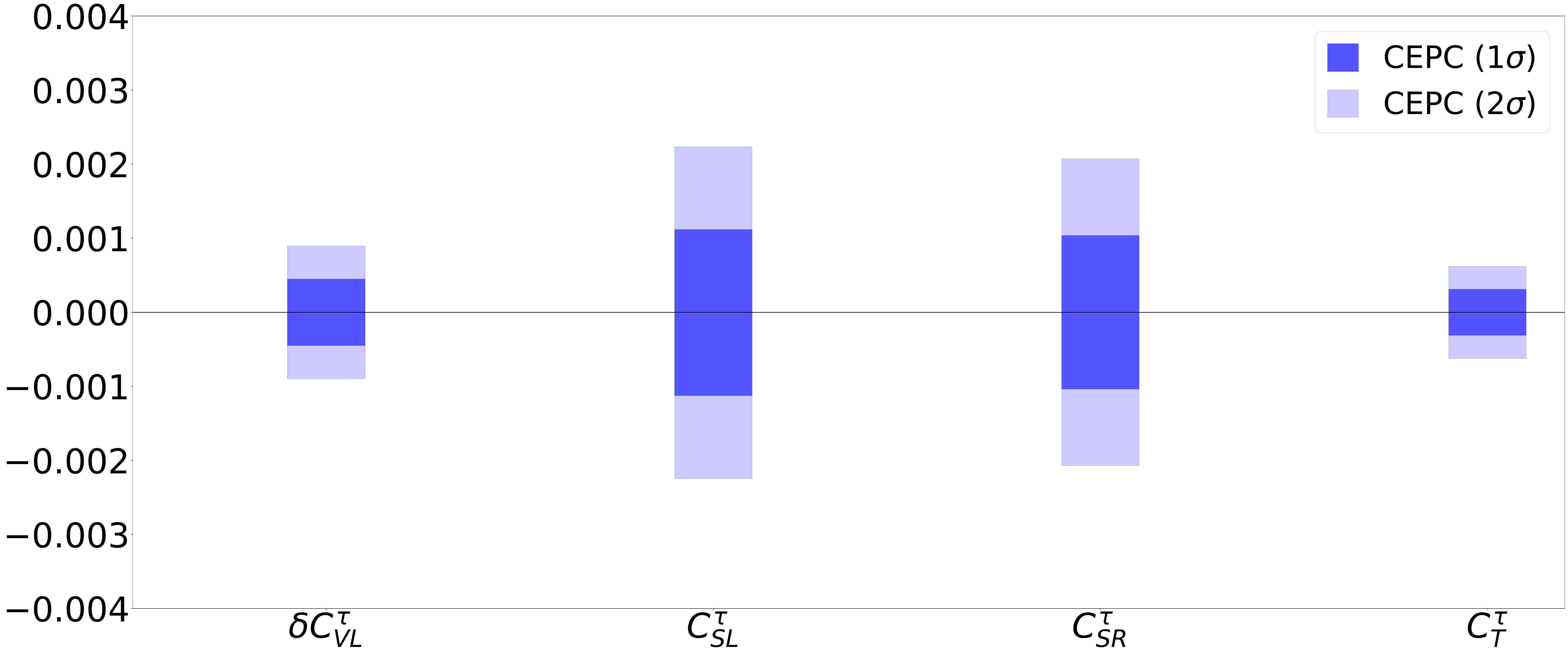}
    \caption{Marginalized constraints on the Wilson coefficients of $b\to c\tau \nu$ LEFT at the CEPC, with $\delta C_{V_L}^{\tau} = C_{V_L}^{\tau} - \delta C_{V_L,{\rm SM}}^{\tau}$. This plot is taken from Ref.~\cite{Ho:2022ipo}.}
    \label{fig:FCCCLEFT}
\end{figure}

Note that these measurements cover a variety of $b\to c \tau \nu$ transitions: such as the ones from pseudoscalar ($B_{s,c}$) to vector ($D_s^\ast$, $J/\psi$) or pseudoscalar ($D_s$); those from baryon ($\Lambda_b$) to another baryon ($\Lambda_c$); and the decays of a pseudoscalar ($B_{c}$) to a pair of fermions. Consequently, they can be employed to constrain different LEFT operators that can induce $b\to c \tau \nu$ transitions. Following the approach in Ref.~\cite{Ho:2022ipo}, we present in Figure~\ref{fig:FCCCLEFT} the marginalized constraints on the Wilson coefficients of $b\to c\tau \nu$ LEFT at the CEPC, based on the results of~\cite{Zheng:2020ult,Ho:2022ipo}. In this context, these Wilson coefficients can be universally constrained to a level of $\mathcal{O}(10^{-3})$.\footnote{In this analysis, the operator $O_{V_R}$ has been turned off, as it cannot be generated by UV physics respecting the $SU(2)_L$ gauge symmetry at a dimension-6 level.} 

\begin{table}[t]
\centering
\fontsize{10pt}{12pt}\selectfont
\begin{tabular}{cc}
\toprule[1pt]
Process & Observable \\
\hline
%$b\to cl\nu$,$b\to ul\nu$ & LFU\\ \hline
$b\to cl\nu$ & $R_{H_{c}}(R_{J/\psi},R_{D_s^{(*)}},R_{\Lambda_c})$\\
\hline
$B_c\to \tau \nu$ &$|V_{cb}|$\\ \hline
$B\to \tau \nu$ &$|V_{ub}|$\\ 
\bottomrule[1pt]
\end{tabular}
\caption{List of benchmark FCCC semileptonic and leptonic $b$-decay channels that can be investigated at CEPC. }
\label{tab:obervable1}
\end{table}

Additionally, several unexplored topics of FCCC physics deserve attention. Firstly, in view of the scientific significance of testing LFU, it is necessary to establish the CEPC sensitivity for a full list of $R_{H_c}$ measurements including the traditional $R_D$ and $R_{D^\ast}$, higher-resonant $R_{D^{\ast\ast}}$~\cite{Bernlochner:2017jxt}, remaining baryonic modes such as $R_{\Xi_c}$, etc., and their corresponding differential measurements. Also, to provide an LFU test for all three generations, it is natural to extend studies to the measurement of ${\text{BR}(b\to c\mu\nu)}/{\text{BR}(b\to ce\nu)}$, where it is crucial to reduce the systematics to a level comparable to the statistical errors. The relevant benchmark channels that can be investigated at CEPC are listed in Table~\ref{tab:obervable1}. 
Secondly, the superior precision of measuring the $B$ meson flight distance at the CEPC creates a new opportunity for the measurement of time-dependent CP-violation in semileptonic $b\to c \ell \nu$ decays. With this approach, the $CP$-violating markers in $B^0_{(s)}-\bar{B}^0_{(s)}$ mixing, which are encoded as $\mathcal{A}^{d}_{\rm SL}$ and $\mathcal{A}^{s}_{\rm SL}$~\cite{Charles:2020dfl,Grossman:2021xfq} respectively, can be extracted by measuring the $B^0$ and $B_s^0$ decays. 
As these measurements can contribute significantly to the global constraints on the parameters $\beta$ and $\beta_s$~\cite{Artuso:2015swg,Chang:2017wpl}, where the current experimental precision remains far from the SM predictions, it is of high value to perform a more dedicated sensitivity analysis with either fast or full simulations.

\section{FCNC \texorpdfstring{$\boldsymbol{b}$}{b}-Hadron Decays}
\label{sec:FCNC}

\begin{figure}[t]
\centering
\includegraphics[width=\textwidth]{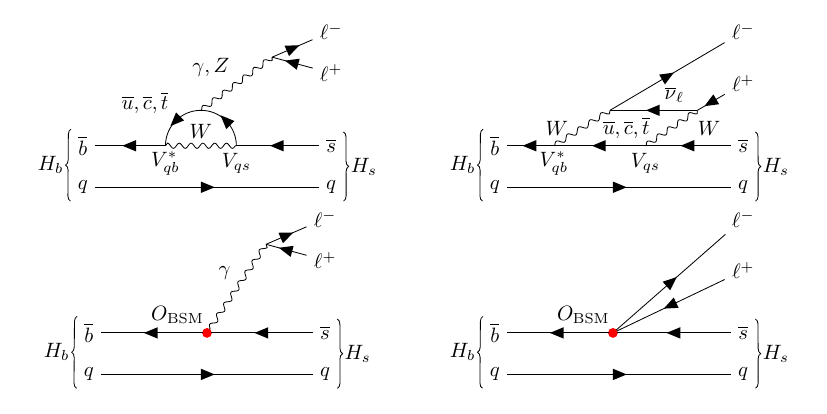}
    
    \caption{Illustrative Feynman diagrams for the transition $H_b\to H_s\ell^+\ell^-$. \textbf{UPPER}: SM examples. \textbf{BOTTOM}: BSM examples.}
    \label{fig:FCNC_Hb_Feynman}
\end{figure}

FCNC transitions are prohibited at tree level in the SM. While being enabled by EW penguin or box diagrams (see Figure~\ref{fig:FCNC_Hb_Feynman}), these transitions are subject to a joint suppression by off-diagonal CKM matrix elements and loop factors, and thus are rare. 
Because of this feature, the FCNC processes emerge uniquely sensitive to weak NP effects that may otherwise evade detection. Given a relative deviation of $\delta_{\rm rare}$ in signal rate from the SM prediction, the energy scale probed can reach~\cite{Altmannshofer:2022hfs} 
\begin{equation}
    \Lambda_{\rm NP}^{\rm rare} \sim \Big(\frac{\alpha}{4\pi} \frac{m_t^2}{m_W^2}G_F|V_{tb}V_{ts}^\ast|\delta_{\rm rare}\Big)^{-\frac{1}{2}}  \sim (30~\text{TeV})\times \delta^{-\frac{1}{2}}_{\rm rare}~
\end{equation}
and 
\begin{equation}
    \Lambda_{\rm NP}^{\rm rare} \sim \Big(\frac{\alpha}{4\pi} \frac{m_t^2}{m_W^2}G_F|V_{tb}V_{td}^\ast|\delta_{\rm rare}\Big)^{-\frac{1}{2}}  \sim (67~\text{TeV})\times \delta^{-\frac{1}{2}}_{\rm rare}~
\end{equation}
for the $b\to s$ and $b\to d$ transitions, respectively. 
Notably, while the FCNC processes are rarer than the FCCC ones in the SM, $\Lambda_\text{NP}^{\rm rare}$ can be comparable to, or even higher than, $\Lambda_\text{NP}^{\rm SL}$ as long as $\delta_{\rm rare} \lesssim 100 \delta_{\rm SL}$ is achieved.

Similar to the $b\to c\ell \nu$ transitions investigated in Section~\ref{sec:FCCC}, we have the dimension-6 LEFT Hamiltonian to parametrize the $b\to s$ transitions:
\begin{align}
    \mathcal{H}^{\rm eff}_{b\to s} &= - \frac{4G_F}{\sqrt{2}}V_{tb}V_{ts}^{\ast} \frac{\alpha}{4\pi} \sum_j(C_j O_j + C_j^\prime O_j^\prime) + (C_L O_L + C_R O_R) + \textrm{h.c.}, 
    \label{eq:FCNC-LEFT}
\end{align}
where the operators of interest include
\begin{equation}
\begin{aligned}
    O_{S}^{(\prime)} &= m_b (\bar{s} P_{R(L)} b)(\bar{\ell} \ell) , \qquad & 
    O_{P}^{(\prime)} &= m_b (\bar{s} P_{R(L)} b)(\bar{\ell} \gamma^5 \ell), \\
    O_{9}^{(\prime)} &= (\bar{s}\gamma^\mu P_{L(R)} b)(\bar{\ell} \gamma_\mu \ell), \qquad & 
    O_{10}^{(\prime)} &= (\bar{s}\gamma^\mu P_{L(R)} b)(\bar{\ell} \gamma_\mu \gamma^5 \ell), \\
    O_{T(T5)} &= (\bar{s} \sigma_{\mu\nu} b)(\bar{\ell} \sigma^{\mu\nu}(\gamma^5) \ell), \qquad &
    O_{7}^{(\prime)} &= \frac{1}{e} m_b(\bar{s}\sigma^{\mu\nu} P_{R(L)} b) F_{\mu\nu}, \\
    O_{L(R)} & = (\bar{s} \gamma^\mu P_{L(R)} b)(\bar{\nu}\gamma_\mu P_L \nu).
\end{aligned} 
\end{equation}
Among these operators, the first five encode the scalar-, vector-, and tensor-mediated $b\to s$ transitions with a pair of charged leptons and may violate LFU. The presence and absence of a ``prime'' denote the $b\to s$ currents which are subject to the left- and right-handed chiral projections respectively, while the opposite convention applies to the dipole operators $O_{7}^{(\prime)}$. $O_{L(R)}$ encodes the vector-mediated $b\to s$ transitions with a pair of neutrinos. $O_{7}^{(\prime)}$ is an EM dipole operator which can either yield decays with an on-shell photon or mediate $b\to s \ell\ell$ transitions (see the bottom left panel in Figure~\ref{fig:FCNC_Hb_Feynman}). Note that, when the strange-quark and lepton masses are neglected, the SM contributes to $O_{9}$, $O_{10}$, $O_{L}$ and $O_{7}$ only.

\begin{figure}[t]
\centering
\includegraphics[width=1\textwidth]{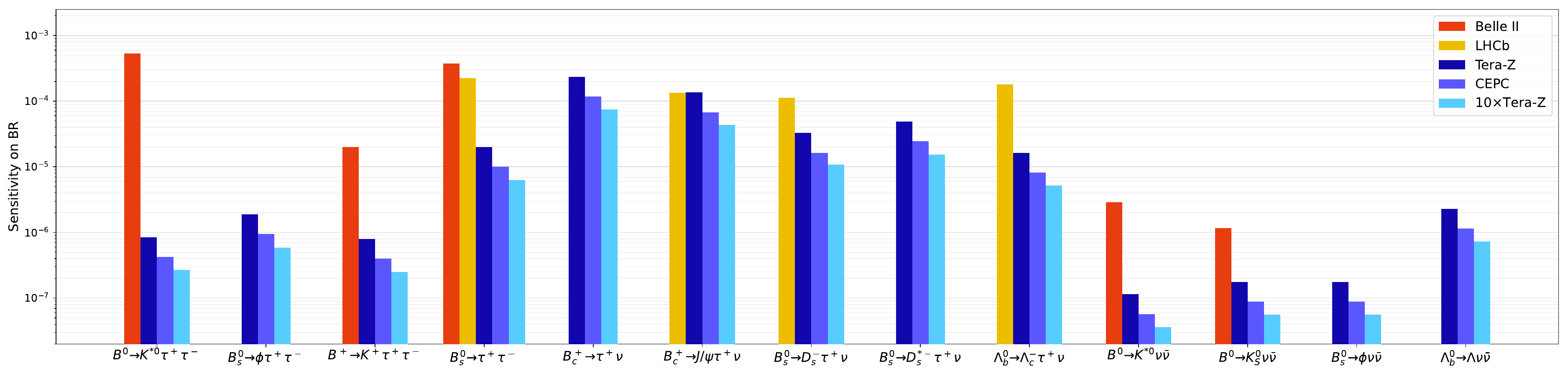}
\caption{Projected sensitivities of measuring the $b\to s\tau\tau$~\cite{Li:2020bvr}, $b\to s\nu\bar{\nu}$~\cite{Li:2022tov,Amhis:2023mpj} 
and $b\to c\tau\nu$~\cite{Zheng:2020ult,Ho:2022ipo} transitions at the $Z$ pole. The sensitivities at Belle II @ 50 $\text{ab}^{-1}$~\cite{Belle-II:2018jsg,Belle-II:2022cgf} and LHCb Upgrade II~\cite{Bediaga:2018lhg,Bifani:2018zmi} have also been provided as a reference. Note that LHCb sensitivities are generated by combining the analyses of $\tau^+ \to \pi^+\pi^-\pi^-(\pi^0)\nu$ and $\tau \to \mu \nu \bar \nu$. This plot is taken from Ref.~\cite{Ho:2022ipo}, with additional $b\to s\nu\bar{\nu}$ modes included. 
}
\label{fig:overall}
\end{figure}

In this section, we will mostly focus on the measurements of $b\to s \tau\tau$, $b \to s \nu \bar\nu$ and $b\to s \gamma$ transitions. The CEPC offers a great platform for these studies, particularly during its $Z$ pole run. The extraordinarily high luminosity delivered by the CEPC ensures considerable signal statistics for even the most elusive decay modes with BRs typically $\lesssim 10^{-5}$. Moreover, as compared to the LHCb detector, the planned detectors of the CEPC are better suited for the reconstruction of $\tau$ leptons and thus the measurement of $b\to s \tau\tau$, for the measurement of missing energy and hence of $b \to s \nu \bar\nu$, and for photon identification as needed for the measurement of $b\to s \gamma$. A combination of these advantages yields an enhanced sensitivity for both testing the SM and probing NP effects. The CEPC thus represents an ideal facility for investigating these rare FCNC decays and the underlying physics. 
It is worth noting that both $b \to s \nu \bar\nu$ and, especially, $b \to s \tau\tau$ transitions, for which we have very poor experimental information so far, are extremely sensitive to test a wide class of motivated NP models with new dynamics coupled mainly to the third generation~\cite{Calibbi:2015kma,Allwicher:2023shc}. 
For the convenience of the discussion below, we summarize the projected sensitivities to $b\to s\tau\tau$ and $b\to s\nu\bar{\nu}$ transitions, together with the $b\to c\tau\nu$ processes discussed in Section~\ref{sec:FCCC}, in Figure~\ref{fig:overall}. At the end of this section, we will extend the discussions to the possibilities of testing the SM global symmetries with forbidden $b$-hadron decays.

\subsection{Di-lepton Modes}

\begin{figure}[t]
    \centering
    \includegraphics[height=5.cm]{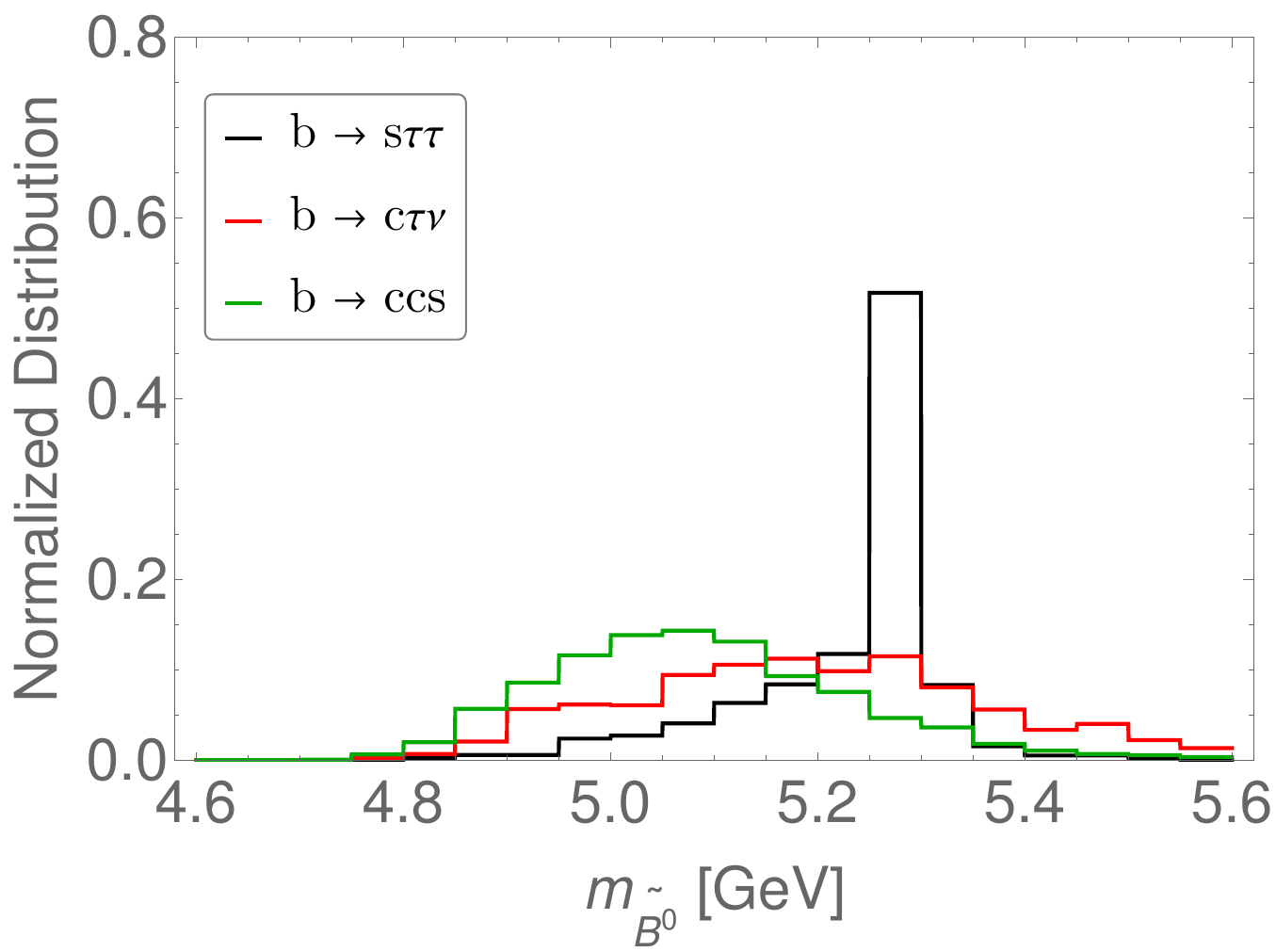}
    \hfill
    \includegraphics[height=5.cm]{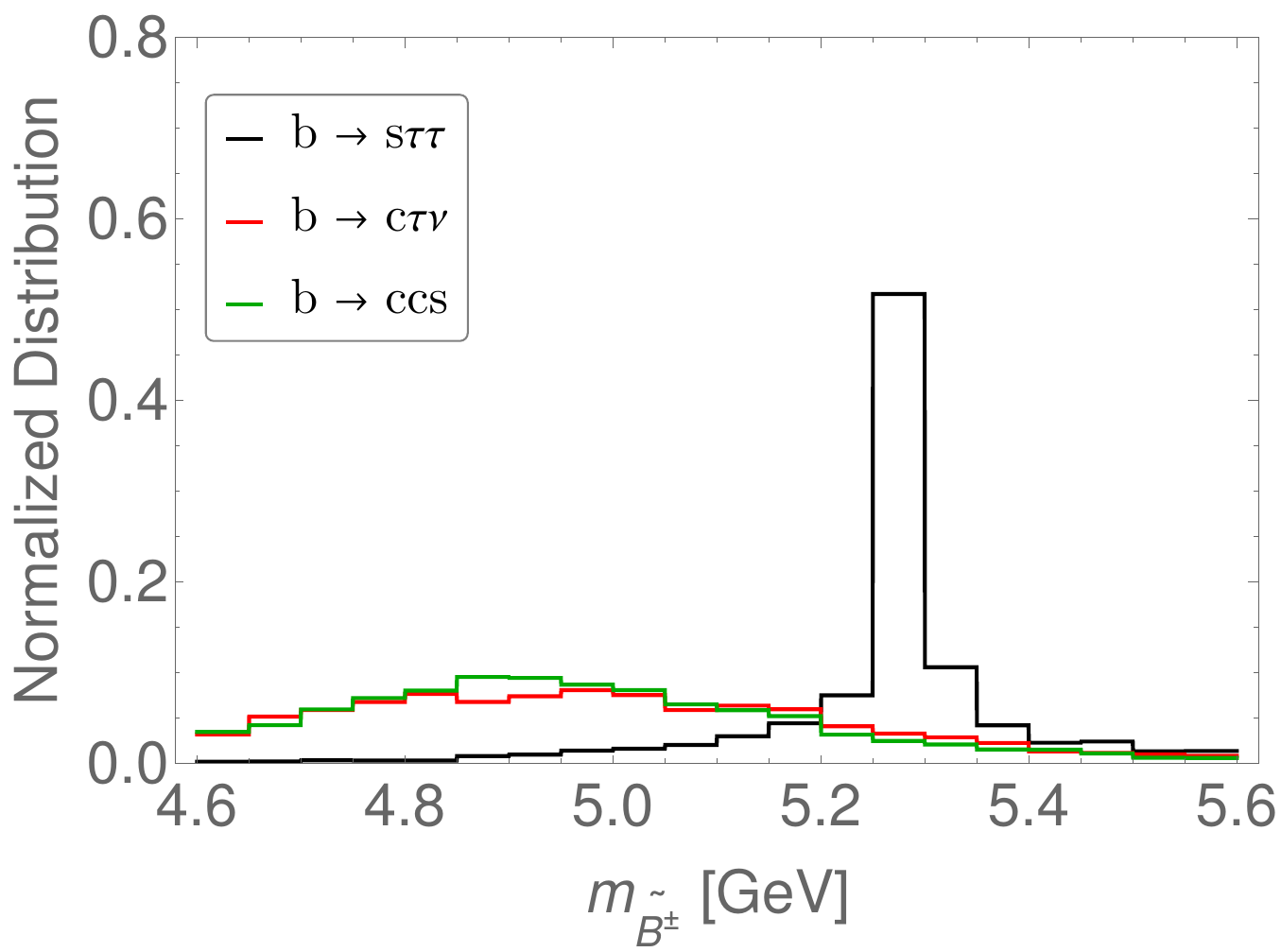}
    \caption{Mass reconstruction for the signal $b$-mesons in the measurements of $b\to s\tau\tau$ at the $Z$ pole, with $\tau^\pm \to \pi^\pm\pi^\pm\pi^\mp\nu$~\cite{Li:2020bvr}. \textbf{LEFT:} $B^0\to K^{\ast 0} \tau^- \tau^+$. 
    \textbf{RIGHT:} $B^+ \to K^+ \tau^- \tau^+$. The major backgrounds arise from the $b\to c \tau \nu$ and $b\to c c s $ transitions and are both reconstructed.} 
    \label{fig:FCNCdilep}
\end{figure}

In general, the reconstruction of $b\to s \tau\tau$ is more involved compared to the reconstruction of $b\to s ee, s\mu\mu$. 
As the $\tau$ decays result in neutrino production, the $b\to s \tau\tau$ events are not fully visible to a detector. This difficulty, however, can be well-addressed at a machine like the CEPC. In a recent study~\cite{Li:2020bvr} (for discussions on $B^0\to K^{\ast 0} \tau^- \tau^+$, also see~\cite{Kamenik:2017ghi}), the sensitivity for measuring a set of benchmark $b\to s\tau\tau$ transitions, including $B^0\to K^{\ast 0} \tau^- \tau^+$, $B_s^0\to\phi \tau^- \tau^+$, $B^+ \to K^+ \tau^- \tau^+  $ and $B_s^0 \to \tau^- \tau^+$, at the $Z$ pole has been systematically analyzed. To utilize the machine's capability, a tracker-based scheme to reconstruct the signal $B$ mesons that works for these $b\to s\tau\tau$ channels has been developed, achieved by using the decay modes of $\tau^\pm\to \pi^\pm\pi^\pm\pi^\mp\nu$. Such a tracker-based scheme also benefits from the particle kinematics at the $Z$ pole. Due to their boost, the signal $b$ hadrons tend to travel further (compared to, {\it e.g.}, Belle II) before their decay, which benefits the relevant tracking measurements. The predominant backgrounds for these measurements are the Cabibbo-favored $b\to c+X$ processes. Recall that both $D^\pm$ and $D_s^\pm$ mesons have masses and lifetimes comparable to those of $\tau$ leptons and thus may decay to a vertex of $\pi^\pm\pi^\pm\pi^\mp$ with extra particles. Therefore, they can fake the $\tau$ leptons in the signal. In Figure~\ref{fig:FCNCdilep} we demonstrate the mass reconstruction for the signal $b$-mesons in the measurements of $B^0\to K^{\ast 0} \tau \tau$ and $B^+ \to K^+ \tau^- \tau^+$ at the $Z$ pole. These two channels involve the decay of $b$-mesons into vector and pseudoscalar mesons respectively. They are sensitive to the LEFT in approximately orthogonal ways and thus are complementary in probing NP~\cite{Li:2020bvr}.

\begin{figure}[t]
    \centering
    \includegraphics[width=0.8\textwidth]{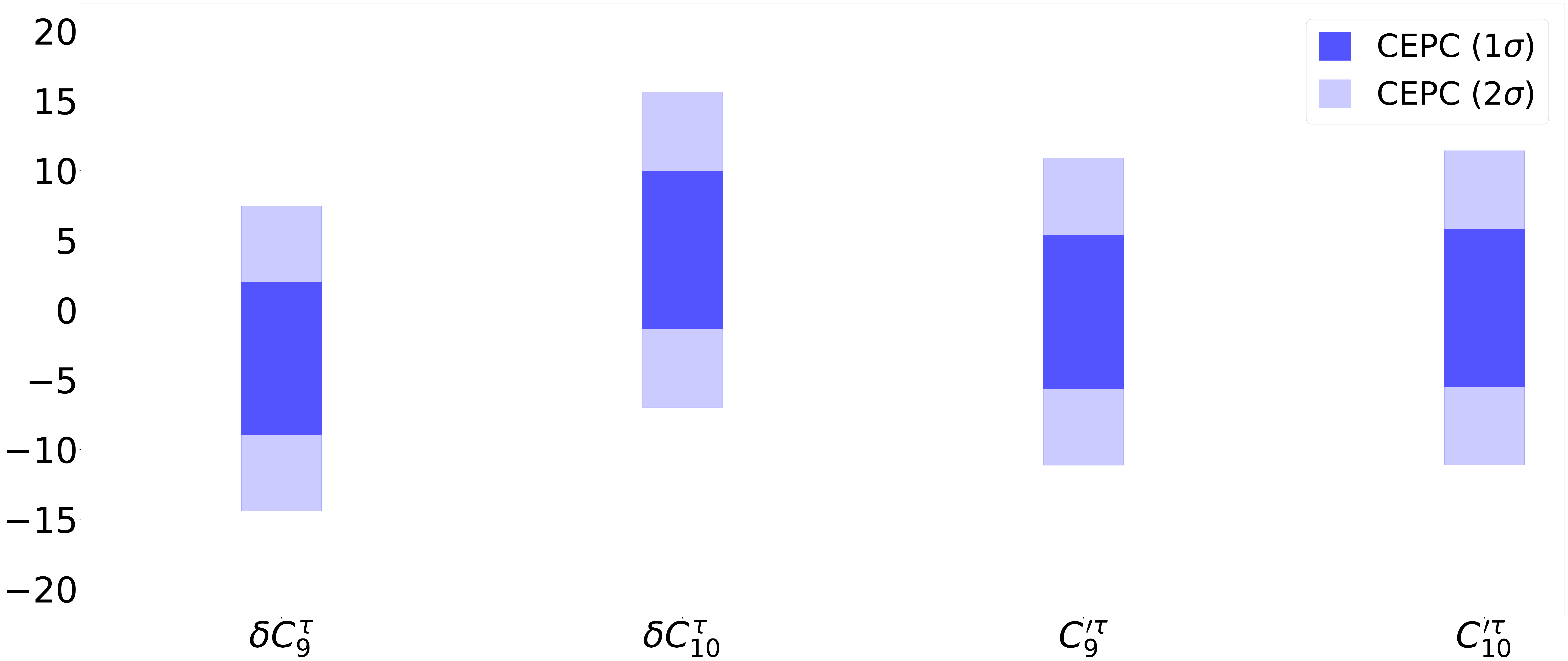}
    \caption{Marginalized constraints on the Wilson coefficients of $b\to s\tau \tau$ LEFT (vector current only) at the CEPC, with $\delta C_9^\tau = C_9^\tau - C_{9,{\rm SM}}^\tau$ and $\delta C_{10}^\tau = C_{10}^\tau - C_{10,{\rm SM}}^\tau$. This plot is adapted from Ref.~\cite{Li:2020bvr}.}
    \label{fig:FCNCLEFT}
\end{figure}

As illustrated in Figure~\ref{fig:overall}, the Tera-$Z$ and $10\times$Tera-$Z$ machines would be able to measure the BRs of $B^0\to K^{\ast 0} \tau^- \tau^+$, $B_s^0\to\phi \tau^- \tau^+$ and $B^+ \to K^+ \tau^- \tau^+$ with an absolute precision of $\mathcal O(10^{-7} - 10^{-6})$, as well as BR($B_s^0 \to \tau^- \tau^+$) with an absolute precision of $\mathcal O(10^{-6} - 10^{-5})$. In comparison, Belle~II and LHCb either have no sensitivity to these measurements or can only yield a sensitivity that is one to two orders of magnitude weaker. With the baseline luminosity, this indicates that the CEPC will be able to identify $\sim \mathcal{O}(1)$ deviations from the SM predictions and further probe the $b\to s\tau \tau$ LEFT operators.  Figure~\ref{fig:FCNCLEFT} shows the marginalized constraints on the corresponding Wilson coefficients in the presence of the vector-mediated operators only.

In spite of this progress, the study of FCNC $b$ rare decays at CEPC should be extended in multiple directions. Firstly, the CEPC constraints on the LEFT operators in Eq.~(\ref{eq:FCNC-LEFT}) should be improved.  Currently, the sensitivity to BR($B_s\to \tau^- \tau^+$) is too weak to probe unconstrained LEFT parameter space. BR($B^0\to K^{\ast 0} \tau^- \tau^+$) and BR($B_s^0\to\phi \tau^- \tau^+$) are both pseudoscalar to vector transitions and have a similar dependence on the NP parameters. 
% This leaves a few directions poorly constrained in the LEFT parameter space. 
To improve the constraints on the relevant LEFT coefficients, one can consider: (i) introducing differential observables, such as forward-backward asymmetry and $\tau$ polarimetry~\cite{Kamenik:2017ghi}; and (ii) incorporating $b\to s\tau\tau$ transitions of different nature, such as the baryonic decay $\Lambda_b \to \Lambda\tau^- \tau^+$. 
Interestingly, within the context of an $SU(2)_L$-invariant EFT, sizable NP contributions to the $b\to s\tau\tau$ transitions are often accompanied with large effects on the left-handed vector current NP operators that contribute to the LFU observables $R_{D^{(*)}}$, which currently exhibit some tension with the SM predictions~\cite{Capdevila:2023yhq,HFLAV2024winter}.

A second area of improvement would be to advance the study on LFU tests at the CEPC. The CEPC analysis in Ref.~\cite{Li:2020bvr} focuses on the di-$\tau$ mode of $b\to s$ transitions. To paint a full picture in this context, it is of high value to extend the analysis to $b\to s \ell \ell$. The measurements of, {\it e.g.}, $R_{K^{(\ast)}}$, $R_{pK}$~\cite{Aaij:2019bzx}, $R_\phi$~\cite{Aaij:2021pkz}, $R_{f_2^\prime(1525)}$~\cite{Aaij:2021pkz} and even $R_{\Lambda}$ could provide important insights regarding LFU. For some of these measurements, the systematic uncertainties induced by PID could be dominant. The superior electron- and muon-ID capabilities of future detectors are anticipated to offer an edge over  LHCb. Notably, the luminosity advantage of the CEPC in measuring the $b\to s \tau\tau$ transitions could be extended to ultra-rare channels such as $B_s^0 \to \mu^+ \mu^-$. The measurement of $\text{BR}(B_s^0 \to \mu^+ \mu^-)$ in the SM is known to be statistically limited, due to its tiny value of around $\sim 3.0 \times 10^{-9}$~\cite{LHCb:2017rmj}. With a yield of $\sim 1.2 \times 10^{11}$ for $B_s^0$ mesons at the CEPC, about 360 $B_s^0 \to \mu^+ \mu^-$ events are expected to be produced, which provides a good opportunity to improve the precision of its measurements.

Finally, sensitivity studies should be extended to $b\to d \ell^+\ell^-$ transitions at the CEPC. The $b\to d \ell^+\ell^-$ transitions represent another independent category of FCNC rare $b$-decays, and hence play a role complementary to the $b\to s \ell^+\ell^-$ transitions in exploring flavor physics. 
The measurements of these channels including both signal rate and $CP$ violation~\cite{Descotes-Genon:2020tnz,Fleischer:2022klb} may share difficulties similar to those of $b\to s\ell^+\ell^-$ decays, and hence would impose similar requirements for the detector performance at the CEPC. All of these issues deserve further detailed examinations.

\subsection{Neutrino Modes}

The $b\to s\nu\bar{\nu}$ decay is immune to non-factorizable charm-loop corrections and photonic-penguin contributions. Therefore, the theoretical calculation for its SM rate is cleaner than that for the $b\to s\ell\ell$ transitions, which yields BR($B_s^0 \to \phi\nu\bar{\nu})_{\mathrm{SM}}=(9.93\pm 0.72)\times 10^{-6}$~\cite{Li:2022tov}. The $b\to s\nu\bar{\nu}$ decay can be used to probe light dark sectors, such as dark photons, sterile neutrinos, axions/axion-like-particles (ALPs), or neutral scalars, which may significantly alter the kinematics of visible particles~\cite{Batell:2009jf,Dror:2017nsg,Hou:2024vyw}, (for discussions on the light dark sectors at CEPC, also see Section~\ref{sec:BSM}). Also, due to the constraints of electroweak gauge symmetry, the impacts of NP on the $b \to s\nu\bar{\nu}$ and $b \to s\ell^+\ell^-$ decays could be interconnected. Thus, the measurement of $b\to s\nu\bar{\nu}$ offers a complementary probe to look into the underlying physics~\cite{Bhattacharya:2014wla,Calibbi:2015kma}.

\begin{figure}[t]
\centering

\includegraphics[width=\textwidth]{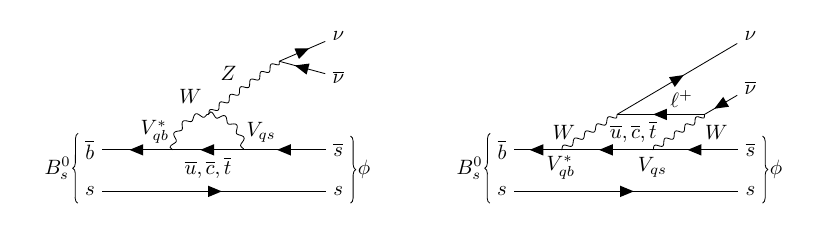}
    \caption{Illustrative Feynman diagrams for the transition $B_s^0\to \phi\nu\overline{\nu}$ in the SM. \textbf{LEFT}: EW penguin diagram. \textbf{RIGHT}: EW box diagram.}
    \label{fig:FCNC_Bs_Feynman}
\end{figure}

\begin{figure}[t]
    \centering
    \includegraphics[height=5.4cm]{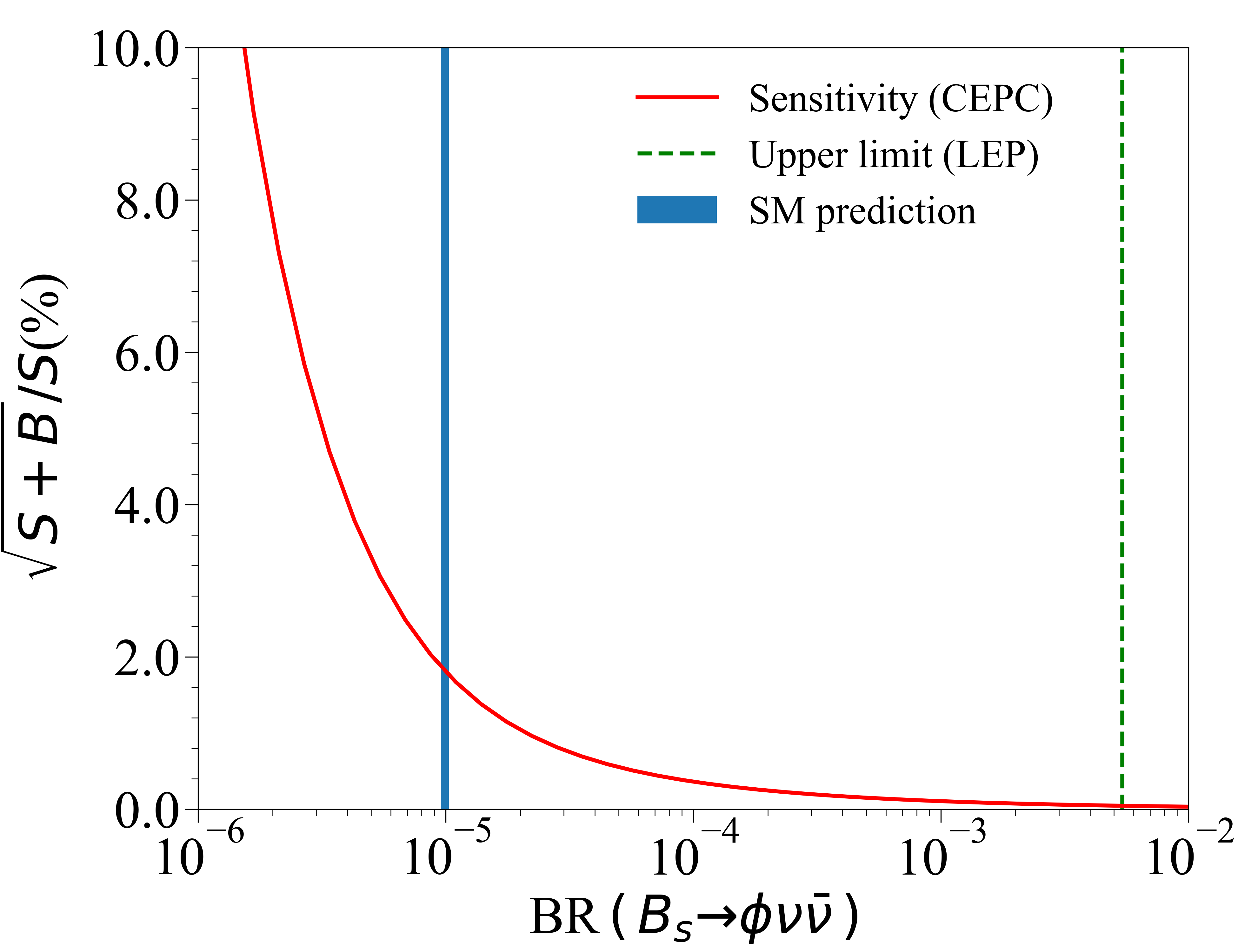}
    \hfill
    \includegraphics[height=5.4cm]{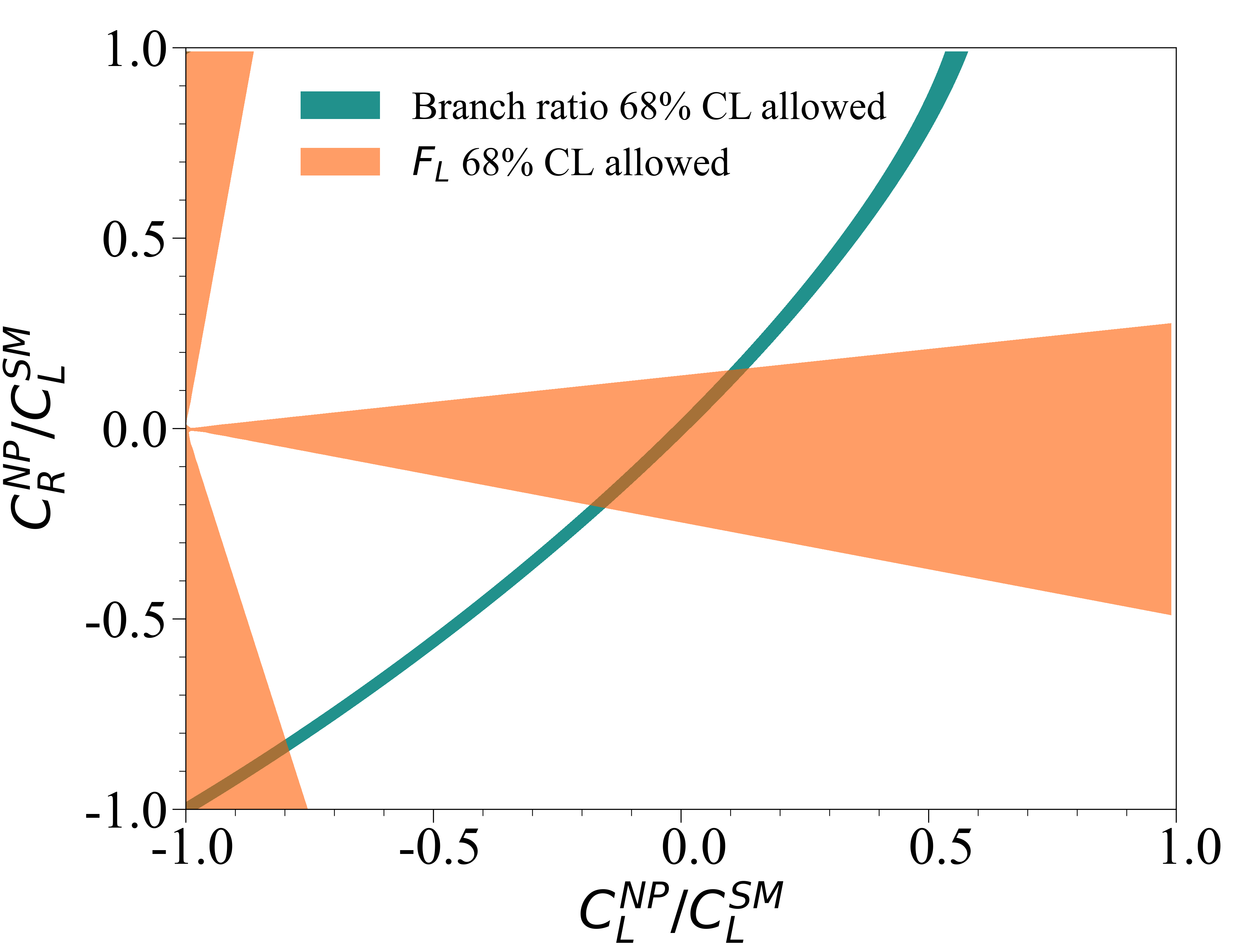}
    \caption{\textbf{LEFT}: Relative precision for measuring the signal strength of $B_s^0 \to \phi\nu\bar{\nu}$ at Tera-$Z$, as a function of its BR. \textbf{RIGHT}: Constraints on the LEFT coefficients $C_L^{\rm NP} \equiv C_{L}-C_L^{\rm SM}$ and $C_R$ with the measurements of the overall $B_s^0 \to \phi\nu\bar{\nu}$ decay rate (green band) and the $\phi$ polarization $F_L$ (orange regions). These plots are taken from Ref.~\cite{Li:2022tov}.}
    \label{fig:BsPhiNuNu_Accuracy}
\end{figure}

A dedicated study of the $B_s^0\to\phi\nu\bar{\nu}$ decay (see Figure~\ref{fig:FCNC_Bs_Feynman}) at the $Z$ pole has been  conducted, using full simulation samples aligned with the CEPC detector profile~\cite{Li:2022tov}. This study, facilitated by the large $B_s^0$ statistics at the CEPC (see Table~\ref{tab:BYield}), suggests that a precise measurement of such a rare decay is possible. 
Explicitly, the accurate $\phi$ and $B_s^0$ reconstructions in this analysis reduce the $Z\to q\bar{q}$ events by a factor $\sim \mathcal{O}(10^{-8})$, with a signal efficiency $\sim 3\%$, leaving primarily the $Z\to b\bar{b}$ events as the backgrounds. As a result, a relative precision $\lesssim 2\%$ can be achieved for measuring the SM $B_s^0\to\phi\nu\bar{\nu}$ signal, as shown in the left panel of Figure~\ref{fig:BsPhiNuNu_Accuracy}. Particularly, with a high  signal-to-background ratio of $\simeq 77\%$, the robustness of this measurement against potential systematic uncertainties is largely assured. This study has also shown that the constraints obtained from this measurement can contribute pivotally to the global determination of NP effects, {\it e.g.}, the ones encoded in the LEFT, (see the right panel of Figure~\ref{fig:BsPhiNuNu_Accuracy}).

In addition to the $B_s^0\to\phi\nu\bar{\nu}$ decay, there exist a set of other physics processes that can be applied to study the $b\to s\nu\bar{\nu}$ transitions at the CEPC, for example $B^+ \to K^+\nu\bar{\nu}$, $B^+ \to K^{+\ast} \nu\bar{\nu}$, and $B^0 \to K^{0\ast} \nu\bar{\nu}$. Interestingly, the Belle~II collaboration has recently performed a search for the rare $B^+ \to K^+\nu\bar{\nu}$ decay using an inclusive tagging approach, and obtained a branching fraction of $(2.7\pm0.7)\times 10^{-5}$~\cite{Belle-II:2023esi}, with a significance of $3.5$ standard deviation with respect to the background-only hypothesis. This measurement also shows a $2.9$ standard deviation departure from the SM expectation~\cite{Buras:2014fpa,Becirevic:2023aov}. The expected precision of the branching ratios for $B\to K^{(*)}\nu\bar\nu$ with $50~\text{ab}^{-1}$ by combining the charged and neutral $B$ decay modes are of the order of $10\%$~\cite{Belle-II:2022cgf}.
Yet, by leveraging its advantages in reconstructing the missing energy and producing the $b$-hadrons, the CEPC may push this precision to a much higher level. 
Such expectations have been confirmed by a recent study at FCC-$ee$~\cite{Amhis:2023mpj}.

\iffalse
\begin{figure}[t]
    \centering
    \includegraphics[width=0.9\textwidth]{fig/bsnn_fcc.png}
    \caption{Plots from Ref.~\cite{Amhis:2023mpj}. The sensitivity estimates of the $B^0 \to K^{0\ast} \nu\bar{\nu}$ and the $B_s^0\to\phi\nu\bar{\nu}$ processes at FCC-$ee$, with a $Z$ pole dataset of $6\times10^{12}$ $Z$ bosons.}
    \label{fig:bsnnfcc}
\end{figure}
\fi

Furthermore, probes of other decay modes involving long-lived $s$-hadrons, such as $B^0 \to K^0_S\nu \bar\nu$, $\Lambda_b \to \Lambda \nu\bar{\nu}$ and $\Xi_b^\pm \to \Xi^\pm \nu\bar{\nu}$ could also help pin down the $b\to s\nu \bar\nu$ transition. The decays of the intermediate neutral particles in general give rise to vertices with a displacement of $\mathcal{O}(10)$\,cm. Therefore the precision of these channels highly depends on the reconstruction and resolution of these significantly displaced vertices. From a preliminary estimate~\cite{Aleksan:2021fbx}, it is possible to achieve an $80\%$ reconstruction efficiency for the $K_S^0$ and $\Lambda$ vertices at a CEPC environment, opening up the opportunity to perform a combined constraint of $b s\nu\bar{\nu}$ effective interactions with all the aforementioned decay modes.
In particular, the baryonic processes such as $\Lambda_b \to \Lambda \nu\bar{\nu}$ and $\Xi_b^\pm \to \Xi^\pm \nu\bar{\nu}$ are unique opportunities at the CEPC as they are above the production threshold of the Belle II experiment. Since form factors of these baryonic modes are different from those of the mesonic modes, studies of these channels will bring independent information to understand the dynamics underlying 
the $b\to s\nu\bar{\nu}$ transition in a global fit.

\subsection{Radiative Modes}

\begin{figure}[t]
\centering
    \includegraphics[width=\textwidth]{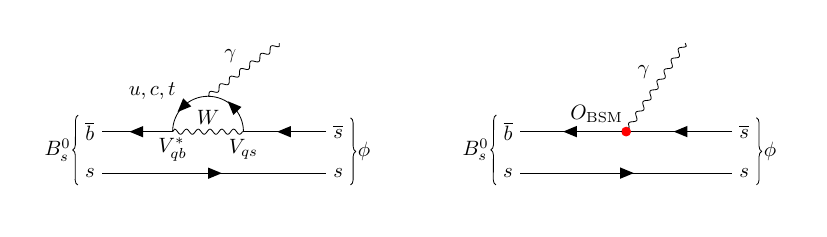}
    \caption{Illustrative Feynman diagrams for the decay $B_s^0\to \phi \gamma$. \textbf{LEFT}: SM example. \textbf{RIGHT}: BSM example.}
    \label{fig:Radiative_Feynman}
\end{figure}

The third category of FCNC rare $B$ decays consists of radiative ones, such as $b \to s\gamma, d\gamma$. These modes are sensitive to the EM dipole operators $O_7$ and $O_7^\prime$. A wealth of data, including the inclusive $B \to X_{s,d}\gamma$ decays, as well as the direct $CP$ violation $A_{CP}$ and time-dependent $CP$ violation $S_{CP}$ in various $b\to s\gamma$ decays, has yielded complementary insights into the corresponding Wilson coefficients $C_7$ and $C_7^\prime$. At the CEPC, however, the reach for FCNC radiative modes is yet to be fully explored, despite their scientific significance~\cite{Talk1}. One such example is the $B_s^0 \to \phi\,(\to K^+ K^-) \,\gamma$ decay, illustrated in Figure~\ref{fig:Radiative_Feynman}. Achieving a high accuracy in reconstructing the signal $B_s^0$ meson necessitates superior photon angular and energy resolution. For the LHCb Upgrade II, it was found that BR$(B_s^0\to\phi\gamma)$ could be measured with a statistical uncertainty $\sim 0.1\%$, and the $CP$ parameters can also be well measured~\cite{Bediaga:2018lhg, Aaij:2019pnd}. These sensitivities are expected to be further improved at the CEPC due to the potentially high performance of its ECAL.  
This study can be extended to baryonic radiative decays of the $b\to s\gamma$ type, such as $\Lambda_b \to \Lambda\gamma$ and $\Xi_b\to \Xi \gamma$, again with an expected sensitivity better than the LHCb~\cite{Aaij:2019hhx}. The study can also be extended to $b \to d\gamma$ decays, which can broaden our understanding of the FCNC transition amplitudes and potentially refine the CKM matrix determinations. Finally, if the ECAL of the CEPC allows an efficient reconstruction of $\pi^0, \eta \to \gamma\gamma$~\cite{Wang:2022nrm}, the double-radiative decays of $B_{s,d} \to \gamma\gamma$ could be measured~\cite{Bobeth:2011st}.
Theoretical studies show that the $\Lambda_{\rm QCD}/m_b$ power corrections in these channels are well under control, 
making them new
benchmark probes of non-standard dynamics~\cite{Qin:2022rlk,Shen:2020hfq}. The SM predictions for their BRs are given by~\cite{Qin:2022rlk,Shen:2020hfq} 
\begin{equation}
    \text{BR}(B_s^0\to\gamma\gamma)=(3.8^{+1.9}_{-2.1})\times10^{-7}\;, \; \text{BR}(B^0\to\gamma\gamma)=(1.9^{+1.1}_{-1.0})\times10^{-8} \; .
\end{equation} 
Belle II has assessed its sensitivities to be respectively $\sim 23\%$ and $\sim 10\%$~\cite{Belle-II:2018jsg} relative to the theoretical estimates in Ref.~\cite{Bosch:2002bv} that, we notice, are a factor of few larger than those provided above. Recently, an analysis combining the Belle and available Belle II data sets an upper limit of $\text{BR}(B^0\to\gamma\gamma)<6.4\times 10^{-8}$ at $90\%$ confidence level~\cite{Belle:2024bsw}. %\ACR{what about $B->Xs\gamma\gamma$?}

\subsection{Tests of SM Global Symmetries}

An important class of observables include $b$-hadron decays that are forbidden because of the global symmetries of the SM. Aside from gauge symmetries, the SM respects or approximately respects a series of global symmetries, yielding, at different levels, the conservation of lepton family numbers, lepton and baryon numbers. The only-known breaking effects for these symmetries are highly suppressed in collider environments: lepton family numbers in the charged lepton sector are only violated through neutrino mixing and thus suppressed by the small neutrino mass differences; lepton and baryon numbers are only violated by the non-perturbative $SU(2)_L$ sphaleron which breaks both the lepton number and baryon number but conserves their difference exactly. The observation of Lepton Flavor Violation (LFV) in the charged lepton sector, as well as Lepton Number Violation (LNV), Baryon Number Violation (BNV) in any perturbative processes thus would be an indisputable evidence for BSM physics. 
Interestingly, LFV and LFU violation (LFUV) receive contributions respectively from the flavor off-diagonal and  diagonal components of the same classes of EFT operators and thus are often correlated in UV-complete NP models. The modes that are forbidden in the SM often yield striking signals that are dramatically distinct from the background events. Just like the LFU tests, the CEPC with its large statistics and clean environment can play a significant role in examining these global symmetries.

Some of the FCNC studies presented in previous subsections can be extended to the null tests of SM global symmetries, in a straightforward way. For example, one can investigate the LFV effects in the $b$-hadron decays~\cite{Chrzaszcz:2021nuk}, such as $H_b \to H_{d/s}\tau \ell$, where $\ell$ denotes an electron or a muon. 
These decays are significant for testing current anomalies in semi-leptonic $b$-hadron decays~\cite{Capdevila:2023yhq} and, more in general, heavy NP coupling preferably to the third generation~\cite{Calibbi:2015kma,Allwicher:2023shc}. In the past, experimental efforts have primarily focused on the modes $B^{+} \to K^+ \tau\ell$, yielding $\mathcal{O}(10^{-5})$ upper limits on their branching ratios~\cite{BaBar:2012azg,Belle:2022pcr}.  Topological reconstruction techniques, employing a fast parametric simulation with momentum reconstruction resolutions and vertex detector performance, have been implemented to simulate LFV signal events for $B^{0} \to K^{*0}\mu\tau$ as well. Initial explorations have demonstrated the detector requirements, offering guidance for future design and optimization goals for the vertex detector of the CEPC. 
As for LFV two-body decays, preliminary studies have shown that  -- while the CEPC constraints on the decays such as $B^0_{(s)} \to \mu^{\pm} e^{\mp}$ and $B^0_{(s)} \to \tau^{\pm} \mu^{\mp}$ can at most match the LHCb sensitivity~\cite{Bediaga:2018lhg} -- an improvement in the sensitivity to $B^0_{(s)} \to \tau^{\pm} e^{\mp}$ could be achieved at the CEPC due to the expected excellent electron identification.

The CEPC also provides a platform for testing LNV and BNV in $b$-hadron decays. For instance, LNV can be tested by measuring the same-sign di-lepton decay $B^{+} \to \pi^{-}(K^{-}) \ell^{+} \ell^{+}$, where the sensitivities are  primarily influenced by statistics and lepton charge identification. Unlike the LHCb analysis which has focused on the di-muon mode~\cite{Aaij:2011ex,Aaij:2014aba}, the CEPC may have a good sensitivity for the same-sign di-electron mode also, given its low misidentification rates for electrons. The BNV measurements may feature the signals such as forbidden baryon-antibaryon oscillations~\cite{Aaij:2017inn} and explicit BNV decays. One example in the latter case is $\Lambda_b^{0} \to h^{-}(h^{0})\ell^{+}$, which arise from the dimension-6 BNV operators $qq^{\prime}q^{\prime\prime}\ell$ where $B-L$ is conserved.

Interestingly, BNV is one of the three Sakharov conditions~\cite{Sakharov:1967dj} required for dynamically generating the baryon asymmetry of the Universe (BAU). Hence, the measurement of BNV modes may provide valuable clues for resolving this long-standing cosmological puzzle. For example, introducing a dark matter candidate carrying baryon number, the $B$-mesogenesis model~\cite{Elor:2018twp} predicts the BNV separately in the visible sector and dark matter sector, simultaneously achieving baryogenesis and the correct dark matter relic abundance. This model can be tested by measuring invisible decays of neutral bottom baryons such as $\Lambda_b^0$  -- for further discussions on its collider phenomenology, see~\cite{Elahi:2021jia,Alonso-Alvarez:2021oaj,Alonso-Alvarez:2021qfd}.
In a recent study~\cite{Zheng:2024tkj}, it has been shown that the important constraints on the model parameters can be obtained at the $Z$ pole run of the CEPC.

\section{\texorpdfstring{$\boldsymbol{CP}$}{CP} Violation in \texorpdfstring{$\boldsymbol{b}$}{b}-Hadron Decays}
\label{sec:CPV}

In the SM, the flavor properties of quarks are mainly encoded in the CKM matrix, including what concerns the phenomena involving $CP$ violation. The independent entries include three Euler angles entangling the three generations and one CKM phase as the only source of $CP$ violation in the SM~\cite{Kobayashi:1973fv}. 
Yet, addressing the puzzle of BAU dynamically requires additional $CP$ violation, as one of the Sakharov conditions. 
This consideration has motivated extensive explorations in last decades. $b$-hadron decays provide a handle particularly suitable for this study. Theoretically, it has been demonstrated in Ref.~\cite{Liu:2011jh} that the $CP$ violation in $B$ meson systems can drive the BAU generation though EW baryogenesis. Experimentally, the heavy-flavor measurements represent one of the most important tasks in flavor physics. At the CEPC, such measurements are expected to greatly benefit from high statistics, low backgrounds, efficient hadron ID, and extreme displacement resolution. The observables, handled by proper analysis of amplitudes, can be fed into the global fit of the CKM matrix. Any deviation from the CKM unitarity would be a smoking-gun signature for NP including new $CP$ violation.

Generally, there are three categories of observables for $CP$ violation: $CP$ violation in decay (direct $CP$ violation), $CP$ violation in mixing (indirect $CP$ violation) and $CP$ violation through the interference between mixing and decay.\footnote{It was suggested recently~\cite{Shen:2023nuw} that double-mixing $CP$ violation is possible in cascade decays involving two neutral mesons in the decay chain, induced by the interference of different meson oscillating paths. Such double-mixing $CP$ violation may occur in specific channels such as $B_s^0\to \rho^0K\to \rho^0(\pi^-\ell^+\nu)$ and $B^0\to D^0K \to D^0(\pi^+\ell^-\bar{\nu})$ and the measurement of $CP$ asymmetry depends on oscillation time of both $B^0_{(s)}$ and $K$.} The $CP$ violation in decay can be measured through a process, where the initial particle does not mix with its $CP$ conjugate and the final state is not a $CP$ eigenstate, and its $CP$ conjugate. The $CP$ violation is then manifested as a time-integrated asymmetry in statistics between these two processes. The effective statistics is determined by both of the overall signal rate and the efficiency of tagging initial heavy-flavored particles. As introduced in Section~\ref{sec:DetectorFeature}, the effective tagging efficiency $\epsilon_{\rm eff}$ can be estimated as $ \epsilon_{\rm tag} (1 - 2\omega)^2$ for some specific processes, where $\epsilon_{\rm tag}$ and $\omega$ are the raw tagging efficiency and mistagging rate, respectively~\cite{Moser:1996xf}. 
Regarding the application for determining the CKM parameters, one example is related to measuring the time-integrated $CP$ asymmetry in the $B^+\to D^{(\ast) 0} K^{(\ast) +}$ decay~\cite{HeavyFlavorAveragingGroup:2022wzx,Aleksan:2021fbx}.
Ref.~\cite{Aleksan:2021fbx} exploits high acceptance and excellent reconstruction of $K^0_S$ from $D^0\to K^0_S \pi^0 $ to study $B^\pm \to D^0(\bar{D}^0) K^\pm$, assuming a crystal ECAL for FCC-$ee$, and finds that the $\gamma_s$ parameter for the $bs$ unitarity triangle could be determined with a precision $\sim \mathcal{O}(1^\circ)$.

The observations of $CP$ violations in mixing and interference between mixing and decay involve decaying processes of neutral particles which as flavor eigenstates are not identical with their mass eigenstates. In the former case, the decays are flavor-specific. The $CP$ violation is often measured as a time-integrated asymmetry for semi-leptonic decays like $M_0\to l^- X$ and $\bar M_0\to l^+ X$. Differently, the latter case requests the decay products to form a $CP$ eigenstate such that an interference can occur between the amplitudes with and without a mixing. $B^0$ and $B_s^0$ as neutral heavy-flavored mesons are especially relevant here. Because of the oscillations between them and  their $CP$-conjugate before decay, the $CP$ asymmetry generically demonstrates a time dependence which can be leveraged for detecting the $CP$ violation. General pattern holds for this time-dependence despite the diversity of possible decaying processes. The asymmetry is proportional to the oscillatory factors with the period determined by the mass gap ($\Delta m$) between the mass eigenstates of initial particles and non-oscillatory factors caused by the decay-width difference ($\Delta \Gamma$) of these mass eigenstates. Because $\Delta m \gg \Delta \Gamma$ for the $B^0$ and $B_s^0$ mesons, the oscillatory factors are relatively more relevant for their $CP$ violation measurements~\cite{HeavyFlavorAveragingGroup:2022wzx}. The mistagging probability $\omega$ becomes significant in this case, as the algorithm must determine the charge of initial $b$ quarks after the $b-\bar{b}$ oscillation happens. Another factor affecting the overall precision is the decay time determination, which is mainly limited by the vertex resolution of the tracking system.

The charge determination of initial $b$ quarks is primarily affected by the mixing-induced oscillations. One way to address this difficulty is to utilize the information of the companion $b$-hadron. If the companion $b$ quark hadronizes into non-oscillatory species such as $B^\pm$ and is subsequently identified, then the charge of the original signal $b$ quark can be identified. Alternatively, one can employ the products of QCD shower and hadronization, as they manifest the original $b$-quark charge before the oscillation occurs. For example, the $B_s^0$ meson is often accompanied by a collimated $K^+$ meson, where the strange quarks are pair-produced. Recent study in~\cite{Li:2022tlo} suggests that an $\epsilon_{\rm eff}$ value of $\gtrsim 20\%$ can be achieved at the CEPC, much higher than $\sim 5\%$ at LHCb~\cite{Fazzini:2018dyq}. This result is also consistent with another CEPC study which combines leading charged particle in a jet and momentum-weighted jet charge~\cite{Cui:2023kqb}, yielding an $\epsilon_{\rm tag} \sim 39\%$ and $20\%$ for inclusive $c$ or $b$ jets respectively. Notably, utilizing the method of jet origin identification and the ParticleNet algorithm developed in Ref.~\cite{Liang:2023wpt}, the  jet charge flip rates could be controlled to 19\% and 7\% for inclusive $b$ and $c$ jets, corresponding to an effective tagging power of 37\% and 54\%, respectively.  
More details can be found in Section~\ref{sec:CEPC}.

The decay-time measurements at the CEPC are expected to benefit from its clean physics environment and well-designed tracking system. The full simulation in Ref.~\cite{Li:2022tlo} reports a CEPC resolution of $\lesssim 5$\,fs for measuring the 4-prong decay $B_s^0\to J/\psi \phi \to \mu^+\mu^-K^+K^-$, which is much better than the typical LHCb level of $\sim 20-30$\,fs. This will bring great benefits to the measurements of time-dependent $CP$ violation and also, for the role of $\Delta m$ and $\Delta \Gamma$ as basic inputs, the global CKM fit. Additionally, a study in the FCC-$ee$ context~\cite{Aleksan:2021gii} suggests a relative uncertainty of $\lesssim 3\times 10^{-5}$ for the $\Delta m$ measurement of $B_s^0$ meson, which is about one order of magnitude better than the current level. We hope that dedicated studies in the future could help validate such results 
and reveal the full potential of the CEPC in measuring these basic flavor physics parameters.

\begin{figure}[t]
    \centering
    \includegraphics[height=4.7cm]{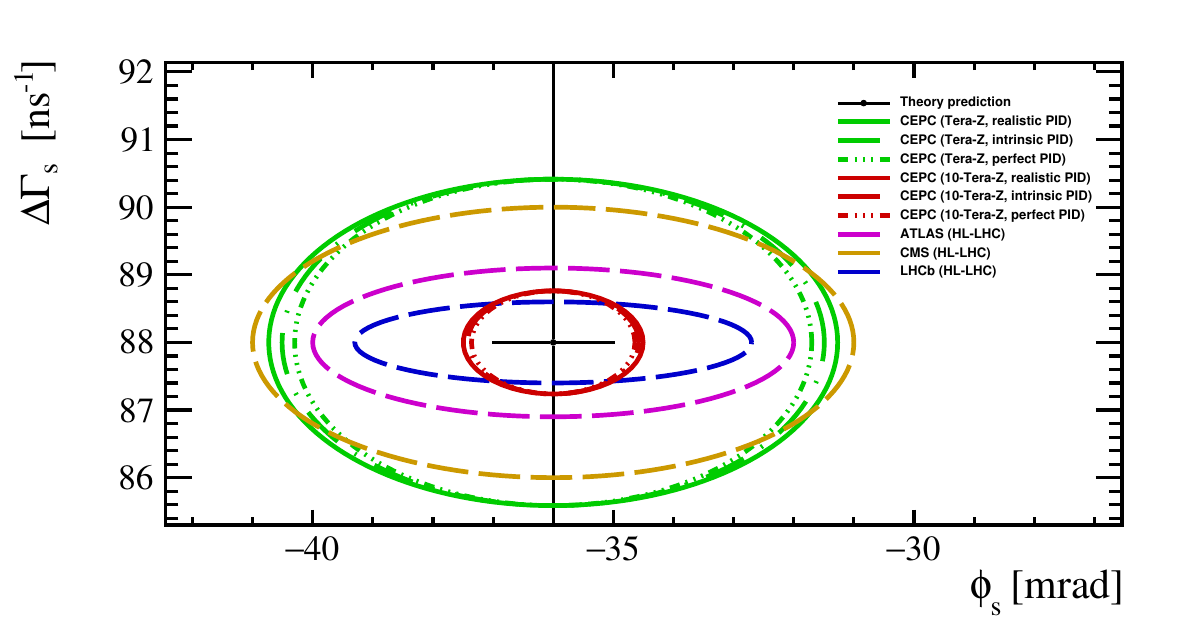}
    \hfill
    \includegraphics[height=4.3cm]{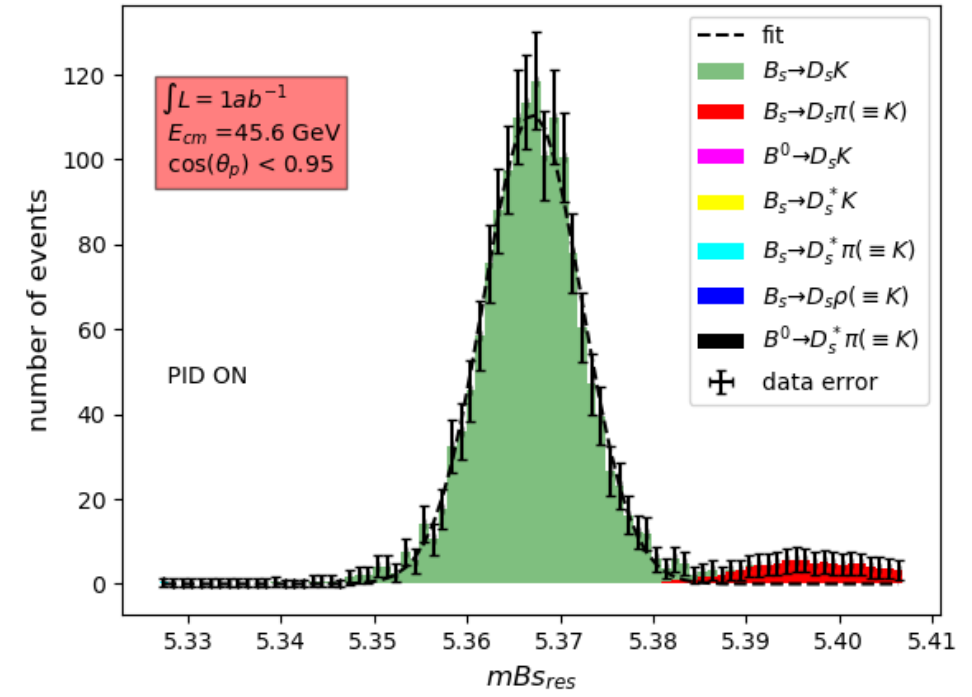}
    \caption{\textbf{LEFT:} Projected 68\% confidence level (CL) sensitivities of measuring the parameters $\Delta\Gamma_s$ and $\phi_s \approx -2 \beta_s$ at the CEPC~\cite{Li:2022tlo}, through the time-dependent $CP$ violation in the decay $B_s^0 \to J/\psi(\to \mu^+\mu^-) \phi (\to K^+ K^-)$. \textbf{RIGHT:} $B_s^0$ mass reconstruction in the decays $B_s^0 \to D^\pm_s (\to \phi \pi^\pm \to K^+K^- \pi^\pm) K^\mp$ at the $Z$ pole of FCC-$ee$~\cite{Aleksan:2021gii}. } 
    \label{fig:time_depedent}
\end{figure}

The time-dependent $CP$ measurements can be also applied to test the $bs$ unitarity triangle. The decay of $B_s^0\to J/\psi \phi \to \mu^+\mu^-K^+K^-$ has been widely used for this purpose~\cite{Dighe:1998vk,Faller:2008gt}.
Figure~\ref{fig:time_depedent} displays in its left panel the projected CEPC sensitivities of measuring the parameters $\Delta\Gamma_s$ and $\phi_s \approx -2 \beta_s$ in this channel~\cite{Li:2022tlo}. 
The performed full simulation indicates that the CEPC could reduce the uncertainty for  
$\beta_s$ to $\sim 2.3 {\rm mrad} \sim 0.13^\circ$ ~\cite{Li:2022tlo}, 
improving the existing precision by several times. 
FCC-$ee$ also reported its study on the time-dependent $CP$ measurements in the same decay mode, and additionally $B_s^0\to D_s^\pm K^\mp$~\cite{Aleksan:2021gii}, with fast simulation. The right panel of Figure~\ref{fig:time_depedent} shows the mass reconstruction of $B_s^0$ mesons achieved in this study. 
Most combinatoric and misidentification-induced backgrounds can be removed with the PID algorithm, yielding a sharp peak of signal events. In this context, 
the triangle parameter $\alpha_s$ and $\beta_s$ can be measured with a precision of $0.4^\circ$ and $0.035^\circ$, respectively~\cite{Aleksan:2021gii}. The CEPC results are weaker than those of FCC-$ee$. However, considering the recent advancement of jet origin ID at the CEPC, comparable sensitivities could be finally achieved for both machines.

\begin{figure}[t]
    \centering
    \includegraphics[width=0.4\textwidth]{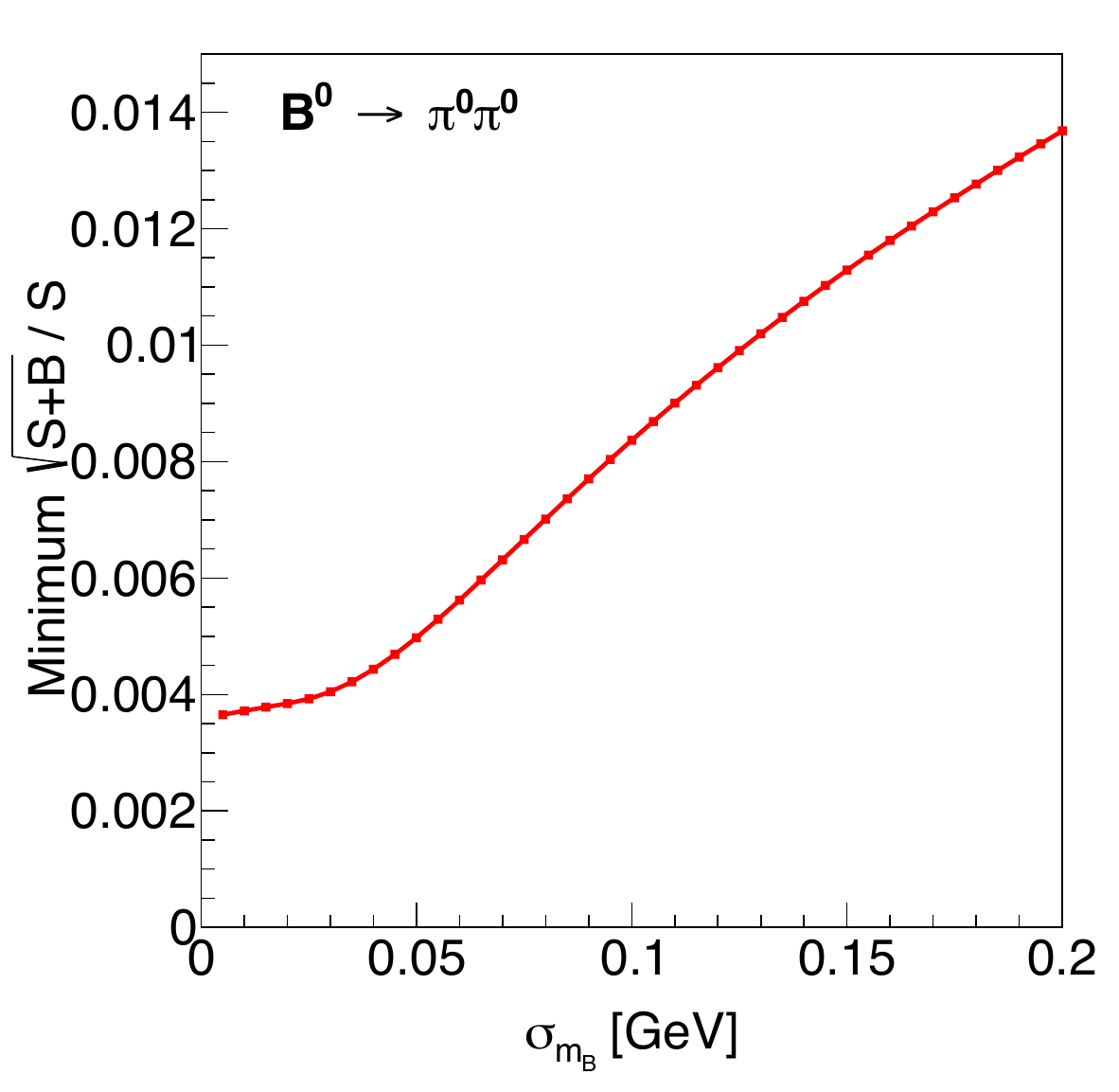}
    \hspace{1cm}
    \includegraphics[width=0.4\textwidth]{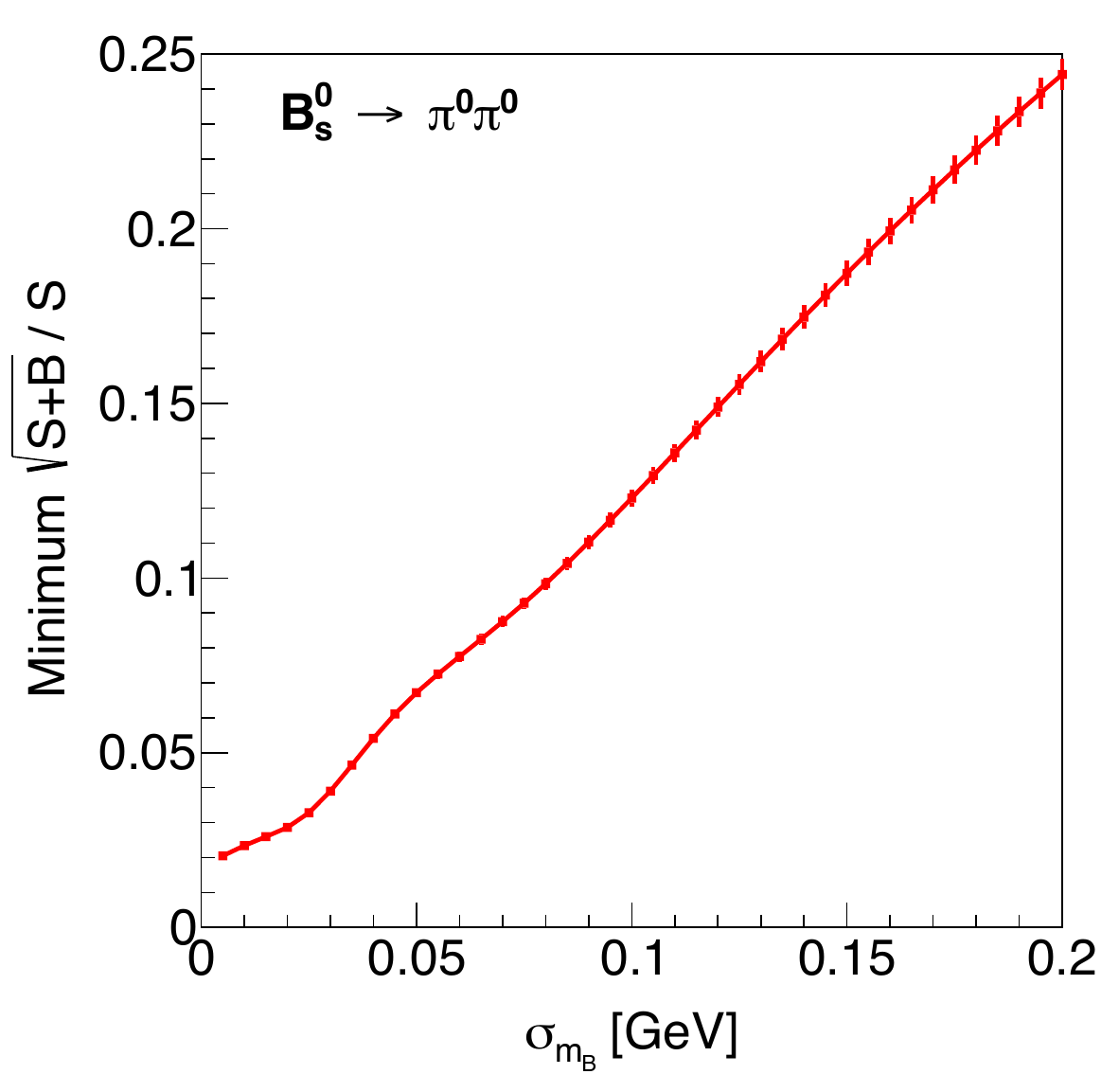}
    \caption{Relative uncertainties (statistical only) of measuring BR($B^{0} \to \pi^{0}\pi^{0} \to 4\gamma$) (\textbf{LEFT}) and BR($B^{0}_{s} \to \pi^{0}\pi^{0} \to 4\gamma$) (\textbf{RIGHT}) at the CEPC as a function of the $B$-meson mass resolution $\sigma_{m_B}$. The plots are taken from Ref.~\cite{Wang:2022nrm}. }
    \label{fig:AccuracyB0sTo2Pi0}
\end{figure}

\begin{figure}[t]
    \centering
    \includegraphics[width=7.5cm]{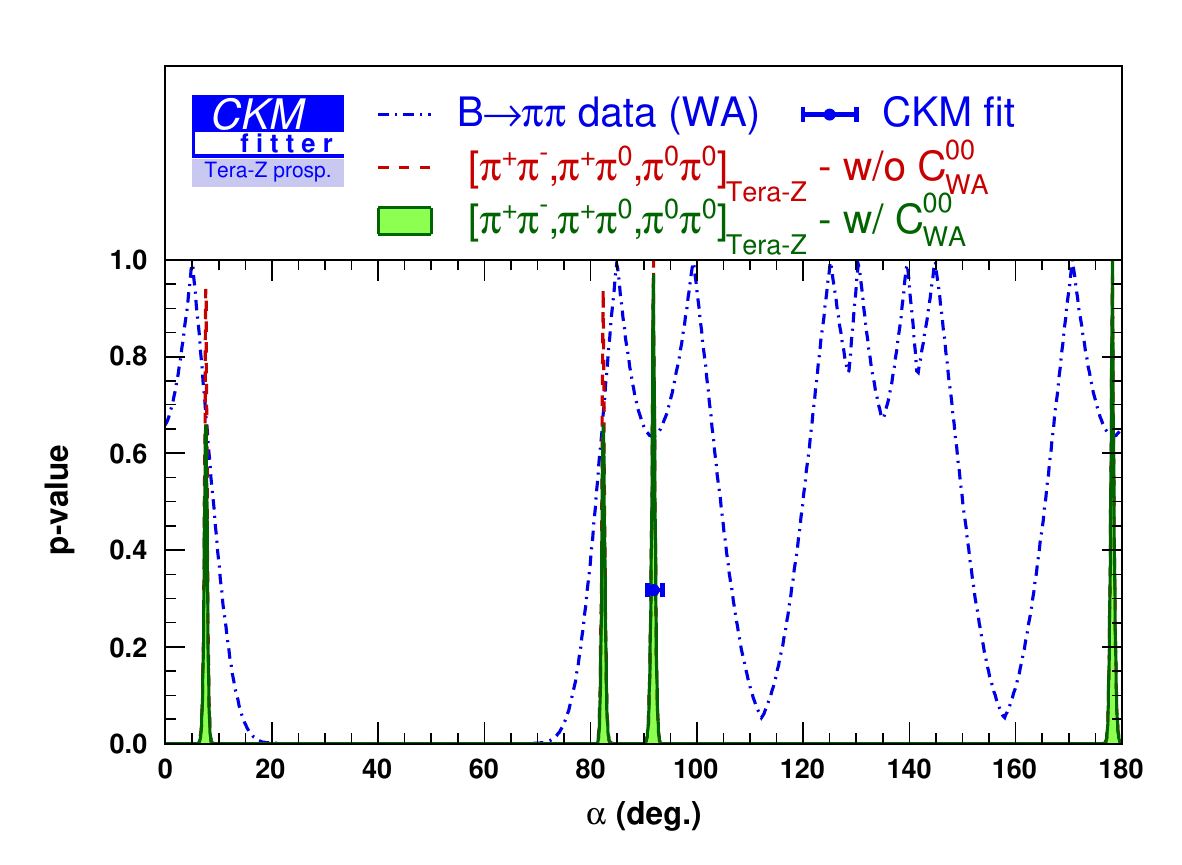}
    \hfill
    \includegraphics[width=7.5cm]{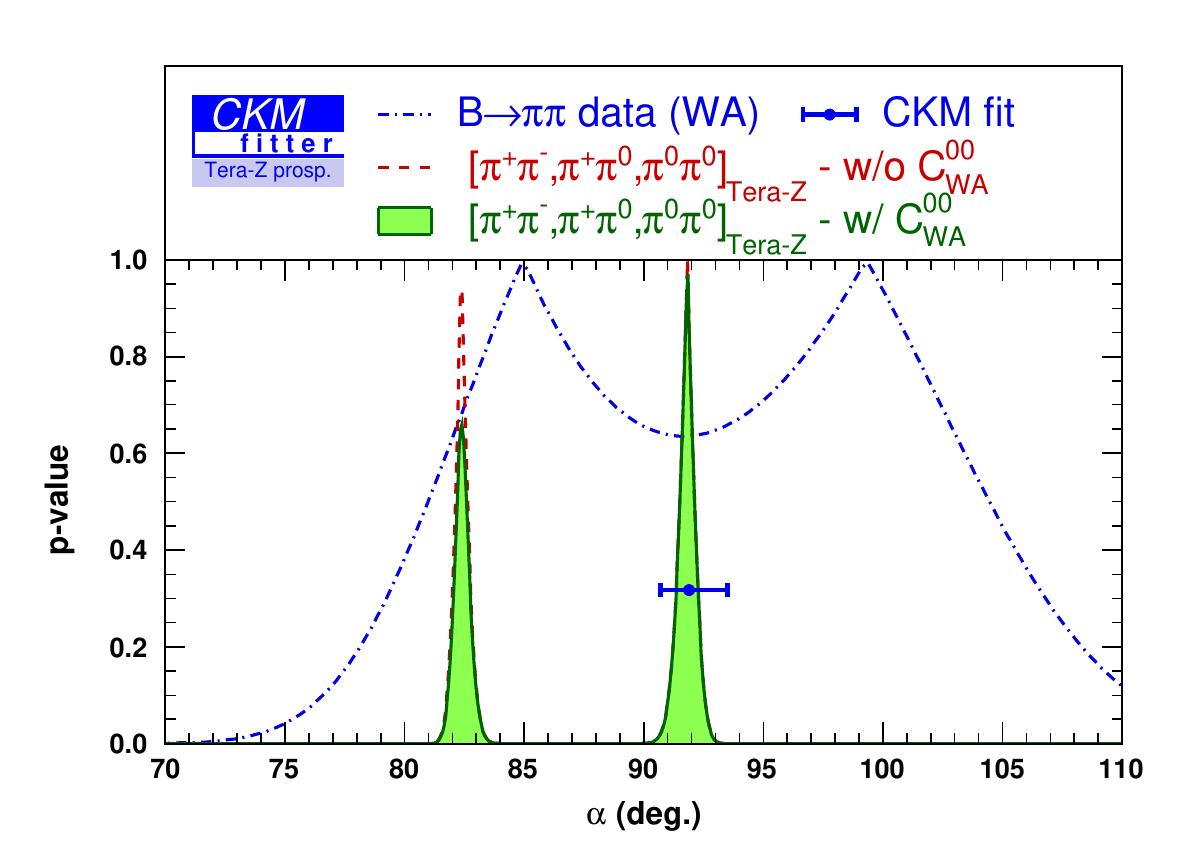}
    \caption{$p$-value for the determination of the CKM angle $\alpha$~\cite{Wang:2022nrm}. With the current $B^0\to\pi\pi$ measurements (dotted-dashed blue) and global CKM fitting (blue error bars) as a reference, we demonstrate two different scenarios of the CEPC measurements as a Tera-$Z$ factory, where the CEPC data is used alone (dashed red) or combined with the current world average of $B$-factory measurements (filled green). \textbf{LEFT}: Scan over the whole range of $\alpha$.  \textbf{RIGHT}: Scan around the value favored by the global CKM fit. 
}
\label{fig:CKMAlpha_TeraFull}
\end{figure}

Yet, the oscillating effects of neutral $B$ mesons are not always trackable. One example is the decay of $B^0_{(s)}\to \pi^0\pi^0\to 4\gamma$, where the tracker loses its power and reconstructing $B^0_{(s)}$ decay time becomes extremely challenging. One can perform time-integrated measurements only for such decays. A sensitivity study on this case has been taken with the CEPC fast simulation in Ref.~\cite{Wang:2022nrm}. Figure~\ref{fig:AccuracyB0sTo2Pi0} displays the obtained relative uncertainties (statistical only) as a function of the $B$-meson mass resolution $\sigma_{m_B}$. For BR($B^{0} \to \pi^{0}\pi^{0} \to 4\gamma$) and BR($B^{0}_{s} \to \pi^{0}\pi^{0} \to 4\gamma$), the Tera-$Z$ precisions are expected to be $\lesssim \mathcal O(1\%)$ and $\lesssim \mathcal O(10\%)$, respectively. Here the magnitude of $\sigma_{m_B}$ is significantly influenced by the ECAL performance. The benchmark presented reflects an ECAL resolution of $\sim 3\%/\sqrt{E\text{(GeV)}}$, which could be achieved with a fully crystal ECAL~\cite{Lucchini:2020bac}. The pair of $B$ mesons produced at $Z$ pole are not entangled, unlike the entangled $B$ production by $\Upsilon(4S)\to 2B$ decays in $B$-factories. Consequently, the time-integrated observables for $CP$ violation at the CEPC are slightly different from their $B$-factory counterparts. Combining the future CEPC and Belle II results of measuring $B\to \pi\pi$, the CKM angle $\alpha$ could be constrained~\cite{Charles:2017evz} to a level as small as  0.4$^\circ$ if theoretical errors are resolved. 
These projected results are illustrated in Figure~\ref{fig:CKMAlpha_TeraFull}, which indicate that the CEPC measurements can constrain $\alpha$ much stronger than the current data. The measurement of time-dependent $CP$ violation in the $B^0\to \pi^0\pi^0$ decay, using the $\pi^0 \to e^-e^+\gamma$ Dalitz mode, has been explored by Belle II collaboration~\cite{Belle-II:2018jsg}. The sensitivity relies on the quality of the $\pi^0 \to e^-e^+\gamma$ decay vertex reconstruction, which is yet to be studied at the CEPC.

%~~~~~~~~~~~~~~~~~~~~~~~~~~~~~~~~~~~~~~~~~~~
%~~~~~~~~~~~~~~~~~~~~~~~~~~~~~~~~~~~~~~~~~~~

\begin{figure}[t]
\centering
    \includegraphics[width=\textwidth]{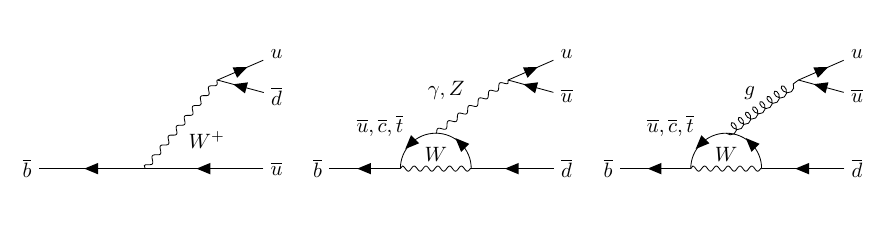}
    
    \caption{Illustrative Feynman diagrams for the transition $\bar{b}\to \bar{u} u \bar{d}$. \textbf{LEFT}: tree level. \textbf{MIDDLE}: EW penguin diagram. \textbf{RIGHT}:  QCD penguin diagram.}
    \label{fig:CPV_Feynman}
\end{figure}

Despite the analyses discussed above, many studies regarding the $CP$ violation at the $Z$ pole and the relevant physics have yet to be explored. For example, the $\beta$ angle is known to be primarily determined by the measurements of the $b\to c\bar{c} s$ transitions such as $B^0 \to J/\psi K^{0}$ and their time-dependent $CP$ violation~\cite{HeavyFlavorAveragingGroup:2022wzx}. A dedicated simulation is needed to validate the projected $Z$-factory sensitivities in Ref.~\cite{Abada:2019lih}. Also, the $CP$ violation in the $b\to u\bar{u} d$ transitions (see Figure~\ref{fig:CPV_Feynman}) such as $B\to \rho\rho$ and $B\to \rho\pi$, can be relevant for the determination of the $\alpha$ angle. These studies also echo the recent report of the first evidence from LHCb for direct CP violation arising from the $b\to c\bar{c} q$  transitions~\cite{LHCb:2024exp}, where $q = s, d$. But, digging out the potential of a $Z$ factory for the CKM global fit demands systematic sensitivity studies on these measurements of $CP$ violation. 

Another recent achievement is the first definitive observation of $CP$ violation in the decays of baryons, a class of particles that had remained experimentally elusive despite decades of study~\cite{LHCb:2025ray}. Using the full Run 1 and Run 2 dataset of the LHCb, the analysis focuses on the four-body decay $\Lambda_b^0 \to pK^-\pi^+\pi^-$ and its $CP$-conjugate process $\bar{\Lambda}_b^0 \to \overline{p}K^+\pi^-\pi^+$, comprising over 80,000 reconstructed events. The global $CP$ asymmetry is directly measured to be $\mathcal{A}_{CP} = (2.45 \pm 0.46 \pm 0.10)\%$, establishing this effect with 5.2$\sigma$ statistical significance after careful control of systematic uncertainties. This discovery is particularly significant as $CP$ violation had previously only been observed in meson systems, despite both quark-level transitions being theoretically predicted to show similar effects. In view of the rich statistics of $\Lambda_b^0$ and $\bar{\Lambda}_b^0$ and their boost kinematics at $Z$ pole (see Table~\ref{tab:BYield}), it is natural to extend the studies of $CP$ violation in baryon systems from LHCb to CEPC, as we have done for measuring the $b\to c \tau \nu$ transitions and also testing the LFU (see Section~\ref{sec:FCCC}).  

Finally, more opportunities for studying $CP$ violation beyond the currently well-established observables at the CEPC are also expected due to the unique detector and kinematic conditions of this machine. However, additional theoretical inputs are needed to make specific recommendations.

\section{Charm and Strange Physics}
\label{sec:charmstrange}

The branching ratio of $Z$ boson decays to a pair of charm quarks is BR($Z\to c\bar{c})\simeq 12\%$ in the SM, which suggests that the CEPC $Z$-pole operation mode could also serve as a charm factory.  Given the CEPC's high luminosity, low background, and excellent detector performance, CEPC may significantly enhance the precisions of certain studies in charm physics. The charm quark may carry the information on NP, while the recent observation of $CP$ violation in charm decays~\cite{LHCb:2011osy,CDF:2012ous,Aaij:2019kcg} further strengthens the need for its study. 

One benchmark case for charm physics at the CEPC, akin to the discussion in Section~\ref{sec:FCCC} and~\ref{sec:FCNC}, could be semileptonic $c$-hadron decays. 
The FCNC charm decays are rare in the SM. Different from the down-type FCNC, where the quarks in the loops are dominated by top, the up-type FCNC is dominated by the loops of $b$, $s$ and $d$. The mass hierarchy for down-type quarks is relatively small compared to up-type quarks, yielding an even stronger GIM suppression. The sensitivity of rare charm decays to the NP is thus expected to be high~\cite{Burdman:2001tf,deBoer:2015boa,Fajfer:2015mia}. 
Nevertheless, due to large resonance contributions, 
it is much more challenging to estimate the hadronic effects in charm decays. The heavy quark expansion method usually adopted for estimating rare $b$ decays also becomes less reliable here. The short-distance physics in rare charm decays is thus difficult to probe through the BRs. Instead, we may consider the observables with a symmetry-protected suppression in the SM and essentially free of hadronic uncertainties. For example, we can test LFU in semileptonic $c\to u \ell^+\ell^-$ decays~\cite{Bause:2019vpr} and search for LFV decays such as $D \to \pi e\mu$ and $D_s \to K e\mu$~\cite{Guadagnoli:2022oxk}. We can also examine angular distributions in semileptonic $c\to u \ell^+\ell^-$ decays~\cite{DeBoer:2018pdx,Golz:2022alh} as well as di-neutrino decays,{\it e.g.}, $D \to \pi \nu \bar{\nu}$ and $D_s \to K \nu \bar{\nu}$~\cite{Bause:2020xzj,BESIII:2021slf}. Any observation of a non-standard effect in these measurements would be an evidence for the NP.

Moreover, it is important to examine hadronic $c$-hadron decays for charm physics. 
A preliminary qualitative estimate of the potential for studying charm physics at the CEPC can be made by estimating the charm particle yields. Table~\ref{tab:charm} shows the number of $D^0$'s and related fully hadronic final state decay modes collected by the LHCb experiment during its Run-2 period (approximately 6 fb$^{-1}$),  the expected data to be collected over the entire lifetime of the LHC and LHCb (approximately 300 fb$^{-1}$), as well as the number of corresponding decay modes expected to be collected at the CEPC in the 50MW SR power beam Z-pole operation mode. Additionally, we compared the number of relevant decay modes reconstructed in certain physics analyses at LHCb, and estimated number of selected events at CEPC. 
According to Ref.~\cite{DELPHI:1993gqe}, the efficiency for reconstructing and selecting charm meson decays at a typical electron-positron collider detector operating at Z-pole is at the level of $10\%$. Here, we assume for all decay channels at CEPC, the reconstruction and selection efficiencies are $10\%$.

\begin{table}[h]
\centering
{\begin{tabular}{cccc}
\toprule[1pt]
Decays & LHCb ( 6 fb$^{-1}$) & LHCb ( 300 fb$^{-1}$)  & CEPC (4 Tera $Z$) \\% &  Comments \\
\midrule
$D^{*+}$ & $4.7\times 10^{12}$ & $2.4\times 10^{14}$ & $4.6 \times 10^{11}$ \\
$D^0$ from $D^{*+}$ & $3.2\times 10^{12}$ & $1.6\times 10^{14}$ & $3.1 \times 10^{11}$ \\
\midrule
$D^{*+}\rightarrow (D^0 \to K^- K^+)\pi^+ $ & $1.6\times 10^{10}$ & $6.5\times 10^{11}$ & $1.3\times 10^9$ \\
$D^{*+}\rightarrow (D^0 \to \pi^- \pi^+)\pi^+ $ & $4.6\times 10^{9}$ & $2.3\times 10^{11}$ & $4.5 \times 10^8$ \\
$D^{*+}\rightarrow (D^0 \to K^- \pi^+)\pi^+ $ & $1.6\times 10^{11}$ & $6.3\times 10^{12}$ & $1.2\times 10^{10}$ \\
$D^{*+}\rightarrow (D^0 \to \pi^- \pi^+ \pi^0)\pi^+$ & $4.8\times 10^{10}$ & $2.4\times 10^{12}$ & $4.6 \times 10^{9}$ \\
$D^{*+}\rightarrow (D^0\to K^- \pi^+ \pi^0)\pi^+$ & $4.6\times 10^{11}$ & $2.3\times 10^{13}$ & $4.4 \times 10^{10}$ \\
\midrule
Reco. \& Sel. $D^0 \to K^- K^+ $ & $5.8\times 10^{7}$\,\cite{LHCb:2021vmn} & $2.9\times 10^{9}$ & $1.3\times 10^8$ \\
Reco. \& Sel. $D^0 \to \pi^- \pi^+ $ & $1.8\times 10^{7}$\,\cite{LHCb:2021vmn} & $9\times 10^{8}$ & $4.5 \times 10^7$ \\
Reco. \& Sel. $D^0 \to K^- \pi^+ $ & $5.2\times 10^{8}$\,\cite{LHCb:2021vmn} & $2.6\times 10^{10}$ & $1.2\times 10^{9}$ \\
Reco. \& Sel. $D^0 \to \pi^- \pi^+ \pi^0 $ & $2.5\times 10^{6}$\,\cite{LHCb:2023mwc} & $1.2\times 10^{8}$ & $4.6 \times 10^{8}$ \\
Reco. \& Sel. $D^0\to K^- \pi^+ \pi^0 $ & $1.9\times 10^{7}$\,\cite{LHCb:2023mwc} & $9.6\times 10^{8}$ & $4.4 \times 10^{9}$ \\
\bottomrule[1pt]
\end{tabular}}
\caption{ The number of ($D^0$) and related fully hadronic final state decay modes produced at the LHCb experiment during its Run-2 period (approximately 6 fb$^{-1}$) and the expected data to be produced over the entire lifetime of the LHC and LHCb (approximately 300 fb$^{-1}$), as well as the number of corresponding decay modes expected to be produced at the CEPC Z-pole operation mode. The total yields at LHCb is estimated using the cross-section measured by Ref.~\cite{LHCb:2015swx}, the reconstructed and selected events from LHCb are obtained from  Ref.~\cite{LHCb:2021vmn,LHCb:2023mwc}, while the reconstruction and selection efficiency at CEPC is assumed to be 10\%. }
\label{tab:charm}
\end{table}

As an experiment at a hadron collider, LHCb has the advantage of a high production cross-section for $D^0$ particles, which is a level unattainable by the CEPC in its Z-pole running mode. Therefore, despite the lower reconstruction efficiency, LHCb has a significant statistical advantage over CEPC for $D^0$ decays to fully charged hadronic final states. However, from the above comparison, it can be concluded that the LHCb experiment has particularly low efficiency for reconstructing $\pi^0$ particles, and for decay modes with $\pi^0$ final states, LHCb does not have a statistical advantage over CEPC in terms of reconstructed decay events. Therefore, conducting flavor physics research involving $\pi^0$ particles at the CEPC, such as searching for CP violation in the $D \to \pi \pi \pi^0$ decay, is promising in achieving measurement results comparable to LHCb's precision.

The $c$-hadron decays with a final state of $CP$ eigenstate, such as $D^0 \to K_S^0\pi^0$, $K_S^0\omega$ and $K_S^0\phi$, are valuable for extracting the $CP$ violation parameter values of $B\to DK$ decays and are hence important for determining the CKM angle $\gamma$~\cite{HeavyFlavorAveragingGroup:2022wzx} (see discussions in Section~\ref{sec:CPV} also). Regarding direct $CP$ violation in charm decays, one important target is to measure $\Delta \mathcal{A}_{\rm CP}\equiv \mathcal{A}_{\rm CP}(K^+K^-) -\mathcal{A}_{\rm CP}(\pi^+\pi^-)$~\cite{Cheng:2012xb,Cheng:2012wr,Chala:2019fdb,Dery:2019ysp,Bause:2020obd,Lenz:2023rlq}. The current experimental precision on this observable is $3\times 10^{-4}$~\cite{LHCb:2019hro}, which is expected to be improved to $\sim 3\times 10^{-5}$ at the LHCb Upgrade II~\cite{Bediaga:2018lhg}. The CEPC potential for this measurement, as well as its possible extension to channels such as $D^0\to K^{+}  K^{\ast -}$ and $\pi^+\rho^- $ or $a_0^+\pi^-$~\cite{Cheng:2019ggx,Cheng:2021yrn,Cheng:2024zul}, remains to be accessed. Finally, we would mention that the investigation of hadronic $c$-hadron decays may also benefit the study on $b$ physics, as $b$-hadrons decay significantly through the $b\to c+X$ EW transition, where the intermediate charm reconstruction is often necessary for the full event reconstruction.

Furthermore, for semileptonic or fully leptonic final state decays, especially those involving neutrino final states, the CEPC is expected to yield better results than hadronic colliders. Semileptonic or fully  leptonic decays of $D^\pm$ and $D_s^\pm$ mesons are among the simplest and best-understood probes of $c \to d$ and $c \to s$ quark flavor-changing transitions. 
The amplitude of such decays consists of the annihilation of the initial quark-antiquark pair ($c\bar{d}$ or $c\bar{s}$) into a virtual $W^\pm$ that subsequently materializes as an antilepton-neutrino pair ($l^+\nu_l$),
therefore can be used to determine the CKM matrix elemets $|V_{cd}|$ and $|V_{cs}|$. 

The Standard Model branching fraction of purely leptonic $D^\pm$ and $D_s^\pm$ decays is given by
\begin{equation}
   B(D^+_q \to l^+\nu_l) = \frac{G^2_F}{8\pi}\tau_{D_q} f^2_{D_q} | V_{cq}|^2 m_{D_q} m_{l}^2 \bigg(1- \frac{m_l^2}{m_{D_q}^2}\bigg)^2 
\end{equation}
where $m_{D_q}$ is the $D_q$ meson mass, $\tau_{D_q}$ is its lifetime, $m_l$ is the charged lepton mass, $G_F$ is the Fermi coupling constant. The parameter $f_{D_q}$ is the $D_q$ meson decay constant and parametrizes the overlap of the wave functions of the constituent quark and anti-quark, and $|V_{cq}|$ is the magnitude of the relevant CKM matrix element. With $f_{D_q}$ calculated precisely by theories like lattice QCD, $|V_{cq}|$ can be determined by measuring the  branching fraction of such decays. The current uncertainty is dominated by experimental uncertainties in these measurements, therefore CEPC has the potential to increase the precision given it may increase the yields of relevent decays by  several orders of magnitudes compare to current electron-positron experiments. 

Similarly, semileptonic decays $D_q \to \pi l^+\nu_l$ and  $D_q \to K l^+\nu_l$ can also be used to determine of $|V_{cq}|$. The precisions of branching fraction measurements are related to the experimental yields and theoretical calculation of the form factor. 
Nowadays, the dominant uncertainties of $|V_{cq}|$ measurements are from the theoretical calculation of the form factor, therefore, even CEPC can have several orders of magnitude higher yields of relevant decays, the ability to increase the  $|V_{cq}|$ precision through semileptonic decays is limited.\footnote{Hadronic $W$ decay could also play an important role when determining $|V_{cq}|$, see Sec.~\ref{sec:Vcb_W} for more details.}

A strange physics program can be also developed at the CEPC, as its tracking system is compatible with the lifetime of approximately $\mathcal {O}(100)$\,ps for many strange hadrons. A full-simulation study in~\cite{Zheng:2020qyh} has showcased promising reconstruction quality
for $K_S^0$ and $\Lambda$ decaying into a pair of charged tracks at the CEPC, featuring an efficiency $\gtrsim 80\%$ and a purity $\sim 95\%$. Differently, the higher-intensity experiments such as kaon factories prioritize the detection of longer-lived $K^\pm$ and $K_L^0$ particles, including the planned upgrades~\cite{NA62:2017rwk,KOTO:2018dsc,Goudzovski:2022vbt}. 
One benchmark of $K_S^0$ or $\Lambda$ decays at the CEPC is $K_S^0\to \mu\mu$. Currently, its BR is constrained to be $\mathcal{O}(100)$ times greater than its SM prediction $\sim 5\times 10^{-12}$~\cite{Dery:2021mct}. However, as this decay is rare, the NP may induce a sizable deviation from the SM prediction for its BR.  
With more than $10^{12}$ $K_S^0$ produced in hadronic $Z$ decays, the CEPC shall be sensitive to detect such kind of NP. 
Additionally, for the events of $Z\to s\bar{s}$, tagging the sign of strange quark  prior to the $K^0-\bar{K}^0$ mixing could be achieved. This is analogous to the $b$ or $c$ sign tagging. The measurements of $CP$ violation from the interference between $K_S^0$ and $K_L^0$ decays is thus possible, allowing the extrapolation of $|V_{td}V_{ts}|\sin(\beta+\beta_s)$~\cite{Dery:2021mct,Dery:2022yqc}. The CEPC sensitivities could be extended to rare decays with additional neutral particles, such as $K_S^0\to \mu\mu\gamma$ or $K_S^0\to \mu\mu\pi^0$. Due to the small rates for these channels, systematics should be evaluated carefully in simulation, which will be left to future study.

\section{\texorpdfstring{$\boldsymbol{\tau}$}{t} Physics}
\label{sec:tau}

With BR$(Z\to \tau^- \tau^+) \simeq 3\%$~\cite{ParticleDataGroup:2024cfk}, the CEPC is anticipated to yield $\simeq 1.2 \times 10^{11}$ $\tau^{+}\tau^{-}$ pairs~\cite{CEPCStudyGroup:2018ghi} -- see Table~\ref{tab:BYield}. The machine could thus produce five orders of magnitude more $\tau$ leptons than the LEP~\cite{ALEPH:2005ab}. 
The absence of accompanying particle showers and large boosts ($\gamma_{\tau} \simeq 26$) in $\tau$ production at the $Z$ pole renders these events particularly favorable for precise measurements and searches for rare or forbidden processes. The amount of $\tau$ events at the CEPC is nearly triple that expected at Belle~II ($\simeq 4.5\times 10^{10}$ $\tau$ pairs)~\cite{Belle-II:2018jsg,Banerjee:2022xuw}, while the reconstruction efficiency of the $\tau$ leptons and the identification of some particular decay modes could be significantly better due to the larger boost and the particle flow oriented detector design at CEPC. Similarly, the $\tau$ event yield at the CEPC is anticipated to be several times more than those at the proposed STCF project ($\simeq 3.5\times 10^{10}$ $\tau$ pairs in 10 years)~\cite{Shi:2020nrf,Achasov:2023gey}. These attributes make the CEPC an excellent environment for $\tau$ physics which could significantly contribute to the future of the field. The preliminary study in Ref.~\cite{Yu:2020bxh} investigated the tagging efficiency of inclusive $\tau$ hadronic modes using full simulation, obtaining an efficiency times purity of approximately $70\%$, ascertained from $W^{+}W^{-}$ events. Concurrently, research is being undertaken to scrutinize the exclusive tagging of prominent $\tau$ decay modes with the dual-readout calorimeter at the $Z$ pole~\cite{Talk3}. Preliminary results suggest that the average $\tau$-tagging accuracy of seven common decay modes is around $90\%$. Detector performances of $\tau$-tagging at the $Z$ pole with the aid of machine learning (ML) algorithms were also investigated in Ref.~\cite{Giagu:2022gmq}, where it was shown that deep learning models applied to the IDEA detector design can classify different $\tau$ decay modes with an average accuracy of 91\% and discriminate $\tau$ jets from QCD jets with an accuracy larger than 95\%.

Recent $\tau$ physics projections and potential measurements at the $Z$ pole of an $e^-e^+$ collider have been comprehensively summarized in Refs.~\cite{Dam:2018rfz,Tau_Talk_2023,lusiani_note}. These analyses, predominantly founded on rapid simulations within the FCC-$ee$ context, provide valuable benchmarks. These comprehensive studies focus on precision decay time and mass measurements, LFU tests in leptonic $\tau$ decays, and LFV searches in $\tau$ decays.

\begin{table}[t]
\centering
{\begin{tabular}{ccccc}
\toprule[1pt]
Measurement & Current & Belle II & FCC & CEPC prelim. \\% &  Comments \\
\midrule
Lifetime [sec] & $(2903 \pm 5)\times 10^{-16}$ &   & $\pm~6\times 10^{-18}$ & $\pm~7\times 10^{-18}$ \\
BR($\tau \to e \nu\bar\nu$)   & $(17.82\pm 0.04)\%$ &   & $\pm~0.003\%$  & $\pm~0.003\%$ \\ 
BR($\tau \to \mu \nu\bar\nu$)   & $(17.39\pm 0.04)\%$ &   & $\pm~0.003\% $ & $\pm~0.003\%$ \\ 
% $m_\tau$ [MeV]  & $ 1776.93 \pm 0.09$  &   & \makecell{$\pm$ 0.0023 (stat.)\\ $\pm$ 0.025 (syst.)} & \\ % old lusiani_note
$m_\tau$ [MeV]  & $ 1776.93 \pm 0.09$  &   & \makecell{$\pm$ 0.0016 (stat.)\\ $\pm$ 0.018 (syst.)} & \\ % new lusiani_2024_57pxj-6xd43
\midrule
BR($\tau \to \mu\mu\mu$) & $< 2.1\times 10^{-8}$ &  $3.6\times 10^{-10}$ & $1.4\times 10^{-11}$ &  $10^{-10}$  \\
BR($\tau \to \mu\gamma$) &    $<4.4\times 10^{-8}$ &  $6.9\times 10^{-9}$   & $1.2\times 10^{-9}$ & $10^{-10}$ \\
\bottomrule[1pt]
\end{tabular}}
\caption{Current~\cite{ParticleDataGroup:2024cfk} and projected sensitivities at Belle~II~\cite{Belle-II:2018jsg,Banerjee:2022vdd,Banerjee:2022xuw}, FCC-$ee$~\cite{Dam:2018rfz,Tau_Talk_2023,lusiani_2024_57pxj-6xd43} and CEPC~\cite{Talk_Dan}, for some $\tau$ physics measurements. For other LFV leptonic modes $\tau\to \ell^{(\prime)}\ell\bar\ell$, for which dedicated studies are still missing, we expect that the CEPC can achieve a sensitivity similar to that estimated for $\tau \to \mu\mu\mu$. Similarly, a sensitivity for $\tau \to e\gamma$ of the same order of magnitude as that for $\tau \to \mu\gamma$ can be plausibly reached.}
\label{tab:tau1}
\end{table}

%%%%
\subsection{LFV in \texorpdfstring{${\tau}$}{t} Decays}

LFV ${\tau}$ decays are complementary to LFV observables at higher energy scales (see Section~\ref{sec:Forbidden}), which highlights the theoretical importance of these modes in discriminating among different NP models~\cite{Celis:2014asa,Calibbi:2017uvl,Cornella:2021sby}.
Table~\ref{tab:tau1} displays current limits and FCC-$ee$ projections from Refs.~\cite{Dam:2018rfz,Tau_Talk_2023,lusiani_note} and CEPC preliminary estimates from Ref.~\cite{Talk_Dan} for the LFV leptonic $\tau$ decay mode $\tau\to \mu\mu\mu$ and  radiative one $\tau\to\mu\gamma$. At the CEPC, the former search is expected to be background free due to the excellent muon identification and momentum reconstruction. The LFV radiative $\tau$ decays are subject to a background from $Z\to\tau\tau\gamma$ followed by ordinary leptonic $\tau$ decays, which can be alleviated by precise measurements of photon momenta.
Given the excellent electron identification performance anticipated at the CEPC~\cite{Yu:2021pxc}, we expect that a sensitivity similar to the one displayed in Table~\ref{tab:tau1} for $\tau \to \mu\mu\mu$ could be achieved for other LFV leptonic decay modes, such as $\tau\to eee$, $\tau\to \mu ee$, , $\tau\to e \mu\mu$. Similarly, we expect the CEPC sensitivity to $\tau\to e\gamma$ to be comparable to that of $\tau\to\mu\gamma$. 
The CEPC prospects should be also compared with the future reach of Belle~II. Based on projections from the existing Belle results, the prospects for over 50 distinct LFV $\tau$ decay modes have been presented in Ref.~\cite{Belle-II:2018jsg} and recently revised in Ref.~\cite{Banerjee:2022xuw,Banerjee:2022vdd}. With 50~$\text{ab}^{-1}$ of collected data, Belle~II is expected to set limits in the $10^{-10}-10^{-9}$ range for most decay modes with a notable exception of the radiative decays, $\tau \to \ell\gamma$. %
The BRs for these decays can not be constrained much below than the $10^{-8}$ level, as a consequence of the difficult background from initial-state-radiation photons affecting $e^-e^+$ colliders running at energies around the $\Upsilon(\text{nS})$ resonances.
As we can see, a Tera-$Z$ factory can play a crucial role in discovering or constraining $\tau$ LFV by searching for radiative modes --- and, more in general, it will be complementary to Belle~II measurements, reaching a comparable sensitivity for the leptonic modes as shown in Table~\ref{tab:tau1}.

The CEPC sensitivity to LFV $\tau$ decays can be interpreted in terms of constraints on EFT operators. For instance, the limit $\text{BR}(\tau\to \mu\gamma) < 10^{-10}$ would imply a lower bound $\Lambda > 2800$~TeV on the energy scale of the LFV dipole operators \mbox{$\frac{1}{\Lambda^2}(\bar \mu \sigma^{\mu\nu} P_{L,R} \tau) \Phi F_{\mu\nu}$}, where $\Phi$ is the Higgs field and $F_{\mu\nu}$ is the EM field tensor. Similarly, \mbox{$\text{BR}(\tau\to \mu\mu\mu) < 10^{-10}$} would translate into the constraint $\Lambda > 44$~TeV on the scale of four-lepton LFV operators of the kind
\mbox{$\frac{1}{\Lambda^2}(\bar \mu \gamma^\mu P_{L,R} \tau) (\bar \mu \gamma_\mu P_{L,R} \mu)$}.

To achieve the sensitivities displayed in Table~\ref{tab:tau1}, the ECAL/PFA performance will be crucial, especially when the LFV final states have one or more neutral components. 
Besides the radiative decays, other examples of such a situation include
$\tau \to \ell h^0$ with $h^0=\pi^0(\to\gamma\gamma),~\eta(\gamma\gamma),~\eta^\prime(\pi^+\pi^-\eta)$, etc. Additionally, since LFV $\tau$  decays do not feature neutrinos, the $m_\tau$ invariant mass reconstruction plays a crucial rule in suppressing large backgrounds from ordinary $\tau$ decays. For explicit discussions of the $\tau \to \ell \gamma$ phenomenology at Tera-$Z$ factories, see~\cite{Dam:2018rfz,Talk_Dan}, while studies of the prospects for hadronic LFV $\tau$  decays, such as $\tau \to \ell\pi$ or $\tau\to \ell\rho$, are still lacking and will require future dedicated efforts. Finally, we notice that, in presence of a light NP boson $a$ with LFV couplings to SM leptons, decays such as $\tau\to\ell a$ can also occur. We will discuss such exotic LFV $\tau$ decay modes in Section~\ref{sec:BSM}.

%%%%
\subsection{LFU of \texorpdfstring{${\tau}$}{t} Decays}
\label{sec:tauLFU}

\begin{figure}[t]
    \centering
    \includegraphics[width=\textwidth]{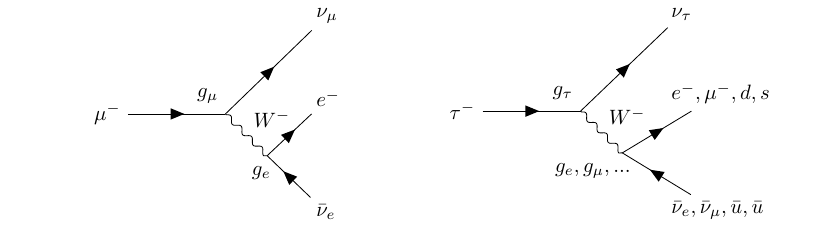}
    \caption{Illustrative Feynman diagrams for the muon and tau decays. In the SM, $g_e = g_\mu = g_\tau$ is predicted. 
    }
    \label{fig:tau}
\end{figure}

In Table~\ref{tab:tau1}, we report current accuracy and Tera-$Z$ prospects of measurements of the $\tau$ mass, lifetime, and the BRs of standard leptonic $\tau$ decays.
These are the crucial quantities to perform tests of the LFU in $\tau$ and $\mu$ decays. The SM predicts LFU of weak charged currents, that is, that the three lepton families couple with the same strength to $W^\pm$ bosons, {\it i.e.}, $g_e = g_\mu = g_\tau = g$, where $g = e /\sin\theta_{W}$ is the $SU(2)_L$ gauge coupling, cf.~Figure~\ref{fig:tau}. Inspecting the processes in this figure, one can see that the LFU prediction can be tested by measuring the following quantities:
\begin{align}
\label{eq:gmu/ge}
\left(\frac{g_\mu}{g_e}\right)^2 &= \frac{\text{BR}(\tau \to \mu \nu\bar\nu)}{\text{BR}(\tau \to e \nu\bar\nu)} \frac{f(m_e^2/m_\tau^2)}{f(m_\mu^2/m_\tau^2)} \frac{R_W^{\tau e}}{R_W^{\tau \mu}}\, ,
\\
\left(\frac{g_\tau}{g_{e/\mu}}\right)^2 & = \frac{\tau_\mu}{\tau_\tau} \left(\frac{m_\mu}{m_\tau}\right)^5
\frac{\text{BR}(\tau \to \mu/e\, \nu\bar \nu)}{\text{BR}(\mu \to e \nu\bar\nu)}
\frac{f(m_e^2/m_\mu^2)}{f(m_{\mu/e}^2/m_\tau^2)} 
\frac{R_W^{\mu e} R_\gamma^\mu}{R_W^{\tau \mu/e} R_\gamma^\tau}\,,%\quad (\ell=e,\mu),
\label{eq:gtau/gell}
\end{align}
where $\tau_{\tau/\mu}$ is the decaying lepton lifetime, $f(x)=1-8x+ 8x^3 -x^4-12 x^2\log x$ is a phase-space factor, $R_W^{\ell^\prime\ell} = 1 + \frac{3}{5} \frac{m_{\ell^\prime}^2}{m_W^2} + \frac{9}{5}\frac{m_{\ell}^2}{m_W^2}$ and $R_\gamma^\ell = 1 + \frac{\alpha(m_\ell)}{2\pi} \left(\frac{25}{4} -\pi^2\right)$ are EW and QED radiative corrections respectively~\cite{HeavyFlavorAveragingGroup:2022wzx,Pich:2013lsa}.\footnote{Numerically one obtains $R_\gamma^\mu/R_\gamma^\tau -1\simeq 8.0 \times 10^{-5}$~\cite{HeavyFlavorAveragingGroup:2022wzx}, $R_W^{\tau e}/R_W^{\tau \mu} -1 \simeq  -\frac{9}{5} \frac{m_\mu^2}{m_W^2} \simeq -3.1\times 10^{-6}$ and $R_W^{\mu e}/R_W^{\tau \ell} -1 \simeq  -\frac{3}{5} \frac{m_\tau^2}{m_W^2} \simeq -2.9\times 10^{-4}$.}
Using the purely leptonic processes in Figure~\ref{fig:tau}, the current experimental determination of the coupling ratios results to be compatible with LFU at the per mil level~\cite{Tau_poster_2024,HeavyFlavorAveragingGroupHFLAV:2024ctg}: 
\begin{align}
    \frac{g_\mu}{g_e}  = 1.0002\pm 0.0011\,,~~
    \frac{g_\tau}{g_e}  = 1.0018\pm 0.0014\,,~~
    \frac{g_\tau}{g_\mu}  = 1.0016\pm 0.0014\,.
    \label{eq:LFUtest}
\end{align}   

As muon physics quantities are known with high precision, the above uncertainties mainly stem from the measurements of $\tau$ leptonic BRs, lifetime and mass. The present relative uncertainties on $\text{BR}(\tau \to e \nu\bar\nu)$ and $\text{BR}(\tau \to \mu \nu\bar\nu)$ are respectively 2.2$\permil$ and 2.3$\permil$~\cite{ParticleDataGroup:2024cfk}, which yield an impact of $1.1\permil$ on the measurement of coupling ratios. As one can see, they constitute the source of largest uncertainty at the moment. The impact of $\tau_\tau$ on the uncertainty of $g_\tau/g_\ell$ is at a comparable level, namely $0.9\permil$, given its current $1.7\permil$ relative precision~\cite{ParticleDataGroup:2024cfk}. The current world average for $m_\tau$ is substantially more precise, with a relative error of $5\times 10^{-5}$~\cite{ParticleDataGroup:2024cfk}, which contributes to the uncertainty of $g_\tau/g_\ell$ only at the $0.2\permil$ level.

\begin{figure}[t]
    \centering
    \includegraphics[width=0.6\textwidth]{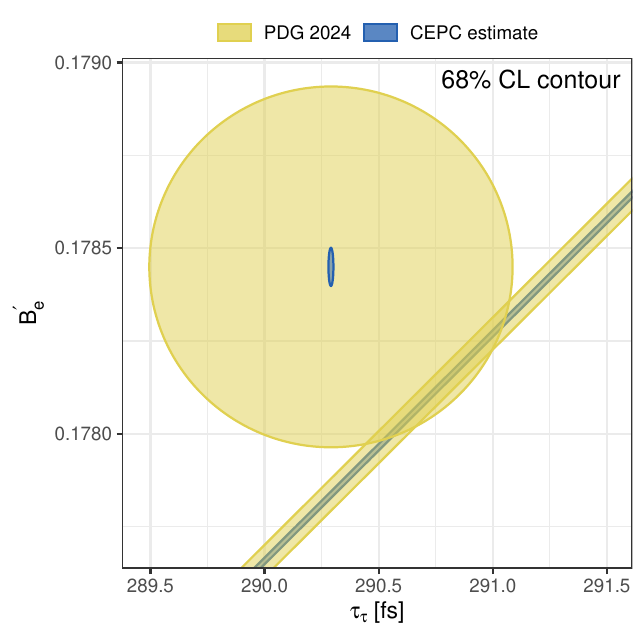}
    \caption{Expected precision of testing LFU by measuring the SM properties of the $\tau$ lepton in the CEPC era.
    The yellow (blue) areas correspond to the present (future) 68\%~CL allowed regions. The ellipse shows the measured value of the $\tau$ lifetime and $\mathcal{B}_e^\prime$. $\mathcal{B}_e^\prime$ is defined as the average of the measured value of $\text{BR}(\tau \to e\nu \bar\nu)$ and its predicted value obtained by setting $g_\mu = g_e$ in Eq.~(\ref{eq:gmu/ge}). The diagonal band displays the SM prediction, based on taking $g_\tau = g_\ell$ in  Eq.~(\ref{eq:gtau/gell}). The width of the band is due to the experimental uncertainty on $m_\tau$.
    This plot is based on Ref.~\cite{Tau_poster_2024,HeavyFlavorAveragingGroupHFLAV:2024ctg} -- see also~\cite{lusiani_note,Dam:2018rfz,Tau_Talk_2023}.}
    \label{fig:LFU_tau_cepc}
\end{figure}

As shown in Table~\ref{tab:tau1}, an improvement by a factor of few for the precision of the $m_\tau$ measurement is possible at Tera-$Z$ factories, such that $m_\tau$ would be known precisely enough to allow to perform the LFU test in Eq.~(\ref{eq:gtau/gell}) with an uncertainty at the $0.1\permil$ level or below.  
Moreover, substantial improvements on the determination of $m_\tau$ are also to be expected at BESIII~\cite{BESIII:2020nme}, Belle~II~\cite{Belle-II:2018jsg} -- which recently released the most precise single measurement, $m_\tau = 1777.09 \pm 0.08(\text{stat}) \pm 0.11(\text{syst})$~MeV~\cite{Belle-II:2023izd} -- and at STCF~\cite{Achasov:2023gey}. Therefore, $m_\tau$ is not expected to be a limiting factor for an improved LFU test. 
As suggested by Table~\ref{tab:tau1}, Tera-$Z$ factories can play a major role for what concerns the measurements of the BRs and lifetime. Actually, the current world average for $\text{BR}(\tau \to \ell \nu\bar\nu)$ is dominated by the LEP measurements that are statistically limited, although the systematic errors are typically just a factor of two smaller than the statistical ones~\cite{ParticleDataGroup:2024cfk}.\footnote{A precise measurement of the ratio ${\text{BR}(\tau \to \mu \nu\bar\nu)}/{\text{BR}(\tau \to e \nu\bar\nu)}$ has been recently published by Belle~II~\cite{Belle-II:2024vvr}, which contributed to the improved $g_\mu/g_e$ measurement displayed in Eq.~(\ref{eq:LFUtest}).} 
Differently, the measurements of $\tau_\tau$ at the LEP have comparable statistical and systematic uncertainties, which are respectively twice and three times larger than those of the most precise measurement of $\tau_\tau$ from the  Belle experiment~\cite{Belle:2013teo}. The $\tau_\tau$ measurements however are simpler at a Tera-$Z$ factory than those at Belle, given the large boost stemming from $m_Z \gg m_\tau$, while the statistics is not going to be a concern at the CEPC also. So the main challenge will be to control the systematics on $\tau_\tau$ and $\text{BR}(\tau \to \ell \nu\bar \nu)$ at a level better than the LEP ones.

To achieve such a goal is possible. As the systematics at the LEP are mainly caused by its sample size, with much higher luminosities, the CEPC may reduce the systematics by one order of magnitude for the $\text{BR}(\tau \to \ell \nu\bar \nu)$ measurements and to a level comparable to the statistical ones for the $\tau_\tau$ analyses~\cite{Dam:2018rfz, Tau_Talk_2023, lusiani_note}. 
The LFU test summarized in Eq.~(\ref{eq:LFUtest}) thus may reach a precision level of $\pm 10^{-4}$. Figure~\ref{fig:LFU_tau_cepc} illustrates the impact of measuring the SM properties of the $\tau$ lepton with such a precision.
Reaching this level of precision would make the CEPC very sensitive to LFUV NP scenarios, such as those discussed in the literature addressing the $R_{D^{(\ast)}}$ anomaly~\cite{HeavyFlavorAveragingGroup:2022wzx}
and, more generally, to models with new dynamics coupled mainly to the third generation~\cite{Allwicher:2023shc}. 
As shown, {\it e.g.}, in Refs.~\cite{Feruglio:2017rjo,Allwicher:2021ndi}, the tests of LFU in the $\tau$ sector are already providing important constraints on such models. 

As another example of the discovery potential of these measurements, we can consider the operator $ \frac{1}{\Lambda^2} i(\vpjt)(\bar L_3 \tau^I \gamma^\mu L_3)$ (with $L_3 \equiv (\nu_\tau ,~\tau_L)^T$), which only involves (left-handed) $\tau$ leptons and is  flavor-conserving. 
The presence of such an operator would induce a shift $g_\tau = g\left(1+ \frac{v^2}{\Lambda^2}\right)$~\cite{Jenkins:2017jig}, where $v \simeq 246$~GeV is the vacuum expectation value of the Higgs field, without affecting the couplings to electrons and muons, $g_e = g_\mu = g$. A precision of 0.1$\permil$ level in the determination of $g_\tau/g_\ell$ would then test a NP sector generating such an operator up to $\Lambda \approx 20$~TeV.

%%%%
\subsection{Opportunities with Hadronic \texorpdfstring{$\tau$}{t} Decays}

Hadronic $\tau$ decays represent an important branch of $\tau$ physics. Currently, many leading constraints on the branching fractions of the various exclusive hadrnoic $\tau$ decay channels are set up by the LEP~\cite{Ackerstaff:1998yj,ALEPH:2005qgp}, including $\tau^- \to \pi^- K_L^0 K_S^0 \nu_\tau $~\cite{ALEPH:1997trn}, $\tau^- \to K^- 3\pi^0 \nu_\tau$~\cite{L3:1994hjr}, $\tau^- \to \pi^-\pi^0 \nu_\tau$~\cite{ALEPH:2005qgp} and so forth. The CEPC's performance in these measurements, especially for the processes with relatively high hadron multiplicity ({\it e.g.}, 3 and 5 hadrons) and in a large hadron invariant mass region, is expected to exceed the LEP. It is promising that CEPC has a good opportunity to provide more precise measurements for a significant portion of inclusive and exclusive hadronic $\tau$ decay channels~\cite{ParticleDataGroup:2024cfk}, which highlights the advantage of high-energy $e^-e^+$ colliders over other flavor factories in this field.

Hadronic $\tau$ decays bring in new physical opportunities while many of them are yet to be explored, especially at the CEPC. For example, inclusive hadronic decays are crucial for extracting the strong coupling constant $\alpha_s(m_\tau)$~\cite{Pich:2013lsa,Pich:2020qna}, which is currently limited by uncertainties in the large-recoil region. Precise measurements on the invariant-mass distributions of the hadronic systems in exclusive $\tau$ decays can tightly constrain the properties of different types of hadron resonances, which will in turn provide valuable inputs to study the $CP$ violation in hadronic $\tau$ decay processes. Another example is the measurement of $\tau \to K(+X)$ decays, which is useful for the determination of $|V_{us}|$. Then such measurements can offer an alternative important way to address the Cabbibo anomaly~\cite{HeavyFlavorAveragingGroup:2022wzx}, {\it i.e.}, the unitarity violation of the first row of the CKM matrix. Additionally, polarization measurements of the $\tau$ leptons produced via $Z$ decays can provide additional tests of the LFU and relevant inputs for the EWPOs global fit~\cite{DELPHI:2000wje,L3:2000vgx,Tenchini:2016puu}. These measurements are often performed in the hadronic decays $\tau\to \rho\nu$ and $\tau\to \pi\nu$. For more theoretical insights and details on hadronic $\tau$ decays, see~\cite{Pich:2013lsa,Pich:2020qna}. 

Hadronic $\tau$ decays could be also employed to improve the measurements of the currently weakly-constrained  $\tau$ anomalous magnetic moment ($a_\tau$) and electric dipole moment ($d_\tau$), along the lines taken for Belle~II in, {\it e.g.}, Refs.~\cite{Chen:2018cxt,Crivellin:2021spu}.
Recently the CMS reported the best limit so far on the $\tau$ magnetic moment, {\it i.e.}, $a_\tau = 0.0009^{+0.0032}_{-0.0031}$~\cite{CMS:2024skm}.
Before this progress, the tightest constraint of $-0.052 < a_\tau < 0.013$ at 95\%~CL was obtained by the DELPHI experiment that used $\tau$ lepton pairs produced from the photon-photon collisions off the $Z$~pole~\cite{DELPHI:2003nah}. These experimental limits are still a factor of few away from the SM prediction, {\it i.e.}, $a_\tau^\textsc{sm}= 0.00117721(5)$~\cite{Eidelman:2007sb}, while it has been shown that $a_\tau$ could be tested at the level of $\sim 10^{-6}$ in the Belle~II experiment~\cite{Chen:2018cxt,Crivellin:2021spu}. The potential role of future $e^-e^+$ colliders in this endeavor needs to be studied. 

Furthermore, the large number of $\tau^- \tau^+$ pairs produced at the $Z$-pole and the improved reconstruction efficiency of $\tau$ leptons, both of which are expected for the CEPC, will allow to efficiently constrain $\tau$ weak-electric dipole moment ($d^{w}_\tau$) defined in, {\it e.g.}, Ref.~\cite{Bernabeu:1993er}. In the SM, this $CP$-violating observable is predicted to arise at two-loop level and its value is hence negligibly small~\cite{Bernreuther:1988jr}.
Any experimental observation of a non-vanishing $d_{\tau}^{w}$ value would be a clear NP signal. Using the tau polarization method~\cite{Bernabeu:1993er,Stiegler:1992sy}, the ALEPH has provided the best limit on $d_{\tau}^{w}$ so far, with $\mathrm{Re}[d_{\tau}^{w}] < 5.0\times 10^{-18} \, e\,\mathrm{cm}$ and $\mathrm{Im}[d_{\tau}^{w}] < 1.1 \times 10^{-17}\, e\,\mathrm{cm}$ at 95\%\,CL~\cite{ALEPH:2002kbp}. With $1.2\times 10^{11}$ $\tau^-\tau^+$ pairs at the CEPC and optimal observables introduced in Refs.~\cite{Davier:1992nw,Diehl:1993br,Bernreuther:1993nd,Bernreuther:2021elu}, a preliminary analysis indicates that the statistical uncertainty for measuring $\mathrm{Re} [d^{w}_{\tau}]$ and $\mathrm{Im} [d^{w}_{\tau}]$ could be reduced to $\sim 10^{-21}\, e\, \mathrm{cm}$, significantly superior to the current best limit.

Hadronic $\tau$ decays can be employed to study other $CP$ violation observables~\cite{Tsai:1996ps,Kuhn:1996dv,Kimura:2009pm,Bigi:2012kz,Kiers:2012fy}. One benchmark mode is the decay $\tau\to K_S^0\pi\nu_\tau$. 
 It has been shown~\cite{Bigi:2005ts,Grossman:2011zk} that the well-established $CP$ violation in $K^0-\bar{K}^0$ mixing can induce an $\mathcal{O}(10^{-3})$ asymmetry  between the rates of $\tau^+$ and $\tau^-$ decays. Furthermore, NP may provide contribution interfering with the SM amplitudes.  
 Assuming that the hadronic $\tau$ decays indeed receive additional contributions from NP degrees of freedom, which carry different weak and strong phases from that of the SM contribution, one can then construct $CP$-violating observables in terms of the interference between the SM and NP amplitudes. Due to the linear dependence on the NP amplitude, these observables may have a sensitivity to NP comparable to processes that are forbidden or strongly suppressed within the SM, such as $\tau\to \mu\gamma$ and the electric dipole moment of leptons, which are usually quadratic in the NP amplitude~\cite{Bigi:2012kz}. 
Searches for $CP$ violation in the decay $\tau\to K_S^0\pi\nu_\tau$ have been performed in several experiments. After initial null results from CLEO~\cite{CLEO:1998lwx,CLEO:2001lhp} and Belle~\cite{Belle:2011sna}, the BaBar collaboration reported in Ref.~\cite{BaBar:2011pij} the observation of anomalous $CP$ violation based on the difference between $\tau^+$ and $\tau^-$ decay rates. This measurement is in tension with the SM prediction~\cite{Bigi:2005ts,Grossman:2011zk,Chen:2019vbr} with a significance of $2.8\,\sigma$. The result prompted a number of NP explanations involving, {\it e.g.}, the introduction of non-standard tensor interactions~\cite{Cirigliano:2017tqn,Rendon:2019awg,Chen:2019vbr}. 

$CP$ violation in $\tau\to K_S^0\pi\nu_\tau$ can be also measured through angular distributions of its decay products, 
even if their rest frame can not be exactly reconstructed~\cite{Kuhn:1996dv}. 
The observable is defined as~\cite{Belle-II:2018jsg}
\begin{equation}\label{eq:ACP_i}
A_i^{CP}=\frac{\int_{s_{1,i}}^{s_{2,i}}\int_{-1}^{1} \cos\alpha \left[\frac{d^2 \Gamma(\tau^-\to K_S^0\pi^-\nu_\tau)}{ds\,d\cos\alpha} -\frac{d^2 \Gamma(\tau^+\to K_S^0\pi^+\bar{\nu}_\tau)}{ds\,d\cos\alpha}\right] ds\,d\cos\alpha}{\frac{1}{2}\int_{s_{1,i}}^{s_{2,i}}\int_{-1}^{1}\left[\frac{d^2 \Gamma(\tau^-\to K_S^0\pi^-\nu_\tau)}{ds\,d\cos\alpha}+\frac{d^2 \Gamma(\tau^+\to K_S^0\pi^+\bar{\nu}_\tau)}{ds\,d\cos\alpha}\right] ds\,d\cos\alpha}\,,
\end{equation}
which is the difference between the angular differential decay widths of $\tau^-$ and $\tau^+$ weighted by $\cos\alpha$, where $\alpha$ is the angle between the directions of the $K$ and $\tau$ momenta in the $K\pi$ rest frame. This observable can be analyzed in individual bins of the $K\pi$ invariant mass squared ($s$), with the $i$-th bin defined by an interval $[s_{1,i}, s_{2,i}]$~\cite{Belle:2011sna}. 
As discussed above, $CP$ violation in $K^0-\bar{K}^0$ mixing induces a non-vanishing effect for this observable~\cite{Chen:2020uxi,Chen:2021udz}. Direct $CP$ violation then arises from, {\it e.g.},     
the interference between an S-wave from exotic scalar-exchange diagrams and a P-wave from SM $W$-exchange diagrams, provided that the couplings of the exotic scalars with fermions are complex. This possibility has been studied for both polarized and unpolarized beams~\cite{Tsai:1996ps,Kuhn:1996dv}. While still plagued by large experimental uncertainties, the current constraints could be significantly improved with more precise measurements expected to be performed at Belle II~\cite{Belle-II:2018jsg}, as well as at future Tera-$Z$~\cite{Pich:2020qna} and STCF~\cite{Sang:2020ksa} facilities.

\section{Flavor Physics in \texorpdfstring{$\boldsymbol{Z}$}{Z} Boson Decays} 
\label{sec:Zdecay}

The LFV and LFU can be tested in multiple ways at the CEPC, which vary from heavy-flavored fermion to $Z$ and Higgs boson decays. Among them the $Z$ boson decays are of particular interest for the Tera-$Z$ events expected in the CEPC $Z$-pole run. In addition to these effects, the $Z$ boson decays can be also applied for examining QCD factorization theorem and investigating hadron inner structure. We will explore these topics in Section~\ref{sec:Forbidden} and Section~\ref{sec:HadronicZ}, respectively.

\subsection{LFV and LFU}
\label{sec:Forbidden}

 Let us consider first the searches for LFV in $Z$ boson decays. In Table~\ref{tab:ZLFV}, we summarize the current limits on $Z\to\ell\ell^\prime$ and the projected sensitivities at the high-luminosity run of the LHC (HL-LHC) and at the FCC-$ee$ and CEPC $Z$ factories. While the current limits can be improved at HL-LHC, such an improvement is expected to be within one order of magnitude as a consequence of the large background from $Z\to \tau\tau$. This background is difficult to deal with at hadron colliders, however it could be well addressed at a machine like CEPC due to its expected excellent identification of $\tau$ leptons.   
 Moreover, for an $e^-e^+$ collider the precise knowledge on the initial state kinematics, such as the constraint on the di-lepton invariant mass, $m^2_{\ell \ell^\prime} = m_Z^2$, is only limited by the beam energy spread, in contrast to hadronic machines where this constraint is instead limited by the large width of the $Z$ boson. With the expected 
 high accuracy in measuring the momenta of the tracks and good control of the beam energy, this may benefit a lot the event reconstruction. Finally, the sensitivity to BR($Z\to\mu e$) is mainly limited by the background from $Z\to\mu\mu$ with one of the muons being misidentified as an electron in the ECAL~\cite{Dam:2018rfz}. Hence, the expected precise PID at the CEPC could be another important advantage.

As demonstrated in Refs.~\cite{Calibbi:2021pyh,Altmannshofer:2023tsa}, although the allowed rate of $Z \to \mu e$ generally lies well below the expected sensitivity,\footnote{Barring unlikely accidental cancellations among different contributions, searches for LFV in muon decays set the indirect constraint ${\rm BR}(Z \to \mu e) \lesssim 10^{-12}$~\cite{Calibbi:2021pyh}.} a Tera-$Z$ factory, with its $\mathcal{O}(10^{12})$ $Z$ decays, holds promise for $Z \to \tau \ell$ decays. Their rate can be as large as $\text{BR}(Z \to \tau \ell) \approx 10^{-7}$ without violating the indirect limits set by the LFV measurement in $\tau$ decays~\cite{Calibbi:2021pyh}.

\begin{figure}[t]
    \centering
    \includegraphics[width=0.9\textwidth]{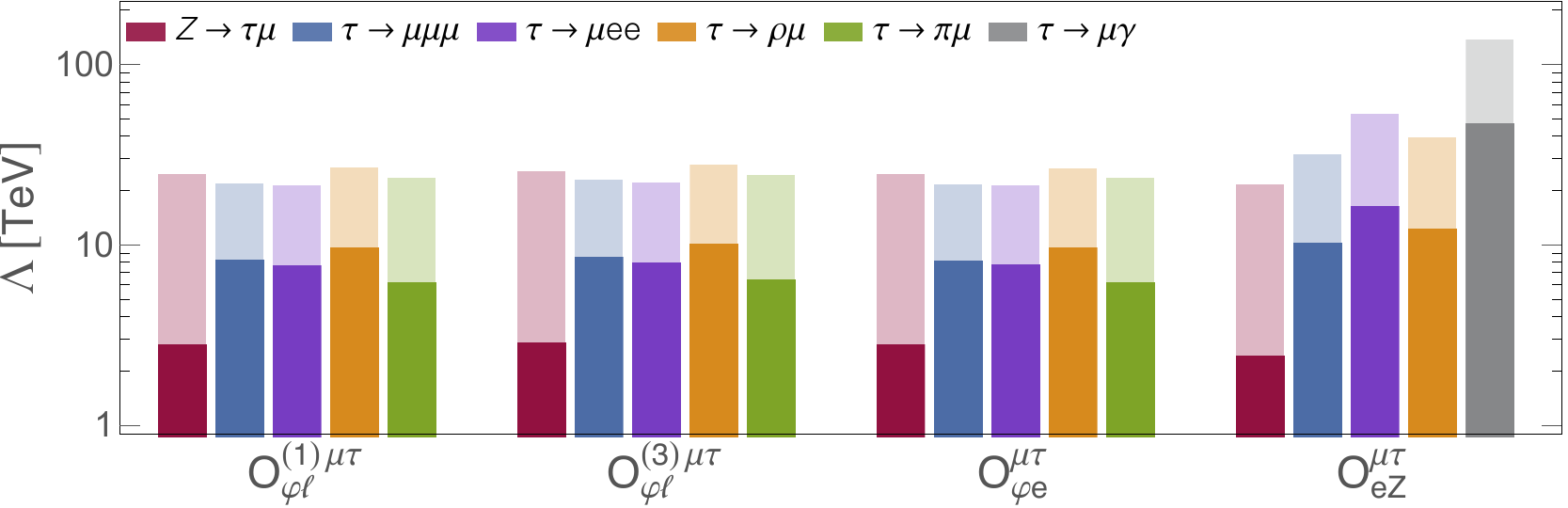}
    \caption{Sensitivity reach for probing the NP scale of the LFV operators in Eq.~(\ref{eq:HLops}) and Eq.~(\ref{eq:dipoleops}). Here the current bounds (dark-colored bars) are set by ATLAS~\cite{ATLAS:2021bdj} ($Z\to\tau\mu$) and $B$ factories~\cite{ParticleDataGroup:2024cfk} (LFV $\tau$ decays),
        and the projected sensitivities (light-colored bars) are based on searches for $Z\to\tau\mu$ at the CEPC $Z$ pole run with 100~$\text{ab}^{-1}$ 
    and $\tau \to \mu$ transitions at Belle II with 50~$\text{ab}^{-1}$~\cite{Belle-II:2018jsg}, see Tables~\ref{tab:tau1} and \ref{tab:ZLFV}. The Wilson coefficients have been set equal to one uniformly. This plot is taken from Ref.~\cite{Calibbi:2021pyh}.} 
    \label{fig:ZLFV}
\end{figure}

In Ref.~\cite{Calibbi:2021pyh}, the present and future limits on LFV $Z$ decays have been interpreted as constraints on the NP energy scale within the dimension-6 SM EFT (SMEFT)  $\mathcal{H}_\textsc{smeft} \supset \frac{1}{\Lambda^2}\sum_a C_a O_a$ \cite{Buchmuller:1985jz,Grzadkowski:1995rx}, where
\begin{align}
 O_{\vp \ell }^{(1)} \equiv i(\vpj)(\bar L\gamma^\mu L)\,,\ 
O_{\vp \ell}^{(3)} \equiv i(\vpjt)(\bar L \tau^I \gamma^\mu L)\,, \ 
 O_{\vp e} \equiv i(\vpj)(\bar E \gamma^\mu E) 
 \label{eq:HLops}
\end{align}
are Higgs current operators and 
\begin{eqnarray}
 O_{eZ} \equiv \left(\sw O_{eB} + \cw O_{eW}\right)\,
\label{eq:dipoleops}
\end{eqnarray}
is a linear combination of the dipole operators 
\begin{align}
O_{eB} \equiv (\bar L \sigma^{\mu\nu} E) \Phi B_{\mu\nu}\,,\quad
O_{eW} \equiv (\bar L \sigma^{\mu\nu} E) \tau^I \Phi W_{\mu\nu}^I\,.
\end{align}
Here $L$ and $E$ are, respectively, the SM doublet and singlet lepton fields (with flavor indices omitted), $\Phi$ is the Higgs doublet, $B_{\mu\nu}$ and $W^I_{\mu\nu}$ ($I=1,2,3$) are, respectively, the $U(1)_Y$ and $SU(2)_L$ field strengths, $\tau^I$ are the Pauli matrices, and $\vpj$ is defined as $\Phi^\dag (D_\mu \Phi)-(D_\mu \Phi)^\dag \Phi$.
In Figure~\ref{fig:ZLFV}, we illustrate the NP scale associated to these LFV operators that the CEPC can reach by searching for $Z \to \tau \mu$ if a sensitivity such as in Table~\ref{tab:ZLFV} is achieved. As one can see, NP scales up to $20-30$~TeV can be probed at the CEPC. Such a performance is comparable with that of Belle II through searches for LFV $\tau$ decays -- assuming an integrated luminosity of 50~$\text{ab}^{-1}$. Searches for $Z \to \tau e$ are expected to deliver similar sensitivities~\cite{Calibbi:2021pyh}.

\begin{table}[t!]
\centering
{\begin{tabular}{cccccc}
\toprule[1pt]
Measurement & Current & HL-LHC & FCC & CEPC prelim. \\% &  Comments \\
\midrule
 BR($Z \to \tau \mu $) & $<6.5\times 10^{-6}$ & $1.4\times 10^{-6}$ &  $10^{-9}$ & $10^{-9}$ \\% &  \multirow{2}{*}{$\tau\tau$ bkg,  $\sigma(p_{\rm track})$ \& $\sigma(E_{\rm beam})$ limited} \\
 BR($Z \to \tau e $) & $<5.0\times 10^{-6}$ & $1.1\times 10^{-6}$  & $10^{-9}$ &  & \\
 BR($Z \to \mu e $) & $<2.62\times 10^{-7}$ & $5.7\times 10^{-8}$  & $10^{-8}-10^{-10}$ & $10^{-9}$ \\% &  PID limited\\
\bottomrule[1pt]
\end{tabular}}
\caption{Current 95\%~CL limits on LFV in $Z$ decays~\cite{ATLAS:2021bdj,ATLAS:2022uhq} and projected sensitivities at HL-LHC and the FCC-$ee$~\cite{Dam:2018rfz} and CEPC~\cite{Talk_Dan} $Z$ factories (see also \cite{Altmannshofer:2023tsa}). For HL-LHC, we naively scaled the current limits, which were obtained by ATLAS employing $139~\text{fb}^{-1}$ of data~\cite{ATLAS:2021bdj,ATLAS:2022uhq}, to the target luminosity $3000~\text{fb}^{-1}$.}
\label{tab:ZLFV}
\end{table}

The study in Ref.~\cite{Altmannshofer:2023tsa} considers an alternative probe at future electron-positron colliders: the non-resonant production of $\tau\mu$, and examines the CEPC's expected sensitivity to its signals. This signal exhibits a characteristic dependence on the center-of-mass energy, depending on the nature of the dominant LFV operator.  
The contributions of operators containing the $Z$ boson, Eq.~(\ref{eq:HLops}) and  Eq.~(\ref{eq:dipoleops}), are resonantly enhanced on the $Z$ pole. At higher energies, dipole interactions as in Eq.~(\ref{eq:dipoleops}) yield a cross section that remains constant for large values of the center-of-mass energy squared $s$, while the Higgs current interactions in Eq.~(\ref{eq:HLops}) result in a cross section that decreases as $1/s$ for large $s$. In contrast, contributions to the non-resonant $e^{+}e^{-} \to \tau\mu$ cross section from contact interactions -- {\it i.e.}, 4-fermion operators such as $(\bar e \gamma_\mu P_X e)(\bar \mu \gamma^\mu P_Y \tau)$ ($X,Y=L,R$) -- increases linearly with $s$. 
Overall, the Tera-$Z$ factories can test NP scales up to  $\mathcal{O}(10)~\text{TeV}$, rivaling the sensitivities of searching for the LFV tau decays at Belle~II. The framework provided by this study enables a disentanglement of contributions from different operators, exploiting the complementarity of searches at various center-of-mass energies. Additional diagnostic measures could be provided also by measurements of forward-backward asymmetry or $CP$ violation.

The searches for flavor violation in the $Z$ boson decays can be extended to the quark sector also. The flavor-violating hadronic $Z$ decays are absent at tree level in the SM and thus can serve as an efficient probe for the NP that significantly enhances such decays. Given the nominal yields of 4 Tera $Z$ boson at the CEPC and employing the method for particle ID  in~\cite{Liang:2023wpt}, we expect the 95\% CL upper limits to reach $10^{-7}$ for the $Z\to bs$ and $Z\to bd$ decays, $3\times 10^{-7}$ for $Z\to cu$, and $7\times 10^{-7}$ for $Z\to sd$, in statistics. These limits are orders of magnitudes stronger than the current ones, and especially for the $Z\to bs$ mode, only twice larger than the SM prediction. Calibration and systematic control will be the major challenges in achieving the expected precision for these measurements. This remains an open question that requires dedicated efforts to address.

The LFU tests have been discussed in the FCCC and FCNC $b$-hadron decays in Section~\ref{sec:FCCC} and Section~\ref{sec:FCNC}. These tests can be also performed in $Z$ decays. Currently, the LFUV for the $Z$ boson couplings have been constrained to per mil level~\cite{ALEPH:2005ab}: 
\begin{align}
\frac{{\rm BR}(Z\to \mu^+\mu^-)}{{\rm BR}(Z\to e^-e^+)} = 1.0009 \pm 0.0028\,, \quad
\frac{{\rm BR}(Z\to \tau^- \tau^+)}{{\rm BR}(Z\to e^-e^+)} = 1.0019 \pm 0.0032\,.
\end{align}
While the used data sets are old ($1.7\times 10^7$ $Z$ events at LEP, and $6\times 10^5$ $Z$ events with polarized beams at SLC), these constraints have strongly limited the space for NP models aiming to address the anomalies in FCCC and FCNC semileptonic $B$ decays~\cite{Feruglio:2017rjo}.
In addition, an enhanced rate of $Z\to\mu^+\mu^-$ is predicted within a wide class of NP models addressing the muon $g-2$ anomaly~\cite{Crivellin:2021rbq}. 
In the future, reaching a precision of $\mathcal O(10^{-4})$ in the measurements of ${\rm BR}(Z\to \ell^+\ell^-)$ will allow to probe the scale $\Lambda$ of the flavor-conserving components of the operators in Eq.~(\ref{eq:HLops}) involving $\tau$ leptons up to $\approx 20$~TeV.  
Similarly, a $Z$ LFU test with such a level of precision would reach $\Lambda \approx 10$~TeV for the semileptonic operator $(\bar Q_3 \gamma_\mu Q_3)(\bar L_3 \gamma_\mu L_3)$ only comprising third generation fermions, which can also provide relevant contributions to other LFU observables such as $R_{D^{(*)}}$~\cite{Feruglio:2017rjo}, cf.~Eq.~(\ref{eq:RHc}).  
Notably, in a Tera-$Z$ factory these prospected measurements would be mainly limited by systematics, while statistical and systematic errors are of the same order of magnitude at the LEP. Hence, further scrutiny on these systematic uncertainties is necessary to assess the CEPC capability of performing the tests of LFU in $Z$ decays. 
The theoretical uncertainties of the SM prediction also need to be estimated.

\subsection{Factorization Theorem and Hadron Inner Structure}
\label{sec:HadronicZ}

During the $Z$-factory phase of CEPC, one can also explore exclusive hadronic $Z$ decays, such as $Z \to J/\psi\gamma$ and $Z\to \pi^+\pi^-$, which have never been observed before. Different from heavy-flavor physics on the bottom and charm mass scales, these decays occur at the EW scale and usually have a better convergence behavior. This may greatly benefit the examination of QCD factorization theorem and the investigation of hadron inner structure.

The factorization formalism for exclusive decays~\cite{Beneke:1999br,Beneke:2000ry,Keum:2000wi,Lu:2000em} is standard. 
Its application to $B$ decays however is hindered by large power corrections of $\mathcal{O}(\Lambda^n_\text{QCD}/m_b^n)$ where the convergence is slow due to the smallness of $b$ quark mass. This theorem, however, can be circumvented for exclusive $Z$ decays, as the large $Z$ mass yields a more efficient suppression for these power corrections. The exclusive $Z$ decays thus 
can serve as a touchstone for examining the factorization formalism. 
The benchmark channel $Z\to J/\psi\,\gamma$, with a BR $\sim 10^{-7}$~\cite{Grossmann:2015lea}, could be measured at the CEPC~\cite{Talk_Dan} with a precision much higher than the current limit of $<1.4\times 10^{-6}$~\cite{Sirunyan:2018fmm}. The two-meson-only $Z$ decays have an even smaller BR of $\lesssim 10^{-11}$~\cite{Cheng:2018khi,Bergstrom:1990bu}. While a discovery would be unattainable at both the LHC and the CEPC, the CEPC is expected to establish much more stringent upper limits for their event rates.

\begin{table}[t]
    \centering
    \resizebox{1\textwidth}{!}{\begin{tabular}{ccccccc}
    \toprule[1pt]
 Measurement & SM Prediction & Current Limits~\cite{ParticleDataGroup:2024cfk} & CEPC prelim. \\  
    \midrule
BR($Z\to \pi^+\pi^-$)  & $(8.3\pm0.5)\times10^{-13}$~\cite{Cheng:2018khi} & - & $\mathcal{O}(10^{-10})$ \\% & $\sigma(\vec{p}_{\rm track})$ limited, good PID\\
BR($Z\to \pi^+\pi^-\pi^0$) & - &  - & $\mathcal{O}(10^{-9})$ \\
BR($Z\to J/\psi \gamma $) & $(8.02\pm0.45)\times10^{-8}$~\cite{Grossmann:2015lea} & $<1.4\times 10^{-6}$  & $10^{-9}-10^{-10}$ \\
BR($Z\to \rho \gamma $) & $(4.19\pm0.47)\times10^{-9}$~\cite{Grossmann:2015lea} & $<2.5\times 10^{-5}$ \\
\bottomrule[1pt]
    \end{tabular}}
    \caption{Preliminary estimates on the Tera-$Z$ sensitivities for measuring exclusive hadronic $Z$ decays~\cite{Talk_Dan}, with the CEPC full simulation samples. The exact results and systematic effects remain to be explored. 
    }
    \label{tab:Zexclusive}
\end{table}

The radiative decay $Z\to M\gamma$ can serve as a tool to investigate the internal structure of light mesons. Its information is a crucial theoretical input for factorization formulae, typically formulated as light-cone distribution amplitudes (LCDAs). While the parton-distribution function (PDF) can be precisely determined by high-energy inclusive processes, a comparable comprehensive experimental determination of meson LCDAs is still lacking. The $Z\to M\gamma$ decay provides an ideal platform for extracting the leading-power LCDAs of mesons. This is not only due to the involvement of only one meson in the process, but also because the large $Z$ mass once again significantly suppresses power corrections, resulting in a clean environment. As indicated in Table~\ref{tab:Zexclusive}, the CEPC is expected to be able to determine the LCDAs of mesons such as $J/\psi$ and $\rho$ by accurately measuring their corresponding radiative decays.

Flavor-specific examples also encompass the Higgs exclusive hadronic decays, believed to be more sensitive to NP, especially to non-standard
Yukawa couplings of the Higgs boson~\cite{Koenig:2015pha}. Such decays can be examined within the Higgs factory mode of the CEPC, and are thus primarily limited by statistics rather than systematic uncertainties. Despite the challenging nature of measuring these rare processes, exclusive decays $h\to V\gamma$ of the Higgs boson at the LHC, the high luminosity run of the LHC and the CEPC could provide the much-needed platform to investigate these processes. These measurements could be vital also for testing the QCD factorization approach and extracting valuable information about the LCDAs of various mesons.

\section{Flavor Physics beyond \texorpdfstring{$\boldsymbol{Z}$}{Z} Pole}
\label{sec:beyondZ}

Similar to the case of $Z$ boson decays, flavor physics can be explored in physical processes of other EW-scale particles such as $W$ boson, Higgs boson and top quark. The productions of these particles are rich in the CEPC runs beyond $Z$ pole including at $WW$ threshold, Higgs factory and also $t\bar t$ threshold (see Table~\ref{tab:Yield_T1}). Such a strategy well-complements the study of heavy flavor physics ($b,c,\tau$), as the probed energy domain and hence the relevant physical effects ({\it e.g.}, QCD effects) could be very different. This will necessarily provide new insights into  fundamental rules in flavor physics. It is thus of high scientific value to extend the flavor-physics study from the heavy-flavored fermions to these EW-scale particles. In this section, instead of a comprehensive study we will demonstrate several benchmark cases involving $W$ boson, Higgs boson and top quark, respectively.

\subsection{Flavor Physics and \texorpdfstring{$W$}{W} Boson Decays}
\label{sec:Vcb_W}

The CEPC is expected to produce $\gtrsim 10^{9}$ $W$ bosons combining 
the $WW$ threshold and Higgs factory operations. 
This large statistics and clean physics environment provides excellent opportunities for investigating flavor physics at a scale much higher than hadron scales. 

\begin{figure}[h!]
\centering
    \includegraphics[width=\textwidth]{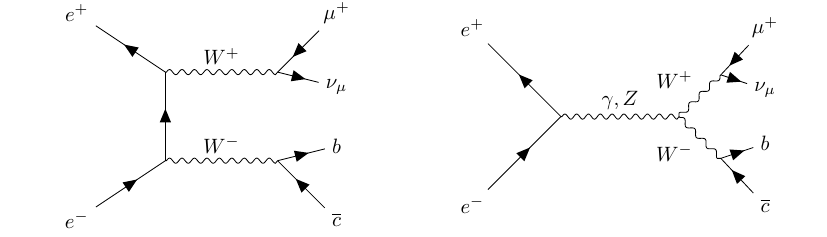}
    \caption{Illustrative Feynman diagrams for the process $e^-e^+\to W^+W^-\to cb \mu \nu$.}
    \label{fig:WW_Vcb_Feynman}
\end{figure}

\begin{figure}[t!]
    \centering
    \includegraphics[width=0.8\textwidth]{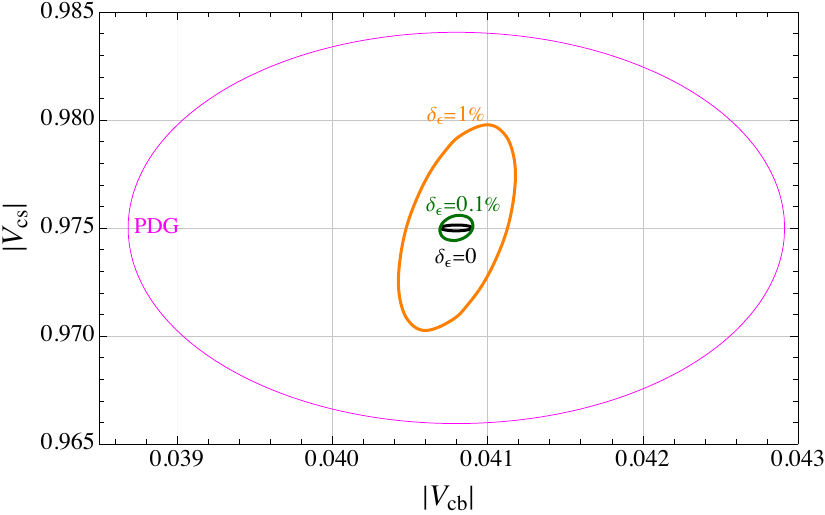}
    \caption{68\% CL $|V_{cs}|$-$|V_{cb}|$ precision contours at different systematic uncertainty scenarios. More details can be found in~\cite{Marzocca:2024mkc}.}
    \label{fig:Vcb_Vcs_Contours}
\end{figure}

One important case is to measure the CKM matrix elements such as $|V_{cb}|$ and $|V_{cs}|$ in the on-shell $W$ boson decays (for illustrative Feynman diagrams, see Figure~\ref{fig:WW_Vcb_Feynman}). Currently, there is a long-standing discrepancy of $\sim 3\sigma$ or $0.0031$ in absolute value on the $|V_{cb}|$ determination between the inclusive and exclusive $B$ meson decays~\cite{HeavyFlavorAveragingGroup:2022wzx}.
This discrepancy, however, is not very indicative for the NP, as both methods rely on semileptonic $b$-hadron decays and consequently are susceptible to theoretical uncertainties from non-perturbative QCD~\cite{HeavyFlavorAveragingGroup:2022wzx,Crivellin:2014zpa}. These QCD effects could be significantly suppressed at a higher energy scale, thereby improving the theoretical predictability~\cite{Descotes-Genon:2018foz}. The precise measurement of $|V_{cs}|$ is also valuable, allowing further investigation of the CKM unitarity. In recent studies at the FCC~\cite{Charles:2020dfl,Marzocca:2024mkc}, it is suggested that the fully hadronic decaying $W$ boson pairs produced from the $WW$ threshold run could be utilized to determine $|V_{cb}|$ and $|V_{cs}|$ simultaneously. Systematic uncertainties, especially the calibration of jet flavor tagging performance, become essential as illustrated in Figure~\ref{fig:Vcb_Vcs_Contours}. In the optimistic case where systematic uncertainty is of $\mathcal{O}(0.1\%)$, the method could yield relative uncertainties as low as $ 0.16\%$ and $0.05\%$ for the $|V_{cb}|$ and $|V_{cs}|$ measurement, respectively. Similarly, by incorporating both semileptonic and fully hadronic decays from $\mathcal{O}(10^9)$ $W$ bosons generated during $WW$ and Higgs operations, and employing advanced jet flavor tagging techniques~\cite{Liang:2023wpt}, CEPC could enhance the relative statistical sensitivity of $|V_{cs}|$ to approximately 0.006\%. Such relative precisions are better than the current ones, $e.g.$, $\gtrsim 1\%$ for $|V_{cb}|$~\cite{HeavyFlavorAveragingGroup:2022wzx}. 
Another dedicated study~\cite{Liang:2024hox} indicates that the Higgs factory operation at the CEPC may provide a comparable or even better sensitivity for measuring $|V_{cb}|$, with a large integrated luminosity ($\sim 20$~ab$^{-1}$).

%Furthermore, the CEPC can capitalize on its large sample of produced $W$ bosons to directly measure the CKM matrix elements through $W$ boson decays. A recent study at the FCC~\cite{Marzocca:2024mkc} demonstrates that, using fully hadronic decays of $6 \times 10^8$ $W$ bosons from the $WW$ threshold scan, statistical uncertainties on $|V_{cs}|$ and $|V_{cb}|$ can reach 0.008\% and 0.14\%, respectively, and the control of systematic uncertainties especially the calibration of jet flavor tagging  performance becomes essential, as illustrated in Figure~\ref{fig:Vcb_Vcs_Contours}. By incorporating both semileptonic and fully hadronic decays from $\mathcal{O}(10^9)$ $W$ bosons generated during $WW$ and Higgs operations, and employing advanced jet flavor tagging techniques~\cite{Liang:2023wpt}, the CEPC could enhance the statistical sensitivity of $|V_{cs}|$ to approximately 0.006\%, nearly two orders of magnitude better than the current precision~\cite{ParticleDataGroup:2024cfk}.
%This offers a complementary and highly precise alternative to measurements derived from $D$-meson decays.
%Additionally, a full simulation study at the CEPC~\cite{Liang:2024hox} explored direct measurements of $|V_{cb}|$ from $W$ boson decays, primarily using semileptonic $WW$ events during Higgs operation, achieving a statistical uncertainty of $\lesssim 0.4\%$ (see Section~\ref{sec:Vcb_W} for details).

\begin{figure}[t]
    \centering
    \includegraphics[width=10 cm]{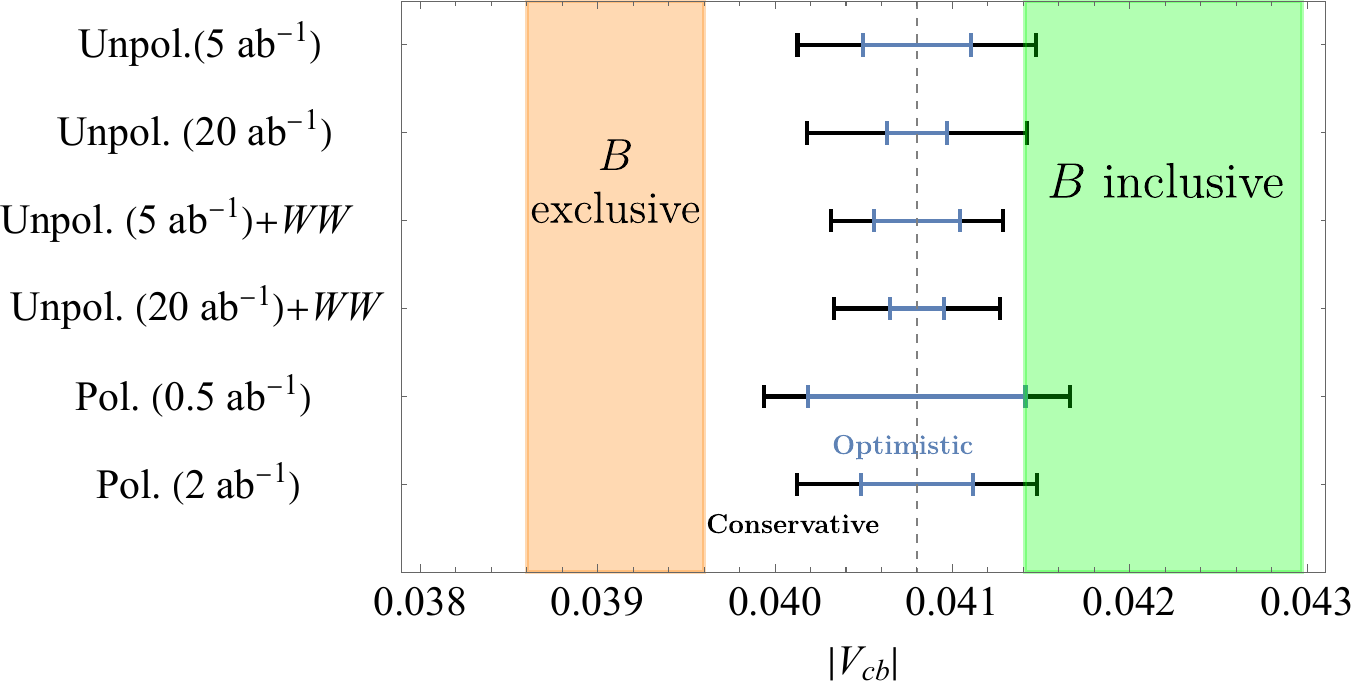}
    \caption{Projected sensitivities of measuring $|V_{cb}|$ in  $W\to cb$ decays in different future lepton collider benchmarks~\cite{Liang:2024hox}. The fourth topdown bar corresponds to the CEPC scenario, given an unpolarized Higgs factory with an extended run and a $WW$ threshold run. Black and blue error bars are based on conservative and optimistic estimates on systematics. For comparison, the current determination of $|V_{cb}|$ from inclusive and exclusive $B$ decays are also shown~\cite{HeavyFlavorAveragingGroup:2022wzx}. The PDG average of $|V_{cb}|$~\cite{ParticleDataGroup:2024cfk} is taken as a nominal central value for all future measurements. 
    }
    \label{fig:VcbWW-2}
\end{figure}

The dedicated Higgs-factory study in~\cite{Liang:2024hox} employed a full simulation of the CEPC CDR detector design~\cite{CEPCStudyGroup:2018ghi}. The signal events of $e^-e^+ \to W^+W^- \to \ell \nu cb$ are distinguished from  major backgrounds including other semileptonic $WW$ events and various processes of $e^-e^+ \to 4(2)$ fermions, through the application of a multivariate classifier. Here the advanced algorithm of jet origin identification~\cite{Liang:2023wpt} was applied for flavor tagging. By combining lepton flavors of $WW\to \ell\nu cb$, the relative statistical uncertainty for measuring $|V_{cb}|$ was found to be $\lesssim 0.4\%$~\cite{Liang:2024hox}, which has a potential to resolve the $|V_{cb}|$ tension. The projected sensitivities for the CEPC and other Higgs factory benchmarks are demonstrated in Figure~\ref{fig:VcbWW-2}. Despite this encouraging outcome, the precision of measuring $|V_{cb}|$ and $|V_{cs}|$ could be further increased by improving the jet flavor tagging with, {\it e.g.}, advanced algorithms and innovative designs for the vertex detector system. Notably, the ultimate precision of measuring $|V_{cb}|$ and $V_{cs}$ relies on also the controlling of systematic uncertainties, especially flavor tagging efficiency and mistag rates.

The measurements of leptonic $W$ boson decays at the CEPC  
also raise new possibilities of testing the LFU in the charged-current processes, in addition to the ones discussed in Sections~\ref{sec:FCCC} and~\ref{sec:tauLFU}. 
Currently the world averages for the width ratios of leptonic $W$ boson decays are~\cite{ParticleDataGroup:2024cfk}:
\begin{align}
&\frac{{\rm BR}(W\to \mu \nu)}{{\rm BR}(W\to e \nu)} = 1.002 \pm 0.006\,, \quad
\frac{{\rm BR}(W\to \tau \nu)}{{\rm BR}(W\to e \nu)} = 1.015 \pm 0.020\,, \nonumber \\
&\frac{{\rm BR}(W\to \tau \nu)}{{\rm BR}(W\to \mu \nu)} = 1.002 \pm 0.020\,,
\end{align}
which are consistent with the SM predictions at percent or even sub-percent levels. These results are based on the LHC  measurements~\cite{ATLAS:2016nqi,ATLAS:2020xea,CMS:2022mhs}, and are more precise than those of the combined LEP analyses by a factor about two~\cite{ALEPH:2013dgf}. With $\sim 10^4$ times larger statistics than that of the LEP and improved control of systematic errors, the CEPC is expected to be in an excellent position to substantially improve the LFU tests in the $W$ boson decays.

\subsection{Flavor-Violating Higgs Boson Decays}
\label{sec:Higgsfcnc}

\begin{figure}[t]
    \centering
    \includegraphics[width=0.45\textwidth]{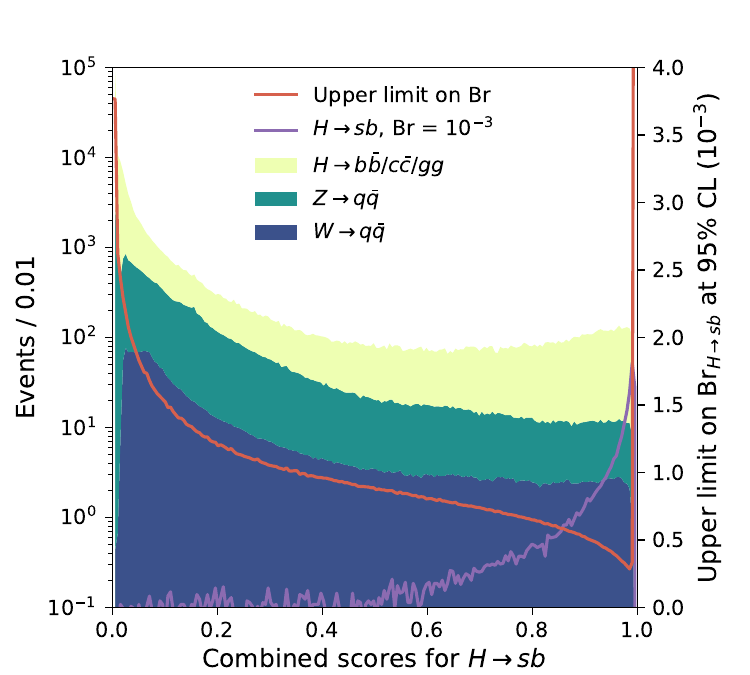}
    \includegraphics[width=0.45\textwidth]{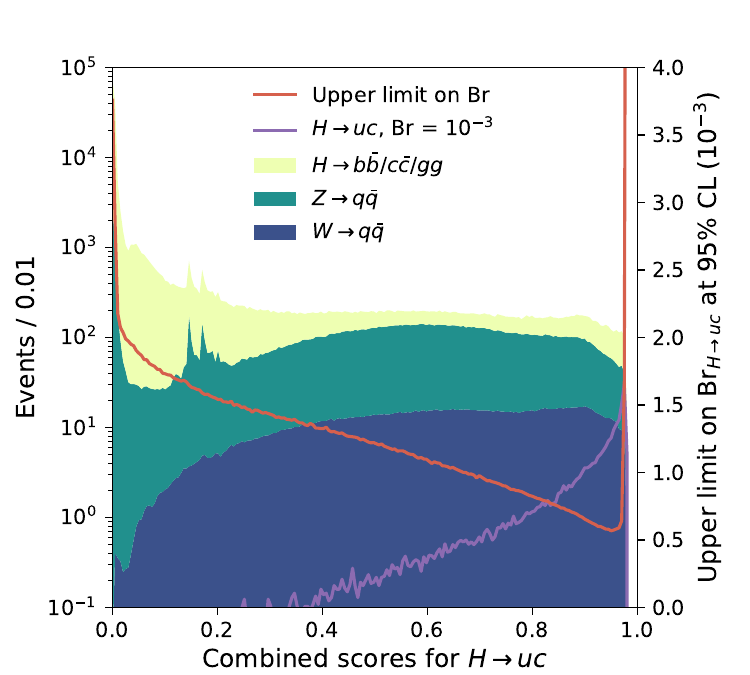}
    \caption{Projected sensitivities for measuring the flavor off-diagonal Higgs decays $H\to sb$ (\textbf{LEFT}) and $H\to uc$ (\textbf{RIGHT})  
    in the $\nu\bar{\nu} H$ process at the CEPC~\cite{Liang:2023wpt}.} 
    \label{fig:Hqq}
\end{figure}

\begin{figure}[t]
    \centering
    \includegraphics[width=0.465\textwidth]{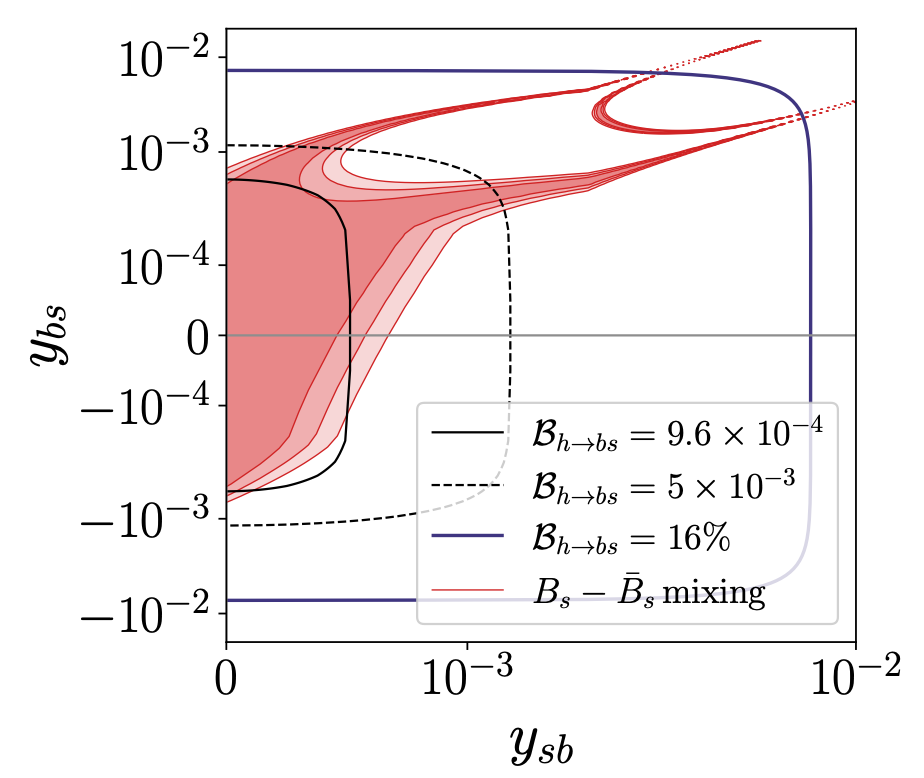}
    \includegraphics[width=0.44\textwidth]{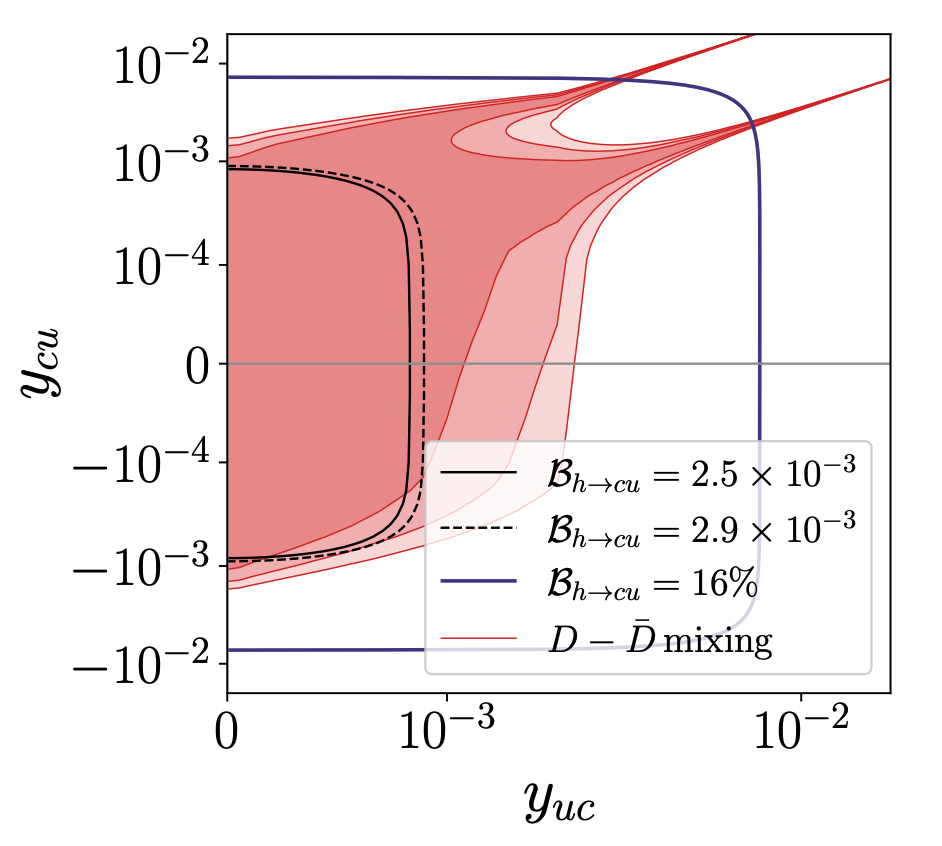}
    \caption{Projected limits on the flavor off-diagonal Yukawa couplings $y_{ij}$. The 16\% limit is derived from the current upper limits on the undetermined Higgs decays at the LHC~\cite{2022cms,2022atlas}. The black lines denote the expected limits to be achieved at the FCC-$ee$ Higgs factory. The red shaded regions, from dark to light, represent the constraints at $1\sigma$, $2\sigma$, $3\sigma$ CLs, respectively, interpreted from the current limits on the $B_s^0 - \bar{B}_s^0$ (\textbf{LEFT}) and $D^0 - \bar{D}^0$ (\textbf{RIGHT}) mixings. The plots are taken from Ref.~\cite{kamenik2023flavor}.
}
    \label{fig:hfcnc_fcc}
\end{figure}

With a yield of $4.3\times 10^6$ Higgs bosons, the study on flavor-violating physics can be naturally extended from the CEPC $Z$ pole to its Higgs factory, by investigating the Higgs hadronic decays $H \to q_i q_j$, $l_i l_j$ with $i\neq j$.\footnote{Here $q_j$ denotes $\bar{q}_j$, and similarly $l_j$ denotes $\bar {l}_j$.}
These flavor-violating Higgs boson decays are forbidden at tree level in the SM, and have a tiny BR up to $\mathcal O(10^{-7})$ due to loop suppression. The NP arising from, {\it e.g.}, multiple Higgs doublets models, however, could enhance the BRs of these decay modes by orders of magnitude~\cite{PhysRevD.73.055015, Crivellin:2017upt}.

For the measurements of flavor-violating hadronic Higgs boson decays, the tagging of quark flavor is crucial and can be addressed using the method of jet origin recognition developed in Ref.~\cite{Liang:2023wpt}. As shown in Figure~\ref{fig:Hqq}, the BRs for the decays $H\to sb$ and $uc$ can be measured at the CEPC with an upper limit $\sim 0.03\%$ and $0.08\%$ at 95\% CL, respectively. A study at the FCC-$ee$~\cite{kamenik2023flavor} indicates comparable sensitivities of measuring $\text{BR}(H\to bs)$ and $\text{BR}(H\to cu)$, estimating the upper limits to be  $\sim \mathcal{O}(10^{-3})$.

The flavor off-diagonal Yukawa couplings $y_{ij}$ also contribute to low-energy-scale observables, such as the $B_s^0 - \bar{B}_s^0$ and $D^0 - \bar{D}^0$ mass splittings and the $B_s^0 \to \mu^+\mu^-$ and $B_s^0 \to \tau^- \tau^+$ decay rates. Measuring these observables thus can yield constraints also on the rate of flavor-violating hadronic Higgs decays. A comparison between the limits obtained by these methods on $y_{ij}$ is demonstrated in Figure~\ref{fig:hfcnc_fcc}. As shown by this figure, the limits obtained from measuring the Higgs decays at the FCC-$ee$ are expected to be comparable with the ones set by the current measurements of $B_s^0 - \bar{B}_s^0$ and $D^0 - \bar{D}^0$ mixing. The best limits of CEPC indicated in Figure~\ref{fig:Hqq} are stronger than the ones set by the black-solid curve in Figure~\ref{fig:hfcnc_fcc} by several times.

\begin{figure}[t]
    \centering
        \includegraphics[width=0.65\textwidth]{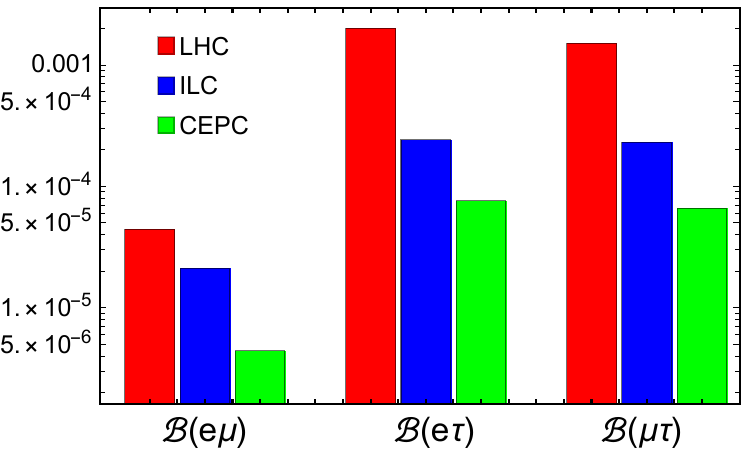}
    \caption{Projected upper limits on the LFV Higgs decays at the LHC, ILC and CEPC. The figure is updated from~\cite{Qin:2017aju}.}
    \label{fig:higgsfcnc_lep}
\end{figure}

The CEPC may yield even stronger limits on the LFV Higgs decays, namely $H\to \ell^+_i\ell^-_j$~($i\neq j$), since the charged leptons could be identified with a higher purity and efficiency compared to the jets. A study regarding this possibility has been performed in~\cite{Qin:2017aju}. As shown in  Figure~\ref{fig:higgsfcnc_lep}, with the CEPC TDR setup of 4.3 million Higgs bosons, the BR can be constrained statistically to a level of $10^{-5}-10^{-4}$ for $H\to e\tau$ and $\mu\tau$, where leptonic decays have been considered for $\tau$ reconstruction~\cite{Qin:2017aju}, and of $\mathcal{O}(10^{-6})$ for the decay mode of $H\to \mu e$.\footnote{However, it is worthwhile to note that, barring fine-tuned cancellations, $\text{BR}(H\to \mu e)$ is indirectly constrained to $\lesssim 10^{-8}$ by current limits on the LFV muon decays~\cite{Harnik:2012pb}.} 
The limits on BR$(H\to \ell \tau)$ can be further improved by including the hadronic $\tau$ decay modes in the analysis.

\subsection{FCNC Top Quark Physics}
\label{sec:topfcnc}

Top quark may carry key information on the dynamics of EW symmetry breaking (see, {\it e.g.},~\cite{Hill:2002ap}). The CEPC program provides opportunities to probe top-quark-related FCNC processes through both anomalous single top production below the top pair production threshold and top decays in the $t\bar t$ events at $\sqrt{s}=360$~GeV. Below we will show a study on the FCNC top production in the Higgs-factory run performed in Ref.~\cite{Shi:2019epw}. The FCNC top quark decays at the top pair threshold of an $e^-e^+$ collider however has been much less studied.

\begin{figure}[t]
    \centering
    \includegraphics[width=\textwidth]{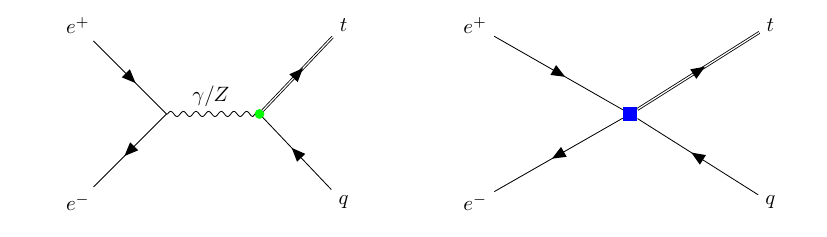}
    \caption{Illustrative Feynman diagrams for the FCNC single top production $e^-e^+\to t(\bar{t})j$. The green dot and blue square represent two-fermion FCNC and four-fermion (two-lepton two-quark) contact operators, respectively.}
    \label{fig:topfcnc_diagram}
\end{figure}

The LHC TOP Working Group~\cite{Aguilar-Saavedra:2018ksv} provides a systematic SMEFT description on FCNC top quark physics. The single top production with a light jet ``$j$'', {\it i.e.}, $e^-e^+\to t(\bar{t})j$, while being suppressed by the GIM mechanism in the SM, can be enhanced by the NP-induced two-fermion FCNC operators 
\begin{equation}
\label{eq:topFCNC2f}
\begin{aligned}
	O_{\varphi q}^{1(ij)} = i\left(\vpj\right) \left( \bar Q_i \gamma^\mu Q_j \right),  \ \
	& O_{\varphi q}^{3(ij)} = i\left(\vpjt\right) \left( \bar Q_i \gamma^\mu \tau^I Q_j \right), \\
	O_{\varphi u}^{(ij)} =  i&\left(\vpj\right) \left( \bar U_i \gamma^\mu U_j \right),
	\\
	O_{uW}^{(ij)}=\left( \bar Q_i\sigma^{\mu\nu}\tau^I U_j \right)\tilde \Phi W_{\mu\nu}^I, \ \ 
	&O_{uB}^{(ij)}=\left( \bar Q_i\sigma^{\mu\nu} U_j \right)\tilde \Phi B_{\mu\nu},
\end{aligned} 
\end{equation}
and four-fermion contact operators
\begin{equation}
\begin{aligned}
	O_{lq}^{1(ijkl)}=\left( \bar L_i \gamma_\mu L_j \right)
			\left( \bar Q_k \gamma^\mu Q_l \right),  \ \ 
	&O_{lq}^{3(ijkl)}=\left( \bar L_i \gamma_\mu \tau^I L_j \right)
			\left( \bar Q_k \gamma^\mu \tau^I Q_l \right), 
			\\
	O_{lu}^{(ijkl)}=&\left(  \bar L_i \gamma_\mu L_j \right)
			\left( \bar U_k \gamma^\mu U_l \right),
	\\
	O_{eq}^{(ijkl)}=\left( \bar E_i \gamma_\mu E_j \right)
			\left( \bar Q_k \gamma^\mu Q_l \right), \ \ 
	&O_{eu}^{(ijkl)}=\left( \bar E_i \gamma_\mu E_j \right)
			\left( \bar U_k \gamma^\mu U_l \right),
	\\
	O_{lequ}^{1(ijkl)}=\left( \bar L_i E_j \right)\varepsilon
			\left( \bar Q_k U_l \right), \ \ 
	&O_{lequ}^{3(ijkl)}=\left( \bar L_i \sigma_{\mu\nu} E_j \right)\varepsilon
			\left( \bar Q_k \sigma^{\mu\nu} U_l \right).
\end{aligned} 
\end{equation}
Here $i$, $j$, $k$, $l$ are flavor indices and $\Phi$ is the SM Higgs doublet. Their contributions to this physical processes are shown in Figure~\ref{fig:topfcnc_diagram}.

\begin{figure}[t]
    \centering
        \includegraphics[width=0.85\textwidth]{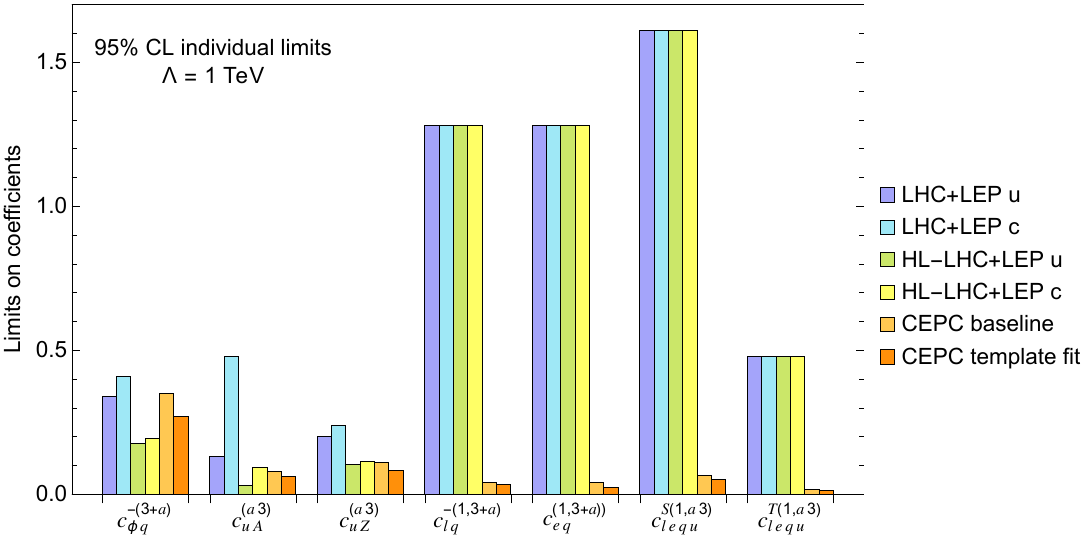}
    \caption{Projected limits on the FCNC top quark operators at the CEPC Higgs factory run with single top production. For comparison, the existing LHC+LEP2 bounds and the expected limits from HL-LHC+LEP2 are also shown. Here $c_{\varphi q}^{-(3+a)}$,  $c_{uA}^{(a3)}$,  $c_{uZ}^{(a3)}$ and $c_{lq}^{-(1,3+a)}$, $c_{eq}^{(1,3+a)}$, $c_{lequ}^{S(1,a3)}$,  $c_{lequ}^{T(1,a3)}$ are linearly combined Wilson coefficients of the two-fermion FCNC operators and the four-fermion contact operators, respectively. These parameters are assumed to be real, with their limits being generated by switching on the correspondent operators individually. The LHC bounds on the four-fermion operators are obtained by recasting the $t\to q \ell\ell$ searching results. The ``CEPC baseline'' shows the baseline analysis by tagging a single top quark decaying leptonically, while the ``CEPC template fit'' exploits additionally $c$-tagging (only for the $a=2$ operators) and top quark scattering angle to enhance signal recognition. This plot is taken from~\cite{Shi:2019epw}.  
    }
    \label{fig:topfcnc_1dlimit}
\end{figure}

Currently, the best constraints on the two-fermion FCNC operators and four-fermion contact operators are set by the LHC~\cite{ATLAS:2015iqc,ATLAS:2018xxe,ATLAS:2018zsq,CMS:2015kek,CMS:2016uzc} and LEP2 data, 
respectively~\cite{OPAL:2001spi,ALEPH:2002wad,L3:2002hbp,DELPHI:2011ab} (see also~\cite{Durieux:2014xla,Cerri:2018ypt}).
The measurements in the latter case are exactly based on the FCNC single top quark production. The prospects for measuring these operators via the same process at the CEPC have been studied in Ref.~\cite{Shi:2019epw}, by assuming an integrated luminosity of 5.6 ab$^{-1}$ at $\sqrt{s}= 240\,$GeV and a CEPC detector profile as presented in~\cite{CEPCStudyGroup:2018ghi}. For the semileptonic top quark decays, the signal signature contains one bottom quark jet, one up or charm quark jet, one charged lepton and missing energy, while the major background is the $WW$ production with one $W$ boson decaying hadronically and the other one leptonically. 
As shown in Figure~\ref{fig:topfcnc_1dlimit}, at the CEPC the current limits for the four-fermion contact operators can be improved by one to two orders of magnitude. 
These constraints could be further improved by exploiting additional kinematic features of the FCNC single top quark production. 
The capacity of tagging light-flavored jets at the CEPC also presents the possibility to distinguish the SMEFT operators with $j=u$ quarks from those of $j=c$.
The Lorentz structure of the operators are reflected in the kinematics of the top quark and hence its decay products. The observables such as differential distributions and forward-backward asymmetries thus may help lift the degeneracy between their Wilson coefficients if an FCNC signal is observed.

Other than the single top production at $\sqrt{s}=240$~GeV, the CEPC is also expected to produce 0.6$\times 10^6$ $t\bar{t}$ events at the $\sqrt{s}=360$~GeV run. This data set can be used to search for FCNC top decays such as $t\to qZ$ and $t\to q H$ with $q=c,u$~\cite{Balazs:1998sb,He:1998ie,Diaz-Cruz:2001igs}. Consider the $t\to q H$ decays as an example. These decays may arise from the dimension-6 Yukawa-type operators\,\cite{Ge:2016zro,Aguilar-Saavedra:2018ksv} 
\begin{eqnarray}
O^{(ij)}_{u\varphi} = (\Phi^\dag \Phi)\bar Q_i^{} \tilde \Phi U_j^{}~,\quad O^{(ij)}_{d\varphi}=  (\Phi^\dag \Phi)\bar Q_i^{} \Phi D_j^{}~.
\end{eqnarray}
In this context, the mass matrix of the  up-type quarks and their couplings with the physical Higgs boson ($\mathcal{L} \supset y_{ij}\bar q_{i}H u_j$) are not aligned, generically yielding the $t\to q H$ decays. The current LHC bound for the $t\!\to\! cH$ decay is  
BR$(t\!\to\! cH)\!<\!4.3\!\times\! 10^{-4}$  at 95\%\,C.L.~\cite{ATLAS:2023ujo}, 
implying $y_{ct}^2\!+\! y_{tc}^2 \!<\!0.0032$. With the expected yield of 0.6$\times 10^6$ $t\bar{t}$ events, the CEPC could improve this limit to the $\mathcal O(10^{-5})$ level and, accordingly, the constraint on $y_{ct}^2\!+\! y_{tc}^2$ by one order of magnitude.

\section{Spectroscopy and Exotics}
\label{sec:exotic}

\begin{figure}
    \centering
    \includegraphics[width=\textwidth]{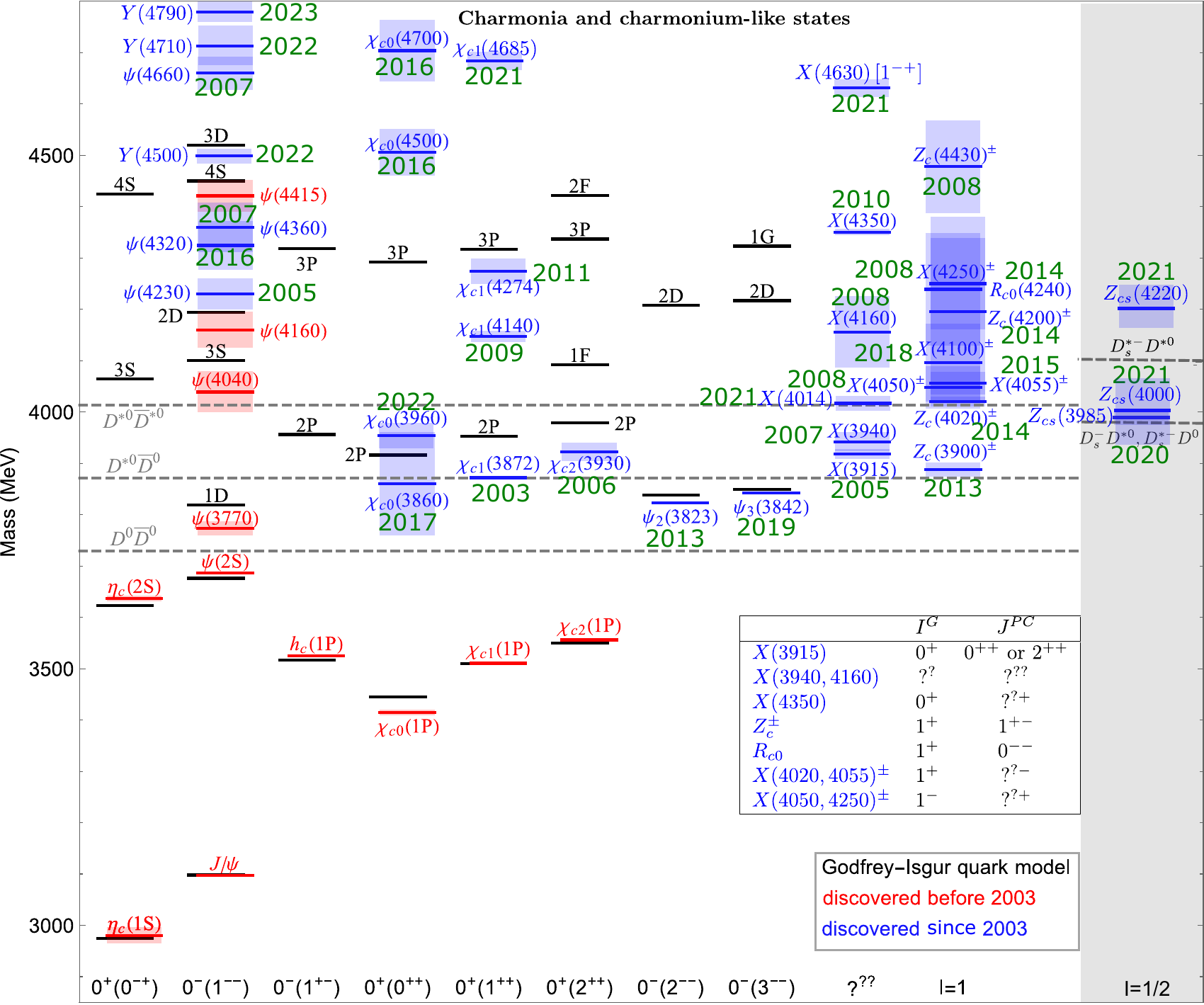}
    \caption{Spectrum of the charmonium and charmonium-like states. Black lines represent the masses in the Godfrey-Isgur quark model~\cite{Godfrey:1985xj}. The red and blue lines represent the states observed experimentally before 2003 and since 2003, respectively. For the latter, the years when the states were observed are labeled in green. The height of each shadow indicates the width of the corresponding state. We also show a few two-body open-charm thresholds as dashed lines.}
    \label{fig:charmoniumspec}
\end{figure}

Spectroscopy of hadrons is critical for understanding the mass generation in QCD, given the persisting mystery of color confinement. Although exotic hadrons, extending beyond conventional quark-antiquark mesons and three-quark baryons, have been postulated since the invention of the quark model, strong evidence for their existence only emerged recently as a result of significant experimental progress. In particular, the discovery of the $D_{s0}^*(2317)$ meson by BaBar~\cite{BaBar:2003oey} and the $X(3872)$ meson, also known as $\chi_{c1}(3872)$~\cite{ParticleDataGroup:2024cfk}, by Belle~\cite{Choi:2007wga}, has resulted in a surge of interest from both experimental and theoretical sides. During the past two decades dozens of exotic states, with a noteworthy characteristic of narrow states located near the threshold for production of a pair of open-flavor hadrons, have been identified. Nevertheless, intriguing resonant structures, that are explicitly exotic, were observed, such as the $Z_c(3900)^\pm$ by BESIII~\cite{BESIII:2013ris} and Belle~\cite{Belle:2013yex}, hidden-charm strange tetraquark $Z_{cs}$ candidates by BESIII~\cite{BESIII:2020qkh} and LHCb~\cite{LHCb:2021uow}, hidden-charm $P_c$ pentaquarks~\cite{LHCb:2015yax,LHCb:2019kea}, double-charm $T_{cc}(3875)^+$ tetraquark~\cite{LHCb:2021vvq}, and fully-charmed tetraquarks ({\it e.g.}, $X(6900)$) by LHCb~\cite{LHCb:2020bwg}, ATLAS~\cite{ATLAS:2023bft} and CMS~\cite{CMS:2023owd}. 
It is evident from Figure~\ref{fig:charmoniumspec} that most of these newly observed states in the charmonium mass region go beyond the charmonium spectrum predicted by  quark models ({\it e.g.}, the Godfrey-Isgur quark model~\cite{Godfrey:1985xj}).
These discoveries spur plenty of efforts in trying to reveal the nature of the new hadrons and to gain deeper understanding of nonperturbative strong interactions. For recent reviews, one may refer to Refs.~\cite{Chen:2016qju,Hosaka:2016pey,Chen:2016spr,Esposito:2016noz,Lebed:2016hpi,Guo:2017jvc,Ali:2017jda,Olsen:2017bmm,Liu:2019zoy,Brambilla:2019esw,Guo:2019twa,Chen:2022asf}. A wide spectrum of potential new resonances and a multitude of observables make hadron spectroscopy a promising avenue for discoveries at CEPC. This is particularly relevant considering that the formation of multiquark exotics would favor the heavy-flavored systems, which can be well treated as non-relativistic systems~\cite{Chen:2016qju,Lebed:2016hpi,Iwasaki:1975pv,Ader:1981db,Zouzou:1986qh,Heller:1985cb,Lloyd:2003yc,Richard:2017vry,Anwar:2017toa,Karliner:2016zzc,Bai:2016int,Berezhnoy:2011xn,Debastiani:2017msn,Esposito:2018cwh,Hughes:2017xie,Wu:2016vtq,Liu:2019zuc}, and the spectra of the fully-heavy exotics, such as $bb\bar{b}\bar{b}$, $cc\bar{c}\bar{c}$, $bb\bar{c}\bar{c}$, $bc\bar{b}\bar{c}$, etc., can be accessed at CEPC. Note that it is still unclear how many and what kinds of exotic multiquark states we should expect, and how these multiquark states can be stablized by the nonperturbative strong interactions. At CEPC, systematic measurements of these heavy-flavored multiquark states should be able to provide  crucial insights into the underlying binding mechanism for these heavy-flavored exotic states.

Despite numerous works and tremendous efforts on the understanding of these novel structures observed in experiment, a comprehensive solution for describing and classifying them remains elusive. Thereby, experimental data are paramount for further theoretical development. 
At CEPC, the production of exotic states from $b$-hadron decays, directly from the $Z$ decays or from initial state radiation is expected. 

For example, the hidden-charm exotic states such as $X(3872)$ and $P_c(4450)$ can be produced at CEPC via $b\to c\bar{c}s$ transitions after $b$-flavored hadrons are formed. Given the abundant production of heavy quark pairs ({\it e.g.},~the branching fraction of $Z\to b\bar b$ is $(15.12 \pm 0.05) \%$~\cite{ParticleDataGroup:2024cfk}), a considerable amount of exotic hadrons, including known ones and new states, can be generated. It should be stressed that this also allows to access a broad spectrum of conventional heavy-flavored mesons and baryons, which can hardly be probed by the present facilities, including  excited states and multi-heavy baryons such as $\Xi_{bb}$.

\begin{figure}
\begin{center}
    
    \includegraphics[width=\textwidth]{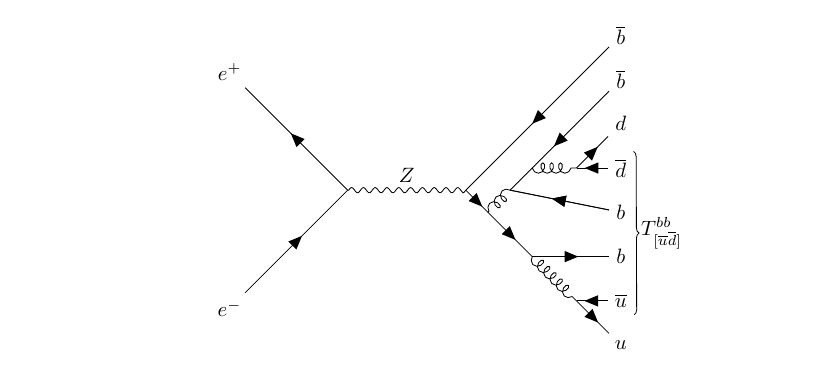}
    \end{center}

    \caption{An illustrative Feynman diagram for the  production of tetraquark state $T_{[\overline{u}\overline{b}]}^{bb}$ from the $Z\to bb\bar{b}\bar{b}$ decay. }
    \label{fig:Teraquark_Feynman}
\end{figure}

At CEPC, another significant source of exotic or multi-flavored hadrons at the $Z$ pole comes from $Z\to q\bar{q}q^\prime\bar{q}^\prime$. The multiple heavy quarks produced, either of the same or opposite signs, could hadronize into various (exotic) species if their relative velocity is low enough. The process is highly relevant to the $B_c$ physics studies since $B_c$ from the $Z$ pole mainly comes from $Z\to b\bar{b}c\bar{c}$ decays~\cite{Chang:1992bb,OPAL:1998gdf,DELPHI:1996vyn,ALEPH:1997oob}. In addition, the measurement of many inclusive rates of new resonances might occur for the first time, and the observation of numerous new decay modes is anticipated. With regards to doubly-heavy baryons ($bbq$, $bcq$ and $ccq$) and doubly-heavy exotic states (for instance, the double-charm tetraquark $T_{cc}(3875)^+$~\cite{LHCb:2021vvq,LHCb:2021auc}, double-bottom tetraquarks~\cite{Ader:1981db,Manohar:1992nd,Karliner:2017qjm,Eichten:2017ffp} and hidden-bottom pentaquarks~\cite{Wu:2010rv}), the high mass threshold necessitates $Z$ inclusive decays as their main production mechanism. An example of Feynman diagrams contributing to the production of a double-bottom tetraquark is shown in Figure~\ref{fig:Teraquark_Feynman}.

Simplified assumptions and parton-level simulations were employed to deduce the inclusive decay rates: BR($Z\to X+T^{cc}_{[\bar{q}\bar{q}^\prime]})\sim \mathcal{O}(10^{-6})$, BR($Z\to X+\Xi_{cc})\sim 5\times 10^{-5}$, and BR($Z\to X+\Omega_{cc})\sim 1\times 10^{-5}$ at the $Z$ pole~\cite{Qin:2020zlg}. Additionally, BR($Z\to X+T^{{bb}}_{[\bar{q}\bar{q}^\prime]})\sim \mathcal{O}(10^{-6})$ was also calculated~\cite{Ali:2018ifm}.
It's worth noting that $T^{{bb}}_{[\bar{q}\bar{q}^\prime]}$ could have a mass lower than the sum of $B$ and $B^*$ meson mass, thus it could only decay via weak interaction - as predicted by various of theoretical and lattice works, resulting in a life time comparable to the $B$ hadrons.
Therefore, the typical decay chain ($T^{{bb}}_{[\bar{q}\bar{q}^\prime]} \to B \to D $) could result in very special event topology, which could be well reconstructed using state-of-the-art vertex detector. 
Preliminary calculation shows that percentage level of accuracy in measuring $T^{{bb}}_{[\bar{q}\bar{q}^\prime]}$ signal strength could be achieved at CEPC. 

\begin{figure}[tb]
    \includegraphics[width=0.495\textwidth]{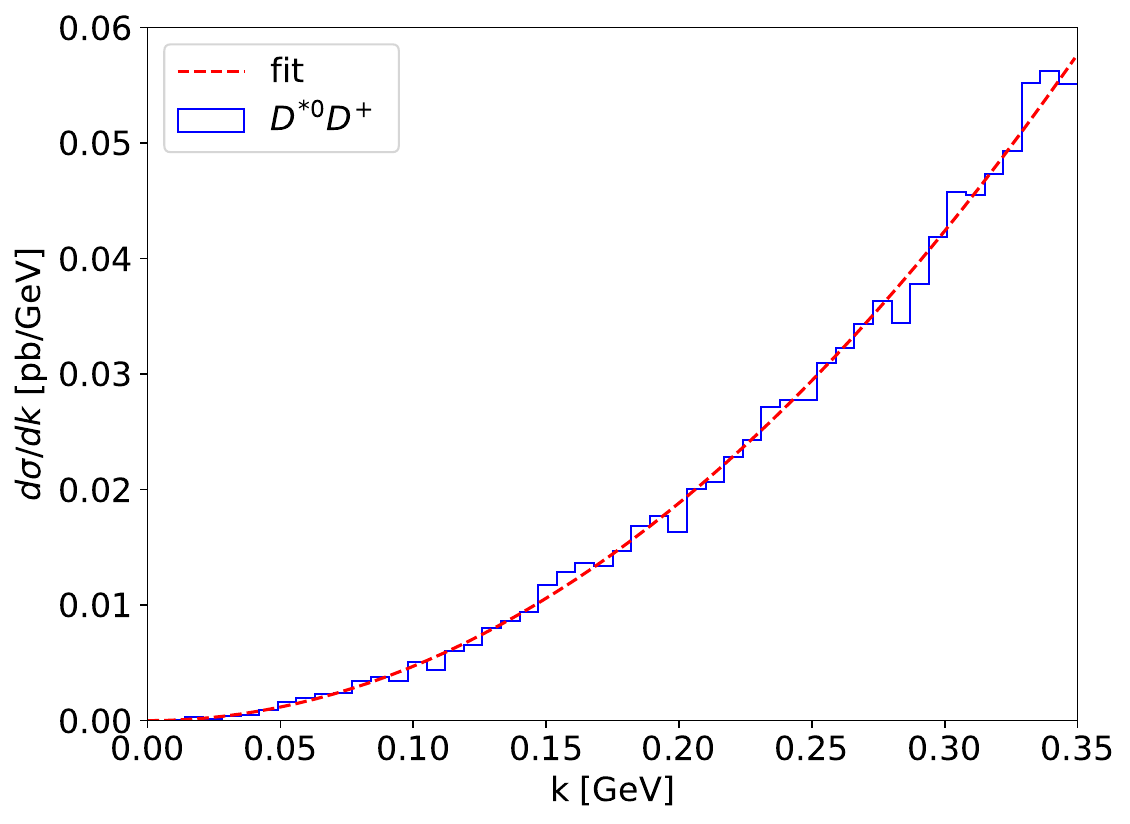} 
    \hfill 
    \includegraphics[width=0.495\textwidth]{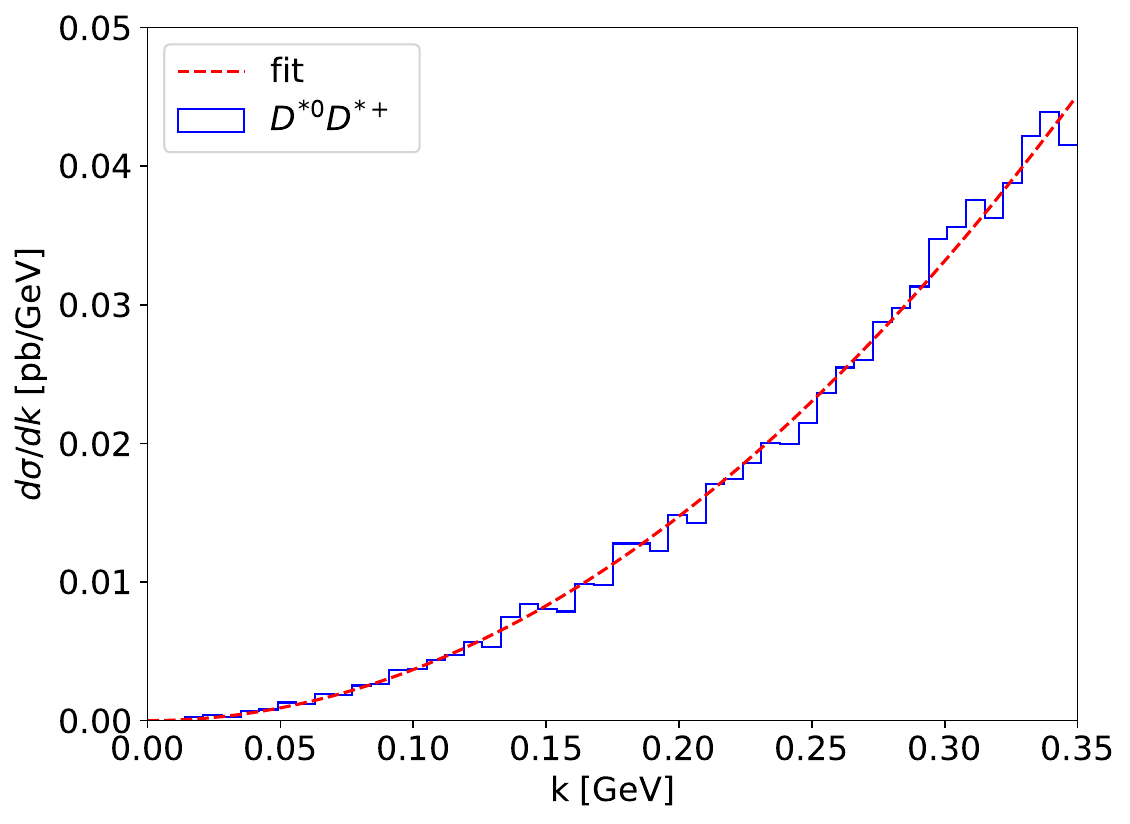} 
   \caption{Differential cross sections of $e^-e^+ \to Z^0 \to D^{*0}D^+$ and $D^{*0}D^{*+}$ generated using Pythia (histograms) and fit with $d\sigma/dk\propto k^2$, where $k$ is the relative momentum between the $D^{*0}$ and $D^+$ meson (dashed curves)~\cite{Jia:2024owm}. 
   }
   \label{fig:diffxsec_tcc}
\end{figure}

One may also estimate the inclusive production cross section of double-charm tetraquarks of the hadronic molecular type (for systematic predictions, see, {\it e.g.}, \cite{Dong:2021bvy}) by combining Monte Carlo event generators and nonrelativistic effective field theory (NREFT). 
Such method can successfully reproduce the inclusive cross section of the $X(3872)$ at hadron colliders~\cite{Artoisenet:2010uu,Guo:2014sca,Albaladejo:2017blx}.
Using Pythia 8.3~\cite{Bierlich:2022pfr} to generate differential distributions of the $D^{(*)}D^*$ pairs with low relative momenta (see Figure~\ref{fig:diffxsec_tcc}) and using NREFT to compute the effective couplings of the $T_{cc}(3875)$ to $DD^*$ and its hypothesized spin partner $T_{cc}'$ to $D^*D^*$~\cite{Du:2021zzh}, one finds that both the inclusive cross section for the $T_{cc}(3875)$ and $T_{cc}'$ at the $Z$ pole are of the order of a few to 10~fb~\cite{Jia:2024owm}.
Given the expected integrated luminosity of 100~ab$^{-1}$ at the $Z$ pole at CEPC (see Table~\ref{tab:Yield_T1}), one expects $10^5-10^6$ $T_{cc}$ and $T_{cc}'$ to be produced, consistent with the estimate in Ref.~\cite{Ali:2018ifm}. Events involving these states can be reconstructed from the $DD\pi(\pi)$ final states or similar ones with the pions replaced by photons.

Due to the high uncertainties in their differential rates and decay final states, performing a MC simulation of such exotic hadron events and reconstructing their resonance is impractical without more advanced theoretical calculations or analysis algorithms. On the other hand, additional recent efforts have been predicted the production of doubly-flavored baryons, {\it i.e.}, $\Xi_{cc}$, $\Xi_{bc}$, and $\Xi_{bb}$, at the $Z$ pole and provided the differential distributions~\cite{Luo:2022jxq,Niu:2023ojf}.

\section{Light BSM States from Heavy Flavors}
\label{sec:BSM}

Light particles are widely predicted in BSM scenarios involving dark sectors and feebly interacting particles~\cite{Antel:2023hkf}, and may couple to lepton and quark  sectors. Candidates for such particles include axions and axion-like-particles $a$~\cite{Bauer:2017ris,Liu:2022tqn,Calibbi:2022izs,Cheung:2023nzg}, dark photons $A'$ and light $Z^\prime$ bosons~\cite{He:2017zzr}, heavy neutral leptons (HNL)~\cite{Blondel:2014bra,Ding:2019tqq,Shen:2022ffi}, hidden valley hadrons such as the dark pion $\hat \pi$~\cite{Cheng:2021kjg}, etc. As a paradigmatic example, let us consider an ALP $a$ that couples with the SM fermions via the dimension-5 operators 
\begin{equation}
\mathcal{L}\supset \frac{\partial_\mu a}{2 f_a} \left(c^A_{f f^\prime}\,\bar{f}\gamma^\mu \gamma^5 f^\prime + c^V_{f f^\prime}\,\bar{f}\gamma^\mu  f^\prime\right)\,, 
\label{eq:ALP}
\end{equation}
where $f$ and $f^\prime$ are SM fermions, $c^{A,V}_{ff^\prime}$ are dimensionless couplings, (with the vector ones $c^V_{ff}$ being unphysical if $f=f^\prime$), and $f_a$ is the ALP decay constant that can be regarded as a measure of the NP energy scale. These light BSM states could thus be explored in flavor-physics experiments if they are radiated from initial or final state particles, or they are produced in lepton/quark decays. Interestingly, the production in the latter case does not conserve lepton flavor and the sensitivity to UV scales is parametrically enhanced by the narrow width of the SM fermions. Owing to their feebly-interacting nature, (so as for them to remain undetected so far), the produced BSM particles tend to be long-lived. They are often subject to displaced decays or they contribute to missing energy directly. Both kinematic features being used as collider signatures of light BSM particles have been widely studied. Note that the heavy-flavored particles in the SM are also long-lived; to enable their identification, detectors have often been designed for reconstructing the tracking/vertexing information with high quality. Even if the light BSM particle in question is invisible, the techniques for reconstructing the missing energy at the $Z$ pole can facilitate the reconstruction of its invariant mass. Therefore, the exploration of light BSM states in this context is naturally expected. Below, let us consider the detection of light BSM states which are produced via the decays of heavy-flavored leptons and quarks, using the ALP and dark pion as respective examples.

\begin{figure}[t]
\centering
    
    \includegraphics[width=\textwidth]{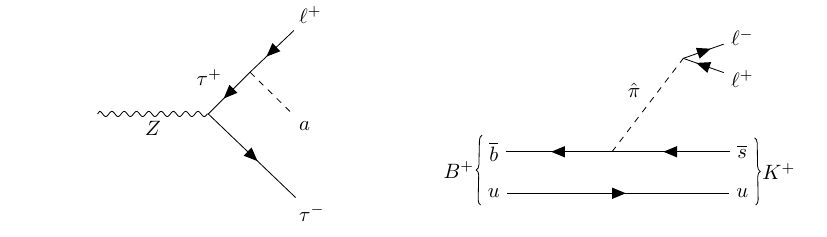}
    \caption{Illustrative Feynman diagrams of light BSM states produced via their couplings with the flavor sector, including the light dark pion $\hat\pi$ and the ALP $a$. \textbf{LEFT}: Illustrative Feynman diagrams for the ALP production in $Z\to \tau^- \tau^+$ events via lepton flavor violating couplings.  \textbf{RIGHT}: $B^+\to K^+\hat\pi(\to\mu^+\mu^-)$. The flavor-changing interaction between the SM quarks and $\hat\pi$ can arise either at the tree level or through an EW loop. 
    }
    \label{fig:FCNC_BSM_Feynman}
\end{figure}

 \subsection{Lepton Sector}

\begin{figure}[t]
    \centering
    \includegraphics[height=5cm]{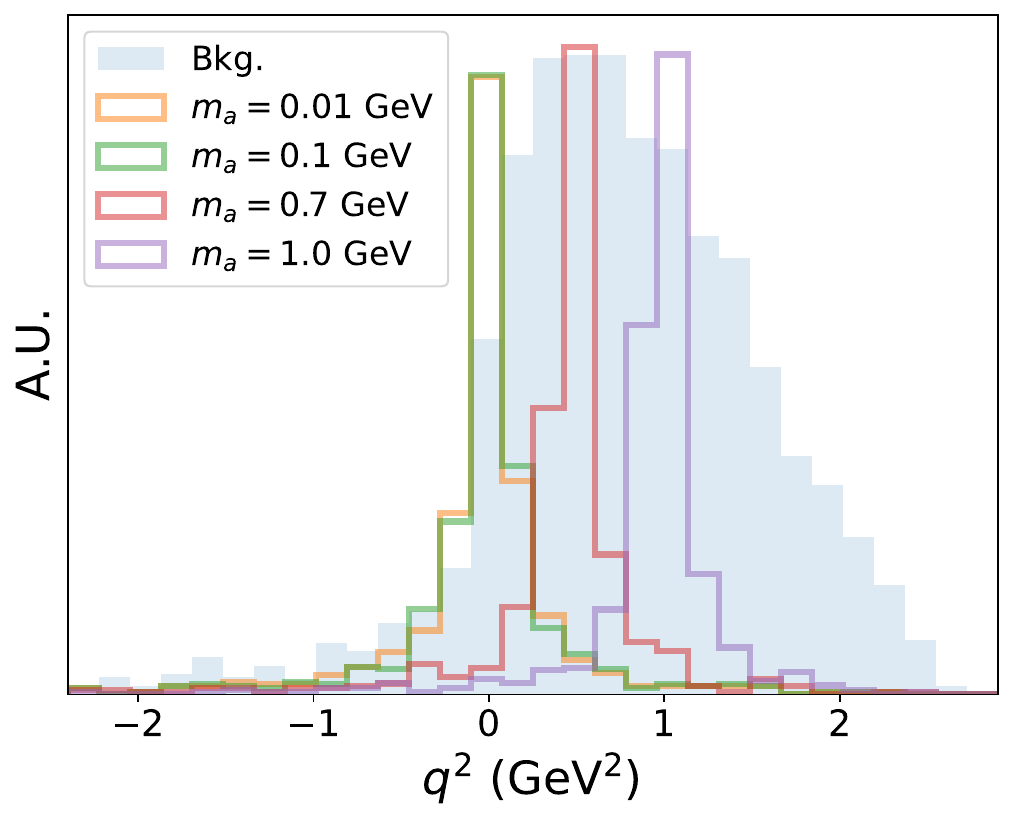}
    \hfill
    \includegraphics[height=5cm]{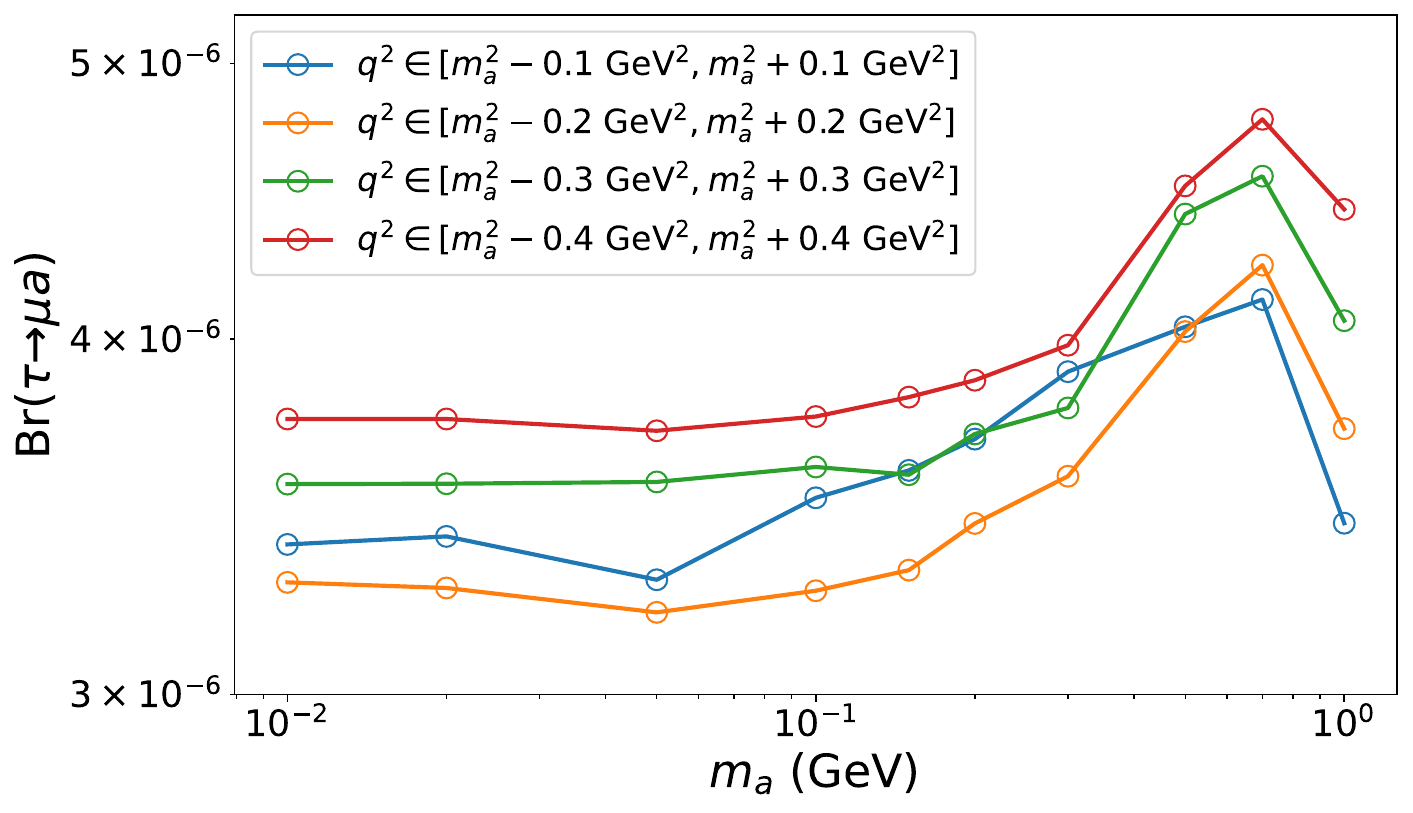}
    \caption{Preliminary sensitivity analysis for searching for an invisible ALP in the $Z \to \tau(\to \mu a)\tau(\to 3 \pi \nu)$ events at the CEPC.  \textbf{LEFT:} Reconstruction of $q^2 \equiv (p_\tau - p_\mu)^2$. 
  \textbf{RIGHT:} Upper limits on BR$(\tau \to \mu a)$ with 95\% CL, where four $q^2$ windows have been considered. The plots are taken from~\cite{KLL}. }
    \label{fig:BSM_tau}
\end{figure}

As discussed in Sections~\ref{sec:FCCC},~\ref{sec:FCNC}, and~\ref{sec:tau}, the CEPC has a strong potential for carrying out $\tau$-related searches, due to the excellent performance of its tracker. A prominent example is the LFV decay $\tau \to \ell a$ (see the left panel of Figure~\ref{fig:FCNC_BSM_Feynman}) with the ALP $a$ being invisible~\cite{Calibbi:2020jvd}. The major backgrounds then arise from the  
$\tau \to \ell \nu \nu$ decays, which share the signal signature of one visible object and missing energy. 
 Let us consider a full reconstruction of the $Z\to \tau\tau$ event.  
 Indeed, the 3-prong decays of the second $\tau$ in the $Z\to \tau\tau$ event can yield an efficient determination for the $\tau$ momentum direction. Combining this result with some other kinematic constraints, such as the $\tau$ mass on-shell condition and energy-momentum conservation, we are able to reconstruct the invisible mass $q^2\equiv (p_\tau-p_\ell)^2 =m_a^2$  accurately. 
The results from a preliminary sensitivity analysis are presented in Figure~\ref{fig:BSM_tau}, where the events are simulated with non-zero spatial beam spread, initial state radiation, and finite tracking/calorimetry resolution. As shown in the left panel, the reconstructed $q^2$ for the signal events sharply peaks at $m_a^2$, in contrast to that of the backgrounds. The right panel shows the expected CEPC 95\% C.L. upper limits on BR$(\tau \to \mu a)$. Compared with the current Belle II bound, {\it i.e.}, BR$(\tau \to \mu a)< 5.9 \times 10^{-4}$ (95\%~CL) for a practically massless ALP~\cite{Belle-II:2022heu}, the estimated CEPC limits are about two orders of magnitude stronger. In terms of the interactions in Eq.~(\ref{eq:ALP}), this implies that a NP scale as high as $f_a/c^{A,V}_{\tau\mu} \sim \mathcal{O}(10^8)$~GeV could be probed at the CEPC. 

The light ALPs can be also searched for by their lepton-flavor-conserving radiation, such as that in the $Z\to \tau\tau a$ process~\cite{Calibbi:2022izs}. 
Currently, the ALP coupling with $\tau$ leptons is essentially yet unconstrained. 
For the case of $Z\to \mu\mu a$, where the dynamics is relatively simple, it has been shown~\cite{Calibbi:2022izs} that the CEPC has the potential to reach $\text{BR}(Z\to \mu\mu a)\lesssim 3\times 10^{-11}$, yielding a limit to the ALP coupling with muons of $f_a/c^A_{\mu\mu} \gtrsim 1$~TeV. 

Moreover, both Dirac and Majorana HNLs can be produced via LFV processes. The HNLs might be responsible for the origin of neutrino mass, the puzzle of dark matter and even the cosmic baryon asymmetry. 
Their mixing with neutrinos allows them to be produced via $\tau$ decays such as $\tau \to \ell \nu N$ and $\tau \to \pi N$, if they are lighter than the $\tau$ lepton. This provides an alternative to the $Z\to \nu N$ decays in searching for HNLs at the $Z$ pole~\cite{Blondel:2022qqo}. Nevertheless, the relevant sensitivity analysis is yet to be explored.

\subsection{Quark Sector}

\begin{figure}[t]
    \centering
    \includegraphics[width=9cm]{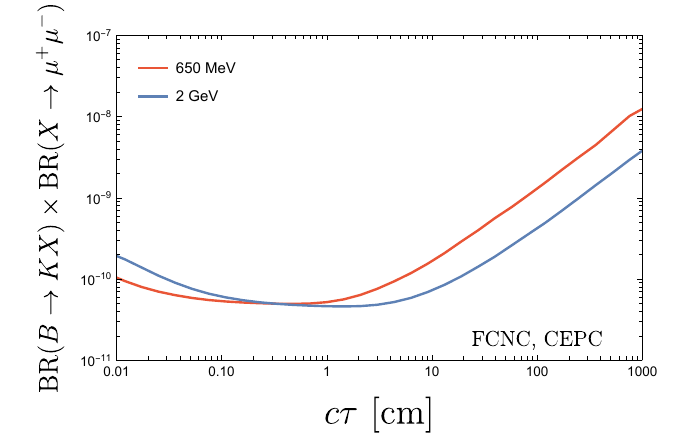}
    \caption{Preliminary expected limits for searching for a long-lived dark pion in $B\to K X(\to \mu\mu$) events at the CEPC as a function of the $\hat{\pi}$ decay length, plot customized from results of~\cite{Cheng:2024aco}.}
    \label{fig:BSM_FCNC}
\end{figure}

Light BSM particles can be also produced in heavy-flavored quark decays~\cite{Batell:2009jf,Kamenik:2011vy,Cheng:2024hvq,Calibbi:2016hwq,Aloni:2019ruo,MartinCamalich:2020dfe,Cheng:2021kjg}. As an example, let us consider a dark pion from the strong dynamics of a hidden sector, where this dark pion also couples with the SM leptons, yielding a signature of a displaced di-lepton vertex from its decay (see the right panel of Figure~\ref{fig:FCNC_BSM_Feynman})~\cite{Cheng:2021kjg}.  
The reconstruction of a narrow di-lepton resonance away from the primary vertex with high quality then allows for the efficient distinction of the signal events from the backgrounds. Figure~\ref{fig:BSM_FCNC} demonstrates preliminary limits for searching for a long-lived particle in $B\to K X(\to \mu\mu$) events at the CEPC~\cite{Cheng:2024hvq}, where $X$ denotes the long-lived new particle. The strongest constraints, namely  BR$(B\to K X(\to \mu\mu)) \lesssim 10^{-10}$, are achieved while the proper lifetime of $X$ is $\sim 0.1-10$~cm. Compared to relevant LHCb limits~\cite{LHCb:2016awg,LHCb:2020ysn}, the CEPC analysis is sensitive to a wider lifetime range and can be generalized to various final states other than $\mu\mu$. It will be convenient to describe such a new light degree of freedom by Eq.~\eqref{eq:ALP} if the new particle is a pseudoscalar since it behaves as an ALP at low energy scales. The BR limit above can then be interpreted as a probe of the decay constant $f_a$ of an ALP through its coupling with SM quarks.
Even when the FCNC couplings are absent at tree level, they will be generated at one loop by EW interactions. In the case where the couplings to all fermions are close to unity ($c^A_{ff}\sim \mathcal{O}(1)$), the constraint on $f_a$ by the CEPC will be up to $\sim \mathcal{O}(10^7)$~GeV~\cite{Cheng:2021kjg}. If a large FCNC coupling $c^V_{bs}\sim 1$ presents at tree level, the constraints on $f_a$ will be even higher, though all such limits will also depend on other parameters that control the dark pion lifetime, such as $m_{\hat{\pi}}$.

Finally, we remark that this strategy can be applied to searching for other long-lived light BSM bosons, if they are produced and decay in a similar way. Also, it is interesting to extend this study to the case where these particles decay outside the detector and hence contribute to the missing energy directly. In the latter case, the CLEO analysis performed about twenty years ago~\cite{CLEO:2001acz} still provides the current strongest constraints on $\text{BR}(B^\pm \to \pi^\pm/K^\pm + X) < 4.9\times 10^{-5}$. 
These constraints can be interpreted as $f_a \gtrsim 10^8$~GeV in the relevant QCD axion scenarios~\cite{MartinCamalich:2020dfe}. However, the sensitivity prospect for such a measurement at the CEPC is still missing.

\section{Detector Performance Requirements}
\label{sec:detectorrequirement}

The CEPC’s extensive flavor physics program consequently imposes stringent and multifaceted requirements on detector performance, which becomes a key challenge in the design and optimization of the CEPC detector.
Many physics benchmark analyses presented in this manuscript serve as references for detector requirement and optimization studies by quantifying the correlations between anticipated precisions and critical detector performance. These studies indicate that a suitable detector for the CEPC flavor physics measurements should be able to:

\begin{itemize}
    \item Provide a large acceptance of nearly 4$\pi$ solid angle coverage, a low momentum threshold for charged tracks, and low energy thresholds for photons and neutral hadrons.    
    In flavor physics, many measurements involve reconstruction of excited heavy hadrons. These excited resonances could decay into their base state together with a photon or a pion with typical energy of $\mathcal{O}(10-100)\,$MeV, as shown in Figure~\ref{fig:low_threshold_benchmark}.
    The low energy/momentum threshold is crucial for identifying these heavy-flavored hadrons. Notably, low-momentum charged pions also contribute to the jet charge measurement.

    \begin{figure}[t]
        \centering
        \includegraphics[width=0.45\textwidth]{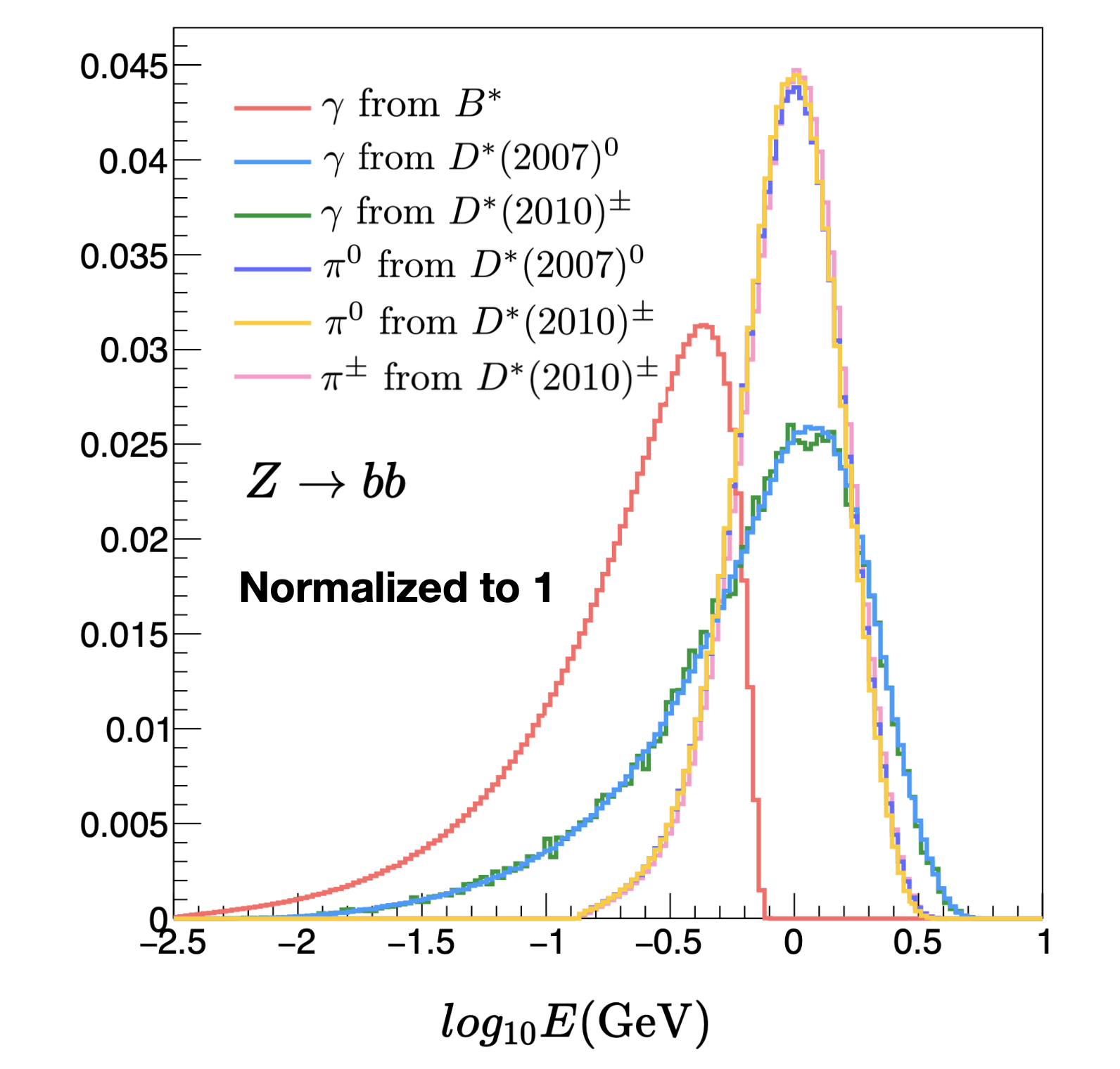}
        \includegraphics[width=0.45\textwidth]{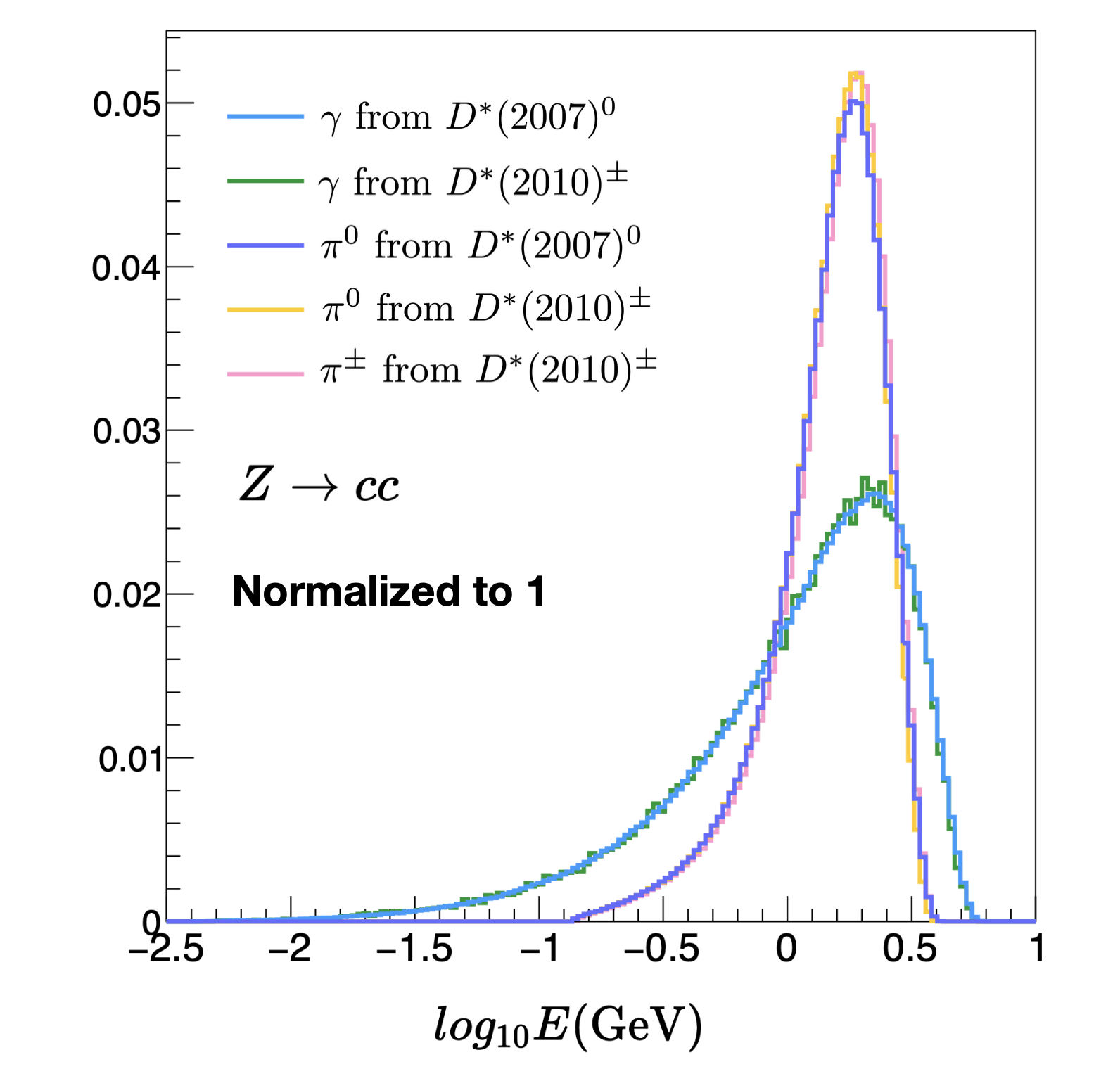}
        \caption{Energy distributions of $\gamma$, $\pi^0$, and $\pi^{\pm}$ generated from the decays of typical excited heavy hadrons in the $Z\to b\bar{b}, c\bar{c}$ ($\sqrt{s}$ = 91.2 GeV) processes.}
        \label{fig:low_threshold_benchmark}
    \end{figure}

    \item Achieve excellent intrinsic resolution. Usually, the intrinsic momentum resolution of the tracker should reach 0.1\% level in the barrel region, while the intrinsic energy resolution of the ECAL is suggested to be better than $3\%/\sqrt{E\text{(GeV)}}$.
    The latter is particularly relevant for distinguishing between $B^0$ and $B^0_s$ when they decay into photons~\cite{Wang:2022nrm}. 
    Moreover, to efficiently reconstruct the decay vertex of $\tau$ lepton and heavy flavor hadrons, the vertex position resolution is suggested to be better than 5~$\upmu$m, with the vertex detector placed sufficiently close to the interaction point~\cite{Vcb_Hao_Talk}.

    \item Provide excellent particle flow reconstruction and PID. The CEPC flavor physics significantly involves analyzing hadronic events at the $Z$ pole. Accurately identifying the decay products (charged particles, photons, and neutral hadrons) of individual heavy-flavored particles such as $b$-hadron and $\tau$ is thus important. 
    Figure~\ref{fig:Pid requirement} demonstrates the reconstruction efficiency and purity of $\phi$ in the decay $B_{s}^{0} \to \phi\nu\bar{\nu}$ and the  anticipated precision of measuring its signal rate as a function of the $K/\pi$ separation power. Such a correlation indicates the necessity of obtaining a $K/\pi$ separation power better than $3\,\sigma$~\cite{Li:2022tov,Li:2022tlo}. 
    The PID can be improved with various technologies. For example, the CEPC CDR detector employs TPC as its main tracker, which could provide $dE/dx$ and $dN/dx$ measurements. If the $dE/dx$ (or $dN/dx$) can be measured with a relative accuracy of 3\%, and considering a TOF measurement of 50~ps at cluster level~\cite{Che:2022dig}, the reconstruction efficiency and purity of inclusive charged kaons in the hadronic $Z$ pole sample could both exceed 95\%. Recently, a concept of one-to-one correspondence reconstruction between visible final state particles and reconstructed particles was developed, by applying ML techniques to the information from the 5-dimensional calorimeter~\cite{Wang:2024eji}.  The potential of identifying nine types of particles simultaneously is demonstrated in left panel of Figure~\ref{fig:PFA_1-1}. For charged particles and photons, identification efficiencies of 97\% to nearly 100\% could be achieved, while for neutral hadrons, efficiencies of 75\% to 80\% are also attainable.

    \begin{figure}[t]
        \centering
        \includegraphics[height=5.5cm]{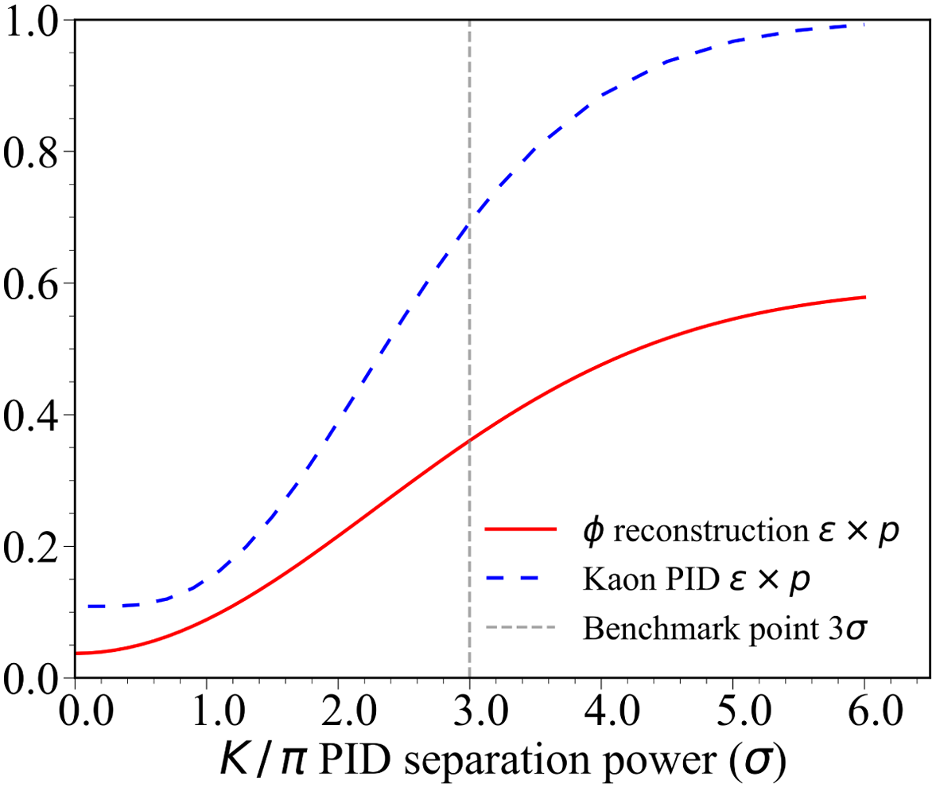}
        \includegraphics[height=5.5cm]{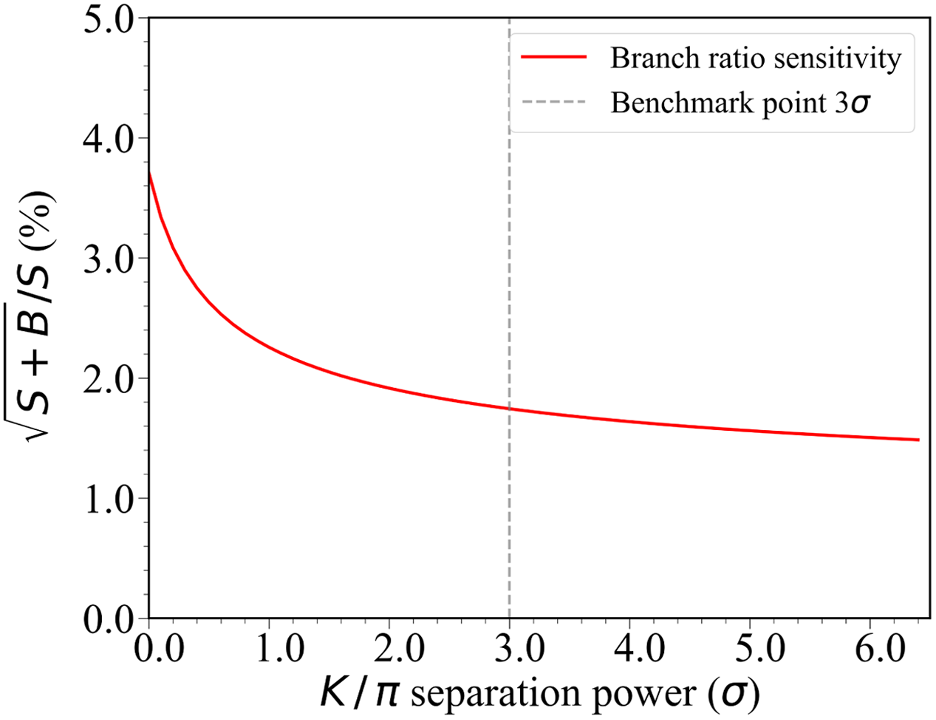}
        % \hfill
        \caption{Reconstruction efficiency and purity of $\phi$ in the decay $B_{s}^{0} \to \phi\nu\bar{\nu}$ (\textbf{LEFT}) and anticipated precision of measuring its signal rate (\textbf{RIGHT}) as a function of the $K/\pi$ separation power. Here, the separation power is defined as $|\mu_1 - \mu_2|/\sqrt{\sigma_{1}^2 +\sigma_{2}^2}$, where $\mu_i$ and $\sigma_i$ denote the mean and standard deviation of Gaussian distributions. The two plots are taken from~\cite{Li:2022tov}.
        }
        \label{fig:Pid requirement}
    \end{figure}

    \item Reconstruct missing energy and momentum excellently.
    The CEPC is expected to offer a unique advantage over hadron collider for the measurements involving missing energy and momentum, such as those of $b$-hadron semi-leptonic decays and potential dark matter production. As these measurements are often based on hadronic events at the $Z$ pole, accurately reconstructing the four-momentum of visible final state particles is essential for meeting this expectation. The PFA is crucial in this regard, by integrating information from various sub-detectors to achieve high precision. 
    As shown in~\cite{Wang:2024eji},  
    using the reconstruction of one-to-one correspondence can improve the BMR by 25\% beyond the CDR performance, achieving a value below 3\% (see right panel of Figure 44).
    A better BMR, and consequently improved missing momentum resolution at the CEPC, will enhance the flavor measurements involving missing particles and may enable new flavor physics measurements that are not feasible in other experiments.

    \item Deliver stable performance over time. The stability of detector response is crucial for minimizing systematic uncertainties. Reliable performance depends on the system's ability to endure the beam environment, so the detector design must be robust enough to withstand beam-induced background while limiting its impact on physics measurements to an acceptable level. Efficient monitoring of various subsystems %such as the magnet and cryogenic systems, 
    is essential for calibrating the detector and mitigating systematic effects. 
    Currently, the machine-detector interface optimization, integration studies and machine protection designs are still in active development.
    Additionally, the accelerator's performance must remain stable, as it directly influences the collision environment, including instantaneous luminosity and collision energy. The accelerator ring may also contribute significantly to machine-induced background, introducing further systematic uncertainties. These discussions are especially relevant for measuring tree-level processes in flavor physics such as the FCCC transitions. As the signal rates are relatively high, in these cases the statistical errors could be much lower than systematic uncertainties.

    \item Realize a scenario of being effectively triggerless and free from pile-ups. The CEPC detector is anticipated to efficiently reconstruct physics events while minimizing noise contamination to an acceptable level. With an event rate of $10^5$ Hz at the $Z$ pole, a dedicated Trigger-DAQ system is essential to meet this expectation, known as the triggerless equivalent scenario. Additionally, online event-building could be complicated due to the high event rate and the varying response times of different subdetectors ({\it e.g.}, TPC and calorimeters may detect neutron-induced hits milliseconds after a collision), leading to overlapping events. This makes it impossible to separate events based solely on time. New reconstruction technologies are thus needed to efficiently and accurately reconstruct low-level physics objects such as tracks and clusters and associate them with different vertices. One potential solution is to use the PFA that incorporates both spatial and temporal information.

\end{itemize}

% ~~~~~~~~~~~~~~~~~~~~~~~~~~~~~~~~~~~~~~~~~~~~~~~~~~
\begin{figure}[!t]
    \centering
    \adjustbox{valign=c}{\includegraphics[scale=0.288]{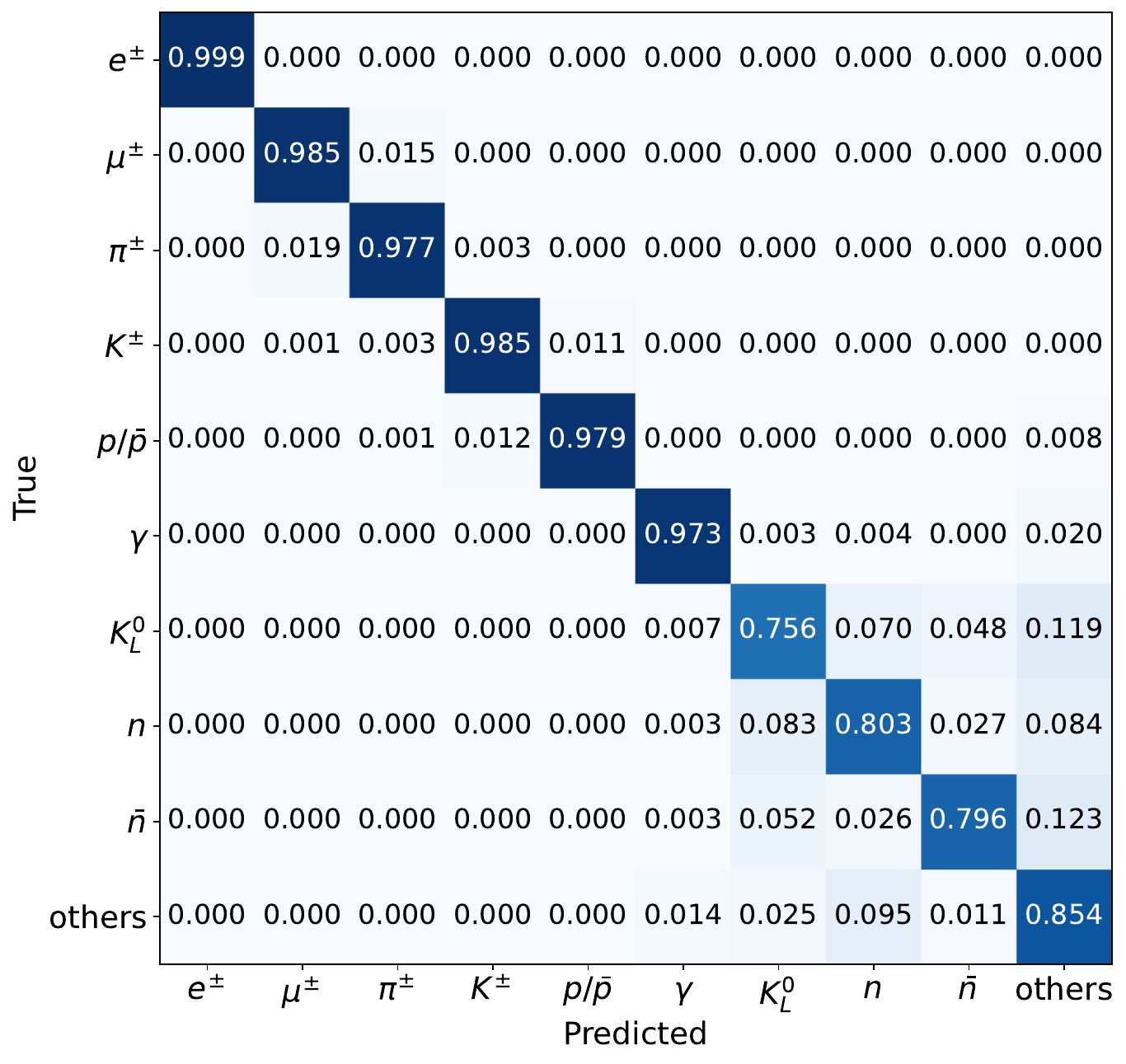}}
    \adjustbox{valign=c}{\includegraphics[scale=0.35]{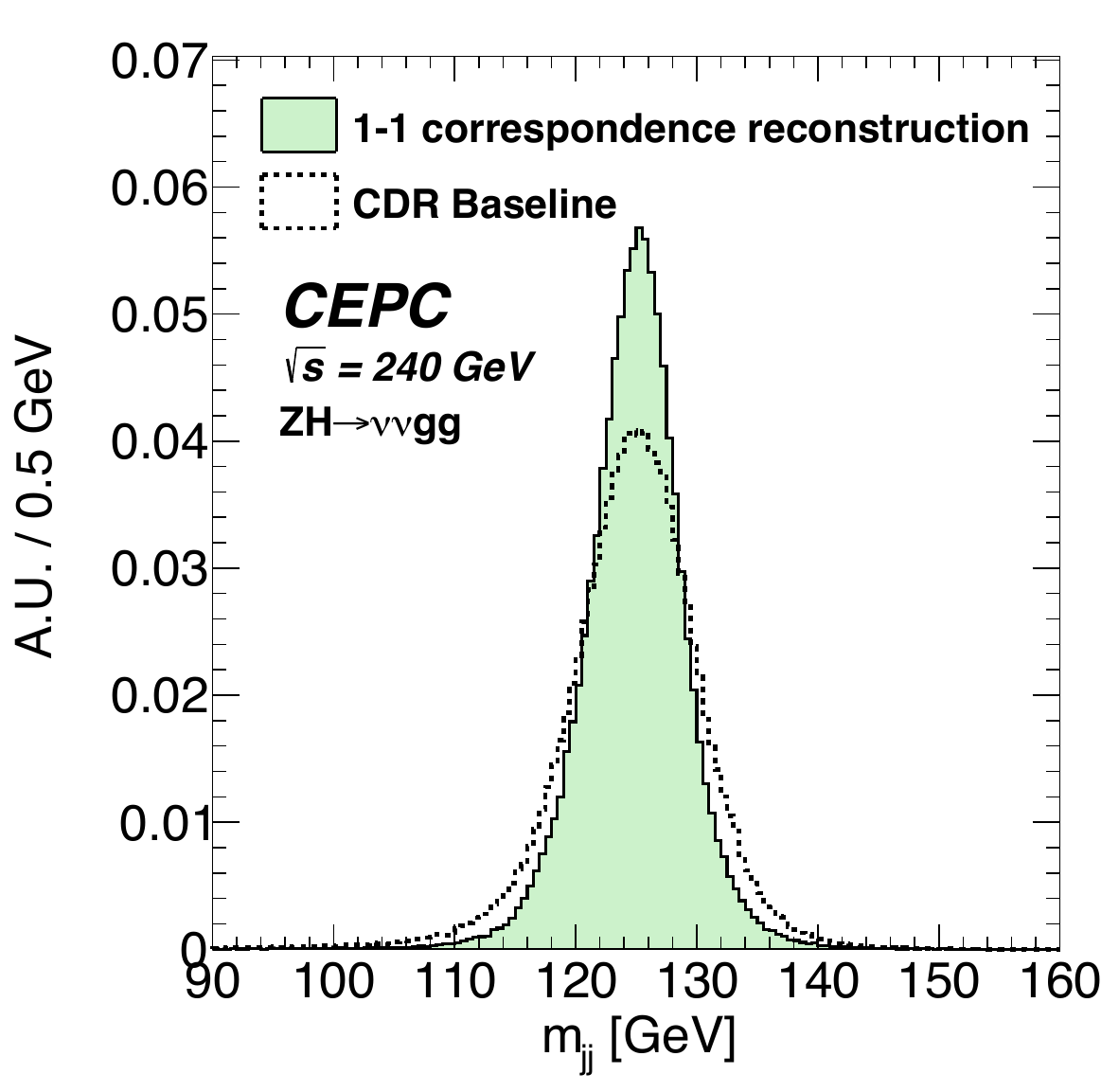}}
    \caption{
       \textbf{LEFT:} Confusion matrix for the identification of 9 types of particles.
    \textbf{RIGHT:} Invariant mass distributions of hadronically decayed Higgs bosons at the CEPC CDR phase and improved by the one-to-one (1-1) correspondence reconstruction.
  Both plots are taken from~\cite{Wang:2024eji}.
    }
    \label{fig:PFA_1-1}
\end{figure}

All of these requirements could be addressed through comprehensive detector design, key technology R$\&$D, and reconstruction algorithm studies. It is crucial to consider them collectively, as many are interconnected and may conflict with each other. For example, while incorporating TOF systems can significantly enhance PID performance, it also introduces additional upstream material that may adversely affect the intrinsic energy resolution of the ECAL.

\section{Summary and Outlook}
\label{sec:summary}

An electron-positron Higgs factory is identified as the highest priority for future collider facilities. 
According to its accelerator TDR~\cite{CEPCStudyGroup:2023quu}, the CEPC is expected to produce 4 million Higgs bosons, 4 trillions of $Z$ bosons, and billions of $W$ bosons during its 13 years of operation across multiple runs. The CEPC's instantaneous luminosity is so high that it could generate the entire statistics of LEP-I in approximately one minute. This facility thus presents an unprecedented opportunity to advance the study of particle physics.

% \newpage
%~~~~~~~~~~~~~~~~~~~~~~~~~~~~~~~~~~~~~~~~~~~
\begin{sidewaystable}[thp]
\centering
    % \scriptsize
    \tiny
    \renewcommand\arraystretch{1.2}
    \resizebox{.95\textwidth}{!}{
    \begin{tabular}[t]{ccccccccc}
    
        % Title
        % -----------------------------------------------------------
        \hline
        No.
        & Process
        & $\sqrt{s}$ (GeV)
        % & \makecell{Parameter\\ of interest}
        & \makecell{Observable/\\ physics parameter}
        & \makecell{Current precision}
        & \makecell{CEPC precision}
        % & \makecell{CEPC Precision\\ 4 Tera-$Z$}
        & \makecell{Estimation\\ method}
        & \makecell{Key performance}
        & \makecell{Relevant\\ section}
        \\

        % |Vcb|, Bc->tau nu
        % -----------------------------------------------------------
        \hline
        1
        & $B_c\to\tau\nu$
        & 91.2
        % & $|V_{cb}|$
        % & $\sigma(\mu)/\mu$
        & \makecell{BR\\ ($|V_{cb}|$)}
        & $\lesssim 30\%$~\cite{Akeroyd:2017mhr}
        % & \makecell{relative (stat. only)\\ $\mathcal{O}(1\%)$~\cite{Zheng:2020ult}}
        & \makecell{relative (stat. only)\\ $\mathcal{O}(0.5\%) ^{*}$~\cite{Zheng:2020ult}}
        & Full simulation
        & \makecell{Tracker\\ Lepton ID\\ Missing energy\\ Jet origin ID}
        & \ref{sec:FCCC}
        \\

        % LFU R(Hc), b->clv, $b\to c\ell \nu$, $R_{H_c}$
        % -----------------------------------------------------------
        \hline
        2
        & $B_c\to J/\psi \tau \nu$
        & 91.2
        % & -
        & $R_{J/\psi}$
        % & $0.71 \pm 0.17 \pm 0.18$~\cite{Aaij:2017tyk}
        & \makecell{$\pm 0.17 \pm 0.18$~\cite{Aaij:2017tyk}\\ relative $\pm 24\% \pm 25\%$}
        % & \makecell{relative (stat. only)\\ $\lesssim 5\%$~\cite{Ho:2022ipo}}
        & \makecell{relative (stat. only)\\ $\lesssim 2.5\% ^{*}$~\cite{Ho:2022ipo}}
        & Fast simulation
        & \makecell{Tracker\\ Vertex}
        & \ref{sec:FCCC}
        \\
        
        % LFU R(Hc), b->clv
        % -----------------------------------------------------------
        \hline
        3
        & $B_s^0 \to D_s \tau \nu$
        & 91.2
        % & -
        & $R_{D_s}$
        & -
        % & \makecell{relative (stat. only)\\ $\lesssim 0.4\%$~\cite{Ho:2022ipo}}
        & \makecell{relative (stat. only)\\ $2.1 \times 10^{-3\,*}$~\cite{Ho:2022ipo}}
        & Fast simulation
        & \makecell{Tracker\\ Vertex}
        & \ref{sec:FCCC}
        \\
        % -----------------------------------------------------------
        \hline
        4
        & $B_s^0 \to D_s^{\ast} \tau \nu$
        & 91.2
        % & -
        & $R_{D_s^{\ast}}$
        & -
        % & \makecell{relative (stat. only)\\ $\lesssim 0.4\%$~\cite{Ho:2022ipo}}
        & \makecell{relative (stat. only)\\ $1.6 \times 10^{-3\,^*}$~\cite{Ho:2022ipo}}
        & Fast simulation
        & \makecell{Tracker\\ Vertex}
        & \ref{sec:FCCC}
        \\
        
        % LFU R(Hc), b->clv
        % -----------------------------------------------------------
        \hline
        5
        & $\Lambda_b \to \Lambda_c \tau \nu$
        & 91.2
        % & -
        & $R_{\Lambda_c}$
        % & 0.242 \pm 0.076$~\cite{LHCb:2022piu}
        & \makecell{$\pm 0.076$~\cite{LHCb:2022piu}\\ relative 31\%}
        % & \makecell{relative (stat. only)\\ $\sim 0.1\%$~\cite{Ho:2022ipo}}
        & \makecell{relative (stat. only)\\ $\sim 0.05\%^{\,*}$~\cite{Ho:2022ipo}}
        & Fast simulation
        & \makecell{Tracker\\ Vertex}
        & \ref{sec:FCCC}
        \\
        
        % & \makecell{$\mathcal{O}(1)$ deviation from SM\\ \color{red}{30 TeV?}}
        % LFU, Anomaly, $b\to s\tau^- \tau^+$
        % -----------------------------------------------------------
        \hline
        6
        & $B^0\to K^{\ast 0} \tau^- \tau^+$
        & 91.2
        % & -
        & BR
        & -
        & $\lesssim \mathcal{O}(10^{-6})$~\cite{Li:2020bvr}
        % & 
        & Fast simulation
        & \makecell{Tracker\\ Vertex\\ Jet origin ID}
        & \ref{sec:FCNC}
        \\
        
        % LFU, Anomaly, b->s tau tau
        % -----------------------------------------------------------
        \hline
        7
        & $B_s^0\to\phi \tau^- \tau^+$
        & 91.2
        % & -
        & BR
        & -
        & $\lesssim \mathcal{O}(10^{-6})$~\cite{Li:2020bvr}
        % & 
        & Fast simulation
        & \makecell{Tracker\\ Vertex\\ Jet origin ID}
        & \ref{sec:FCNC}
        \\

        % LFU, Anomaly, b->s tau tau
        % -----------------------------------------------------------
        \hline
        8
        & $B^+ \to K^+ \tau^- \tau^+$
        & 91.2
        % & -
        & BR
        & $<2.25\times10^{-3}$~\cite{ParticleDataGroup:2024cfk}
        & $\lesssim \mathcal{O}(10^{-6})$~\cite{Li:2020bvr}
        % & 
        & Fast simulation
        & \makecell{Tracker\\ Vertex\\ Jet origin ID}
        & \ref{sec:FCNC}
        \\

        % LFU, Anomaly, b->s tau tau
        % -----------------------------------------------------------
        \hline
        9
        & $B_s^0 \to \tau^- \tau^+$
        & 91.2
        % & -
        & BR
        & $<6.8\times10^{-3}$~\cite{ParticleDataGroup:2024cfk}
        & $\lesssim \mathcal{O}(10^{-5})$~\cite{Li:2020bvr}
        % & 
        & Fast simulation
        & \makecell{Tracker\\ Vertex\\ Jet origin ID}
        & \ref{sec:FCNC}
        \\
        
        % Anomaly, $b\to s\nu\bar{\nu}$, Bs->phivv
        % -----------------------------------------------------------
        \hline
        10
        & $B_s^0\to \phi\nu\bar{\nu}$
        & 91.2
        % & -
        % & $\sigma(\mu)/\mu$
        & BR
        & $< 5.4 \times 10^{-3}$~\cite{ParticleDataGroup:2024cfk}
        % & \makecell{relative (stat. only)\\ $\lesssim 2\%$~\cite{Li:2022tov}}
        & \makecell{relative (stat. only)\\ $\lesssim 1\% ^{*}$~\cite{Li:2022tov}}
        & Full simulation
        & \makecell{Tracker\\ Vertex\\ Missing energy\\ PID}
        & \ref{sec:FCNC}
        \\
        
        % $B_s^0\to J/\psi\phi$, $\phi_s$, $\Gamma_s$, $\Delta \Gamma_s$
        % -----------------------------------------------------------
        \hline
        11
        & $B_s^0\to J/\psi\phi$
        & 91.2
        % & $\phi_s$ (= -2$\beta_s$)
        & $\Gamma_s$, $\Delta \Gamma_s$, $\phi_s$
        
        % & \makecell{$\Gamma_s$ = 657.3 $\pm$ 2.3 ns$^{-1}$~\cite{ParticleDataGroup:2024cfk}\\ $\Delta \Gamma_s$ = $65.7 \pm 4.3 \pm 3.7$ ns$^{-1}$~\cite{ATLAS:2020lbz}\\ $\phi_s$ = $-87 \pm 36 \pm 21$ mrad~\cite{ATLAS:2020lbz}}

        & \makecell{$\sigma(\Gamma_s)$ = $\pm$ 2.3 ns$^{-1}$~\cite{ParticleDataGroup:2024cfk}\\ $\sigma(\Delta \Gamma_s)$ = $\pm 4.3 \pm 3.7$ ns$^{-1}$~\cite{ATLAS:2020lbz}\\ $\sigma(\phi_s)$ = $\pm 36 \pm 21$ mrad~\cite{ATLAS:2020lbz}}
        
        % & \makecell{$\sigma(\Gamma_s)$ = 0.072 ns$^{-1}$\\ $\sigma(\Delta \Gamma_s)$ = 0.24 ns$^{-1}$\\ $\sigma(\phi_s)$ = 4.3 mrad}~\cite{Li:2022tlo}
        & \makecell{$\sigma(\Gamma_s)$ = 0.036 ns$^{-1}{}^{*}$\\ $\sigma(\Delta \Gamma_s)$ = 0.12 ns$^{-1}{}^{*}$\\ $\sigma(\phi_s)$ = 2.2 mrad$ ^{*}$}~\cite{Li:2022tlo}
        
        & Full simulation
        & \makecell{Tracker\\ Vertex\\ Lifetime resolution\\ Jet origin ID}
        & \ref{sec:CPV}
        \\
        
        % B->pipi
        % -----------------------------------------------------------
        \hline
        12
        & \makecell{$B^0\to\pi^0\pi^0$}
        & 91.2
        % & $\alpha$
        % & \makecell{$\sigma(\mu)/\mu$\\ $A_{CP}$}
        & \makecell{BR, $A_{CP}$\\ ($\alpha$)}
        
        % & \makecell{$\text{BR}^{00} = (1.59\pm 0.26)\times 10^{-6}$ (16\%)\\ $C_{CP}^{00} = -0.33 \pm 0.22$}~\cite{ParticleDataGroup:2024cfk}
        & \makecell{$\sigma(\text{BR})/\text{BR}^{00} = 16\%$\\ $\sigma(C_{CP}^{00})$ = $\pm$ 0.22}~\cite{ParticleDataGroup:2024cfk}
        
        % & \makecell{$\sigma(\text{BR})/\text{BR}^{00} = 0.45\%$\\ $\sigma(a_{CP}^{00})$ = $\pm$ (0.014--0.018)}~\cite{Wang:2022nrm}
        & \makecell{$\sigma(\text{BR})/\text{BR}^{00} = 0.25\% ^{*}$\\ $\sigma(a_{CP}^{00})$ = $\pm 0.01 ^{*}$}~\cite{Wang:2022nrm}
        
        & Fast simulation
        & \makecell{ECAL\\ Jet origin ID}
        & \ref{sec:CPV}
        \\
        
        % B->pipi
        % -----------------------------------------------------------
        \hline
        13
        & \makecell{$B^0\to\pi^+\pi^-$}
        & 91.2
        % & $\alpha$
        % & \makecell{$\sigma(\mu)/\mu$\\ $A_{CP}$}
        & \makecell{BR\\ ($\alpha$)}
        
        % & \makecell{$\text{BR}^{+0} = (5.5\pm 0.4)\times 10^{-6}$ (7\%)}~\cite{ParticleDataGroup:2024cfk}
        & \makecell{$\sigma(\text{BR})/\text{BR}^{+0} = 7\%$}~\cite{ParticleDataGroup:2024cfk}
        
        % & \makecell{$\sigma(\text{BR})/\text{BR}^{+0} = 0.19\%$}~\cite{Wang:2022nrm}
        & \makecell{$\sigma(\text{BR})/\text{BR}^{+0} = 0.1\% ^{*}$}~\cite{Wang:2022nrm}
        & Fast simulation
        & \makecell{ECAL\\ Tracker\\ Jet origin ID}
        & \ref{sec:CPV}
        \\
        
        % B->pipi
        % -----------------------------------------------------------
        \hline
        14
        & \makecell{$B^+\to\pi^+\pi^0$}
        & 91.2
        % & $\alpha$
        % & \makecell{$\sigma(\mu)/\mu$\\ $A_{CP}$}
        & \makecell{BR, $A_{CP}$\\ ($\alpha$)}
        
        % & \makecell{$\text{BR}^{+-} = (5.12\pm 0.19)\times 10^{-6}$ (4\%)\\ $C_{CP}^{+-} = -0.314 \pm 0.030$\\ $S_{CP}^{+-} = -0.670 \pm 0.030$}~\cite{ParticleDataGroup:2024cfk}
        
        & \makecell{$\sigma(\text{BR})/\text{BR}^{+-} = 4\%$\\ $\sigma(C_{CP}^{+-})$ = $\pm$ 0.030\\ $\sigma(S_{CP}^{+-})$ = $\pm$ 0.030}~\cite{ParticleDataGroup:2024cfk}
        
        % & \makecell{$\sigma(\text{BR})/\text{BR}^{+-} = 0.18\%$\\ $\sigma(C_{CP}^{+-})$ = $\pm$ (0.004--0.005)\\ $\sigma(S_{CP}^{+-})$ = $\pm$ (0.004--0.005)}~\cite{Wang:2022nrm}
        & \makecell{$\sigma(\text{BR})/\text{BR}^{+-} = 0.1\% ^{*}$\\ $\sigma(C_{CP}^{+-})$ = $\pm 0.003 ^{*}$\\ $\sigma(S_{CP}^{+-})$ = $\pm 0.003 ^{*}$}~\cite{Wang:2022nrm}
        & Fast simulation
        & \makecell{ECAL\\ Tracker\\ Vertex\\ Jet origin ID}
        & \ref{sec:CPV}
        \\

        % LFV tau decays
        % -----------------------------------------------------------
        \hline
        15
        & $\tau \to eee$
        & 91.2
        % & -
        & BR
        & $<2.7\times 10^{-8}$~\cite{ParticleDataGroup:2024cfk}
        & $\lesssim \mathcal{O}(10^{-10})$~\cite{Dam:2018rfz,Talk_Dan}
        % & 
        & Extrapolation
        & \makecell{Tracker\\ Lepton ID}
        & \ref{sec:tau}
        \\
        
        % LFV tau decays
        % -----------------------------------------------------------
        \hline
        16
        & $\tau \to e\mu\mu$
        & 91.2
        % & -
        & BR
        & $<2.7\times 10^{-8}$~\cite{ParticleDataGroup:2024cfk}
        & $\lesssim \mathcal{O}(10^{-10})$~\cite{Dam:2018rfz,Talk_Dan}
        % & 
        & Extrapolation
        & \makecell{Tracker\\ Lepton ID}
        & \ref{sec:tau}
        \\
        
        % LFV tau decays
        % -----------------------------------------------------------
        \hline
        17
        & $\tau \to \mu ee$
        & 91.2
        % & -
        & BR
        & $<1.8\times 10^{-8}$~\cite{ParticleDataGroup:2024cfk}
        & $\lesssim \mathcal{O}(10^{-10})$~\cite{Dam:2018rfz,Talk_Dan}
        % & 
        & Extrapolation
        & \makecell{Tracker\\ Lepton ID}
        & \ref{sec:tau}
        \\
        
        % LFV tau decays
        % -----------------------------------------------------------
        \hline
        18
        & $\tau \to \mu\gamma$
        & 91.2
        % & -
        & BR
        & $<4.4\times 10^{-8}$~\cite{ParticleDataGroup:2024cfk}
        & $\lesssim \mathcal{O}(10^{-10})$~\cite{Dam:2018rfz,Talk_Dan}
        % & 
        & Extrapolation
        & \makecell{Tracker\\ Lepton ID\\ ECAL}
        & \ref{sec:tau}
        \\
        
        % LFV tau decays
        % -----------------------------------------------------------
        \hline
        19
        & $\tau \to e\gamma$
        & 91.2
        % & -
        & BR
        & $<3.3\times 10^{-8}$~\cite{ParticleDataGroup:2024cfk}
        & $\lesssim \mathcal{O}(10^{-10})$~\cite{Dam:2018rfz,Talk_Dan}
        % & 
        & Extrapolation
        & \makecell{Tracker\\ Lepton ID\\ ECAL}
        & \ref{sec:tau}
        \\
        
        % LFV tau decays
        % -----------------------------------------------------------
        \hline
        20
        & $\tau \to \mu\mu\mu$
        & 91.2
        % & -
        & BR
        & $< 2.1\times 10^{-8}$~\cite{ParticleDataGroup:2024cfk}
        & $\lesssim \mathcal{O}(10^{-10})$~\cite{Dam:2018rfz,Talk_Dan}
        % & 
        & Extrapolation
        & \makecell{Tracker\\ Lepton ID}
        & \ref{sec:tau}
        \\
        
        % (FCC)
        % -----------------------------------------------------------
        \hline
        21
        & $\tau \to {\rm incl.}$
        & 91.2
        % & -
        & \makecell{$\tau_{\tau}$ (s)\\ lifetime}
        & $\pm 5\times 10^{-16}$~\cite{ParticleDataGroup:2024cfk}
        & $\pm 1 \times 10^{-18}$~\cite{Dam:2018rfz}
        % & 
        & Extrapolation
        & -
        & \ref{sec:tau}
        \\
        
        % (FCC)
        % -----------------------------------------------------------
        \hline
        22
        & $\tau \to {\rm incl.}$
        & 91.2
        % & -
        & $m_{\tau}$ (MeV)
        & $\pm 0.12$~\cite{ParticleDataGroup:2024cfk}
        & \makecell{$\pm$ 0.004 (stat.)\\ $\pm$ 0.1 (sys.)}~\cite{Dam:2018rfz}
        % & 
        & Extrapolation
        & -
        & \ref{sec:tau}
        \\
        
        % LFU tau decay (FCC)
        % -----------------------------------------------------------
        \hline
        23
        & $\tau \to \ell \nu\bar\nu$
        & 91.2
        % & -
        & BR
        & $\pm 4\times 10^{-4}$~\cite{ParticleDataGroup:2024cfk}
        & $\pm 3\times 10^{-5}$~\cite{Dam:2018rfz}
        % & 
        & Extrapolation
        & \makecell{Tracker\\ Lepton ID\\ Missing energy}
        & \ref{sec:tau}
        \\
        \hline
    \end{tabular}
    }
\end{sidewaystable}
%~~~~~~~~~~~~~~~~~~~~~~~~~~~~~~~~~~~~~~~~~~~

% \newpage
%~~~~~~~~~~~~~~~~~~~~~~~~~~~~~~~~~~~~~~~~~~~
\begin{sidewaystable}[thp]
\centering
    % \footnotesize
    \scriptsize
    \renewcommand\arraystretch{1.3}
    \resizebox{.95\textwidth}{!}{
    \begin{tabular}[t]{ccccccccc}
    
        % Title
        % -----------------------------------------------------------
        \hline
        No.
        & Process
        & $\sqrt{s}$ (GeV)
        % & \makecell{Parameter\\ of interest}
        & \makecell{Observable/physics\\ parameter of interest}
        & \makecell{Current precision}
        & \makecell{CEPC precision}
        % & \makecell{CEPC Precision\\ 4 Tera-$Z$}
        & \makecell{Estimation\\ method}
        & \makecell{Key performance}
        & \makecell{Relevant\\ section}
        \\
        
        % -----------------------------------------------------------
        \hline
        24
        & $Z\to \pi^+\pi^-$
        & 91.2
        % & -
        & BR
        & -
        & $\lesssim \mathcal{O}(10^{-10})$~\cite{Talk_Dan}
        
        & Extrapolation
        & \makecell{Tracker\\ PID}
        & \ref{sec:Zdecay}
        \\
        
        % -----------------------------------------------------------
        \hline
        25
        & $Z\to \pi^+\pi^-\pi^0$
        & 91.2
        % & -
        & BR
        & -
        & $\lesssim \mathcal{O}(10^{-9})$~\cite{Talk_Dan}
        
        & Extrapolation
        & \makecell{Tracker\\ PID\\ ECAL}
        & \ref{sec:Zdecay}
        \\
        
        % -----------------------------------------------------------
        \hline
        26
        & $Z\to \rho \gamma$
        & 91.2
        % & -
        & BR
        & $<2.5\times 10^{-5}$~\cite{ParticleDataGroup:2024cfk}
        & $\lesssim \mathcal{O}(10^{-9})$~\cite{Talk_Dan}
        
        & Extrapolation
        & \makecell{Tracker\\ PID\\ ECAL}
        & \ref{sec:Zdecay}
        \\
        
        % -----------------------------------------------------------
        \hline
        27
        & $Z\to J/\psi \gamma$
        & 91.2
        % & -
        & BR
        & $<1.4\times 10^{-6}$~\cite{ParticleDataGroup:2024cfk}
        & $\lesssim$ $10^{-9}$--$10^{-10}$~\cite{Talk_Dan}
        
        & Extrapolation
        & \makecell{Tracker\\ PID\\ ECAL}
        & \ref{sec:Zdecay}
        \\
        
        % LFV $Z$ decays
        % -----------------------------------------------------------
        \hline
        28
        & $Z \to \tau \mu$
        & 91.2
        % & -
        & BR
        & $<6.5\times 10^{-6}$~\cite{ATLAS:2014vur,ATLAS:2020zlz,ATLAS:2021bdj}
        & $\lesssim \mathcal{O}(10^{-9})$~\cite{Dam:2018rfz,Talk_Dan}
        
        & Extrapolation
        & \makecell{$E_{\rm beam}$\\ Tracker\\ PID}
        & \ref{sec:Zdecay}
        \\
        
        % LFV $Z$ decays
        % -----------------------------------------------------------
        \hline
        29
        & $Z \to \tau e $
        & 91.2
        % & -
        & BR
        & $<5.0\times 10^{-6}$~\cite{ATLAS:2014vur,ATLAS:2020zlz,ATLAS:2021bdj}
        & $\lesssim \mathcal{O}(10^{-9})$~\cite{Dam:2018rfz,Talk_Dan}
        
        & Extrapolation
        & \makecell{$E_{\rm beam}$\\ Tracker\\ PID}
        & \ref{sec:Zdecay}
        \\
        
        % LFV $Z$ decays
        % -----------------------------------------------------------
        \hline
        30
        & $Z \to \mu e$
        & 91.2
        % & -
        & BR
        & $<7.5\times 10^{-7}$~\cite{ATLAS:2014vur,ATLAS:2020zlz,ATLAS:2021bdj}
        & $\lesssim 1\times10^{-9}$~\cite{Altmannshofer:2023tsa}
        
        & Extrapolation
        & \makecell{$E_{\rm beam}$\\ Tracker\\ PID}
        & \ref{sec:Zdecay}
        \\

        % Z Hadronic FCNC
        % -----------------------------------------------------------
        \hline
        31
        & \makecell{$Z\to bs$}
        & 91.2
        % & -
        & BR
        & -
        % & $< 2.3 \times 10{-7} ^{*}$
        & $< 10^{-7\,*}$
        
        & Fast simulation
        & \makecell{Jet origin ID}
        & \ref{sec:Zdecay}
        \\
        % -----------------------------------------------------------
        \hline
        32
        & \makecell{$Z\to bd$}
        & 91.2
        % & -
        & BR
        & -
        & $< 10^{-7\,*}$
        
        & Fast simulation
        & \makecell{Jet origin ID}
        & \ref{sec:Zdecay}
        \\
        % -----------------------------------------------------------
        \hline
        33
        & \makecell{$Z\to cu$}
        & 91.2
        % & -
        & BR
        & -
        & $< 3 \times 10^{-7\,*}$
        
        & Fast simulation
        & \makecell{Jet origin ID}
        & \ref{sec:Zdecay}
        \\
        % -----------------------------------------------------------
        \hline
        34
        & \makecell{$Z\to sd$}
        & 91.2
        % & -
        & BR
        & -
        & $< 7 \times 10^{-7\,*}$
        
        & Fast simulation
        & \makecell{Jet origin ID}
        & \ref{sec:Zdecay}
        \\

        % Higgs LFV
        % -----------------------------------------------------------
        \hline
        35
        & \makecell{$H\to e\mu$}
        & 240
        % & -
        & BR
        & $<4.4 \times 10^{-5}$~\cite{ParticleDataGroup:2024cfk}
        & $<6 \times 10^{-6}$~\cite{Qin:2017aju}
        
        & Fast simulation
        & \makecell{Lepton ID}
        & \ref{sec:beyondZ}
        \\
        % -----------------------------------------------------------
        \hline
        36
        & \makecell{$H\to \mu\tau$}
        & 240
        % & -
        & BR
        & $<1.5 \times 10^{-3}$~\cite{ParticleDataGroup:2024cfk}
        & $<6 \times 10^{-5}$~\cite{Qin:2017aju}
        
        & Fast simulation
        & \makecell{Lepton ID}
        & \ref{sec:beyondZ}
        \\
        % -----------------------------------------------------------
        \hline
        37
        & \makecell{$H\to e\tau$}
        & 240
        % & -
        & BR
        & $<2.0 \times 10^{-3}$~\cite{ParticleDataGroup:2024cfk}
        & $<8 \times 10^{-5}$~\cite{Qin:2017aju}
        
        & Fast simulation
        & \makecell{Lepton ID}
        & \ref{sec:beyondZ}
        \\

        % Higgs FCNC
        % -----------------------------------------------------------
        \hline
        38
        & \makecell{$H\to sb$}
        & 240
        % & -
        & BR
        & $\lesssim 10^{-2}$~\cite{ILYUSHIN2020114921}
        % & $\lesssim$ 0.02\%–-0.1\%~\cite{Liang:2023wpt}
        & $<2.2 \times 10^{-4}$~\cite{Liang:2023wpt}
        
        & Full simulation
        & \makecell{Jet origin ID}
        & \ref{sec:beyondZ}
        \\
        % Higgs FCNC
        % -----------------------------------------------------------
        \hline
        39
        & \makecell{$H\to sd$}
        & 240
        % & -
        & BR
        & -
        % & $\lesssim$ 0.02\%–-0.1\%~\cite{Liang:2023wpt}
        & $<8.6 \times 10^{-4}$~\cite{Liang:2023wpt}
        
        & Full simulation
        & \makecell{Jet origin ID}
        & \ref{sec:beyondZ}
        \\
        % Higgs FCNC
        % -----------------------------------------------------------
        \hline
        40
        & \makecell{$H\to db$}
        & 240
        % & -
        & BR
        & $\lesssim 10^{-2}$~\cite{ILYUSHIN2020114921}
        % & $\lesssim$ 0.02\%–-0.1\%~\cite{Liang:2023wpt}
        & $<2.3 \times 10^{-4}$~\cite{Liang:2023wpt}
        
        & Full simulation
        & \makecell{Jet origin ID}
        & \ref{sec:beyondZ}
        \\
        % Higgs FCNC
        % -----------------------------------------------------------
        \hline
        41
        & \makecell{$H\to uc$}
        & 240
        % & -
        & BR
        & -
        % & $\lesssim$ 0.02\%–-0.1\%~\cite{Liang:2023wpt}
        & $<3.9 \times 10^{-4}$~\cite{Liang:2023wpt}
        
        & Full simulation
        & \makecell{Jet origin ID}
        & \ref{sec:beyondZ}
        \\

        % Top FCNC
        % -----------------------------------------------------------
        \hline
        42
        % & $e^-e^+ \to t(\bar{t})j$
        & $e^-e^+ \to tq$
        & 240
        % & -
        % & FCNC constraint coefficients
        & cross section
        & \makecell{two-fermion, LHC~\cite{ATLAS:2015iqc,ATLAS:2018xxe,ATLAS:2018zsq,CMS:2015kek,CMS:2016uzc}\\ four-fermion, LEP2~\cite{OPAL:2001spi,ALEPH:2002wad,L3:2002hbp,DELPHI:2011ab}}
        & \makecell{1--2 orders of magnitude\\ improvement compared to LEP2}~\cite{Shi:2019epw}
        
        & Fast simulation
        & \makecell{Tracker\\ Missing energy\\ Jet origin ID}
        & \ref{sec:beyondZ}
        \\
        
        % |Vcb|, W decay
        % & \makecell{$(38.9 \pm 0.53) \times 10^{-3}$\\ relative $\sim 1.4\%$}
        % & \makecell{HFLAV, average from\\ $B\to D^{(*)}\ell\nu$, $B_s^0\to D^{(*)}_s\mu\nu$}~\cite{HeavyFlavorAveragingGroup:2022wzx}
        % -----------------------------------------------------------
        \hline
        43
        % & \makecell{$WW \to \mu\nu qq$\\ $WW\to\tau(\to\mu\nu\nu)\nu qq$}
        & $WW \to \ell\nu qq$
        & 240
        % & $|V_{cb}|$
        & $|V_{cb}|$
        
        % & \makecell{$(38.9 \pm 0.53) \times 10^{-3}$\\ relative $\sim 1.4\%$}~\cite{HeavyFlavorAveragingGroup:2022wzx}
        % & $(38.9 \pm 0.53) \times 10^{-3}$~\cite{HeavyFlavorAveragingGroup:2022wzx}
        
        & \makecell{$\pm 0.5 \times 10^{-3}$ (inclusive)\\ $\pm 0.6 \times 10^{-3}$ (exclusive)\\ $\pm 1.2 \times 10^{-3}$ (average)}~\cite{ParticleDataGroup:2024cfk}
        
        % & $\lesssim 0.5\%$~\cite{Liang:2024hox}
        & \makecell{$\lesssim 0.2 \times 10^{-3}$~\cite{Liang:2024hox}\\ $L=20$~ab$^{-1}$}
        
        & Full simulation
        & \makecell{Jet origin ID}
        & \ref{sec:beyondZ}
        \\

        % BSM
        % -----------------------------------------------------------
        % & $f_a/c^A_{\mu\mu} \gtrsim 1$~TeV
        \hline
        44
        % & $Z\to \mu\mu a$ 
        & $Z\to \mu\mu X_{\text{inv}}$ 
        & 91.2
        % & -
        & BR
        & -
        & $\lesssim 3\times 10^{-11}$~\cite{Calibbi:2022izs}
        
        & Fast simulation
        & \makecell{Tracker\\ Missing energy}
        & \ref{sec:BSM}
        \\
        
        % tau, BSM
        % -----------------------------------------------------------
        % & $\sim$ $\mathcal{O}(10^8)$ GeV (for $c^{A,V}_{\tau\mu} \sim \mathcal{O}(1)$)
        \hline
        45
        % & $\tau \to \mu a$
        & $\tau \to \mu X_{\text{inv}}$
        & 91.2
        % & -
        & BR
        & $\lesssim 7 \times 10^{-4}$~\cite{Belle-II:2022heu}
        & $\lesssim$ 3--5 $\times 10^{-6}$
        
        & Fast simulation
        & \makecell{Tracker\\ Missing energy}
        & \ref{sec:BSM}
        \\
        
        % % BSM
        % The typical proper lifetime of the LLP in benchmark No. 32 is $\sim$ 0.1--10 cm.
        % % -----------------------------------------------------------
        % % & $\sim$ $\mathcal{O}(10^7)$ GeV ($c^A_{ff}\sim \mathcal{O}(1)$)
        % \hline
        % 32
        % % & $B\to K \hat\pi(\to \mu\mu)$
        % & $B\to K X_{\text{LLP}}(\to \mu\mu)$
        % & 91.2
        % % & -
        % & BR
        % & -
        % & $\lesssim 10^{-10}$~\cite{Hsin-Chia:2023asd}
        
        % & Fast simulation
        % & \makecell{Tracker\\ Vertex}
        % & \ref{sec:BSM}
        % \\

        \hline
    \end{tabular}
    }
    \caption{Summary of flavor physics benchmarks at the CEPC.
    The relevant observables and physics parameters are listed. 
    The symbol $X_{\rm inv}$ in benchmarks No.~44--45 denotes invisible NP particles.
    The CEPC precision for benchmarks marked with stars (*) is obtained by scaling to the statistic of 4 Tera-$Z$, and for Z FCNC hadronic decays (No.~31--34) is statistical only.}
    \label{tab:Benchmarks}
    
\end{sidewaystable}
%~~~~~~~~~~~~~~~~~~~~~~~~~~~~~~~~~~~~~~~~~~~

\begin{figure}[pt]
    \centering
    \rotatebox{90}{
        \begin{minipage}{\textheight}
            \centering
            \includegraphics[width=1.\textwidth]{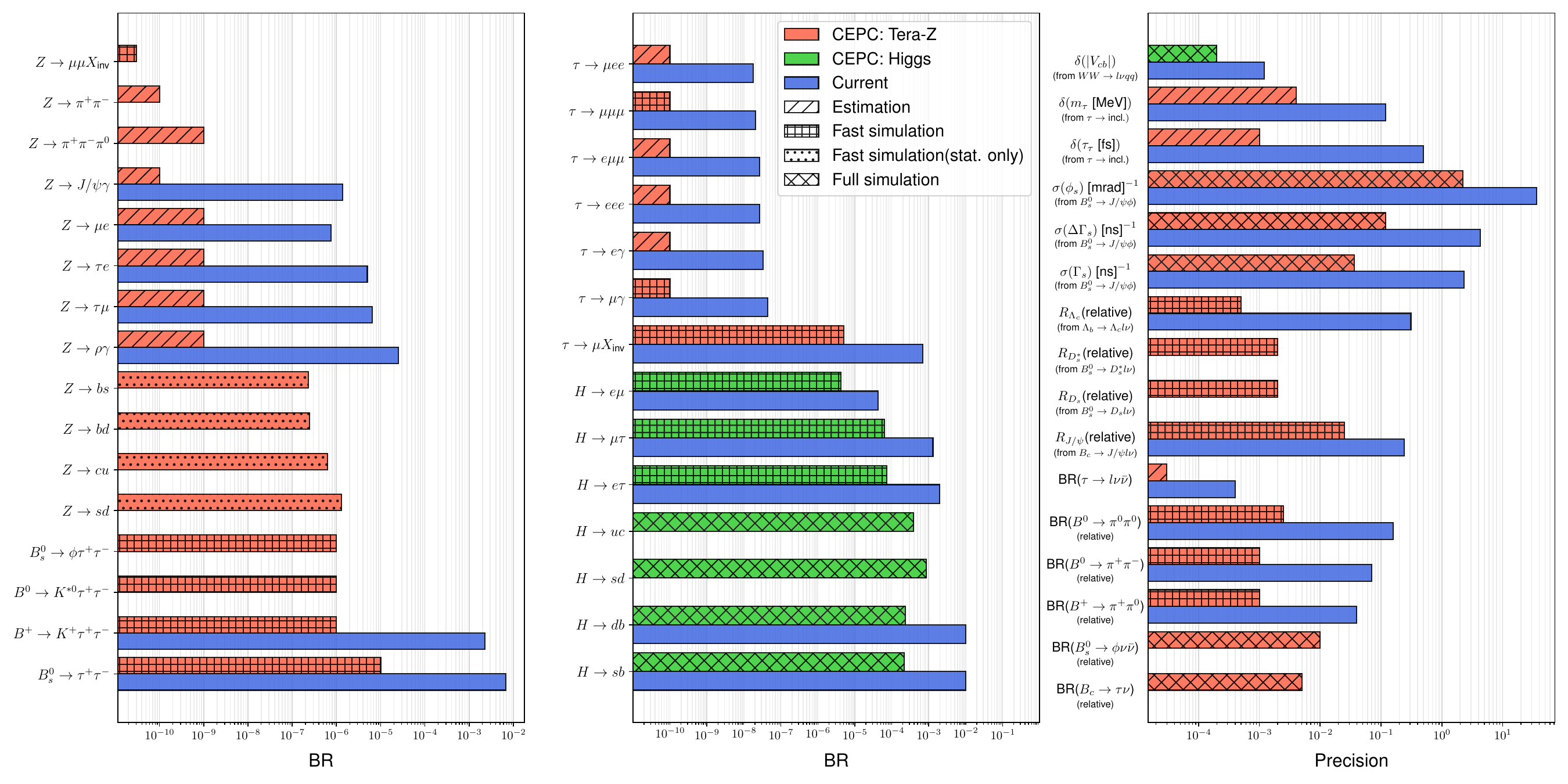}
            \caption{Anticipated upper limits or measurement precisions for the flavor physics benchmarks at the CEPC.
            It should be remarked that the limits of $Z$ hadronic FCNC decays are statistic w.r.t. current performance of jet origin identification, whose calibration remains challenging. A breakthrough is thus needed to control the relevant systematic uncertainty to a level comparable to their statistic ones. 
            }
            \label{fig:BRLimit-Precision}
        \end{minipage}
    }
    \vspace{-8pt}
\end{figure}

This manuscript presents the flavor physics landscape at the CEPC, focusing on heavy-flavored systems particularly $b$-hadrons and $\tau$ leptons, as well as heavy bosons such as $Z$ and $H$. To provide a systematic understanding, the investigation encompasses various physics topics, including FCCC and FCNC transitions, $CP$ violation, LFU, LNV and BNV, exotic states, light BSM particles with a particular emphasis on the $Z$ pole run. The estimated upper limits or measurements' precision for the CEPC benchmarks are summarized in Table~\ref{tab:Benchmarks}, and then visualized as a histogram in Figure~\ref{fig:BRLimit-Precision}.
These benchmarks have been analyzed using various methods for sensitivity estimation, including full simulation, fast simulation based on detector performance modeling, and extrapolations from existing studies. These efforts ensure a comprehensive evaluation of the CEPC's capabilities in exploring flavor physics.

Compared to existing flavor physics platforms, particularly LHCb and Belle II, the CEPC offers significant advantages and unique opportunities for a wide range of measurements. Unlike hadron colliders, the CEPC provides a much cleaner collision environment and a more precise, controllable initial state.
In addition to the favorable collision environment, the PFA-oriented design of the CEPC detector, coupled with the potential implementation of a high-precision calorimeter system, allows for accurate reconstruction of neutral and missing final states. This capability positions the CEPC to excel in measurements involving photons, neutral pions, leptons, and neutrinos, making its results superior to those from LHCb,
and even surpassing those from the upgraded LHCb at the HL-LHC
(see particularly Sections~\ref{sec:FCCC} and \ref{sec:FCNC}).
With a well-defined initial state and reduced event pile-up, the CEPC can effectively access radiative and leptonic decays, thereby enhancing sensitivity of measuring FCNC processes (as discussed in Section~\ref{sec:FCNC}), testing LFV and LFU in $\tau$ decays (see Section~\ref{sec:tau}) and $Z$ boson decays (see Section~\ref{sec:Zdecay}), and searches for rare decay modes.
Moreover, the heavy-flavored hadrons and $\tau$ leptons produced at the CEPC experience a larger boost compared to those generated at $B$ and tau-charm factories~\cite{Belle-II:2018jsg,Achasov:2023gey}. This results in improved precision for measuring lifetimes and secondary vertices, particularly for time-dependent $CP$ asymmetries (as elaborated in Section~\ref{sec:CPV}).
On top of the Tera-$Z$ run, the CEPC will also provide flavor physics measurements at higher center-of-mass energies especially with large integrated luminosity at the Higgs operation, which 
enables precise measurements of flavor-violating Higgs processes and offers direct assessment of the CKM matrix elements through the decays of $W$ bosons (see Section~\ref{sec:beyondZ}). 
The CEPC's wide beam energy range also facilitates the study of hadronic states that cannot be directly produced at Belle II, including $B_c$, $\Lambda_{b}$ and many exotic hadronic states (discussed in  Section~\ref{sec:exotic}).

It is important to emphasize that the flavor physics program at the CEPC is exceptionally rich and diverse, and this paper does not capture the full extent of its potential. Numerous intriguing topics remain to be explored, each offering unique opportunities for discovery. For instance, assessing the impact of the Tera-$Z$ facility, in conjunction with existing experimental setups, on the global CKM fit could refine our understanding of quark mixing parameters. Additionally, extending the study of FCNC from $b\to s \tau\tau$ transition to include the first two generations of leptons would allow researchers to test LFU. Systematically studying $CP$ asymmetry in $B$ and $C$ mesons, and potentially extending it to other meson systems presents exciting avenue for understanding the BAU. Furthermore, physics measurements utilizing $\tau$ lepton pair production at the $Z$ pole can provide critical insights into LFV. The largely unexplored charm and strange quark physics at the CEPC also offer valuable opportunities to investigate strong interactions and flavor symmetries. Lastly, exploring flavor physics beyond the $Z$ pole, {\it i.e.}, flavor-violating top quark decays and searches for light BSM resonances at the $t\bar t$ threshold, could yield significant insights into high-energy processes. Collectively, these research directions will significantly enhance our understanding of fundamental particle interactions and may uncover NP that challenges or extends the current theoretical framework.

To explore the rich flavor physics at the CEPC imposes stringent requirements on detector performance. 
A large geometrical acceptance and low energy/momentum thresholds can reduce the chance of missing visible particles, particularly at the endcap and forward region. Moreover, the efficient separation, reconstruction, and identification of final state particles, where the newly developed method of one-to-one correspondence may play a role, will greatly benefit the reconstruction of hadron events. 
Furthermore, the intrinsic performance of sub-detectors is crucial. For example, flavor physics measurements frequently involve distinguishing mass resonances with small mass differences, such as the $B^0$ and $B^0_s$ mesons. An ECAL with an energy resolution better than $3\%/\sqrt{E\text{(GeV)}}$ is essential in this context. Also, an excellent vertex detector system is mandatory for identifying secondary and tertiary vertices, which are key for characterizing $b$, $c$, and $\tau$ decays, as well as for measuring jet charge. Precise calibration and control of systematic uncertainties require a stable detector system and a high-performance monitoring system for reliable references. Lastly, a highly efficient trigger, DAQ, and event building system are essential for conducting measurements at high event rates, particularly during the CEPC's $Z$-pole operation. Addressing these challenges is imperative to fully exploit the flavor physics potential at the CEPC.

In parallel, the ongoing development and exploration of innovative tools and algorithms are essential for effective data analysis and interpretation in flavor physics research. As the CEPC produces vast amounts of data, traditional analysis methods may struggle to extract meaningful insights, making the application of ML techniques increasingly vital. These algorithms, including supervised and deep learning models, can identify complex patterns within the data, significantly enhancing the accuracy of distinguishing between signal and background events -- an especially critical task in flavor physics, where rare processes often exist amid substantial noise. Moreover, ML can improve measurement precision by refining detector calibrations and enhancing event reconstructions. Several highly relevant developments are jet origin identification, one-to-one correspondence reconstruction, and the application of event-level techniques~\cite{Li:2020vav}. These techniques also facilitate anomaly detection, allowing researchers to flag unusual events that may indicate new physics beyond the Standard Model. The integration of advanced algorithms for data analysis can further enable faster processing times and more efficient data management through approaches like parallel processing and cloud computing. Together, these advancements will be instrumental in maximizing the scientific output of the CEPC, ensuring it remains at the forefront of flavor physics research and empowering researchers to uncover new phenomena, refine theoretical models, and deepen our understanding of fundamental particle interactions.

Given the impressive experimental reach, it is essential to ensure theoretical uncertainties under control with commensurate precision. Especially, most flavor physics measurements are frequently entangled with strong interactions. To match the anticipated experimental precision at the $Z$ factory, high-precision theoretical calculations, particularly those involving QCD, become crucial. Concerning the perturbative QCD effect, this requires higher-order loop calculations based on modern techniques, as reviewed in Ref.~\cite{Buras:2011we}. For some processes like $B_s \to \mu^+\mu^-$, we even need to consider the higher-order EW and QED corrections to match the experimental precision that can be reached at the CEPC. For the nonperturbative QCD effects, on the other hand, we have to employ lattice QCD, various QCD sum-rule techniques, phenomenological fits, and quark model techniques. Especially, the lattice QCD has now been proven to be an indispensable method to determine nonperturbative strong contributions to weak decay processes of $b$ and $c$ quarks. To connect the physics at different energy scales involved in these processes, the methods of effective field theories are also playing a key role, which allow relating them by performing the sequential matching and employing the renormalization group running~\cite{Buras:2020xsm}. 

These different flavor physics facilities, such as LHCb, Belle II, and future $e^-e^+$ colliders, could provide complementary information for flavor physics studies.It is important to combine all these theoretical aspects to provide unambiguous and rigorous interpretations of the experimental data in a global framework, either within the SM or in any BSM scenario. This also makes collaborative interactions between the theory community and experimental collaborations indispensable. Typical examples include the Heavy Flavor Averaging Group that periodically provides updates of properties of heavy-favored hadrons and their transitions~\cite{HeavyFlavorAveragingGroup:2022wzx},%~\cite{Banerjee:2024znd}, 
the Flavour Lattice Averaging Group that periodically provides important lattice inputs for experimental measurements~\cite{FlavourLatticeAveragingGroupFLAG:2024oxs}, and the Muon $g-2$ Theory Initiative~\cite{Aoyama:2020ynm} that is dedicated to a detailed account of recent efforts to improve the calculation of the muon anomalous magnetic moment. At the same time, it would also be beneficial to develop efficient interpretation frameworks capable of combining flavor physics measurements with other measurements, such as those of the Higgs and EW sectors.

To conclude, the flavor physics program at CEPC holds immense scientific promise. Based on its benchmark studies, we conclude that the CEPC could give rise to discoveries of new physical processes,
% such as $b\to s \tau\tau$ transitions, 
boost the precision of many measurements by orders of magnitude, and allow the NP searches to be extended to energy scales of 10 TeV or even higher. However, to fully realize the CEPC potential in physics, dedicated detector design and critical R\&D, as well as theoretical studies, are needed. We hope that the flavor physics studies at the CEPC will not only serve as a reference for evaluating the CEPC physics potential and optimizing its detector design, but also inspire innovative ideas for the development of new technologies, new algorithms and new tools.

\acknowledgments

We would like to thank Roy Aleksan, Patrizia Azzi, Mogens Dam, Christophe Grojean, Michelangelo Mangano, Stephan Monteil, Emmanuel Perez, Wenbin Qian, Huilin Qu, Michele Selvaggi for valuable discussions and their continuous support.

We acknowledge financial support from
the National Natural Science Foundation of China (NSFC) under grants
No.~12125507, No.~12047503,
% Lorenzo (NSFC)
No.~12035008, No.~12211530479,
% Xinqiang Li (NSFC)
No.~12475094, No.~12135006, No.~12075097,
% Qin Qin (NSFC)
No.~12375086, No.~2022YFA1601903,
% Liming Zhang (NSFC)
No. 12061141007,
% Jiayin Gu (NSFC)
No.~12035008, No.~12375091,
% CCAST (NSFC), Yuexin
No. 12342502,
% Q. Zhao (NSFC)
No.~12235018,
% Fusheng Yu (NSFC)
No.~12335003,
% Yadi Wang
No.~12105100,
% Qiang Zhao (NSFC)
No.~12235018,
% Yu Zhang (NSFC)
No.~12475106,
% Yiming Li and Jianchun Wang (NSFC)
No. 11961141015, No.~12188102, No.~12175245, 
% Long Chen (NSFC??)
No.~12205171, No.~12235008, No.~12321005, No.~tsqn202312052, No.~2024HWYQ-005,
% Peng-Cheng Lu (NSFC??)
No.~12405121,
% Xiaokang Du (NSFC)
No.~12447167,
% Jibo He (NSFC)
No.~12061141006,
% Yubo Li (NSFC)
No.~12405102,
% Zong-Guo Si (NSFC)
No.~12321005, No.~12235008,
% Wei Wang (NSFC)
No.~12125503, No.~12335003,
% Wei Su (NSFC)
No.~12305115,
% Fei Wang (NSFC)
No.~12075213, No.~12335005;
the Chinese Academy of Sciences under grants No.~YSBR-101 and No.~XDB34030000; 
the National Key R\&D Program of China under grants No.~2022YFE0116900, No.~2023YFA1606703, and No.~2022YFA1601901;
% Yiming Li and Jianchun Wang
the National Key Research and Development Program of China under grant No.~2023YFA1606300; 
% Cristian Sierra
the Excellent Postdoctoral Program of Jiangsu Province under grant No.~2023ZB891;
% Wei Su 
the Shenzhen Science and Technology Program under grants No.~202206193000001 and No.~20220816094256002; 
% Fei Wang
the Natural Science Foundation for Distinguished Young Scholars of Henan Province under grant No.~242300421046;
% Yadi Wang
the Beijing Municipal Natural Science Foundation under Contracts number JQ22002; 
% Hongkong team
the Area of Excellence under the Grant No.~AoE/P-404/18-3 and the General Research Fund under Grant No.~16304321 (both grants are issued by the Research Grants Council of Hong Kong S.A.R);
% Haijun Yang
the MOST National Key R\&D Program under grant No.~2023YFA1606303;
the Shanghai Key Laboratory for Particle Physics and Cosmology, Key Laboratory for Particle Astrophysics and Cosmology (Ministry of Education), Shanghai Jiao Tong University.

\bibliographystyle{JHEP}
\bibliography{reference}
\end{document}